\documentclass[12pt,a4paper]{report}

\usepackage{graphicx} 
\usepackage{fancyhdr} 
\usepackage[utf8]{inputenc}
\usepackage[T1]{fontenc} 
\usepackage{setspace} 
\usepackage{microtype}
\usepackage{mathptmx} 
\usepackage{slantsc}
\usepackage{titlesec}
\usepackage{mfirstuc} 
\usepackage{calc}
\usepackage{setspace}
\usepackage[acronym, nonumberlist]{glossaries} 
\usepackage[final,unicode]{hyperref} 
\usepackage{pdfpages}
\usepackage{float}
\let\uline\emph
\usepackage[vcentermath]{youngtab}

\definecolor{MyDarkBlue}{rgb}{0.15,0.25,0.45}
\usepackage{amsmath,amssymb}
\usepackage{amsfonts}
\usepackage{mathrsfs}
\usepackage{stackengine} 

\usepackage{bbm}
\usepackage{booktabs}

\usepackage{amsthm}

\usepackage{pgf,tikz}
\usetikzlibrary{cd,decorations.markings,calc}
\usepackage{mathtools}
\usepackage[nosort]{cleveref}

\let\fn\footnote
\renewcommand{\footnote}[1]{\linespread{1.1}\fn{#1}\linespread{1.29}}

\makeatletter\renewcommand{\section}{\@startsection
    {section}{1}{\z@}{-3.5ex plus -1ex minus
        -.2ex}{2.3ex plus .2ex}{\bf }}
\makeatletter\renewcommand{\subsection}{\@startsection{subsection}{2}{\z@}{-3.25ex
        plus -1ex minus
        -.2ex}{1.5ex plus .2ex}{\bf }}
\makeatletter\renewcommand{\subsubsection}{\@startsection{subsubsection}{3}{-2.45ex}{-3.25ex
        plus -1ex minus -.2ex}{1.5ex plus .2ex}{\it }}
\renewcommand{\thesection}{\arabic{section}}
\renewcommand{\thesubsection}{\arabic{section}.\arabic{subsection}}
\renewcommand{\@seccntformat}[1]{\@nameuse{the#1}.~~}

\renewcommand{\theequation}{\thesection.\arabic{equation}}
\makeatletter \@addtoreset{equation}{section}
\def\Ddots{\mathinner{\mkern1mu\raise\p@
        \vbox{\kern7\p@\hbox{.}}\mkern2mu
        \raise4\p@\hbox{.}\mkern2mu\raise7\p@\hbox{.}\mkern1mu}}
\usepackage[toc,page]{appendix}

\renewcommand{\thesection}{\thechapter.\arabic{section}}
\renewcommand{\thesubsection}{\thesection.\arabic{subsection}}

\newtheorem{thm}{Theorem}[section]
\newtheorem{lemma}[thm]{Lemma}
\newtheorem{definition}[thm]{Definition}
\newtheorem{theorem}[thm]{Theorem}
\newtheorem{proposition}[thm]{Proposition}
\newtheorem{corollary}[thm]{Corollary}
\newtheorem{remark}[thm]{Remark}

\renewcommand{\appendices}{
    \section*{Appendix}\label{appendices}\setcounter{subsection}{0}
    \addcontentsline{toc}{section}{Appendix}
    \setcounter{equation}{0}
    \makeatletter
    \renewcommand{\theequation}{\Alph{subsection}.\arabic{equation}}
    \renewcommand{\thesubsection}{\Alph{subsection}}
    \renewcommand{\thethm}{\Alph{subsection}.\arabic{thm}}
    \@addtoreset{equation}{subsection}
    \@addtoreset{thm}{subsection}
    \makeatother
}

\newcommand{\hooklongrightarrow}{\lhook\joinrel\longrightarrow}

\newcommand\cm{\mathrm{cm}}

\def\slasha#1{\setbox0=\hbox{$#1$}#1\hskip-\wd0\hbox to\wd0{\hss\sl/\/\hss}}

\def\periodb#1{\setbox0=\hbox{$#1$}#1\hskip-\wd0\hbox to\wd0{-}}

\newcommand{\unit}{\mathbbm{1}}   			
\newcommand{\id}{\mathrm{id}}   			

\newcommand{\CA}{\mathcal{A}}    			

\newcommand{\CC}{\mathcal{C}}
\newcommand{\CCC}{\mathscr{C}}

\newcommand{\CCG}{\mathscr{G}}

\newcommand{\CP}{\mathcal{P}}

\newcommand{\frg}{\mathfrak{g}}				
\newcommand{\frL}{\mathfrak{L}}				
\newcommand{\frh}{\mathfrak{h}}				
\newcommand{\frA}{\mathfrak{A}}

\newcommand{\frG}{\mathfrak{G}}

\newcommand{\fru}{\mathfrak{u}}

\newcommand{\frv}{\mathfrak{v}}
\newcommand{\frl}{\mathfrak{l}}

\newcommand{\FR}{\mathbbm{R}}     			
\newcommand{\RZ}{\mathbbm{Z}}     			

\newcommand{\dd}{\mathrm{d}}     			
\newcommand{\dpar}{\partial}     			
\newcommand{\de}{\mathrm{e}}     			
\newcommand{\di}{\mathrm{i}}     			
\newcommand{\eps}{{\varepsilon}}			

\newcommand{\sB}{\mathsf{B}}

\newcommand{\sE}{\mathsf{E}}

\newcommand{\eand}{{\qquad\mbox{and}\qquad}}     		
\newcommand{\ewith}{{\qquad\mbox{with}\qquad}}

\newcommand{\der}[1]{\frac{\dpar}{\dpar #1}}   		
\newcommand{\tr}{\,\mathrm{tr}\,}     			

\newcommand{\fra}{\mathfrak{a}}
\newcommand{\fre}{\mathfrak{e}}

\newcommand{\astring}{\mathfrak{string}}

\newcommand{\sU}{\mathsf{U}}     			

\newcommand{\sG}{\mathsf{G}}

\newcommand{\sL}{\mathsf{L}}

\newcommand{\sLie}{\mathsf{Lie}}
\newcommand{\sCE}{\mathsf{CE}}

\newcommand{\sH}{\mathsf{H}}

\newcommand{\sString}{\mathsf{String}}

\newcommand{\acton}{\mathbin\vartriangleright}     			
\newcommand{\tildeacton}{\mathbin{\tilde\vartriangleright}}     			
\renewcommand{\remark}[1]{}     				
\def\tyng(#1){\hbox{\tiny$\yng(#1)$}}			
\def\tyoung(#1){\hbox{\tiny$\young(#1)$}}			

\newlength\tmplength
\def\showmybox{\tmplength=\wd0\relax\box0\kern-\tmplength}
\def\mybox#1{\setbox0=\hbox{\ensurestackMath{#1}}}
\def\SIn#1#2{\stackinset{c}{#1}{c}{#2}}
\def\chint{\mathchoice%
{\mybox{\SIn{1.8pt}{-0pt}{\mathsf C}{\displaystyle\phantom{\int}}}\showmybox}
{\mybox{\SIn{1.5pt}{-0pt}{\scriptstyle\mathsf C}{\textstyle\phantom{\int}}}\showmybox}
{\mybox{\SIn{1.2pt}{-0pt}{\scriptscriptstyle\mathsf C}{\scriptstyle\phantom{\int}}}\showmybox}%
{\mybox{\SIn{1.1pt}{-0pt}{\scalebox{0.7}{$\scriptscriptstyle\mathsf C$}}{%
\scriptscriptstyle\phantom{\int}}}\showmybox}%
\int}

\def\ichint{\mathchoice%
{\mybox{\SIn{9.3pt}{-0pt}{\mathsf C}{\displaystyle\phantom{\int}}}\showmybox}
{\mybox{\SIn{7pt}{-0pt}{\scriptstyle\mathsf C}{\textstyle\phantom{\int}}}\showmybox}
{\mybox{\SIn{5.5pt}{-0pt}{\scriptscriptstyle\mathsf C}{\scriptstyle\phantom{\int}}}\showmybox}%
{\mybox{\SIn{4.4pt}{-0pt}{\scalebox{0.7}{$\scriptscriptstyle\mathsf C$}}{%
\scriptscriptstyle\phantom{\int}}}\showmybox}%
\iint}

\def\chichint{\mathchoice%
{\mybox{\SIn{1.8pt}{-0pt}{\mathsf C}{\displaystyle\phantom{\int}}}\showmybox}
{\mybox{\SIn{1.5pt}{-0pt}{\scriptstyle\mathsf C}{\textstyle\phantom{\int}}}\showmybox}
{\mybox{\SIn{1.2pt}{-0pt}{\scriptscriptstyle\mathsf C}{\scriptstyle\phantom{\int}}}\showmybox}%
{\mybox{\SIn{1.1pt}{-0pt}{\scalebox{0.7}{$\scriptscriptstyle\mathsf C$}}{%
\scriptscriptstyle\phantom{\int}}}\showmybox}%
\mathchoice%
{\mybox{\SIn{9.3pt}{-0pt}{\mathsf C}{\displaystyle\phantom{\int}}}\showmybox}
{\mybox{\SIn{7pt}{-0pt}{\scriptstyle\mathsf C}{\textstyle\phantom{\int}}}\showmybox}
{\mybox{\SIn{5.5pt}{-0pt}{\scriptscriptstyle\mathsf C}{\scriptstyle\phantom{\int}}}\showmybox}%
{\mybox{\SIn{4.4pt}{-0pt}{\scalebox{0.7}{$\scriptscriptstyle\mathsf C$}}{%
\scriptscriptstyle\phantom{\int}}}\showmybox}%
\iint}

\def\ichichint{\mathchoice%
{\mybox{\SIn{1.8pt}{-0pt}{\mathsf C}{\displaystyle\phantom{\iiint}}}\showmybox}
{\mybox{\SIn{1.5pt}{-0pt}{\scriptstyle\mathsf C}{\textstyle\phantom{\iiint}}}\showmybox}
{\mybox{\SIn{1.2pt}{-0pt}{\scriptscriptstyle\mathsf C}{\scriptstyle\phantom{\iiint}}}\showmybox}%
{\mybox{\SIn{1.1pt}{-0pt}{\scalebox{0.7}{$\scriptscriptstyle\mathsf C$}}{%
\scriptscriptstyle\phantom{\iiint}}}\showmybox}%
\mathchoice%
{\mybox{\SIn{9.3pt}{-0pt}{\mathsf C}{\displaystyle\phantom{\iiint}}}\showmybox}
{\mybox{\SIn{7pt}{-0pt}{\scriptstyle\mathsf C}{\textstyle\phantom{\iiint}}}\showmybox}
{\mybox{\SIn{5.5pt}{-0pt}{\scriptscriptstyle\mathsf C}{\scriptstyle\phantom{\iiint}}}\showmybox}%
{\mybox{\SIn{4.4pt}{-0pt}{\scalebox{0.7}{$\scriptscriptstyle\mathsf C$}}{%
\scriptscriptstyle\phantom{\iiint}}}\showmybox}%
\iiint}

\newcommand{\todo}[1]{{\color{red}#1}}

\newcommand{\makecommand}[3]{%
    \foreach \i in #3 {%
        \expandafter\xdef\csname #1\i\endcsname{\noexpand#2{\unexpanded\expandafter{\i}}}%
    }%
}

\makecommand{fr}{\mathfrak}{{c,der,g,h,A,X,gl,sl,o,so,osp,u,su,sp,spin,string,Poisson,td,tb,inn}}

\makecommand{sf}{\mathsf}{{id,String,Lie,Hom,SU,SO,Cat,Set,Spin,Pin,inv,CE,Diff,HSymp,TD,TB,pr,Aut,GL,SL,IO,Mat,Sym,Inn,Out,AUT,alt,MC,USp}}

\makecommand{rm}{\mathrm}{{Ham,diag,im,wk,sk,Im,lp}}

\newcommand{\latinalphabet}{A,a,B,b,C,c,d,D,E,e,F,f,G,g,H,h,I,i,J,j,K,k,L,l,M,m,N,n,O,o,P,p,Q,q,R,r,S,s,T,t,U,u,V,v,W,w,X,x,Y,y,Z,z}
\newcommand{\caplatinalphabet}{A,B,C,D,E,F,G,H,I,J,K,L,M,N,O,P,Q,R,S,T,U,V,W,X,Y,Z}
\makecommand{I}{\mathbbm}{\latinalphabet}
\makecommand{bf}{\mathbf}{\latinalphabet}
\makecommand{bm}{\bm}{\latinalphabet}
\makecommand{ca}{\mathcal}{\caplatinalphabet}
\makecommand{fr}{\mathfrak}{\latinalphabet}
\makecommand{rm}{\mathrm}{\latinalphabet}
\makecommand{sf}{\mathsf}{\latinalphabet}
\makecommand{sc}{\mathscr}{\latinalphabet}

\newcommand{\sfGO}{\sfG\sfO}
\newcommand{\scTD}{\scT\!\scD}

\newcommand{\scGO}{\scG\!\scO}

\definecolor{outrageousorange}{rgb}{1.0, 0.43, 0.29}

\newcommand{\astringsk}{\mathfrak{string}_{\rm sk}}
\newcommand{\astringl}{\mathfrak{string}_{\rm lp}}

\newcommand{\CatCat}{\mathsf{Cat}}

\newcommand{\ainn}{\mathfrak{inn}}
\newcommand{\sInn}{\mathsf{Inn}}
\newcommand{\sW}{\mathsf{W}}

\DeclareMathOperator\Pexp{P\,exp}
\DeclareMathOperator\dom{dom}
\DeclareMathOperator\codom{codom}

\newcommand\tildeotimes{\mathbin{\tilde\otimes}}
\newcommand\sslash{\mathbin{/\!/}}

\def\tyng(#1){\raisebox{0.05cm}{\hbox{\tiny$\yng(#1)$}}}			

\usepackage{stmaryrd}
\DeclareFontFamily{U}{mathx}{\hyphenchar\font45}
\DeclareFontShape{U}{mathx}{m}{n}{
      <5> <6> <7> <8> <9> <10>
      <10.95> <12> <14.4> <17.28> <20.74> <24.88>
      mathx10
      }{}
\DeclareSymbolFont{mathx}{U}{mathx}{m}{n}
\DeclareMathSymbol{\bigoslash}{1}{mathx}{"CD}

\newcommand{\ophLie}{h\mathcal{\caL}ie_2}

\newcommand{\opEilh}{\mathcal{\caE}ilh_2}

\newcommand{\opLeib}{\mathcal{\caL}eib}

\newcommand{\rep}[1]{{\mathbf{#1}}}
\newcommand{\parder}[2][]{%
    \ifthenelse{\equal{#1}{}}{%
        \frac{\partial}{\partial #2}%
    }{%
        \frac{\partial #1}{\partial #2}%
    }%
}

\newcommand{\sst}[1]{{\scriptscriptstyle #1}}
\def\0{{\sst{(0)}}}
\def\1{{\sst{(1)}}}
\def\2{{\sst{(2)}}}
\def\3{{\sst{(3)}}}
\def\4{{\sst{(4)}}}
\def\5{{\sst{(5)}}}
\def\6{{\sst{(6)}}}
\def\7{{\sst{(7)}}}

\fancypagestyle{preliminary}{
    
    \fancyhead[RCL]{}

    \pagenumbering{Roman}
}
\fancypagestyle{chapter}{
    \fancyhead[L]{\normalfont\itshape\fontsize{10pt}{12pt}\selectfont\nouppercase{\leftmark}}
    \fancyhead[R]{}

    \fancyfoot[C]{\thepage}

    \pagenumbering{arabic}
}

\newcommand{\headingBaseline}{12}
\newcommand{\headingBaselineDiv}{10}
\newlength{\chapterFontSize}
\newlength{\sectionFontSize}
\newlength{\subsectionFontSize}
\newlength{\chapterBaseline}
\newlength{\sectionBaseline}
\newlength{\subsectionBaseline}

\titleformat{\chapter}[display]
    {\normalfont\bfseries\fontsize{\chapterFontSize}{\chapterBaseline}\selectfont}{\chaptertitlename\ \thechapter}{14pt}{}
\titlespacing{\chapter}{0pt}{10pt}{25pt}

\titleformat{\section}[hang]
    {\normalfont\bfseries\fontsize{\sectionFontSize}{\sectionBaseline}\selectfont}{\thesection}{5pt}{}
\titlespacing{\section}{0pt}{25pt}{15pt}

\titleformat{\subsection}[hang]
    {\normalfont\bfseries\itshape\fontsize{\subsectionFontSize}{\subsectionBaseline}\selectfont}{\thesubsection}{5pt}{}
\titlespacing{\subsection}{0pt}{20pt}{10pt}

\makenoidxglossaries

\hypersetup{
    colorlinks = true,
    linkcolor = blue, 
    citecolor = blue, 
    urlcolor = blue, 
    filecolor = black 
}

\newcommand{\thesisTitle}{Construction~and application\\of adjusted higher gauge theories}

\newcommand{\authorName}{Hyungrok Kim}
   
\newcommand{\degreeQualification}{Doctor of Philosophy}
\newcommand{\institution}{\phantom{ }}
\newcommand{\school}{School of Mathematics and Computer Science}
\newcommand{\university}{Heriot-Watt University}
\newcommand{\monthDate}{May}
\newcommand{\yearDate}{2023}

\include{I-glossary}
\usepackage{subfiles}
\usepackage[nosort]{cleveref}

\begin{document}

\pagestyle{empty}
\begin{center}
\vspace*{15pt}\par
\setstretch{1}

\begin{spacing}{1.8}
{\Large\bfseries\MakeLowercase{\capitalisewords{\thesisTitle}}}\\
\end{spacing}

\vspace{40pt}\par
\includegraphics[width=140pt]{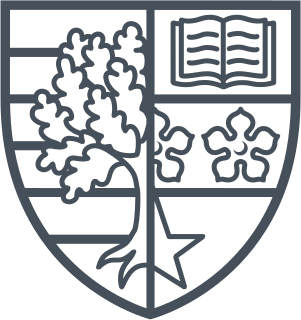}\\
\vspace{40pt}\par

{\itshape\fontsize{15.5pt}{19pt}\selectfont by\\}\vspace{15pt}\par

{
\Large \authorName
}\vspace{55pt}\par

{
\large Submitted for the degree of \\ \vspace{8pt} \Large\slshape\degreeQualification\\
}

\vspace{35pt}\par

{\scshape\setstretch{1.5} \institution\\ \school\\ \university\\
}

\vspace{50pt}\par

{\large \monthDate\ \yearDate}

\vfill

\begin{flushleft}
\setstretch{1.4}\small
The copyright in this thesis is owned by the author. Any quotation from the thesis or use of any of the information contained in it must acknowledge this thesis as the source of the quotation or information.
\end{flushleft}
\end{center}

\clearpage
\pagestyle{preliminary}
\begin{center}
\LARGE\textbf {Abstract}
\end{center}
\vspace{5pt}

\noindent
This thesis investigates several aspects of nonabelian higher gauge theories, which appear in many areas of physics, notably string theory and gauged supergravity. We show that nonabelian higher gauge theory admits a consistent classical nonperturbative formulation insofar as a higher nonabelian parallel transport exists consistently, without requiring certain curvature components (fake curvature) to vanish. 

Next, we explore examples of nonabelian higher gauge theories that naturally appear in high-energy physics. Using a generalisation of \(L_\infty\)-algebras called \(EL_\infty\)-algebras, we show that tensor hierarchies of gauged supergravity naturally admit a formulation in terms of higher nonabelian gauge theories. Furthermore, toroidal compactifications of string theory exhibiting T-duality also naturally contain higher gauge symmetry, which explain several features of nongeometric compactifications (Q- and R-fluxes).


\clearpage
\clearpage
\begin{center}
\LARGE\textbf {Acknowledgements}
\end{center}
\vspace{5pt}

\noindent I wish to thank my supervisor Christian Saemann and my collaborators Leron Borsten, Branislav Jurčo, Tommaso Macrelli, Martin Wolf. I also thank my spouse, Madiha Hussain, for her unwavering support.

\clearpage

{
    \setstretch{1}
    \hypersetup{linkcolor=black}
    \tableofcontents
}


\clearpage

\pagestyle{chapter}

\chapter{Introduction}

This thesis deals with the formulation and application of nonabelian higher gauge theories, which appear in many areas of physics, notably string theory and gauged supergravity. This thesis is based on the papers \cite{Kim:2019owc,Borsten:2021ljb,Kim:2022opr} coauthored in part with Leron Borsten and Christian Saemann.

Let us first explain our interest in higher gauge theories and what they provide.
From a physical standpoint, it is an empirical fact that the universe has four large spacetime dimensions and that its low-energy effective theory is that of general relativity and a non-abelian Yang--Mills theory coupled to matter. Yet the four-dimensionality of spacetime is not obvious: string theory, our best-understood theory of quantum gravity to date, predicts many additional microscopic dimensions, and in more than four dimensions, a generic low-energy theory of a string vacuum instead contains a higher-form generalisation of Yang--Mills theory in which gauge fields of form degree greater than \(1\) couple, in a consistent and nontrivial way, to matter and to gravity. Thus it is of great import to understand the features, constraints, and ansätze allowed by such generic higher gauge theories.

From a mathematical standpoint, the investigation of Yang--Mills theory and string theory have delivered invaluable dividends, and the application of higher-categorical techniques is continuing in steadily conquering diverse areas of mathematics, starting with topology and algebraic geometry.\footnote{Please pardon the capitalistico-imperialistic metaphors. Deconstruct it if you'd like.} Hence, it is a natural mathematical enterprise to marry the two, whence springs higher gauge theory.

Yet the study of higher gauge theories is beset with problems, much as nonabelian Yang--Mills theory requires new and sophisticated mathematical techniques compared to abelian gauge theory. This thesis aims to solve some (we prefer to be modest) of the fundamental problems that arise in the formulation and application of higher gauge theories. First, while the theory as it applies to topologically trivial configurations is describable by a set of differential forms of various degrees with prescribed gauge transformations, any nonperturbative formulation of the theory must include topologically nontrivial configurations, for which the description by mere differential forms is inadequate, much as Yang--Mills theory for an arbitrary principal bundle requires more than just a Lie-algebra-valued 1-form. In particular, one needs a consistent and coherent set of analogues of Wilson loops (in higher gauge theory they must be Wilson surfaces or Wilson (hyper-)volumes, naturally), which suffice to found the theory. This will be accomplished rigorously in this thesis.

Beyond the conceptual problems of the general theory, we also encounter concrete problems in specific higher gauge theories that arise from string theory; after all, formalism is only useful insofar as it is applicable. Perhaps the most prominent nontrivial (i.e.~non-abelian) higher gauge theory in this context is that of tensor hierarchies in gauged supergravity, most of which have natural stringy realisations. In this theory, the challenge we face and address is twofold: one is to make sense of the morass of coupling constants and algebraic identities that arise in tensor hierarchy actions in terms of some suitable algebraic structure, and the second is to use this algebraic structure to formulate (at least in principle) the theory with topologically nontrivial configurations. Our answer to this twofold problem is called \(\ophLie\)-algebras.\footnote{We follow the fine tradition of using inscrutable names for algebraic structures.} We will see that the myriad coupling constants and identities naturally organise themselves via the magic of higher algebraic structures into a simple \(\ophLie\)-algebras, which are homotopy generalisations of Lie algebras; as such, categorical methods then provide a way to integrate them into (higher) Lie groups to formulate topologically nontrivial classical configurations globally.

While gauged supergravity provides a fine example of the nontriviality of a higher gauge Lie algebra, a higher gauge Lie group can also be nontrivial due to global structure rather than locally, just as a Lie group may be nonabelian even though its Lie algebra may be abelian. We argue that such a situation arises in T-duality in string theory. In this case, the \(p\)-forms that appear in toroidal compactifications do not, at first glance, seem to have nonabelian gauge structure. However, if one formulates this theory straightforwardly in a merely abelian way, one finds no trace of the famous \(\sfO(n,n;\IZ)\) symmetry of T-duality. We show that if the higher gauge group is nonabelian in a certain natural manner, this automatically produces \(\sfO(n,n;\IZ)\) automorphisms. Furthermore, we show that this T-duality gauge group extends naturally to include the \(Q\)- and \(R\)-fluxes that appear in nongeometric compactifications: in effect, we trade uncategorified generalised (doubled) geometry for higher categorifications of ordinary geometry.


\section{Parallel transport in higher gauge theory}
After a review of higher gauge theory and mathematical prerequisites, Chapter~4 discusses parallel transport in higher gauge theory. A consistent formulation of nonabelian higher gauge theory on possibly topologically nontrivial higher bundles (e.g.\ higher analogues of instantons) requires a formulation of higher holonomy to which branes can be coupled. Previous approaches to such higher connections on categorified principal bundles require these to be fake-flat. This condition, however, renders them locally gauge equivalent to connections on abelian gerbes. For particular higher gauge groups, for example 2-group models of the string group, this limitation can be overcome by generalising the notion of higher connection. Starting from this observation, we define a corresponding generalised higher holonomy functor which is free from the fake-flatness condition, leading to a truly non-abelian parallel transport. 

    As reviewed in Chapter~2, nonabelian higher gauge theory as conventionally formulated suffers from severe technical problems. Fundamentally, this issue arises from the fact that the field strengths as defined by the ordinary Weil algebra are generically too restrictive in all but the simplest (abelian) situations. In order to rectify this problem, we employ \emph{adjusted Weil algebras} \cite{Saemann:2019dsl} that encode a more general notion of field strengths and demonstrate that it permits a consistent topologically non-trivial formulation in that its Wilson surfaces, which suffice to found the theory classically, are well defined.
   
    We start from the observation that the dual of the Weil algebra of a higher Lie $n$-algebra $\frL$ is isomorphic to the Lie $(n+1)$-algebra of inner derivations $\ainn(\frL)$ of $\frL$; details are found in \cref{app:inner_weil}. For a Lie 2-algebra $\frL$, this leads to the Lie 3-algebra $\ainn(\frL)$, which can be described in two ways: first, as a 3-term $L_\infty$-algebra and second, as a 2-crossed module of Lie algebras, as done in~\cite{Roberts:0708.1741}. The former is directly obtained from the Weil algebra, while the latter contains some additional information and is readily integrated to a 2-crossed module of Lie groups. In \cref{app:Lie3_2_crossed}, we explain in detail the correspondence between Lie 3-algebras and 2-crossed modules of Lie algebras.
    
    The adjustment of the Weil algebra amounts thus to an adjustment of the algebra of inner derivations. One result that certainly deserves further study is that the data required to lift the Lie 3-algebra into a 2-crossed module of Lie algebras is precisely the data encoding the adjustment of the Weil algebra; see sections~\ref{ssec:loop_model_of_Lie_algebra} and~\ref{ssec:adjusted_inner_derivation}. There, we also compute the corresponding integrated adjusted inner automorphism Lie 2-groups in the form of a 2-crossed module of Lie groups and compare them to the unadjusted forms.
    
    For simplicity, we focus our discussion on {\em local} parallel transport over a contractible patch $U$ of the spacetime manifold; gluing the local picture to a global one is mostly a technicality. Local parallel transport in ordinary gauge theory with gauge group\footnote{i.e.~the structure group of an underlying principal bundle} $\sG$ is essentially a functor $\Phi$ from the path groupoid $\CP U$, which has points in $U$ as its objects and paths between these as morphisms, to the one-object groupoid $\sB\sG$ which has $\sG$ as its group of morphisms:
    \begin{equation}
        \Phi\colon \CP U\longrightarrow \sB\sG~,\hspace{2cm}
        \begin{tikzcd}
            \mbox{paths} \arrow[d,shift left] \arrow[d,shift right]\arrow[r]{}{\Phi_1} & \sG \arrow[d,shift left] \arrow[d,shift right]\\
            U \arrow[r]{}{\Phi_0} & * 
        \end{tikzcd}
    \end{equation}
    see section~\ref{ssec:parallel} for technical details. Any path is thus associated to a group element such that constant paths are mapped to $\unit_\sG$ and composition of paths leads to multiplication of the corresponding group elements. These are the axioms of a parallel transport. A connection 1-form is readily extracted from considering infinitesimal paths and conversely, a connection 1-form maps a path to a group element by the usual path-ordered exponential.
    
    An alternative yet equivalent picture is obtained from the short exact sequence of groupoids
    \begin{equation}
        *\longrightarrow \begin{array}{c} \sG\\[-0.75ex] \downdownarrows\\[-0.3ex]\sG\end{array} \hooklongrightarrow \sInn(\sG)\longrightarrow \begin{array}{c} \sG\\[-0.75ex] \downdownarrows\\[-0.75ex]*\end{array}\longrightarrow *~,
    \end{equation}
    where $\sInn(\sG)$ is the Lie 2-group of inner automorphisms. Instead of a functor $\Phi$ from $\CP U$ to $\sB\sG$, we can also consider a 2-functor $\varPhi$ from the 2-groupoid $\CP_{(2)} U$ of points, paths and paths or homotopies between paths to $\sB\sInn(\sG)$. We find that the globular identities\footnote{i.e.~the relations between domain and codomain for morphisms and higher morphisms: the domain of the domain is the domain of the codomain, and the codomain of the domain is the codomain of the codomain} for $\varPhi$ reduce the defining data to the same as for $\Phi$, which simply corresponds to Stokes' theorem at the level of connection 1-forms and curvature 2-forms.
    
    Gauge transformations are encoded in natural transformations between the corresponding functors. In the case of 2-functors $\varPhi:\CP_{(2)} U\rightarrow \sB\sInn(\sG)$, we must restrict the 2-natural transformations to obtain the correct set of gauge transformations, as we explain in detail in section~\ref{ssec:derived_parallel_ordinary}.
    
    In the context of higher gauge theory with Lie 2-group $\CCG$, similarly, a strict 2-functor $\Phi\colon\CP_{(2)}U\rightarrow \sB\CCG$ induces a strict 3-functor $\varPhi\colon\CP_{(3)} U \rightarrow \sB\sInn(\CCG)$. In both cases, the fake curvature condition appears as a necessary condition for the existence of these functors.
    
    If we replace, however, the 3-group of inner automorphisms $\sInn(\CCG)$ with the 3-group of {\em adjusted} inner automorphisms $\sInn_{\rm adj}(\CCG)$ and consider a strict 3-functor
    \begin{equation}
        \varPhi^{\rm adj}\colon\CP_{(3)}U\longrightarrow \sB\sInn_{\rm adj}(\CCG)~,
    \end{equation}
    we obtain a new higher-dimensional parallel transport. The higher-dimensional Stokes' theorem is automatically satisfied and merely enforces the definition of higher curvatures together with the corresponding higher Bianchi identities. This parallel transport is truly non-abelian and underlies the self-interacting field theories constructed from the kinematical data arising from the adjusted Weil algebra. Contrary to the unadjusted parallel transport, this 3-functor only simplifies to a 2-functor if the underlying connection is fake flat.
    
    Altogether, we conclude that while for ordinary (higher) gauge theories there exist two fully equivalent ways of defining parallel transport, this is no longer the case if we aim for an adjusted higher parallel transport, where only one of these definitions is possible. In particular, a general higher parallel transport along $d$-dimensional volumes with underlying gauge $d$-group $\CCG$ which admits an adjusted higher $(d+1)$-group of inner automorphisms $\sInn_{\rm adj}(\CCG)$ is based on a strict $(d+1)$-functor
    \begin{equation}
        \varPhi^{\rm adj}\colon\CP_{(d+1)}U\longrightarrow \sB\sInn_{\rm adj}(\CCG)~.
    \end{equation}
    Gauge transformations arise from appropriately restricted $d$-natural transformations. For $d=1$, $\sInn(\CCG)$ is always adjusted, since there are no higher curvatures, and the 2-functor simplifies to a functor $\Phi:\CP U\rightarrow \sB\CCG$, reproducing the usual higher transport. If an adjustment is not possible for $d>1$, then only an unadjusted parallel transport exists, which is locally gauge equivalent to an abelian one. 

\section{Tensor hierarchies via \(E_2L_\infty\)-algebras}
Chapter~5 introduces a generalisation of \(L_\infty\)-algebras called \(E_2L_\infty\)-algebras, which relax up to homotopy not only the Jacobi identity but also the antisymmetry of the Lie bracket. These provide the natural algebraic framework for the gauge algebras arising in the tensor hierarchies of gauged supergravity.

    Tensor hierarchies are particular forms of higher gauge theories that were introduced in the context of gauging maximal supergravity theories~\cite{deWit:2005hv, deWit:2005ub,Samtleben:2005bp,deWit:2008ta,Bergshoeff:2009ph}. They are constructed using the embedding tensor formalism, introduced in~\cite{Cordaro:1998tx, Nicolai:2000sc, deWit:2002vt, deWit:2003hq}. For comprehensive reviews  see~\cite{Samtleben:2008pe, Trigiante:2016mnt}. Tensor hierarchies are also crucial to, for example, conformal field theories such as the $\mathcal{N}=(1,0)$ superconformal models of~\cite{Samtleben:2011fj, Samtleben:2012mi,Samtleben:2012fb}.
    Although initially applied to gauged supergravity theories,  tensor hierarchies do not  require supersymmetry and appear through the embedding tensor formalism applied to  the gauging of  a  broader class of Einstein--Maxwell-matter  theories, as discussed in~\cite{Bergshoeff:2009ph,Hartong:2009vc}.
    
    For us, tensor hierarchies provide a natural and nontrivial class of examples of nonabelian higher gauge theories. In particular, the field strengths that appear in the literature lie beyond the ansatz of unadjusted Weil algebras and require adjustment for their formulation. 
    
    In Chapter~5, we show that the adjustment data of tensor hierarchy theories are captured by a new kind of homotopy algebras called $E_2L_\infty$-algebras, which can be regarded as a weaker yet quasi-isomorphic form of $L_\infty$-algebras. Our construction generalises previous work on weaker forms of categorified Lie algebras~\cite{Roytenberg:0712.3461,Squires:2011aa}, from where we also borrowed the nomenclature.

    In particular, we show in \cref{thm:firmly_adjust} that there is a particular class of $L_\infty$-algebras that come with a natural adjustment encoded in an $E_2L_\infty$-algebra. This class is precisely the one arising in the tensor hierarchies of gauged supergravity. The latter are thus adjusted higher gauge algebras, employing $E_2L_\infty$-algebras in their construction.
    
    Besides giving the general definition for $E_2L_\infty$-algebra in a fashion that can be readily used for explicit computations, we also develop the general structure theory to some extent:
    \begin{itemize}
        \item[$\Diamond$] The key to most of our discussion is the notion of $\ophLie$-algebras, which are differential graded Leibniz algebras in which the Leibniz bracket fails to be graded antisymmetric up to a homotopy given by an alternator, whose failure to be graded symmetric is controlled by a higher alternator, and so on ad infinitum.
        \item[$\Diamond$] Koszul dual to the concept of $\ophLie$-algebras is that of $\opEilh$-algebras, and we can use semifree $\opEilh$-algebras to define the homotopy algebras of $\ophLie$-algebras, which we call $E_2L_\infty$-algebras. 
        \item[$\Diamond$] $E_2L_\infty$-algebras come with a good notion of homotopy transfer and, correspondingly, with a minimal model theorem.
        \item[$\Diamond$] $L_\infty$-algebras are special cases of $E_2L_\infty$-algebras, and $L_\infty$-algebra morphisms lift to $E_2L_\infty$-algebra morphisms.
        \item[$\Diamond$] An $E_2L_\infty$-algebra can be antisymmetrised to an $L_\infty$-algebra which, when regarded as an $E_2L_\infty$-algebra, is quasi-isomorphic to the original $E_2L_\infty$-algebra. Correspondingly, any $E_2L_\infty$-algebra is quasi-isomorphic to a differential graded Lie algebra.
        \item[$\Diamond$] Any differential graded Lie algebra gives naturally rise to an $\ophLie$-algebra, i.e.~to a hemistrict $E_2L_\infty$-algebra.
        \item[$\Diamond$] All of the above can be made explicit in terms of multilinear maps, at least order by order, and we give explicit formulas that should prove useful in future applications.
    \end{itemize}
    
    We can then show that given an $\ophLie$-algebra originating from a differential graded Lie algebra, adjusted notions of the curvatures of higher gauge theory are naturally found. These adjusted curvatures are precisely the ones of the tensor hierarchies of gauged supergravity for maximal supersymmetry. In the past, these gauge theories have been regarded as gauge theories of Leibniz algebras or various notions of enhanced Leibniz algebras, see our discussion in \cref{sec:comparison}. By the principles of categorification, it is clear that higher gauge algebras always have to be some higher form of Lie algebras, as these are the ones integrating to (higher) symmetry group. We show that the various forms of enhanced Leibniz algebras proposed in the literature are indeed axiomatically incomplete forms of $\ophLie$-algebras or weaker higher Lie algebras.

\section{Higher gauge symmetry of T-duality}
In Chapter~6, we exploit nonabelian higher gauge theory to describe T-duality between general geometric and non-geometric backgrounds as higher groupoid bundles with connections. Our description extends the previous observation by Nikolaus and Waldorf that the topological aspects of geometric and half-geometric T-dualities can be described in terms of higher geometry. We extend their construction in two ways. First, we endow the higher geometries with adjusted connections, which allow us to discuss explicit formulas for the metric and the Kalb-Ramond field of a T-background. Second, we extend the principal 2-bundles to augmented 2-groupoid bundles, which accommodate the scalar fields arising in T-duality along several directions as well as Q- and R-fluxes. This provides a natural way to formulate global aspects of nongeometric string compactifications in terms of the well-understood theory of higher geometry.

    T-duality is a crucial feature of string theory which sets it apart from field theories of point particles. In its simplest form, T-duality relates two string theories whose target spaces are of the form $X\times S^1$ for some Lorentzian manifold $X$, interchanging the momentum modes and the winding modes along the circle $S^1$. More complicated is the case in which a T-duality involves a non-trivial circle bundle that carries in addition a Kalb--Ramond two-form field $B$ describing a topologically non-trivial gerbe. Here, a T-duality can link target spaces with different topologies. Even more generally, we can consider T-dualities along a torus fibration and introduce additional background $p$-form potentials with non-vanishing curvatures corresponding to (higher) gerbes. Such T-dualities can link fully geometric target spaces to non-geometric target spaces. A class of mildly non-geometric target spaces is known as T-folds. These are still locally geometric, but their local data are glued together by an element in the T-duality group $\sfO(n,n;\IZ)$. T-dualities can, however, also produce $R$-spaces, for which there is not even a local geometric description. It is clear that a complete understanding of string theory requires an understanding of these non-geometric backgrounds. Given that T-duality is a duality on topologically non-trivial target spaces, it is particularly important not to work merely locally, as done in much of the literature. One of the aims of this chapter is to provide a clean mathematical description of such non-geometric backgrounds arising in the context of non-trivial topologies.
    
    By now, T-duality has attracted considerable mathematical interest due to its relation to a number of important mathematical constructions such as mirror symmetry and the Fourier--Mukai transform. The observation that T-duality can change the topology of the target space was linked in a formalism usually called topological T-duality to the existence of a Gysin sequence~\cite{Bouwknegt:2003vb,Bouwknegt:2003zg}. The latter provides an explicit relation between different topological classes, e.g.~between the first Chern class of a torus fibration and the Dixmier--Douady class of a gerbe on its total space. Subsequent works have found interpretations of the corresponding open-string versions of these non-geometric backgrounds in terms of non-commutative~\cite{Mathai:2004qq} and non-associative geometries~\cite{Bouwknegt:2004ap}.
    
    A useful geometric description of T-duality called T-correspondences was given in~\cite{Bunke:2005sn,Bunke:2005um}. Here, a T-background is defined as a torus bundle $P$ over a base manifold $X$ together with the Dixmier--Douady class $H$ of an abelian gerbe over $P$. A T-duality between two such pairs $(\check P, \check H)$ and $(\hat P,\hat H)$ is then formulated as a relation between the pullbacks of $\check H$ and $\hat H$ to the correspondence space $\check P\times_X\hat P$; the data $(\check P, \check H)$, $(\hat P,\hat H)$, and the relation are collectively called a T-duality triple. Quite recently, it was observed that T-backgrounds and indeed full T-duality triples can be represented by 2-stacks~\cite{Nikolaus:2018qop}. That is, a topological, geometric T-background can be equivalently seen as a principal 2-bundle or gerbe with a particular structure 2-group $\sfTB^\text{F2}_n$ that encodes both the torus directions as well as the gerbe part. The same holds for a T-duality triple between geometric T-backgrounds, where the structure group is denoted by $\sfTD_n$. This structure group comes with two natural projections to $\sfTB^\text{F2}_n$, and the induced map on principal 2-bundles yields the data of a T-duality triple. Interestingly, even half-geometric T-dualities, which link geometric backgrounds with T-folds, can be captured in terms of principal 2-bundles. This opens up the exciting possibility that non-commutative and perhaps even non-associative geometries can be resolved into ordinary but higher geometries, which would clearly constitute a simplification: higher geometry enters the description of geometric backgrounds anyway in the form of gerbes, and higher geometric objects are more readily derived than their non-commutative and, in particular, non-associative counterparts.
    
    The appealing picture obtained in~\cite{Nikolaus:2018qop} poses three evident questions: First, can we extend the topological constructions to a more complete picture by adding a differential refinement in the form of connections? Second, can we extend the half-geometric correspondences further to the most general case of $R$-spaces? And third, is there a fully $\sfO(n,n;\IZ)$-covariant formulation that manifests the action of the T-duality group on the components of this description? In this chapter, we answer all of these questions affirmatively.
    
    A differential refinement of the description of geometric T-duality triples in terms of principal 2-bundles should be straightforward by abstract nonsense. Unfortunately, this is complicated by the fact that the connections on principal 2-bundles as conventionally defined in the literature are either too general (see e.g.~\cite{Breen:math0106083} and~\cite{Aschieri:2003mw}) or too restrictive because they imply a particular flatness condition on the curvature (see e.g.~\cite{Baez:0511710}).  Instead, one has to work with adjusted connections as explained in Chapters~2~and~4. In particular, the finite form of differentially refined, adjusted cocycles was only identified very recently~\cite{Rist:2022hci}. Using this technology, it is not hard to construct the relevant adjustment and to endow the principal 2-bundles describing geometric T-duality correspondences with adjusted connections.
    
    A differential refinement in the half-geometric case, however, requires some more work. Recall that T-duality can be interpreted as a Kaluza--Klein reduction of the correspondence space, cf.~\cite{Berman:2019biz,Alfonsi:2019ggg,Alfonsi:2020nxu,Alfonsi:2021ymc}. A Kalb--Ramond $B$-field on the correspondence space will thus give rise to scalar fields on the base space $X$ after T-duality, or dimensional reduction, along two directions. These scalar fields then take values in the Narain moduli space~\cite{Narain:1985jj}
    \begin{equation}
        M_n=\sfO(n,n;\IZ)~\backslash~\sfO(n,n;\IR)~/~\big(\sfO(n;\IR)\times \sfO(n;\IR)\big)\eqqcolon \sfO(n,n;\IZ)~\backslash~Q_n~.
    \end{equation}
    This makes it obvious that the principal 2-bundles used in~\cite{Nikolaus:2018qop} need to be extended to principal 2-groupoid bundles with structure 2-groupoid given by the 2-group $\sfTD_n$ fibred over the Narain moduli space $M_n$.
    
    As is often the case, it turns out to be convenient to replace the Narain moduli space by the action groupoid for the action of the T-duality group $\sfO(n,n;\IZ)$,
    \begin{equation}
        \sfO(n,n;\IZ)\ltimes Q_n~\rightrightarrows~Q_n~.
    \end{equation}
    The T-duality group $\sfO(n,n;\IZ)$, however, should also induce some transformation on $\sfTD_n$, and the construction of this transformation is (interestingly) non-trivial. In fact, $\sfO(n,n;\IZ)$ by itself does not act on $\sfTD_n$, and one has to introduce a larger 2-group\footnote{It has been recently shown that a 2-group of the form $\scGO(n,n;\IZ)$ is equivalent to the automorphism 2-group of $\sfTD_n$~\cite{Waldorf:2022lzs}.} $\scGO(n,n;\IZ)$. Once the action is derived, the relevant structure 2-groupoid $\scTD_n$ is obvious, and the explicit cocycle description of principal $\scTD_n$-bundles can be given.
    
    There is an evident continuation of our picture to $R$-spaces: the requirement for $0$-form curvatures corresponding to (non-existent) $(-1)$-forms suggests augmenting the 2-groupoid $\scTD_n$ in the simplicial sense to an augmented 2-quasi-groupoid $\scTD^\text{aug}_n$. The relevant space of $(-1)$-simplices is then identified from the observation that R-fluxes are related to particular embedding tensors. Again, the relevant cocycles can be written down, and we obtain a description of T-duality correspondences between general $R$-spaces.\footnote{The slides~\cite{Waldorf:2019aa} mention a possible extension of the framework of~\cite{Waldorf:2022lzs} to $R$-spaces by using Lie 2-groupoids, but they do not give details; in particular, augmentation, which we see as necessary for a proper description of $R$-fluxes, is not mentioned.}

    Our construction only captures fields that have trivial dependence on the T-duality directions. This is simply due to the fact that we interpret T-duality as a Kaluza--Klein reduction from the correspondence space and ignore any massive modes arising therefrom. The latter encode the part of the geometry and dynamics that is not invariant under translation along the fibres, and these fields do not introduce new gauge symmetries. Since the compactification is toroidal, the truncation to massless fields is consistent.

\chapter{A Short Introduction to Higher Gauge Theories}

\section{Why higher categories?}    
    An experimenter observes the Aharonov--Bohm effect and concludes that nature associates to each path a phase, i.e.~an element of $\sU(1)$. The phases add when paths are concatenated; the phases invert when paths are inverted. One would call the set of paths a group, except that paths only compose when the endpoints match. We instead call it a \emph{groupoid}, or more precisely the \emph{path groupoid} of the space, and regard the points in space as objects and the paths as maps or \emph{morphisms} between their endpoints, called \emph{domain} and \emph{codomain}:
    \begin{equation}
        \begin{gathered}
            \begin{tikzcd}
                \codom(\gamma)=y&[-1.2cm]\bullet &[1cm] \arrow[l,bend right,swap]{}{\textrm{path}~\gamma}\bullet&[-1.2cm]x=\dom(\gamma)
            \end{tikzcd}\\
            \begin{tikzcd}
                x &[-1.2cm]\bullet \arrow[loop,swap]{}{\id_x}
            \end{tikzcd}~~~
            \begin{tikzcd}
                y&[-1.2cm]\bullet \arrow[r,bend left]{}{\gamma^{-1}} &[1cm] \bullet&[-1.2cm]x
            \end{tikzcd}~~~
            \begin{tikzcd}
                \bullet & \ar[l,bend right,swap]{}{\gamma_2}\bullet & \bullet \ar[l,bend right,swap]{}{\gamma_1}\ar[ll,bend left]{}{\gamma_2\circ \gamma_1}
            \end{tikzcd}
        \end{gathered}
    \end{equation}
    The group of phases, $\sU(1)$, can also be regarded as a groupoid with morphisms $\sU(1)$ taking a single object $*$ to itself:
    \begin{equation}
        \begin{tikzpicture}[baseline=(current bounding box.center)]
            \node (a) {$*$};
            \path[scale=2] 
            (a) edge [loop right, "$\de^{\frac{1}{2}\pi\di}$"]
            (a) edge [loop below, "$\de^{\pi\di}$"] (a)
            (a) edge [loop above, "$\unit$"] (a);
            \draw [densely dotted]  (240:1.5em) arc (240:120:1.5em);
            \draw [densely dotted]  (60:1.5em) arc (60:30:1.5em);
            \draw [densely dotted]  (330:1.5em) arc (330:300:1.5em);
        \end{tikzpicture}
    \end{equation}
    We call this groupoid $\sB\sU(1)$. The Berry phase, which mathematicians call \emph{holonomy}, maps the objects and morphism in the path groupoid to objects and morphisms in the groupoid $\sB\sU(1)$ of phases. This generalisation of a group homomorphism is called a \emph{functor}.\footnote{The terminology is borrowed from philosophy: more general groupoids are called \emph{categories}, which Saunders Mac~Lane took from Kant; he also took the term \emph{functor} from Carnap.}
    
    Over time, physicists have discovered two variations on the theme. One, discovered by Yang and Mills, replaces the abelian group of phases $\sU(1)$ with non-abelian ones, as necessary for describing strong and weak nuclear forces. The other variation generalises paths to surfaces and higher-dimensional spaces, as necessary for field theory on higher-dimensional spacetimes. String theory seems to require both variations at the same time in e.g.~stacks of NS5- or M5-branes, which have strings in them with self-interacting higher non-abelian gauge fields. Defining the right generalisation of the underlying phases is important to fundamentally understand this physics.
    
    Both variations are captured by essentially obvious generalisations of the holonomy functor, exemplifying the utility of functorial descriptions of mathematical objects. Replacing the groupoid $\sB\sU(1)$ by the groupoid $\sB\sG$ for a non-abelian gauge group $\sG$ is straightforward. The generalisation to a higher-dimensional parallel transport requires the development of higher-dimensional groupoids containing points, paths between points, paths between paths, etc., but also this construction is not hard, using geometric intuition. Here, a new feature is that higher morphisms compose in multiple ways: e.g.~in the case of the 2-groupoid of points, paths and paths-between-paths, paths-between-paths compose both vertically and horizontally:
    \begin{equation*}
        \begin{tikzcd}
            \bullet & \lar[""{name=A},""'{name=B}] \lar[bend left=90, ""'{name=C}] \lar[bend right=90, ""{name=D}] \bullet \arrow[Rightarrow, from=B, to=D] \arrow[Rightarrow, from=C, to=A] 
        \end{tikzcd}
        \longmapsto 
        \begin{tikzcd}
            \bullet & \lar[bend left=90, ""'{name=C}] \lar[bend right=90, ""{name=D}] \bullet \arrow[Rightarrow, from=C, to=D] 
        \end{tikzcd}
        \quad\text{and}\quad
        \begin{tikzcd}
            \bullet & \lar[bend left=60, ""'{name=A}] \lar[bend right=60, ""{name=B}] \bullet \arrow[Rightarrow, from=A, to=B] & 
            \lar[bend left=60, ""'{name=C}] \lar[bend right=60, ""{name=D}] \bullet \arrow[Rightarrow, from=C, to=D]
        \end{tikzcd}
        \longmapsto
        \begin{tikzcd}
            \bullet && 
            \ar[ll, bend left=60, ""'{name=C}] \ar[ll, bend right=60, ""{name=D}] \bullet \arrow[Rightarrow, from=C, to=D]
        \end{tikzcd}
    \end{equation*}
    This gives rise to the term \emph{higher-dimensional algebra} for higher categories and higher groupoids. Higher-dimensional groupoids with single objects then describe higher analogues of groups, just as $\sB\sG$ describes the group $\sG$. This process of adding morphisms between morphisms is known as \emph{categorification}.
    
    The heavy use of groupoids and their higher-dimensional generalisations is thus due to the ease with which they allow us to reproduce and subsequently generalise relevant mathematical definitions, guaranteeing consistency from the outset.
    
    All terms we use are generalisations or categorifications of the mathematical terms underlying ordinary gauge theories. It is more convenient to describe them by equivalent mathematical objects, and one can easily get lost in the nomenclature.
    
    As described above, the gauge group is categorified to a higher Lie group. We describe Lie 2-groups and Lie 3-groups with the more economical language of \emph{crossed modules of Lie groups} and \emph{2-crossed modules of Lie groups}. 
    Just as Lie groups differentiate to Lie algebras, higher Lie groups differentiate to higher Lie algebras. For Lie $n$-algebras, we use three models: we start from \emph{$n$-term $L_\infty$-algebras},
    which we also describe dually via their (degree-shifted) function algebras, known as \emph{Chevalley--Eilenberg algebras}, and the function algebra of their inner derivations, known as \emph{Weil algebras}.
    The third description is in terms of \((n-1)\)-crossed modules of Lie algebras.
    The precise definitions of these are given in the next chapter.
    
    The higher analogue of a principal bundle is a \emph{principal $n$-bundle} or an \emph{\((n-1)\)-gerbe}.\footnote{Standard nomenclature often assumes gerbes to be abelian while principal $n$-bundles are unrestricted.} Locally, the description of connections is easily described as morphisms from the Weil algebra of the gauge $L_\infty$-algebra to the de~Rham complex of the local patch; see section~\ref{ssec:limitations}. For a gauge Lie $n$-algebra, one obtains 1-, 2-, \dots, $n$-forms valued in particular parts of the Lie $n$-algebra and corresponding 2-, 3-, \dots, $(n+1)$-form components of the total curvature. All of the latter, except for the form with top degree, are known as \emph{fake curvatures}.
    
    For higher-dimensional parallel transport we need \emph{higher groupoids}, which, as clear from the higher path groupoid, are essentially collections of objects, morphisms between objects, and higher morphisms between morphisms, such that all morphisms are invertible. We mostly work with strict higher groupoids, i.e.~those for which composition of morphisms is strictly associative and unital. Strict higher \((n+1)\)-groupoids are readily defined by replacing the group of morphisms of a (1-)category with the \(n\)-groupoid of morphisms.
    
    As mentioned before, a \emph{higher group} is defined by a higher groupoid with a single object.\footnote{Pedantically, a \emph{higher group} is obtained by truncating the single object and instead regarding groupoid 1-morphisms as the objects of the higher group and groupoid \((k+1)\)-morphisms as \(k\)-morphisms of the higher group. This generalises the relation between the \(\sG\) and \(\sB\sG\).}
    Our 3-groups are \emph{Gray groups}, which means that the two different ways of evaluating the diagram
    \begin{equation}
        \begin{tikzcd}
            \bullet & \lar[""{name=Aa},""'{name=Ba}] \lar[bend left=90, ""'{name=Ca}] \lar[bend right=90, ""{name=Da}] \bullet \arrow[Rightarrow, from=Ba, to=Da] \arrow[Rightarrow, from=Ca, to=Aa]
            & \lar[""{name=E},""'{name=F}] \lar[bend left=90, ""'{name=G}] \lar[bend right=90, ""{name=H}] \bullet \arrow[Rightarrow, from=F, to=H] \arrow[Rightarrow, from=G, to=E] 
        \end{tikzcd}
    \end{equation}
    are the same, up to isomorphism. For further details, see the literature cited in the respective sections. 

\section{Higher gauge theory and its complications}
    Higher gauge theory is simply a gauge theory in which the ordinary gauge group of Yang--Mills theory is replaced with a higher group. In the abelian case, this generalisation is well-known and standard: they are \emph{\(p\)-form electrodynamics}, which arise plentifully in higher spacetime dimensions. But just as nonabelian Yang--Mills theory shows much richer phenomena than Maxwell theory, we similarly wish to formulate a nonabelian analogue of such abelian higher gauge theories. Nonabelian higher gauge fields are expected to arise in a number of physical contexts, ranging from six-dimensional conformal field theory over supergravity theories to string/M-theory. Such gauge fields are meant to describe higher holonomies, arising from a parallel transport along higher-dimensional spaces, e.g.~surfaces. 
    
    In particular, the classical string couples to the Kalb--Ramond 2-form field $B$, which is part of the connection of an abelian gerbe. This is the higher analogue of a particle coupling to a Maxwell gauge potential $A$, which is part of the connection on an abelian principal bundle. If we now want to generalise connections on abelian gerbes to potentially self-interacting ones, mimicking the transition from Maxwell fields to Yang--Mills fields, we face a number of problems. Using the appropriate language of 2-categories and functorial definitions of higher principal bundles, of their connections and of the induced parallel transport, most of these\footnote{e.g.~the Eckmann--Hilton type argument forbidding a na\"ive non-abelian higher parallel transport} are readily overcome.
    
    We arrive at a theory of non-abelian gerbes or higher principal bundles with connections~\cite{Breen:math0106083,Aschieri:2003mw,Baez:2004in,Baez:0511710} together with an induced parallel transport~\cite{Baez:2002jn,Girelli:2003ev,Baez:2004in,Schreiber:0705.0452,Schreiber:0802.0663,Schreiber:2008aa}; see also~\cite{Soncini:2014zra} as well as~\cite{Baez:2010ya} for an introduction. Topologically, these non-abelian higher principal bundles are simultaneous generalisations of (non-abelian) principal fiber bundles and abelian gerbes. The connections they carry, however, merely generalise connections of abelian gerbes. Consistency of the underlying differential cocycles requires that a particular curvature component, known as \emph{fake curvature}, vanishes. This fake flatness condition also arises from a higher Stokes' theorem, guaranteeing invariance of the induced higher parallel transport under reparameterisations.
    
    Thus, the fake flatness condition forbids a straightforward interpretation of ordinary principal bundles with connections as non-abelian higher principal bundles with connection. This is surprising because categorification usually implies generalisation: a set is trivially a category, a category is trivially a 2-category, a group is trivially a 2-group, and, indeed, a principal bundle is trivially a principal 2-bundle; but not every principal bundle with connection is a principal 2-bundle with connection in the sense of~\cite{Breen:math0106083,Aschieri:2003mw,Baez:2004in,Baez:0511710}. Even worse, connections on non-abelian higher principal bundles are locally gauge equivalent to connections on abelian ones; see~\cite{Saemann:2019dsl} and also~\cite{Gastel:2018joi}. Locally, thus, the extension to non-abelian higher principal bundles is futile, and we can merely hope to answer some topological questions with these, using higher versions of Chern--Simons theories. Other highly interesting theories, such as six-dimensional superconformal field theories involving the tensor multiplet, require local gauge field interactions, also over topologically trivial spaces, and therefore the conventional non-abelian higher principal bundles are inapt for their description.
    
    For certain higher gauge groups, there is a further generalisation of the notion of connection that lifts this limitation. A higher gauge algebra\footnote{Note that we always follow the physicists' nomenclature and identify the terms \emph{gauge group} and \emph{gauge (Lie) algebra} with the structure group and structure Lie algebra of the principal bundle underlying the gauge theory. The gauge group is thus different from the resulting \emph{group of gauge transformations}.} $\frL$ gives rise to a differential graded algebra $\sW(\frL)$, called its \emph{Weil algebra}. The kinematical data of a higher gauge theory over some local patch $U$ of spacetime is fully encoded in a morphism of differential graded algebras from $\sW(\frL)$ to the de~Rham complex $\Omega^\bullet(U)$. However, the na\"ive generalisation of the notion of Weil algebra of a Lie algebra to the Weil algebra of a higher Lie algebra is problematic: the induced definition of invariant polynomials is not compatible with quasi-isomorphisms, which are the appropriate notion of isomorphisms for higher Lie algebras. For particular higher Lie algebras $\frL$, this incompatibility can be overcome by particular deformations of the Weil algebra $\sW(\frL)$~\cite{Sati:2008eg,Sati:2009ic}.
    
    At the field theory level, the BRST complex describing infinitesimal gauge transformations and their actions on the fields arising from morphisms $\sW(\frL)\rightarrow \Omega^\bullet(U)$ is not closed. It closes only up to equations of motion corresponding to the fake curvature condition. The aforementioned deformations of the Weil algebra also cure this problem. Such deformed Weil algebras that induce a closed BRST complex were called \emph{adjusted Weil algebras} in~\cite{Saemann:2019dsl}.
    
    If the higher gauge group is the \emph{string 2-group}\footnote{more precisely: a 2-group model of the string group}, a higher relative of the spin group, the adjustment leads to \emph{differential string structures}. These are expected to arise in the context of string theory and M-theory; see~\cite{Saemann:2017rjm,Saemann:2017zpd,Saemann:2019dsl}.
    

    \section{Cartan's formalism}

    Henri Cartan~\cite{Cartan:1949aaa,Cartan:1949aab} discovered a particularly elegant and useful description of local connection forms on principal fiber bundles. Let $\frg$ be a Lie algebra\footnote{either a finite-dimensional Lie algebra, or an infinite-dimensional Lie algebra with a suitable notion of dual space} with basis $e_\alpha$ and structure constants $f^\gamma_{\alpha\beta}$, such that
    \begin{equation}
        [e_\alpha,e_\beta] \eqqcolon f^\gamma_{\alpha\beta}e_\gamma~.
    \end{equation}
    Dually, $\frg$ can be regarded as the \emph{(graded-commutative) differential graded algebra}
    \begin{equation}
        \sCE(\frg)\coloneqq \big(\bigodot\nolimits^\bullet \frg[1]^*,Q_\sCE\big)=\big(\CC^\infty_{\rm pol}(\frg[1]),Q_\sCE\big)~,
    \end{equation}
    which consists of polynomials in the coordinate functions $t^\alpha\in \frg[1]^*$ of degree one and whose differential $Q_\sCE$ is the homological vector field 
    \begin{equation}
        Q_\sCE=-\tfrac12 f^\gamma_{\alpha\beta}t^\alpha t^\beta \der{t^\gamma}~,~~~|Q|=1~,~~~Q^2=0~.
    \end{equation}
    We call $\sCE(\frg)$ the \emph{Chevalley--Eilenberg algebra} of $\frg$.
    
    Similarly, the Chevalley--Eilenberg algebra of the grade-shifted tangent bundle $T[1]U$ of a local patch $U$ of some manifold $M$ can be identified with the de~Rham complex of $U$,
    \begin{equation}
        \sCE(T[1]U)=\big(\CC^\infty(T[1]U),\dd\big)=\big(\Omega^\bullet(U),\dd\big)~.
    \end{equation}
    Morphisms of differential graded algebras
    \begin{equation}\label{eq:dga_morph_CE}
        \CA:\sCE(\frg)\to \sCE(T[1] U)
    \end{equation}
    preserve the graded algebra structure and are therefore fixed by the image of $t^\alpha$,
    \begin{equation}
        \CA(t^\alpha) \eqqcolon A^\alpha\in \Omega^1(U)~,
    \end{equation}
    a Lie algebra-valued differential form or local connection 1-form $A\coloneqq A^\alpha e_\alpha$ on $U$. Compatibility with the differentials on $\sCE(\frg)$ and $\sCE(T[1]U)$ enforces flatness of this connection,
    \begin{equation}
        \begin{aligned}
            (\dd\circ \CA)(t^\alpha)&=(\CA\circ Q_\sCE)(t^\alpha)\\
            \dd A^\alpha&=\CA(-\tfrac12 f^\alpha_{\beta\gamma}t^\beta t^\gamma)=-\tfrac12 f^\alpha_{\beta\gamma}A^\beta \wedge A^\gamma\\
            &\implies F\coloneqq \dd A+\tfrac12 [A,A]=0~.
        \end{aligned}    
    \end{equation}
    Gauge transformations are encoded in partially flat homotopies between two morphisms $\CA$ and $\tilde \CA$ of type~\eqref{eq:dga_morph_CE}. 
    
    To describe non-flat connections, we enlarge the Chevalley--Eilenberg algebra $\sCE(\frg)$ to the \emph{Weil algebra}
    \begin{equation}
        \sW(\frg)\coloneqq \left(\bigodot\nolimits^\bullet(\frg[1]^*\oplus \frg[2]^*),Q_\sW\right)=\big(\CC^\infty_{\rm pol}(\frg[1]^*\oplus \frg[2]^*),Q_\sW\big)~,
    \end{equation}
    which consists of polynomials in the coordinate functions $t^\alpha\in\frg[1]^*$ and $\hat t^\alpha=\sigma(t^\alpha)\in \frg[2]^*$, where $\sigma:\frg[1]^*\to \frg[2]^*$ is the shift isomorphism. We extend $\sigma$ trivially to $\frg[1]^*\oplus \frg[2]^*$ by $\sigma(\frg[2]^*)\coloneqq 0$ and as a derivation to $\bigodot\nolimits^\bullet(\frg[1]^*\oplus \frg[2]^*)$. We also extend $Q_\sCE$ to $\bigodot\nolimits^\bullet(\frg[1]^*\oplus \frg[2]^*)$ by demanding that
    \begin{equation}
        Q_\sCE \sigma\coloneqq -\sigma Q_\sCE~.
    \end{equation}
    The homological vector field $Q_\sW$ on $\frg[1]\oplus \frg[2]$ is then defined as
    \begin{equation}
        Q_\sW=Q_\sCE+\sigma~.
    \end{equation}
    Explicitly, we have
    \begin{equation}\label{eq:Weil_of_ordinary_Lie}
        Q_\sW\colon~t^\alpha \mapsto -\tfrac12 f^\alpha_{\beta\gamma} t^\beta t^\gamma + \hat t^\alpha \eand
        \hat t^\alpha\mapsto -f^\alpha_{\beta\gamma} t^\beta \hat t^\gamma~,
    \end{equation}
    where $f^\alpha_{\beta\gamma}$ are again the structure constants of $\frg$.
    
    Without going into further details, we note that the Chevalley--Eilenberg algebra of the tangent Lie algebroid $T[1]U$ can be seen as the Weil algebra of the manifold $U$ regarded as the trivial Lie algebroid over itself, $\sCE(T[1]U)=\sW(U)$.
    
    Non-flat connections are described as morphisms of differential graded algebras
    \begin{equation}\label{eq:dga_morph_Weil}
        \CA\colon \sW(\frg)\to \sW(U)~,
    \end{equation}
    which are fixed by their action on the generators $t^\alpha$ and $\hat t^\alpha$. We define
    \begin{equation}
        \begin{aligned}
            A&\coloneqq A^\alpha \tau_\alpha~,~~~&A^\alpha&\coloneqq \CA(t^\alpha)~,\\
            F&\coloneqq F^\alpha \tau_\alpha~,~~~&F^\alpha&\coloneqq \CA(\hat t^\alpha)~.
        \end{aligned}
    \end{equation}
    Compatibility with the differentials and the graded algebra structure implies that
    \begin{equation}
        F=\dd A+\tfrac12[A,A]\eand \dd F+[A,F]=0~.
    \end{equation}
    We thus recover the definition of the curvature and the Bianchi identity. As stated above, gauge transformations are obtained by partially flat homotopies. Recall that a homotopy between morphisms $\CA,\tilde \CA\colon \sW(\frg)\rightarrow \sW(U)$ is given by a morphism 
    \begin{equation}
        \hat \CA\colon\sW(\frg)\rightarrow \sW(U\times I)~~~\mbox{with}~~~\hat \CA(x,0)=\CA(x)~,~~~\hat \CA(x,1)=\tilde \CA(x)~,
    \end{equation}
    where $I=[0,1]$ and $x\in U$. Let $t$ be the coordinate on $I$. The potential 1-form and the curvature 2-form now naturally decompose into two parts:
    \begin{equation}
        \hat A=\hat A_x+\hat \alpha_t\dd t~,~~~\hat F=\hat F_x+\hat \varphi_t\wedge \dd t~,~~~\hat A_x\left(\der t\right)=\hat F_x\left(\der t\right)=0~.
    \end{equation}
    Partial flatness means $\varphi_t=0$, and we can compute
    \begin{equation}
        \delta A\coloneqq\der{t}\hat A(t)\Big|_{t=0}=\dd_x \hat \alpha_t+[\hat A_x,\hat \alpha_t]\Big|_{t=0}=:\dd_x \alpha+[A,\alpha]~,
    \end{equation}
    and we recover the usual form of infinitesimal gauge transformations.
    
    \section{(Unadjusted) Weil algebras and their limitations}\label{ssec:limitations}
    A particularly nice feature of Cartan's description of gauge potentials in terms of morphisms of differential graded algebras is its generality: one can easily replace both the domain and the codomain with more general differential graded algebras. In this chapter, we are interested in more general domains~\cite{Kotov:2007nr,Sati:2008eg}; see e.g.~\cite{Ritter:2015zur} for more general codomains.
    
    An obvious generalisation of the source $\sCE(\frg)$ is obtained by replacing the graded vector space $\frg[1]$ by a more general, $\RZ$-graded vector space 
    \begin{equation}
        E=\bigoplus_{i\in \RZ} E_i~,
    \end{equation}
    again endowed with a nilquadratic vector field $Q_\sCE$ of degree~1. The resulting differential graded algebras are the Chevalley--Eilenberg algebras of \emph{$L_\infty$-algebras}.\footnote{generalising $E$ to a vector bundle directly yields Chevalley--Eilenberg algebras of \emph{$L_\infty$-algebroids}.} These are graded vector spaces $\frL=E[-1]$ together with a set of \emph{higher products}
    \begin{equation}
        \mu_i\colon\frL^{\wedge i} \to \frL
    \end{equation}
    of degree $|\mu_i|=2-i$. The explicit form of the higher products can be derived from $Q_\sCE$; see \cref{app:L_infinity_defs} for explicit formulas and our conventions. Because $Q_\sCE^2=0$, the higher products $\mu_i$ satisfy a generalisation of the Jacobi identity, the \emph{homotopy Jacobi identity}; see \cref{app:L_infinity_defs}.
    If $\frL$ is an $L_\infty$-algebra with underlying graded vector space of the form
    \begin{equation}
        \frL=\frL_{-n+1}\oplus \frL_{-n+2}\oplus \dotsb \oplus \frL_{-1}\oplus \frL_0~,
    \end{equation}
    we say that $\frL$ is an \emph{\(n\)-term \(L_\infty\)-algebra}; it is a model of a Lie \(n\)-algebra. We call an \(L_\infty\)-algebra \emph{strict} if \(\mu_i=0\) for \(i\ge3\).
    
    Flat higher connections are equivalent to morphisms of differential graded algebras from the Chevalley--Eilenberg algebra $\sCE(\frL)$ of an $L_\infty$-algebra $\frL$ to the Weil algebra $\sW(U)$ of the Lie algebroid of smooth functions on a contractible patch $U$, which turns out to be nothing more than the graded-commutative algebra of differential forms on \(U\):
    \begin{equation}
        \sW(U)\cong\Omega^\bullet(U)~.
    \end{equation}
    That is, the data of a flat higher connection are described by morphisms
    \begin{equation}
        \caA\colon \sCE(\frL)\longrightarrow \sW(U)\cong\Omega^\bullet(U)~.
    \end{equation}
    General connections can be described by morphisms from the Weil algebra of $\frL$ to $\sW(U)$:
    \begin{equation}\label{eq:naive_dga_morphism}
        \caA\colon\sfW(\frL)\longrightarrow \sW(U)\cong\Omega^\bullet(U)~.
    \end{equation}
    The definition of the Weil algebra of an \(L_\infty\)-algebra is a straightforward generalisation of the Weil algebra of a Lie algebra:
    \begin{equation}
        \sW(\frL)\coloneqq \big(\bigodot\nolimits^\bullet(\frL[1]^*\oplus \frL[2]^*),Q_\sW\big)~,~~~Q_\sW\coloneqq Q_\sCE+\sigma~,
    \end{equation}
    where $\sigma$ is the trivial extension of the shift isomorphism $\frL[1]^*\to \frL[2]^*$ to $\frL[1]^*\oplus \frL[2]^*$ and further, as a derivation, to $\bigodot\nolimits^\bullet(\frL[1]^*\oplus \frL[2]^*)$, and where $Q_\sCE$ is the extension of the Chevalley--Eilenberg differential by the rule $Q_\sCE \sigma \coloneqq -\sigma Q_\sCE$.
    
    For definiteness, consider a Lie 2-algebra $\frL=\frL_{-1}\oplus \frL_0$. Morphisms of differential graded algebras $\sW(\frL)\to \sW(U)$, where $U$ is a local patch of some manifold $M$, encode the following kinematical data:
    \begin{subequations}\label{eq:unadjusted_fields}
        \begin{align}
            A&\in \Omega^1(U)\otimes \frL_0~,\\
            B&\in \Omega^2(U)\otimes \frL_{-1}~,\\
            F&=\dd A+\tfrac12 \mu_2(A,A)+\mu_1(B)\in \Omega^2(U)\otimes \frL_0~,\\
            \dd F&=-\mu_2(A,F)+\mu_1(H)~,\notag\\
            H&=\dd B+\mu_2(A,B)+\tfrac{1}{3!}\mu_3(A,A,A)\in \Omega^3(U)\otimes \frL_{-1}~,\label{eq:unadjusted_fields_H}\\
            \dd H&=-\mu_2(A,H)+\mu_2(F,B)-\tfrac12\mu_3(A,A,F)~.\notag
        \end{align}
    \end{subequations}
    The infinitesimal gauge transformations are again induced by partially flat infinitesimal homotopies between two morphisms of differential graded algebras. Here, they are parametrised by
    \begin{equation}\label{eq:gauge_parameters}
        (\alpha,\Lambda)\in (\Omega^0(U)\otimes \frL_0)\times (\Omega^1(U)\otimes \frL_{-1})
    \end{equation}
    and read as 
    \begin{subequations}\label{eq:unadjusted_gauge_transf}
        \begin{align}
            \delta A&=\dd \alpha-\mu_1(\Lambda)+\mu_2(A,\alpha)~,\\
            \delta B&=\dd \Lambda+\mu_2(A,\Lambda)+\mu_2(B,\alpha)-\tfrac12\mu_3(A,A,\alpha)~,\label{eq:unadjusted_B_gauge_transformation}
\\
            \delta F &= \mu_2(F,\alpha)~,\\
            \delta H &=\mu_2(H,\alpha)+\mu_2(F,\Lambda)-\mu_3(F,A,\alpha)~.\label{eq:unadjusted_H_gauge_transformation}
        \end{align}
    \end{subequations}
    The commutator of two infinitesimal gauge transformations is
    \begin{equation}
        [\delta_{\alpha+\Lambda},\delta_{\alpha'+\Lambda'}]=\delta_{\mu_2(\alpha+\Lambda,\alpha'+\Lambda')+\mu_3(A,\alpha+\Lambda,\alpha'+\Lambda')}+\mu_3(F,\alpha,\alpha')~,
    \end{equation}
    and we have run into the following severe limitation. Gauge transformations generically only close if the theory is abelian (and thus $\mu_i=0$ for $i\geq 2$) or if the \emph{fake curvature}\footnote{In a general higher gauge theory, a \emph{fake curvature} is any curvature form other than the top form.} $F$ vanishes. The situation is not improved by restricting to strict $L_\infty$-algebras (for which $\mu_i$ with $i\geq 3$ vanishes), since there the condition $F=0$ reappears when composing finite gauge transformations. 
    
    Fake flatness also arises in the conventional definition of higher parallel transport; see~e.g.~\cite{Schreiber:0705.0452,Schreiber:2008aa}. Section~\ref{ssec:derived_parallel_higher} further discusses this point.
    
    For all these reasons, fake flatness $F=0$ is a fixed part of the conventional definition of connections on principal 2-bundles in the literature, cf.~\cite{Breen:math0106083,Aschieri:2003mw,Baez:2004in,Baez:0511710}.
    
    The fake flatness condition $F=0$ is now highly problematic due to the following theorem~\cite{Saemann:2019dsl}; see also~\cite{Gastel:2018joi} for a detailed analysis of the involved gauges:
    \begin{theorem}
        A connection on a non-abelian principal 2-bundle is locally gauge equivalent to a connection on an abelian principal 2-bundle.
    \end{theorem}
    This is somewhat surprising. Topologically, ordinary principal bundles are easily interpreted as principal 2-bundles. A Lie group $\sG$ is readily seen as a Lie 2-group, e.g.~in the form of the crossed module of Lie groups\footnote{See \cref{app:hypercrossed_modules} for definitions.} $*\to\sG$. The cocycles defining a principal 2-bundle with structure 2-group $*\to\sG$ are precisely those of an ordinary principal bundle. As soon as we endow the principal bundle with a connection, however, this embedding breaks; only flat principal bundles can be 2-bundles.

    We also note that the form of the gauge transformations of $H$ makes it very hard to imagine a covariant equation of motion. In particular, a non-abelian (2,0)-theory would involve the self-duality equation in six dimensions; however, the equation $H=\star H$ is not covariant unless $F=0$.
    
    The above observations are \emph{not} specific to kinematical data derived from Lie 2-algebras, but rather constitute a generic feature of higher gauge theories; see~e.g.~the discussion of homotopy Maurer--Cartan theory in~\cite{Jurco:2018sby}. Thus, higher gauge theory as conventionally defined is fake flat and locally abelian. This is well-known in the context of BRST/BV quantisation, where these higher gauge theories lead to an ``open'' complex, which closes only modulo equations of motion.

\chapter{Background on higher Lie groups and higher Lie algebras}
    
    In this chapter, we review and fix conventions for the relevant mathematical technology used in this thesis. This is probably not a good way to learn the relevant material for readers new to higher categories, who are referred to e.g.\ \cite{Borceux:1994aa}. However, unfortunately higher category theory is one of those areas of mathematics where many different conventions are possible and have been used, so we will go to some lengths to specify the conventions that I use, which largely follow \cite{Jurco:2014mva}. The reader is advised to skip this chapter in a first reading and to come back to it whenever she encounters the relevant concepts.
    
    \section{Lie 2-groupoid basics}\label{app:2-groupoid_basics}
    Weak 2-categories, also known as bicategories, were introduced in \cite{Benabou:1967:1}.
    
    \paragraph{Weak 2-categories.} A \underline{weak 2-category} $\scB$ consists of a collection of \uline{objects} or \uline{0-cells} $\scB_0$ and a category of morphisms $\scB(a,b)$ for every pair of objects $a,b\in \scB_0$. Objects and morphisms in these categories are known as \uline{1-} and \uline{2-cells}, respectively. For each object $a\in \scB_0$, there is an \uline{identity 1-cell} $\sfid_a$. Composition of 2-cells is denoted by $\circ$ and called \uline{vertical composition}. \uline{Horizontal composition}, on the other hand, is denoted by \(\otimes\) and is a collection of bifunctors $\scB(a,b)\otimes \scB(b,c)\rightarrow \scB(a,c)$ for all $a,b,c\in \scB_0$. Horizontal composition is not strict and comes with a set of natural isomorphisms known as left- and right unitors,
    \begin{subequations}
        \begin{equation}
            \sfl\colon x\otimes \sfid_\sfs(x)\xRightarrow{~\cong~} x
            \eand 
            \sfr\colon\sfid_\sfs(x)\otimes x\xRightarrow{~\cong~} x~,
        \end{equation}
        as well as an associator, 
        \begin{equation}
            \sfa\colon(x\otimes y)\otimes z\xRightarrow{~\cong~} x\otimes (y\otimes z)~,
        \end{equation}
        for all 1-cells $x,y,z$. These morphisms satisfy coherence conditions known as the pentagon and triangle identities, see~\cite{Jurco:2014mva}. We will exclusively work with ``unital'' weak 2-categories that come with unital horizontal composition, reducing the coherence condition to the pentagon identity for the associator:
        \begin{equation}
            \begin{tikzcd}
                & ((x\otimes y)\otimes u)\otimes v \arrow[rd,Rightarrow,"\sfa"] \arrow[ld,Rightarrow,"\sfa\otimes \sfid",swap] & 
                \\
                (x\otimes(y\otimes u))\otimes v  \arrow[d,Rightarrow,"\sfa",swap] & & (x\otimes y)\otimes (u\otimes v) \arrow[d,Rightarrow,"\sfa"]
                \\
                x\otimes((y\otimes u)\otimes v) \arrow[rr,Rightarrow,"\sfid\otimes \sfa"] & & x\otimes(y\otimes(u\otimes v))
            \end{tikzcd}
        \end{equation}
    \end{subequations}    
    
    \paragraph{2-functors.} Given two weak 2-categories $\scB$ and $\tilde \scB$, a \uline{unital lax 2-functor} $\Phi\colon\scB\rightarrow \tilde \scB$ consists of a function
    \begin{subequations}
        \begin{equation}
            \Phi_0\colon\scB_0\rightarrow \tilde \scB_0~,
        \end{equation}
        a collection of functors
        \begin{equation}
            \Phi_1^{ab}\colon\scB(a,b)\rightarrow \tilde \scB(\Phi_0(a),\Phi_0(b))~,
        \end{equation}
        and a collection of natural transformations
        \begin{equation}
            \Phi_2^{abc}\colon\Phi_1^{ab}(-)\tildeotimes \Phi_1^{bc}(-)~\xRightarrow{~~}\Phi_1^{ac}(-\otimes -)
        \end{equation}
        for all $a,b,c\in \scB_0$. The latter satisfy a coherence condition amounting to the commutative diagram
        \begin{equation}\label{eq:2-functor_associator_coherence}
            \begin{tikzcd}
                & \Phi_1^{ac}(x\otimes y)\,\tildeotimes\, \Phi_1^{cd}(z) \arrow[dr,Rightarrow,"\Phi_2^{acd}"] & 
                \\
                (\Phi_1^{ab}(x)\,\tildeotimes\,\Phi_1^{bc}(y))\,\tildeotimes\, \Phi_1^{cd}(z) \arrow[ur,Rightarrow,"\Phi_2^{abc}\otimes \sfid"]  \arrow[d,Rightarrow,"\tilde \sfa"] & & \Phi_1^{ad}((x\otimes y)\otimes z) \arrow[d,Rightarrow,"\Phi_1^{ad}(\sfa)"]
                \\
                \Phi_1^{ab}(x)\,\tildeotimes\,(\Phi_1^{bc}(y)\,\tildeotimes\, \Phi_1^{cd}(z))\arrow[dr,Rightarrow,"\sfid\otimes \Phi_2^{bcd}"] &  & \Phi_1^{ad}(x\otimes(y\otimes z))
                \\
                & \Phi_1^{ab}(x)\,\tildeotimes\,\Phi_1^{bd}(y\otimes z)\arrow[ur,Rightarrow,"\Phi_2^{abd}"] & 
            \end{tikzcd}
        \end{equation}
    \end{subequations}    
    If the natural transformations are natural isomorphisms, we speak of a \uline{weak 2-functor}.
    
    We note that 2-functors $\Psi\colon\scB_1\rightarrow \scB_2$ and $\Phi\colon\scB_2\rightarrow \scB_3$ compose as 
    \begin{equation}\label{eq:composition_2_functors}
        \begin{gathered}
            \Xi=\Phi\circ \Psi~,
            \\
            \Xi_0=\Phi_0\circ \Psi_0~,~~~
            \Xi^{ab}_1=\Phi^{\tilde a\tilde b}_1\circ \Psi^{ab}_1~,
            \\
            \Xi^{abc}_2(x,y)=\Phi^{\tilde a\tilde c}_1(\Psi^{abc}_2(x,y))\circ \Phi^{\tilde a\tilde b\tilde c}_2(\Psi^{ab}_1(x),\Psi^{bc}_1(y))~,
        \end{gathered}
    \end{equation}
    where $\tilde a=\Psi_0(a)$, etc.
    
    \paragraph{Lie 2-groupoids.} A \uline{(weak) 2-groupoid} is a weak 2-category in which all morphisms are equivalences. That is, all 2-cells are strictly invertible, and all 1-cells are invertible up to isomorphisms. A \uline{Lie 2-groupoid} is then a 2-groupoid internal to a suitable category of smooth manifolds.\footnote{Note that the naive choice of smooth manifolds and smooth maps between these does not contain all pullbacks, which leads to problems in the composition. This is a well-known technicality, which can be resolved by working with diffeological spaces and which we ignore here.}
    
    \section{Higher groups}\label{app:higher_groups}\label{ssec:BUTTERFLY}
    
    \paragraph{2-groups.} A 2-group is a categorified group. In its most general form, a \uline{(weak) 2-group} is a weak monoidal small category in which all morphisms are invertible and all objects are weakly invertible, cf.~e.g.~\cite{Baez:0307200}. Equivalently, we can regard it as a monoidal category of morphisms contained in a pointed\footnote{Although the pointing is unique, one should use pointed (2-)functors between pointed (2-)groupoids in order to get the correct automorphisms, cf.~the $n$Lab page \url{https://ncatlab.org/nlab/show/looping}}  Lie 2-groupoid with a single 0-cell.
    
    If the associator in a 2-group is trivial, then we obtain a \uline{strict 2-group}, which can be regarded as a group object internal to $\CatCat$, the category of small categories. As shown in~\cite{Brown:1976:296-302}, strict 2-groups are equivalent to crossed modules of groups. 
    
    \paragraph{Crossed modules of groups.} A \uline{crossed module of groups} $\caG=(\sfH\xrightarrow{~\sft~}\sfG,\acton)$ is a pair of groups $\sfG,\sfH$ together with a group homomorphism $\sft\colon\sfH\rightarrow \sfG$ and an action of $\sfG$ on $\sfH$ by automorphisms $\acton\colon\sfG\times \sfH\rightarrow \sfH$ such that, for all $g\in\sfG$ and for all $h_{1,2}\in\sfH$, we have
    \begin{equation}
        \sft(g\acton h_1)\ =\ g\sft(h_1)g^{-1}
        \eand
        \sft(h_1)\acton h_2\ =\ h_1h_2h_1^{-1}~.
    \end{equation}
    This has an evident specialization to crossed modules of Lie groups.
    
    Crossed modules are models of strict Lie 2-groups, which are special cases of weak Lie 2-groups --- monoidal categories in which every object has a weak inverse and every morphism has an inverse~\cite{Baez:0307200}. Here, we will use the conventions of~\cite{Jurco:2014mva}, under which the monoidal category $\underline{\caG}$ corresponding to a crossed module $\caG=(\sfH\xrightarrow{~\sft~}\sfG,\acton)$ is defined as follows:
    \begin{equation}\label{eq:mon_cat_from_2_group}
        \begin{aligned}
            \begin{tikzcd}
                \sfG\ltimes \sfH \arrow[r,shift left] 
                \arrow[r,shift right] & \sfG
            \end{tikzcd}~,~&~~
            \begin{tikzcd}[column sep=2.0cm,row sep=large]
                \phantom{\sft(h)} g & \sft(h^{-1})g\arrow[l,bend left,swap,out=-20,in=200]{}{(g,h)}
            \end{tikzcd}~,
            \\
            (g_1,h_1)\circ (\sft(h_1^{-1})g_1,h_2)&\coloneqq(g_1,h_1h_2)~,
            \\
            (g_1,h_1)\otimes (g_2,h_2)&\coloneqq(g_1g_2,(g_1\acton h_2)h_1)~,
            \\
            \sfinv(g_1,h_1)&\coloneqq (g_1^{-1},g_1^{-1}\acton h_1^{-1})~.
        \end{aligned}    
    \end{equation}    
    
    \paragraph{Morphisms.} A \uline{strict morphism} of crossed modules of groups $\Phi\colon\caG\rightarrow \tilde \caG$ is simply a map
    \begin{equation}
        \Phi~~:~~(\sfH\xrightarrow{~\sft~}\sfG,\acton)~~\longrightarrow~~(\tilde \sfH\xrightarrow{~\tilde \sft~}\tilde \sfG,\tilde \acton)
    \end{equation}
    consisting of a pair of group homomorphisms $\Phi_0\colon\sfG\rightarrow \tilde \sfG$ and $\Phi_1\colon\sfH\rightarrow \tilde \sfH$ that are compatible with $\sft$ and $\acton$ in evident ways.
    
    A \uline{very weak morphism} of crossed modules of groups $\Phi\colon\caG\rightarrow \tilde \caG$, also known as a \uline{butterfly}, cf.~\cite{Aldrovandi:0808.3627}, is a commutative diagram of groups
    \begin{equation}\label{eq:butterfly}
        \begin{tikzcd}
            \sfH_1 \arrow[dd,"\sft_1",swap] \arrow[dr,"\lambda_1"]& & \sfH_2 \arrow[dl,"\lambda_2",swap] \arrow[dd,"\sft_2"]
            \\
            & \sfE \arrow[dl,"\gamma_1",swap] \arrow[dr,"\gamma_2"]& 
            \\
            \sfG_1 & & \sfG_2
        \end{tikzcd}
    \end{equation}
    where $\sfE$ is a group, $\lambda_{1,2}$ and $\gamma_{1,2}$ are group homomorphisms, the NE--SW diagonal is a short exact sequence (i.e.~a group extension), and the NW--SE diagonal is a complex. A butterfly from $\CCG$ to $\tilde \CCG$ is {\em flippable} if it is also a butterfly from $\tilde \CCG$ to $\CCG$.
    
    Between these two notions lies the notion of a \uline{weak morphism} of crossed modules, which is induced by a lax 2-functor of the corresponding two one-object Lie 2-groupoids whose categories of morphisms are $\underline{\caG}$ and $\underline{\tilde{\caG}}$, respectively, cf.~\cite{Jurco:2014mva}, as defined in~\cref{app:2-groupoid_basics}.\footnote{In this paper, we refrain from using the notion of crossed intertwiners developed in~\cite{Nikolaus:2018qop} for practical reasons.} Such a morphism $\Phi$ is thus encoded in a functor and a natural transformation,
    \begin{subequations}\label{eq:weak_morphism}
        \begin{equation}
            \Phi_1\colon\underline{\caG}\rightarrow \underline{\tilde{\caG}}
            \eand 
            \Phi_2\colon\Phi_1(-)\tildeotimes \Phi_1(-)\Rightarrow \Phi_1(-\otimes -)~.
        \end{equation}
        Besides the naturality condition 
        \begin{equation}
            \Phi_2(g_1,g_2)\tilde \circ (\Phi_1(g_1,h_1)\tildeotimes \Phi_1(g_2,h_2))=\Phi_1((g_1,h_1)\otimes (g_2,h_2))\tilde \circ \Phi_2(\sft(h_1^{-1})g_1,\sft(h_2^{-1})g_2)
        \end{equation}
        for all $g_{1,2}\in \sfG$ and $h_{1,2}\in \sfH$, we have the coherence condition~\eqref{eq:2-functor_associator_coherence} with trivial associators, resulting in
        \begin{equation}
            \Phi_2(g_1\otimes g_2,g_3)\tilde \circ(\Phi_2(g_1,g_2)\tildeotimes \sfid_{\Phi_1(g_3)})
            =
            \Phi_2(g_1,g_2\otimes g_3)\tilde \circ (\sfid_{\Phi_1(g_1)}\tildeotimes \Phi_2(g_2,g_3))~.
        \end{equation}
        Strict morphisms are evidently included here as weak morphisms with $\Phi_2$ trivial.
    \end{subequations}
    Two weak morphisms of Lie 2-groups $\Psi\colon\underline{\caG}_1\rightarrow \underline{\caG}_2$ and $\Phi\colon\underline{\caG}_2\rightarrow \underline{\caG}_3$ compose into a morphism $\Xi=\Phi\circ \Psi$ with
    \begin{equation}
        \begin{aligned}
            \Xi_1(g,h)&=\Phi_1(\Psi_1(g,h))~,
            \\
            \Xi_2(g_1,g_2)&=\Phi_1(\Psi_2(g_1,g_2))\circ \Phi_2(\Psi_1(g_1),\Psi_1(g_2))
        \end{aligned}
    \end{equation}
    for all $(g,h)$ in the morphisms of $\underline{\caG}_1$, cf.~\eqref{eq:composition_2_functors}.
    
    Weak morphisms of crossed modules are particularly useful for our discussion as they can be readily postcomposed with the lax 2-functors defining principal 2-bundles, cf.~\ref{app:higher_principal_bundles}.
    
    \paragraph{2-group actions.} Any 2-group $\scH$ comes with a 2-group of automorphisms (or equivalences) $\sfAut(\scH)$, having 2-group endomorphisms that are equivalences of categories as its objects and natural 2-transformations between these as its morphisms. An action of a (weak) 2-group $\scG$ on another 2-group $\scH$ is then readily defined as a homomorphism of 2-groups $\Phi\colon\scG\rightarrow \sfAut(\scH)$~\cite{Breen:1992:465-514,Carrasco:1996:4059-4112}. 
    
    Here, we will use the reformulation of~\cite[Prop.~3.2]{Garzn:2001aa} for unital such actions. That is, a unital action of a 2-group $\scG$ on a 2-group $\scH$ is given by a bifunctor
    \begin{subequations}
        \begin{equation}
            \acton\colon\scG\times \scH\rightarrow \scH
        \end{equation}
        and natural isomorphisms
        \begin{equation}
            \begin{aligned}
                \Upsilon_\scG\colon (g_1\otimes g_2)\acton h \xrightarrow{~\cong~} g_1\acton(g_2\acton h)~,
                \\
                \Upsilon_\scH\colon g\acton(h_1\otimes h_2)\xrightarrow{~\cong~} (g\acton h_1)\otimes (g\acton h_2)
            \end{aligned}
        \end{equation}
    \end{subequations}    
    for all objects $g,g_{1,2}\in \scG$ and $h,h_{1,2}\in \scH$. These natural isomorphisms have to satisfy the coherence conditions listed in~\cite[Prop.~3.2]{Garzn:2001aa}. We write $\scG\curvearrowright \scH$ for such an action. We also note that the proof of this proposition gives a helpful definition of the bifunctor $\acton$ in terms of the homomorphism $\scG\rightarrow \sfAut(\scH)$.
    
    \paragraph{Semidirect products.} We further take~\cite[Def.~3.4]{Garzn:2001aa} as our definition of a semidirect product of 2-groups. Given two weak 2-groups $\scG$ and $\scH$ with a unital action $\scG\curvearrowright\scH$, we define the semidirect product $\scG\ltimes \scH$ as the 2-group with underlying Lie groupoid $\scG\times \scH$ and monoidal product
    \begin{subequations}
        \begin{equation}
            (G_1,H_1)\otimes (G_2,H_2) \coloneqq (G_1\otimes G_2,H_1\otimes (G_1\acton H_2))
        \end{equation}
        for all morphisms $G_{1,2}$ in $\scG$ and $H_{1,2}$ in $\scH$. The unit object is 
        \begin{equation}
            \unit_{\scG\ltimes \scH}=(\unit_\scG,\unit_\scH)~,
        \end{equation}
        and the associator is given by
        \begin{equation}\label{eq:associator_semidirect_product}
            \begin{aligned}
                &\sfa(g_1,h_1;g_2,h_2;g_3,h_3)\coloneqq
                \\
                &\hspace{1cm}\Big(\sfa(g_1,g_2,g_3),\big(\sfid_{h_1}\otimes \Upsilon_\scH^{-1}(g_1,h_2,g_2\acton h_3)\big)\circ \big(\sfid_{h_1}\otimes (\sfid_{g_1\acton h_2}\otimes \Upsilon_\scG(g_1,g_2,h_3))\big)\\
                &\hspace{8cm}\circ\sfa(h_1,g_1\acton h_2,(g_1\otimes g_2)\acton h_3)\Big)
            \end{aligned}
        \end{equation}
    \end{subequations}    
    for all objects $g_{1,2,3}\in \scG$ and $h_{1,2,3}\in \scH$, where the inverse of $\Upsilon_\scH$ is with respect to vertical composition. We note that the definitions of group action and semidirect product we use here subsume those used in~\cite{Nikolaus:2018qop}. 
    
    \section{Higher Lie algebras}\label{app:higher_Lie_algebras}    \label{app:L_infinity_defs}
    In this section, we give definitions for \(L_\infty\)-algebras and explain our conventions. We only need to work over the field of real numbers. The original references on \(L_\infty\)-algebras are~\cite{Zwiebach:1992ie,Lada:1992wc,Lada:1994mn}; we follow the conventions in~\cite{Jurco:2018sby}, which may also be helpful.
    
    \paragraph{$L_\infty$-algebras.} An \uline{$L_\infty$-algebra} $\frL$ is a $\IZ$-graded vector space $\frL=\bigoplus_{i\in\IZ} \frL_i$ together with totally antisymmetric multilinear maps $\mu_k\colon\frL^{\wedge k}\rightarrow \frL$ of degree $|\mu_k|=2-k$ satisfying the homotopy Jacobi identity
    \begin{equation}\label{eq:hom_Jac_rel}
        \sum_{i+j=n}\sum_{\sigma\in \bar S_{i|j}}\chi(\sigma;\ell_1,\ldots,\ell_{n})(-1)^{j}\mu_{j+1}(\mu_i(\ell_{\sigma(1)},\ldots,\ell_{\sigma(i)}),\ell_{\sigma(i+1)},\ldots,\ell_{\sigma(n)})=0~,
    \end{equation}
    where the sum runs over all $(i,j)$-unshuffles and $\chi$ denotes the (graded) Koszul sign of the permutation of the arguments.
    
    The rather involved identities~\eqref{eq:hom_Jac_rel} are in fact simply an alternative way of writing the nilquadraticity of the homological vector field $Q$ of the Chevalley--Eilenberg algebra. To translate between both, let $\tau_A$ be a basis of $\frL$ and $\xi^A$ dual coordinate functions on $E=\frL[1]$. Then
    \begin{equation}
        \begin{aligned}
            Q(\xi^A\otimes \tau_A)&=-\sum_{i\geq 1} \tfrac{1}{i!}\mu_i\big(\xi^{A_1}\otimes \tau_{A_1},\dotsc,\xi^{A_i}\otimes \tau_{A_i}\big)\\
            &=-\sum_{i\geq 1} \tfrac{1}{i!}\zeta(A_1,\dots,A_i)\xi^{A_1}\dotsm \xi^{A_i}\otimes\mu_i\big(\tau_{A_1},\dotsc,\tau_{A_i}\big)~,
        \end{aligned}
    \end{equation}
    where the Koszul sign $\zeta(A_1,\dots,A_i)=\pm 1$ arises from permuting odd elements $\xi^{A_j}$ past odd elements $\tau_{A_k}$ or taking them out of odd higher products $\mu_k$. Expanding $Q^2=0$ then reproduces the homotopy Jacobi identities~\eqref{eq:hom_Jac_rel}.
    
    The relation between (differential graded) Lie algebras and (differential graded) commutative associative algebras is the prototype of a long and fruitful story known as \emph{Koszul duality}. Roughly, two classes \(A\), \(B\) of algebraic structures are Koszul dual if a differential graded \(A\)-algebra can be encoded by a differential semi-free \(B\)-algebra whose differential is at most quadratic; surprisingly, this turns out to be a duality, that is, \(B\)-algebras can be so described by semi-free \(A\)-algebras with quadratic differentials. Among others, Koszul duality provides a natural construction of homotopy versions of algebras: homotopy \(A\)-algebras are then given by semi-free \(B\)-algebras with arbitrary nilquadratic differentials, and vice versa.
    For background on Koszul duality, see~\cite{Ginzburg:0709.1228,0821843621,Loday:2012aa} as well as~\cite{Vallette:1202.3245}. 
    
    A \emph{strict \(L_\infty\)-algebra}, such as the loop model of the string Lie 2-algebra, is one in which \(\mu_i = 0\) for \(i\ge3\). That is, it is simply a differential graded Lie algebra, and the formidable homotopy Jacobi identities~\eqref{eq:hom_Jac_rel} simply reduce to the following:
    \begin{itemize}
        \setlength\itemsep{0em}
        \item[$\triangleright$] the differential \(\mu_1\) is nilquadratic;
        \item[$\triangleright$] the differential \(\mu_1\) acts as a graded derivation with respect to the graded bracket \(\mu_2\);
        \item[$\triangleright$] the graded bracket \(\mu_2\) satisfies the Jacobi identity.
    \end{itemize}

    \paragraph{Semistrict Lie~2-algebras.} We will be particularly interested in the case of $L_\infty$-algebras concentrated in degrees $-1$ and $0$ that form models for semistrict Lie 2-algebras. They consist of two vector spaces $\frL=\frL_{-1}\oplus \frL_0$ together with maps
    \begin{equation}
        \begin{gathered}
            \mu_1\colon\frL_{-1}\rightarrow \frL_0~,
            \\
            \mu_2\colon\frL_{0}\wedge \frL_0\rightarrow \frL_0~,~~~            
            \mu_2\colon\frL_{-1}\wedge \frL_0\rightarrow \frL_{-1}~,~~~            
            \mu_2\colon\frL_{0}\wedge \frL_{-1}\rightarrow \frL_{-1}~,
            \\
            \mu_3\colon\frL_0\wedge \frL_0\wedge \frL_0\rightarrow \frL_{-1}
        \end{gathered}
    \end{equation}
    satisfying~\eqref{eq:hom_Jac_rel}.
    
    A morphism $\phi\colon\frL\rightarrow \tilde \frL$ is given by linear maps
    \begin{equation}
        \phi_0\colon\frL_0\rightarrow \frL_0~,~~~\phi_1\colon\frL_{-1}\rightarrow \frL_{-1}~,~~~\phi_2\colon\frL_{0}\wedge \frL_{0}\rightarrow \frL_{-1}
    \end{equation}
    such that
    \begin{equation}\label{eq:Lie_2_algebra_morph}
        \begin{aligned}
            0 &= \phi_1(\mu_1(v_1)) - \tilde\mu_1(\phi_1(v_1))~,\\
            0 &= \phi_0(\mu_2(w_1,w_2)) - \tilde\mu_1(\phi_2(w_1,w_2))-\tilde\mu_2(\phi_0(w_1),\phi_0(w_2))~,\\
            0 &= \phi_1(\mu_2(w_1,v_1)) +\phi_2(\mu_1(v_1),w_1) - \tilde\mu_2(\phi_1(w_1),\phi_1(v_1))~,\\
            0 &= \phi_1(\mu_3(w_1,w_2,w_3)) -\phi_2(\mu_2(w_1,w_2),w_3) + \phi_2(\mu_2(w_1,w_3),w_2)\\
            &\phantom{{}={}} - \phi_2(\mu_2(w_2,w_3),w_1) - \tilde\mu_3(\phi_1(w_1),\phi_1(w_2),\phi_1(w_3)) \\
            &\phantom{{}={}} + \tilde\mu_2(\phi_1(w_1),\phi_2(w_2,w_3))- \tilde\mu_2(\phi_1(w_2),\phi_2(w_1,w_3))\\
            &\phantom{{}={}}+\tilde\mu_2(\phi_1(w_3),\phi_2(w_1,w_2))
        \end{aligned}
    \end{equation}
    for all $v_1\in \frL_{-1}$ and $w_{1,2,3}\in \frL_0$.
    
    \paragraph{Crossed modules of Lie algebras.} Applying the tangent functor to a crossed module of Lie groups $\caG=(\sfH\xrightarrow{~\sft~}\sfG,\acton)$, we obtain a crossed module of Lie algebras $\frG=(\frh\xrightarrow{~\sft~}\frg,\acton)$, where $\frg$ and $\frh$ are the Lie algebras of $\sfG$ and $\sfH$, respectively. Such a crossed module is equivalent to a strict 2-term $L_\infty$-algebra $\frL=\frL_{-1}\oplus \frL_0$ under the relation
    \begin{equation}
        \begin{gathered}
            \frg= \frL_0~,~~~\frh= \frL_{-1}~,
            \\
            [w_1,w_2]=\mu_2(w_1,w_2)~,~~~w\acton v=\mu_2(w,v)~,~~~[v_1,v_2]=\mu_2(\mu_1(v_1),v_2)~,
        \end{gathered}
    \end{equation}    
    cf.~also~\cite{Baez:2003aa}. Applying the tangent functor to a morphism of Lie 2-groups yields a morphism of strict 2-term $L_\infty$-algebras where $\phi_0$, $\phi_1$ and $\phi_2$ are the linearizations or differentials of $\Phi_0$, $\Phi_1$, and $\Phi_2$. The required properties of $\phi$ follow from those of $\Phi$.
    
    \section{Principal 2-bundles with adjusted connection}\label{app:higher_principal_bundles}
    
    \paragraph{Čech groupoid.} Consider a surjective submersion $\sigma\colon Y\rightarrow M$, which defines the \uline{Čech groupoid} 
    \begin{subequations}
        \begin{equation}\label{eq:def_Cech_groupoid}
            \check\scC(Y\rightarrow M)\ \coloneqq\ \left(\begin{tikzcd}
                Y^{[2]}\arrow[r,shift left] 
                \arrow[r,shift right] & Y
            \end{tikzcd}\right)~,~~~
            \begin{tikzcd}[column sep=2.0cm,row sep=large]
                y_1 & y_2\arrow[l,bend left,swap,out=-20,in=200]{}{(y_1,y_2)}~,
            \end{tikzcd}
        \end{equation}
        where $Y^{[2]}$ is the fibered product
        \begin{equation}
            Y^{[2]}=\{(y_1,y_2)\in Y\times Y~|~\sigma(y_1)=\sigma(y_2)\}~.
        \end{equation}
    \end{subequations}    
    This groupoid trivially extends to a higher Lie $n$-groupoid with trivial $k$-morphisms for $k\geq 2$. For most purposes, one can restrict $Y$ to be an ordinary cover given in terms of open subsets of $\IR^n$. In certain cases, it is more convenient to replace the Čech groupoid by a more general, higher groupoid giving rise to hypercovers, cf.~e.g.~\cite{Aldrovandi:0808.3627}.
    
    \paragraph{Cocycle description.} The cocycles of a higher principal bundle with higher structure group $\scG$ subordinate to the submersion $\sigma$ are then given by a higher functor from $\check\scC(Y\rightarrow M)$ to the one-object groupoid $\sfB\scG$ whose (higher) category of morphisms is $\scG$. These cocycles can be differentially refined to allow for a connection~\cite{Jurco:2014mva}.
    
    The introduction of a general connection\footnote{We simply use the term connection to refer to what is sometimes in the abelian gerbe literature called a connective structure and a curving.} requires a so-called \uline{adjustment}, cf.~\cite{Sati:2009ic,Saemann:2019dsl,Kim:2019owc} and in particular \cite{Rist:2022hci} for details. For $\scG$ a 2-group given by a crossed module of Lie groups $\caG\coloneqq (\sfH\xrightarrow{~\sft~}\sfG,\acton)$ with corresponding Lie 2-algebra given by the crossed module of Lie algebras $\frG\coloneqq (\frh\xrightarrow{~\sft~}\frg,\acton)$, this amounts to the following data~\cite{Rist:2022hci}: 
    \begin{subequations}\label{eq:adjusted_cocycles}
        \begin{equation}
            \begin{aligned}
                h\ &\in\  C^\infty(Y^{[3]},\sfH)~,
                \\
                (g,\Lambda)\ &\in\  C^\infty(Y^{[2]},\sfG)\oplus\Omega^1(Y^{[2]},\frh)~,
                \\
                (A,B)\ &\in\ \Omega^1(Y,\frg)\oplus\Omega^2(Y,\frh)
            \end{aligned}
        \end{equation}
        such that\footnote{We note that our conventions are related to those in~\cite{Saemann:2012uq} by the map $B\mapsto -B$ and $H\mapsto -H$.}
        \begin{equation}
            \begin{aligned}
                h_{ikl}h_{ijk}\ &=\ h_{ijl}(g_{ij}\acton h_{jkl})~,
                \\
                g_{ik}\ &=\ \sft(h_{ijk})g_{ij}g_{jk}~,
                \\
                \Lambda_{ik}\ &=\ \Lambda_{jk}+g_{jk}^{-1}\acton\Lambda_{ij}-g_{ik}^{-1}\acton(h_{ijk}\nabla_i h_{ijk}^{-1})~,
                \\
                A_j\ &=\ g^{-1}_{ij}A_ig_{ij}+g^{-1}_{ij}\rmd g_{ij}-\sft(\Lambda_{ij})~,
                \\
                B_j\ &=\ g^{-1}_{ij}\acton B_i+\rmd\Lambda_{ij}+A_j\acton \Lambda_{ij}+\tfrac12[\Lambda_{ij},\Lambda_{ij}]-\kappa(g_{ij},F_i)
            \end{aligned}
        \end{equation}
        for all appropriate $(i,j,\ldots)\in Y^{[n]}$, where $\kappa$ is the contribution of the adjustment. The curvature of this principal 2-bundle is the sum of a 2-form $F$ and a 3-form $H$ and given by
        \begin{equation}
            \begin{aligned}
                F\ &\coloneqq\ \rmd A+\tfrac12[A,A]+\sft(B)\ \in\ \Omega^2(Y,\sfG)~,
                \\
                H\ &\coloneqq\ \rmd B+A\acton B-\kappa(A,F)\ \in\ \Omega^3(Y,\sfH)~.
            \end{aligned}
        \end{equation}
        The adjustment function $\kappa$ is induced by a map
        \begin{equation}
            \kappa\colon\sfG\times \frg\rightarrow \frh
        \end{equation}
        such that generically the equation
        \begin{equation}\label{eq:gluing_condition}
            \begin{aligned}
                (g_2^{-1}g_1^{-1})\acton (h^{-1}(X\acton h))
                &+g_2^{-1}\acton\kappa(g_1,X)
                \\
                &+\kappa(g_2,g_1^{-1}X g_1-\sft(\kappa(g_1,X)))-\kappa(\sft(h)g_1g_2,X)=0
            \end{aligned}
        \end{equation}
        holds for all $g_{1,2}\in \sfG$, $h\in \sfH$, and $X\in \frg$. This condition implies directly that gauge transformations of the $B$-field glue together consistently.
        
        The choice $\kappa=0$ leads to the usual non-abelian gerbes with connection defined in~\cite{Breen:math0106083,Aschieri:2003mw}. When gluing together gauge transformations, $X$ in~\eqref{eq:gluing_condition} is replaced by $F$; therefore, the choice $\kappa=0$ generically requires the fake curvature condition $F=0$.
        
        Examples of adjusted connections are found in~\cite{Rist:2022hci} and (in infinitesimal form) in~\cite{Saemann:2019dsl}.
    \end{subequations}
    
    \paragraph{Bundle isomorphisms.} Coboundaries are encoded in natural isomorphisms between two functors given in terms of the above cocycle data. They are encoded in maps
    \begin{subequations}\label{eq:adjusted_coboundaries}
        \begin{equation}
            \begin{gathered}
                b\ \in\  C^\infty(Y^{[2]},\sfH)~,
                \\
                (a,\lambda)\ \in\  C^\infty(Y,\sfG)\oplus\Omega^1(Y,\frh)~,
            \end{gathered}
        \end{equation}
        and two cocycles $(h,g,\Lambda,A,B)$ and $(\tilde h,\tilde g,\tilde \Lambda,\tilde A,\tilde B)$ are equivalent if
        \begin{equation}
            \begin{aligned}
                \tilde h_{ijk}\ &=\ a_i^{-1}\acton(b_{ik}h_{ijk}(g_{ij}\acton b_{jk}^{-1})b_{ij}^{-1})~,
                \\
                \tilde g_{ij}\ &=\ a_i^{-1}\sft(b_{ij})g_{ij}a_j~,
                \\
                \tilde \Lambda_{ij}\ &=\ a^{-1}_j\acton\Lambda_{ij}+\lambda_j-\tilde{g}^{-1}_{ij}\acton\lambda_i+(a_j^{-1}g_{ij}^{-1})\acton(b_{ij}^{-1}\nabla_ib_{ij})~,                    
                \\
                \tilde{A}_i\ &=\ a_i^{-1}A_ia_i+a_i^{-1}\rmd a_i-\sft(\lambda_i)~,
                \\
                \tilde{B}_i\ &=\ a_i^{-1}\acton B_i+\rmd\lambda_i+\tilde{A}_i\acton\lambda_i+\tfrac12[\lambda_i,\lambda_i]-\kappa(a_i,F_i)~.
            \end{aligned}
        \end{equation}
    \end{subequations}
    
    \paragraph{Postcomposition with 2-group morphisms.} Consider two crossed modules $\caG=(\sfH\xrightarrow{~\sft~}\sfG,\acton)$ and $\tilde\caG=(\tilde\sfH\xrightarrow{~\tilde\sft~}\tilde\sfG,\tilde\acton)$ as well as a principal $\caG$-bundle $\scP$ with cocycles $(g,h)$. Then a 2-group morphism $\Phi\colon\underline{\caG}\rightarrow \underline{\tilde \caG}$ yields a principal $\tilde \caG$-bundle with cocycles given by 
    \begin{subequations}\label{eq:cocycles_under_morphisms}
        \begin{equation}
            \tilde g_{ij}=\Phi_1(g_{ij})\eand \tilde h_{ijk}=\Phi^{\tilde \sfH}_1(h_{ijk})\Phi^{\tilde \sfH}_2(g_{ij},g_{jk})~,
        \end{equation}
        where $\Phi^{\tilde \sfH}_1(h_{ijk})$ is the component of $\Phi_1$ in $\tilde \sfH$. This follows abstractly from the interpretation of $\scP$ and $\Phi$ as weak 2-functors and the fact that 2-functors compose nicely, cf.~\eqref{eq:composition_2_functors}.
        
        For the connection part, we can generalise the discussion in~\cite{Rist:2022hci} to apply the $L_\infty$-algebra morphism induced by the morphism $\Phi$ to the local connection forms:
        \begin{equation}
            \begin{aligned}
                A_i\mapsto \tilde A_i=\phi_0(A)~,~~~B_i\mapsto \tilde B_i=\phi_1(B)+\tfrac12\phi_2(A,A)~.    
            \end{aligned}
        \end{equation}
    \end{subequations}    This construction is familiar from homotopy Maurer--Cartan theory, cf.~e.g.~\cite{Jurco:2018sby} for more details.
    
    Interestingly, as observed in~\cite{Rist:2022hci}, this fully defines $\sft(\Lambda_{ij})$ via the cocycle condition~\eqref{eq:adjusted_cocycles} relating $A_i$ to $A_j$. It remains to lift this to a full map $\Lambda\in\Omega^1(Y^{[2]},\sfH)$, which is best done on a case-by-case basis.
    
    Note that, in the case of morphisms $\Phi\colon\caG\rightarrow \tilde \caG$ that are given by butterflies, the induced morphism of principal 2-bundles is more complicated and may require a refinement of the cover, cf.~\cite[\S~4.2]{Aldrovandi:0909.3350}. One such example is discussed in detail in~\cite{Rist:2022hci}. 
    
    \section{Quasi-groupoids and augmentation}\label{app:quasi-groupoids}
    
    In the following, we briefly summarise some basic material on quasi-groupoids; for a detailed review in our conventions, see~\cite{Jurco:2016qwv} and references therein.
    
    \paragraph{Simplicial manifolds.} Recall that a \uline{simplicial object} in a category $\scC$ is a $\scC$-valued presheaf $X\colon\triangle^{\operatorname{op}}\to\scC$ on the simplex category $\triangle$, which is (the skeleton of) the 1-category of finite non-empty well-ordered sets and order-preserving functions. Every finite well-ordered set is isomorphic to the ordinal $\underline n=\{0,1,\dotsc,n-1\}$, and the image of $\underline n$ is the set of $(n-1)$-simplices: $X_n\coloneqq X(\underline{n+1})$. The images under $X$ of injective order-preserving maps $\underline n\to\underline{n+1}$ give the \uline{face maps} $\sff^n_0,\dotsc,\sff^n_n\colon X_n\to X_{n-1}$, and the images under $X$ of surjective order-preserving maps $\underline{n+2}\to\underline{n+1}$ give the \uline{degeneracy maps} $\sfd^n_0,\dotsc,\sfd^n_n\colon X_n\to X_{n+1}$.
    
    A \uline{simplicial set} is then simply a simplicial object in $\sfSet$, and a \uline{simplicial manifold} is a simplicial object in a suitable category of smooth manifolds. Notice that every simplicial set can be trivially regarded as a discrete simplicial manifold. The \uline{standard simplicial $n$-simplex} $\Delta^n$ is the simplicial set $\triangle^\text{op}\rightarrow \sfSet$ represented by $\underline{n+1}$.
    
    \paragraph{Lie quasi-groupoids.} An \uline{$(n,i)$-horn} in $\Delta^n$ is the union of all faces of $\Delta^n$ except for the $i$th one. The $(n,i)$-horns of a simplicial manifold $\scM$ are the images of the $(n,i)$-horns of $\Delta^n$ in $\scM$. If every horn in $\scM$ can be filled to a simplex, and the face maps from the simplices in $\scM$ to the horns in $\scM$ are surjective submersions, we say that $\scM$ is a Kan simplicial manifold or a \uline{Lie quasi-groupoid}. If all the fillers for $(m,i)$-horns are unique for every $m>n$ and $0<i<m$, then we say that $\scM$ is a \uline{Lie $n$-quasi-groupoid}.
    
    \paragraph{From (higher) groupoids to quasi-groupoids.} The nerve of any groupoid $\scG$, i.e.~the simplicial manifold whose 0-simplices are the objects of $\scG$, whose 1-simplices are the morphisms of $\scG$, and whose 2-simplices are the pairs of composable morphisms of $\scG$, etc., is a 1-quasi-groupoid. In the case of 2-groupoids, the definition of a nerve is slightly more involved but similarly feasible. A common choice is the \uline{Duskin nerve}~\cite{Duskin02simplicialmatrices}, which also forms a 2-quasi-groupoid. A problem in this treatment of 2-groupoids is that the explicit expression of horizontal composition is replaced by the existence of a set of horn fillers. This, together with a vast redundancy of information in quasi-groupoids, is the reason for us not working with quasi-groupoids from the outset.
    
    \paragraph{Augmentation.} In the definition of simplicial objects, we restricted ourselves to non-empty well-ordered sets, but we can naturally extend this to all finite well-ordered sets, including $\underline{-1}=\varnothing$; we denote the resulting category by $\triangle_+$. An \uline{augmented simplicial object} in $\scC$ is then a $\scC$-valued presheaf on $\triangle_+$. Concretely, an augmented simplicial object $X^+_\bullet$ is nothing but a triple $(X_\bullet,X_{-1},\sff^0_0)$ where $X_\bullet$ is an (unaugmented) simplicial object, $X_{-1}$ is an object in $\scC$, and $\sff^0_0\colon X_0\to X_{-1}$ is a morphism in $\scC$ such that
    \begin{equation}
        \sff^0_0\circ \sff^1_0=\sff^0_0\circ \sff^1_1~.
    \end{equation}
    We define an \uline{augmented quasi-groupoid} in a corresponding way. A particularly useful example is that of the augmented Čech groupoid~\eqref{eq:augmented_Cech_nerve}.

\tikzset{Rightarrow/.style={double equal sign distance,>={Implies},->},triple/.style={-,preaction={draw,Rightarrow}},quad/.style={preaction={draw,Rightarrow,shorten >=0pt},shorten >=1pt,-,double,double distance=0.2pt}}    
    
    \section{Categorified groups and hypercrossed modules}\label{app:hypercrossed_modules}
    
    Below, we describe Lie 3-groups in terms of 2-crossed modules of Lie groups, which are special cases of hypercrossed modules.
    There are several, obvious categorifications of crossed modules of Lie groups. Here, we focus on 2-crossed modules~\cite{Conduche:1984:155,Conduche:2003}, which encode semistrict Lie 3-groups called {\em Gray groups}, i.e.~Gray groupoids with a single object; see~\cite{Kamps:2002aa}.
    
    A {\em 2-crossed module of Lie groups} is a triple of Lie groups, arranged in the normal complex\footnote{that is, a chain complex of groups (i.e.~\(\sft\circ\sft=0\)) such that the image of every \(\sft\) is a normal subgroup}
    \begin{equation}
        \sL\ \xrightarrow{~\sft~}\ \sH\ \xrightarrow{~\sft~}\ \sG~,
    \end{equation}
    and endowed with smooth $\sG$-actions on $\sH$ and $\sL$ by automorphisms such that the maps $\sft$ are $\sG$-equivariant:
    \begin{equation}
        \sft(g\acton \ell)=g\acton \sft(\ell)\eand \sft(g\acton h)=g\sft(h)g^{-1}
    \end{equation}
    for all $g\in\sG$, $h\in\sH$, and $\ell\in\sL$. The Peiffer identity of crossed modules of Lie groups is violated, but this violation is controlled by the {\em Peiffer lifting}, which is a $\sG$-equivariant smooth map
    \begin{equation}
        \begin{gathered}
            \{-,-\}\colon\sH\times \sH\to \sL~,\\
        \end{gathered}
    \end{equation}
    satisfying the following relations:
    \begin{subequations}
        \begin{align}
            \sft(\{h_1,h_2\})&=h_1 h_2 h_1^{-1}(\sft(h_1)\acton h_2^{-1})~, \label{eq:global_peiffer_lifting_identity}\\
            \{\sft(\ell_1),\sft(\ell_2)\}&=\ell_1\ell_2\ell_1^{-1}\ell_2^{-1}~, \\
            \{h_1 h_2,h_3\}&=\{h_1,h_2h_3h_2^{-1}\}(\sft(h_1)\acton\{h_2,h_3\})~,\\
            \{h_1,h_2h_3\}&=\{h_1,h_2\}\{h_1,h_3\}\{\sft(\{h_1,h_3\})^{-1},\sft(h_1)\acton h_2\}~,\\
            \ell_1 \left(\sft(h_1)\acton \ell^{-1}_1\right)&=\{\sft(\ell_1),h_1\}\{h_1,\sft(\ell_1)\}
        \end{align}
    \end{subequations}
    for all $h_i\in \sH$ and $\ell_i\in \sL$. 
    
    Given a 2-crossed module of Lie groups \(\sL\to\sH\to\sG\), we can construct a monoidal 2-category 
    \begin{subequations}\label{eq:mon_cat_from_3_group}
        \begin{equation}
            \CCC(\sL\to\sH\to\sG) \coloneqq (
            \sL \rtimes \sH \rtimes \sG \rightrightarrows  \sH \rtimes \sG \rightrightarrows  \sG  )~,
        \end{equation}
        whose globular structure is
        \begin{equation}
            \begin{tikzcd}[column sep=2.5cm,row sep=large]
                \sft(h)g & \ar[l, anchor=center,bend left=45, "{(h,g)}", ""{name=U,inner sep=1pt,above}] \ar[l,anchor=center, bend right=45, "{(\sft(\ell)h,g)}", swap, ""{name=D,inner sep=1pt,below}] 
                g \arrow[Rightarrow,from=U, to=D, "{(\ell,h,g)}",swap]
            \end{tikzcd}
        \end{equation}
    \end{subequations}
    see e.g.~\cite[Section 1.4]{Kamps:2002aa}. Shifting the degrees of all morphisms by one, we define the 3-groupoid $\sB(\CCC(\sL\to\sH\to\sG))$, which is a Gray groupoid. 
    
    Conversely, given a monoidal 2-category $\CCG$ encoding a 3-group, we denote the corresponding 2-crossed module of Lie groups by $\CCG_{\rm \cm}$.
    
    The infinitesimal counterpart of a 2-crossed module of Lie groups is a {\em 2-crossed module of Lie algebras}, which consists of a triple of Lie algebras arranged in the complex
    \begin{equation}
        \frl\ \xrightarrow{~\sft~}\ \frh\ \xrightarrow{~\sft~}\ \frg~.
    \end{equation}
    Additionally, we have $\frg$-actions $\acton$ onto $\frh$ and $\frl$ by derivations. The maps $\sft$ are equivariant with respect to these actions,
    \begin{equation}
        \sft(a\acton c)=a\acton\sft(c)\eand \sft(a\acton b)=[a,\sft(b)]
    \end{equation}
    for all $a\in \frg$, $b\in \frh$, and $c\in \frl$. The violation of the Peiffer identity is controlled by a differential version of the Peiffer lifting, which is a $\frg$-equivariant bilinear map
    \begin{equation}
        \{-,-\}\colon \frh\times \frh\to \frl~,
    \end{equation}
    which also satisfies the following relations:
    \begin{subequations}\label{eq:ax:2-crossed_Lie}
        \begin{align}
            \sft(\{b_1,b_2\})&=[b_1,b_2]-\sft(b_1)\acton b_2~, \label{eq:2cm_axiom2}\\
            \{\sft(c_1),\sft(c_2)\}&=[c_1,c_2]~, \\
            \{b_1,[b_2,b_3]\}&=\{\sft(\{b_1,b_2\}),b_3\}-\{\sft(\{b_1,b_3\}),b_2\}~,\\
            \{[b_1,b_2],b_3\}&=\sft(b_1)\acton\{b_2,b_3\}+\{b_1,[b_2,b_3]\}-\sft(b_2)\acton\{b_1,b_3\}-\{b_2,[b_1,b_3]\}~,\\
            -\sft(b_1)\acton c_1&=\{\sft(c_1),b_1\}+\{b_1,\sft(c_1)\} \label{eq:2cm_axiom6}
        \end{align}
    \end{subequations}
    for all $b_1,b_2,b_3\in \frh$ and $c_1,c_2\in \frl$.
    
    Given a 2-crossed module of Lie algebras $\frl\xrightarrow{~\sft~}\frh \xrightarrow{~\sft~}\frg$, the subcomplexes $\frl\xrightarrow{~\sft ~}\frh$ with action
    \begin{equation}
        b\acton c\coloneqq -\{\sft(c),b\}~,~~~b\in \frh~,~~~c\in \frl
    \end{equation}
    as well as $\sft(\frl)\setminus\frh \xrightarrow{~\sft~}\frg$ with the unmodified action of $\frg$ on $\sft(\frl)\setminus\frh$ also form crossed modules of Lie algebras. 
    
    We explain the relationship between Lie 1-, 2-, and 3-algebras and certain hypercrossed modules of Lie algebras in \cref{app:Lie3_2_crossed}.

    \section{Path and loop groups}\label{app:path_groups}
    
    The construction of the strict 2-group model of the string group~\cite{Baez:2005sn} requires a particular technical choice of path groups and loop groups. In short, path groups are smooth and based; loop groups are based, and consist of loops that are smooth everywhere except at the base point, where they are merely continuous.
    
    Given a finite-dimensional Lie group \(\sG\), the \emph{path group} \(P_0\sG\) is the Fréchet--Lie group of smooth paths \(\gamma\colon [0,1]\to \sG\) such that \(\gamma(0) = \unit_\sG\). The group operation is pointwise multiplication. The \emph{loop group} \(L_0\sG\) is the subgroup of those paths \(\gamma\) such that \(\gamma(0)=\gamma(1)\). We do \emph{not} require any further smoothness at the base point. Thus there is a non-split short exact sequence
    \begin{equation}
        * \to L_0\sG \to P_0\sG \overset\partial\to \sG \to *~,
    \end{equation}
    where \(\partial\colon P_0\sG \to \sG\) is the endpoint evaluation map.
    Given the Lie algebra \(\frg\) of \(\sG\), the corresponding Lie algebras are \(P_0\frg\) and \(L_0\frg\), with obvious definitions and the corresponding non-split short exact sequence
    \begin{equation}
        * \to L_0\frg \to P_0\frg \overset\partial\to \frg \to *~.
    \end{equation}
    
    The Fréchet–Lie group \(\hat L_0\sG\) is the usual Kac--Moody central extension of \(L_0\sG\). Its Lie algebra is
    \begin{equation}
        \hat L_0\frg = L_0\frg \oplus \mathbb R~,
    \end{equation}
    where \(\mathbb R\) is the 1-dimensional abelian Lie algebra and \(\oplus\) is a direct sum of Lie algebras.
    While at the level of Lie algebras \(\hat L_0\frg\) is just a trivial direct sum, at the level of Lie groups \(\hat L_0\sG\) is a nontrivial principal $\sU(1)$-bundle over \(L_0\sG\). We thus have the exact sequences
    \begin{equation}
        * \to \operatorname \sU(1) \to \hat L_0\sG \to P_0\sG \overset\partial\to \sG \to *
    \end{equation}
    and
    \begin{equation}
        * \to \mathbb R\to \hat L_0\frg \to P_0\frg \overset\partial\to \frg \to *~.
    \end{equation}

    \section{Strict Lie 3-algebras and 2-crossed modules of Lie algebras}\label{app:Lie3_2_crossed}
    
    Semistrict Lie 3-algebras can be described both by 2-crossed modules of Lie algebras as well as 3-term \(L_\infty\)-algebras. For our purposes, the precise relation between these is important. Because we could not find the relevant statements in the literature, we give them below.
    
    We first mention the comparison theorems between \(n\)-term \(L_\infty\)-algebras and \((n-1)\)-crossed modules of Lie algebras for \(n\le 2\):
    \begin{theorem}
        A 1-term \(L_\infty\)-algebra is the same thing as a 0-crossed module of Lie algebras (i.e.~a Lie algebra).
    \end{theorem}\begin{proof}
        Trivial.
    \end{proof}
    
    \begin{theorem}
        A strict 2-term \(L_\infty\)-algebra is the same thing as a (1-)crossed module of Lie algebras.
    \end{theorem}\begin{proof}
        Given a strict Lie 2-algebra
        \begin{equation}
            \frL=\big(\frL_{-1} \xrightarrow{~\mu_1~} \frL_0\big)~,
        \end{equation}
        we can construct the crossed module of Lie algebras 
        \begin{equation}
            \begin{matrix}
                \frh & \overset{\sft}\longrightarrow & \frg \\
                b & \overset{\sft}\longmapsto & \mu_1(b)
            \end{matrix}
        \end{equation}
        with \(\frg=\frL_0\) and \(\frh=\frL_{-1}\) and
        \begin{equation}
            [a_1,a_2]_\frg=\mu_2(a_1,a_2)~,~~~
            a_1\acton b_1=\mu_2(a_1,b_1)~,~~~
            [b_1,b_2]_\frh=\mu_2(\mu_1(b_1),b_2)
        \end{equation}
        for all $a_1,a_2\in\frg$ and $b_1,b_2\in \frh$. The inverse construction is also evident.
    \end{proof}
    The next step up in the categorification process turns out to be a bit more complicated.
    \begin{theorem}\label{thm:2cm_to_L_infty}
        The complex of Lie algebras underlying a 2-crossed module of Lie algebras comes with a strict 3-term \(L_\infty\)-algebra structure.
    \end{theorem}
    \begin{proof}
        Given a 2-crossed module of Lie algebras
        \begin{equation}
            (\frl \xrightarrow{~\sft~} \frh \xrightarrow{~\sft~}\frg,\acton,\{-,-\})~,
        \end{equation}
        there is a strict 3-term \(L_\infty\)-algebra
        \begin{equation}
            \begin{matrix}
                \frL\quad=\quad(&\frL_{-2}&\overset{\mu_1}\longrightarrow&\frL_{-1}&\overset{\mu_1}\longrightarrow&\frL_0&)~,\\
                &c&\overset{\mu_1}\longmapsto&\sft(c)\\
                &&&b&\overset{\mu_1}\longmapsto&\sft(b)
            \end{matrix}
        \end{equation}
        where \(\frL_{-2}=\frl\) and \(\frL_{-1}=\frh\) and \(\frL_0=\frg\),
        with non-trivial higher products
        \begin{subequations}
            \begin{align}
                \mu_2(a_1,a_2) &\coloneqq [a_1,a_2]_\frg~, \\
                \mu_2(a_1,b_1)&\coloneqq a_1\acton b_1~, \\
                \mu_2(a_1,c)&\coloneqq a_1\acton c~,\\
                \mu_2(b_1,b_2)&\coloneqq -\{b_1,b_2\}-\{b_2,b_1\}\label{eq:Peiffeb_sym_para_mu2_relation}
            \end{align}
        \end{subequations}
        for all $a_1,a_2\in \frg$, $b_1,b_2\in \frh$, and $c\in \frl$. One readily verifies that the homotopy Jacobi identity is satisfied for these higher products, as in Table~\ref{table:2cm_Lie3_correspondence}.
        \begin{table}
            \begin{center}
                \begin{tabular}{cc}
                    \toprule
                    2-crossed module &homotopy Jacobi identity \\ \midrule
                    \(\sft\circ\sft = 0\) & \(\mu_1(\mu_1(\frL_{-2}))\) \\
                    \(\mathfrak g\)-equivariance of map \(\sft\colon\mathfrak h\to\mathfrak g\) & \(\mu_2(\frL_0,\mu_1(\frL_{-1}))\) \\
                    \(\mathfrak g\)-equivariance of map \(\sft\colon\mathfrak l\to\mathfrak h\) & \(\mu_2(\frL_0,\mu_1(\frL_{-2}))\) \\
                    symmetric part of~\eqref{eq:2cm_axiom2} & \(\mu_2(\mu_1(\frL_{_-1}),\frL_{-1})\) \\
                    \eqref{eq:2cm_axiom6}&\(\mu_2(\mu_1(\frL_{-1}),\frL_{-2})\) \\
                    Jacobi identity for \(\mathfrak g\) Lie bracket & \(\mu_2(\mu_2(\frL_0,\frL_0),\frL_0)\) \\
                    \(\mathfrak g\)-action on \(\mathfrak h\) & \(\mu_2(\mu_2(\frL_0,\frL_0),\frL_{-1})\) \\
                    \(\mathfrak g\)-action on \(\mathfrak l\) & \(\mu_2(\mu_2(\frL_0,\frL_0),\frL_{-2})\) \\
                    symmetric part of \(\mathfrak g\)-equivariance of Peiffer lifting & \(\mu_2(\mu_2(\frL_{-1},\frL_{-1}),\frL_0)\) \\
                    \bottomrule 
                \end{tabular}
            \end{center}
            \caption{Proof that a 2-crossed module of Lie algebras defines an \(L_\infty\)-algebra}\label{table:2cm_Lie3_correspondence}
        \end{table}
    \end{proof}
    \begin{theorem}\label{thm:L_infty_to_2cm}
        A strict Lie 3-algebra \(\frL = \frL_{-2} \oplus \frL_{-1} \oplus \frL_0\) equipped with a choice of graded-symmetric (i.e.~antisymmetric) bilinear map
        \begin{subequations}\label{eq:add_map}
            \begin{equation}
                \llbracket-,-\rrbracket\colon \frL_{-1} \times \frL_{-1} \to \frL_{-2}
            \end{equation}
            which satisfies the identities
            \begin{equation}\label{eq:req_identities}
                \begin{aligned}
                    \llbracket  b_2,\mu_2( b_3,\mu_1( b_1))\rrbracket-\llbracket  b_3,\mu_2( b_2,\mu_1( b_1))\rrbracket+\mu_2(\mu_1( b_1),\llbracket  b_2, b_3\rrbracket)&=0~,\\
                    \llbracket  b_1,\mu_1(\llbracket  b_2, b_3\rrbracket)\rrbracket-
                    \llbracket  b_2,\mu_1(\llbracket  b_1, b_3\rrbracket)\rrbracket+
                    \llbracket  b_3,\mu_1(\llbracket  b_1, b_2\rrbracket)\rrbracket-~~&\\
                    -\tfrac14 \mu_2( b_1,\mu_2( b_2,\mu_1( b_3)))
                    +\tfrac14 \mu_2( b_3,\mu_2( b_2,\mu_1( b_1)))&=0
                \end{aligned}
            \end{equation}
        \end{subequations}
        for all $ b_1,b_2,b_3\in \frL_{-1}$ comes with the structure of a 2-crossed module on its underlying graded vector space, where the Peiffer lifting reads as
        \begin{equation}\label{eq:peiffer_thm}
            \{ b_1, b_2\}=\llbracket b_1, b_2\rrbracket-\tfrac12\mu_2( b_1, b_2)
        \end{equation}
        for all $ b_1, b_2\in \frL_{-1}$.
    \end{theorem}
    \begin{proof}
        Given a 3-term \(L_\infty\)-algebra $\frL=\frL_{-2}\oplus \frL_{-1}\oplus \frL_0$, we construct the complex underlying the 2-crossed module of Lie algebras
        \begin{equation}
            (\frl \xrightarrow{~\sft~} \frh \xrightarrow{~\sft~}\frg,\acton,\{-,-\})
        \end{equation}
        with 
        \begin{equation}
            \frl=\frL_{-2}~,~~~\frh=\frL_{-1}~,~~~\frg=\frL_0~,\eand \sft=\mu_1~.
        \end{equation}
        The Lie bracket on $\frg$ is given by 
        \begin{equation}
            [-,-]_\frg=\mu_2\colon\frg\wedge \frg\to \frg~,
        \end{equation}
        and the actions of $\frg$ on $\frh$ and $\frl$ read as 
        \begin{equation}
            a\acton b\coloneqq \mu_2(a,b)\eand a\acton c\coloneqq \mu_2(a,c)
        \end{equation}
        for $a\in \frg$, $b\in \frh$, $c\in \frl$. The Peiffer lifting~\eqref{eq:peiffer_thm} fixes the Lie brackets on $\frh$ and $\frl$ as
        \begin{subequations}
            \begin{align}
                \label{eq:h_Lie_bracket}
                [b_1,b_2]_{\frh} &\coloneqq \mu_1(\llbracket b_1,b_2\rrbracket) + \tfrac12\big(\mu_2(\mu_1(b_1),b_2)-\mu_2(\mu_1(b_2),b_1)\big)~,\\
                \label{eq:l_Lie_bracket}
                [c_1,c_2]_{\frl} &\coloneqq  \{\mu_1(c_1),\mu_2(c_2)\} = \llbracket\mu_1(c_1),\mu_1(c_2)\rrbracket
            \end{align}
        \end{subequations}
        for all \(b_1,b_2\in\frh\) and $c_1,c_2\in \frl$. Straightforward but lengthy algebraic computations show that these structures satisfy the axioms of a 2-crossed module of Lie algebras~\eqref{eq:ax:2-crossed_Lie} if and only if~\eqref{eq:req_identities} are satisfied.
    \end{proof}
    
    \begin{corollary}\label{cor:trivial_peiffer_lifting}
        Under the correspondence given by theorems~\ref{thm:2cm_to_L_infty} and~\ref{thm:L_infty_to_2cm}, the class of 2-crossed modules of Lie algebras with vanishing Peiffer lifting corresponds precisely to the class of 3-term \(L_\infty\)-algebras with vanishing \(\mu_2\colon \frL_{-1}\wedge\frL_{-1} \to \frL_{-2}\).
    \end{corollary}
    \begin{proof}
        This follows from theorems~\ref{thm:2cm_to_L_infty} and~\ref{thm:L_infty_to_2cm} with the observation that in these cases, the identities~\eqref{eq:req_identities} are trivial.
    \end{proof}
    Thus, we see that this class of 3-term \(L_\infty\)-algebras can be canonically integrated, unlike the general case with non-vanishing \(\mu_2\); this is related to the ambiguity of (and need for) adjustments in the general case.
    
    Altogether, we conclude that 2-crossed modules of Lie algebras readily restrict to strict 3-term $L_\infty$-algebras, but strict 3-term $L_\infty$-algebras can only be extended to 2-crossed modules, if they allow for maps~\eqref{eq:add_map}.

    \section{Inner automorphisms and the Weil algebra}\label{app:inner_weil}
\subsection{Inner automorphisms of Lie groups}\label{ssec:inner_group}
    
    The Weil algebra $\sW(\frg)$ of a Lie algebra $\frg$ encodes a 2-term \(L_\infty\)-algebra with underlying graded vector space $\frg\oplus\frg[1]$, which is isomorphic to the 2-term \(L_\infty\)-algebra of inner derivations, $\ainn(\frg)$; see~\cref{app:inner_weil}. The latter sits in the short exact sequence of graded vector spaces 
    \begin{equation}
        *\longrightarrow 
        \begin{array}{c} *\\[-0.75ex] \downarrow\\[-0.75ex]\frg \end{array}\hooklongrightarrow \ainn(\frg)\longrightarrow 
        \begin{array}{c} \frg[1]\\[-0.3ex] \downarrow\\[-0.75ex]* \end{array}\longrightarrow *~.
    \end{equation}
    For a Lie group \(\sG\) integrating $\frg$, this sequence is the infinitesimal version of the short exact sequence of groupoids,
    \begin{equation}\label{ses:groupoids_Lie}
        *\longrightarrow \begin{array}{c} \sG\\[-0.75ex] \downdownarrows\\[-0.3ex]\sG\end{array} \hooklongrightarrow \sInn(\sG)\longrightarrow \begin{array}{c} \sG\\[-0.75ex] \downdownarrows\\[-0.75ex]*\end{array}\longrightarrow *~.
    \end{equation}
    Here, $\sG\rightrightarrows \sG$ is the Lie group $\sG$, trivially regarded as a groupoid, while $(\sG\rightrightarrows *)=\sB\sG$ is the one-object groupoid with $\sG$ as the group of morphisms. Moreover, $\sInn(\sG)$ is the action groupoid of $\sG$ onto itself by left-multiplication. This is a 2-group, the {\em 2-group of inner automorphisms} of $\sG$. The embedding in the sequence~\eqref{ses:groupoids_Lie} is in fact a morphism of Lie 2-groups, while the second map is merely a groupoid morphism; the groupoid $\sG\rightrightarrows *$ does not admit a 2-group structure unless $\sG$ is abelian.
    
    These structures have important topological interpretations. The geometric realisation $|\sB\sG|$ of the nerve of $\sB\sG$ is the classifying space of $\sG$. Applying the same operations to $\sInn(\sG)$, we recover the universal bundle $|\sE\sG|$ of $\sG$ over $|\sB\sG|$. Also, the action groupoid $\sInn(\sG)=(\sG\rtimes \sG\rightrightarrows \sG)$ is equivalent to the trivial 2-group $(*\rightrightarrows *)$; equivalently, $\ainn(\frg)=(\frg[1]\xrightarrow{~\id~} \frg)$ is quasi-isomorphic to the 0-term \(L_\infty\)-algebra\footnote{by the minimal model theorem; see~\cref{app:Equivalence}}. This corresponds to the universal bundle \(|\sE\sG|\) being contractible.

    \subsection{Inner automorphisms of strict Lie 2-groups}\label{ssec:inner_derivations}
    
    The generalisation to the case of a strict Lie 2-group is discussed in detail in~\cite{Roberts:0708.1741}. The inner automorphisms of a strict Lie 2-group $\CCG$ with corresponding crossed module of Lie groups $\CCG_{\rm \cm}=(\sH\xrightarrow{~\tilde \sft~}\sG,\tildeacton)$ form a Lie 3-group, which is conveniently encoded in the following 2-crossed module\footnote{see again \cref{app:hypercrossed_modules} for the definition} of Lie groups \((\sInn_{\rm \cm}(\CCG),\acton,\{-,-\})\):
    \begin{equation}
        \begin{matrix}
            \sInn_{\rm \cm}(\CCG)&=&(&\sH&\xrightarrow{~\sft~}&\sH\rtimes \sG&\xrightarrow{~\sft~}&\sG&)\\
            &&&h&\xmapsto{~\sft~}&( h^{-1},\tilde\sft(h))\\
            &&& & &(h,g)&\xmapsto{~\sft~}&\tilde \sft(h)g
        \end{matrix}
    \end{equation}
    where the products and actions are evident, in particular
    \begin{equation}
        (h_1,g_1)(h_2,g_2)=\big(h_1(g_1\tildeacton h_2),g_1g_2\big)~,~~~(h_1,g_1)^{-1}=(g_1^{-1}\tildeacton h_1^{-1},g_1^{-1})~,
    \end{equation}
    and the Peiffer lifting is
    \begin{equation}
        \{(h_1,g_1),(h_2,g_2)\}=(g_1g_2g_1^{-1}\tildeacton h_1) h_1^{-1}
    \end{equation}
    for all $g_1,g_2\in \sG$ and $h_1,h_2\in \sH$.    
    
    There is now a higher analogue of sequence~\eqref{ses:groupoids_Lie} involving 2-groupoids. The crossed module of Lie groups $\CCG_{\rm \cm}$ corresponds to a monoidal category $\CCG=(\sH\rtimes \sG\rightrightarrows \sG)$ (see equation~\eqref{eq:mon_cat_from_2_group}), which is trivially regarded as a strict 2-category with only identity 2-morphisms. Moreover, $\sInn_{\rm \cm}(\sH\xrightarrow{\tilde \sft}\sG)$ corresponds to a monoidal 2-category $\sInn(\CCG)$ encoding a 3-group\footnote{more precisely, a \emph{Gray group}; see equation~\eqref{eq:mon_cat_from_3_group}}. We present its globular structure for use in section~\ref{sec:parallel_transport}.
    \begin{subequations}
    \begin{equation}
        \sInn(\CCG)\coloneqq\big(\sH\rtimes((\sH\rtimes \sG)\rtimes \sG)  \rightrightarrows  (\sH\rtimes \sG)\rtimes \sG  \rightrightarrows \sG\big)
    \end{equation}
    \begin{equation}
        \begin{tikzcd}[column sep=4.0cm,row sep=large]
            \tilde \sft(h^2_2)g_2^2g_2^1 & \ar[l, bend left=45, "{\big((h_2^2,g_2^2),g_2^1\big)}", ""{name=U,inner sep=1pt,above}] \ar[l, bend right=45, "{\big((h_2^2(h_2^1)^{-1}\,,\,\tilde \sft(h_2^1)g_2^2),g_2^1\big)}", swap, ""{name=D,inner sep=1pt,below}] g_2^1
            \arrow[Rightarrow,from=U, to=D, "{\big(h_2^1,(h_2^2,g_2^2),g_2^1\big)}",swap]
        \end{tikzcd}
    \end{equation}
    \end{subequations}
    Finally, we have the 2-groupoid\footnote{The component \(\sH\times\sG\) in~\eqref{eq:2-groupoid_BG} is merely a manifold, not a Lie group, since \(\sB\CCG\) is not a 3-group, but merely a 2-groupoid.}
    \begin{equation}\label{eq:2-groupoid_BG}
        \sB\CCG=\big((\sH\times \sG)  \rightrightarrows \sG \rightrightarrows *\big)~.
    \end{equation}
    These three 2-groupoids now fit in the short exact sequence
    \begin{subequations}\label{eq:ses_2-groupoids}
        \begin{equation}
            * \longrightarrow \CCG \xrightarrow{~\Upsilon~} \sInn(\CCG) \xrightarrow{~\Pi~} \sB\CCG \longrightarrow *~,
        \end{equation}
        whose components are as follows:
        \begin{equation}
            \begin{tikzcd}
                \sH\rtimes\sG \rar["\Upsilon_2"] \dar[shift left] \dar[shift right] & \sH\rtimes\big((\sH\rtimes\sG)\rtimes\sG\big) \rar["\Pi_2"] \dar[shift left] \dar[shift right] & \sH\times \sG \dar[shift left] \dar[shift right] \\
                \sH\rtimes\sG \rar["\Upsilon_1"] \dar[shift left] \dar[shift right] & (\sH\rtimes\sG)\rtimes\sG \rar["\Pi_1"] \dar[shift left] \dar[shift right] & \sG \dar[shift left] \dar[shift right] \\
                \sG \rar["\Upsilon_0"] & \sG \rar["\Pi_0"] & * 
            \end{tikzcd}
        \end{equation}
        where the strict 2-functors $\Upsilon$ and $\Pi$ are given by
        \begin{equation}\label{eq:3-group_short_exact_sequence_components}
            \begin{aligned}
                \Upsilon_2\colon&&(h,g) &\mapsto (\unit_\sH,h,g,\unit_\sG)~,~~~&\Pi_2\colon&&(h_1,h_2,g_1,g_2)&\mapsto (h_1,g_2)~,\\
                \Upsilon_1\colon&&(h,g) &\mapsto (h,g,\unit_\sG)~,~~~&\Pi_1\colon&&(h,g_1,g_2)&\mapsto g_2~,\\
                \Upsilon_0\colon&&g &\mapsto g~,~~~&\Pi_0\colon&&g&\mapsto *~.
            \end{aligned}
        \end{equation}
    \end{subequations}
    Again, $\Upsilon$ is also a morphism of strict 3-groups.
    
    At an infinitesimal level, $\CCG$ (and, more evidently, $\CCG_{\rm \cm}$) differentiates to the crossed module of Lie algebras $(\frh\xrightarrow{~\tilde \sft~}\frg)$, where $\frg$ and $\frh$ are the Lie algebras of $\sG$ and $\sH$. Its 2-crossed module of inner derivations has the underlying complex~\cite{Roberts:0708.1741}
    \begin{equation}\label{eq:complex_inn_2cm}
        \begin{array}{ccccccccc}
            \ainn(\frh\xrightarrow{\tilde \sft}\frg)&=&(& 
            \frh &\xrightarrow{~\sft~}&\frh \rtimes \frg&\xrightarrow{~\sft~}&\frg&)~,\\
            &&&b &\xmapsto{~\sft~} & \big(-b,\tilde\sft(b)\big)\\
            &&&&& (b,a) &\xmapsto{~\sft~} & \tilde\sft(b)+a
        \end{array}
    \end{equation}
    with the $\frg$-actions 
    \begin{equation}
        a\acton b \coloneqq  a \tildeacton b\eand a_1 \acton (b,a_2)\coloneqq (a_1\tildeacton b,[a_1,a_2])
    \end{equation}
    and the usual Lie bracket on $\frh \rtimes \frg$, viz. 
    \begin{equation}
        [(b_1,a_1),(b_2,a_2)]\coloneqq  \big([b_1,b_2]+a_1\tildeacton b_2-a_2\tildeacton b_1,[a_1,a_2]\big)~,
    \end{equation}
    leading to the Peiffer lifting
    \begin{equation}
        \{(b_1,a_1),(b_2,a_2)\}\coloneqq a_2\tildeacton b_1
    \end{equation}
    for all $a_1,a_2\in\frg$, $b_1,b_2\in \frh$.
    
    The infinitesimal version of the short exact sequence of 2-groupoids~\eqref{eq:ses_2-groupoids} is the following short exact sequence of graded vector spaces:
    \begin{equation}\label{eq:ses_3-vector}
        \begin{tikzcd}
            * \rar["\upsilon_2"] \dar & \frh\rar["\pi_2"] \dar & \frh \dar \\
            \frh\rar["\upsilon_1"] \dar & \frh\rtimes \frg \rar["\pi_1"] \dar & \frg \dar\\
            \frg \rar["\upsilon_0"] & \frg \rar["\pi_0"] & * 
        \end{tikzcd}
        ~,\hspace{2cm}
        \begin{aligned}
         \upsilon_2\colon&&*&\mapsto 0~,~~~&\pi_2\colon&&b&\mapsto b~,\\[0.5cm]
         \upsilon_1\colon&&b&\mapsto (b,0)~,~~~&\pi_1\colon&&(b,a)&\mapsto a~,\\[0.5cm]
         \upsilon_0\colon&&a&\mapsto a~,~~~&\pi_0\colon&&a&\mapsto *~.
        \end{aligned}
    \end{equation}
    
    Every 2-crossed module of Lie algebras defines a strict 3-term \(L_\infty\)-algebra, while a strict 3-term \(L_\infty\)-algebra \emph{almost} determines a 2-crossed module, with the missing data being the antisymmetric part \(\llbracket-,-\rrbracket\) of the Peiffer lifting \(\{-,-\}\); see \cref{app:Lie3_2_crossed}. Are there 2-crossed modules corresponding to the unadjusted and adjusted Weil algebras? In both cases, the answer is yes. The unadjusted Weil algebra corresponds to the inner derivation 2-crossed module; see \cref{app:inner_weil}. The case of the adjusted Weil algebra is treated in section~\ref{ssec:adjusted_inner_derivation}.

    \subsection{Simplification by coordinate transformation}\label{ssec:simplification}
    
    It is convenient to slightly simplify the description of the inner automorphism 3-group and related Lie 3-algebras. This does not change the definitions, but merely the descriptions.
    
    In the semidirect product \(\frh\rtimes\frg\) in the complex~\eqref{eq:complex_inn_2cm}, we define the Lie subalgebras
    \begin{align}
        \frh' &\coloneqq \operatorname{im}\sft \subseteq\frh\rtimes\frg~,&\frg' &\coloneqq \ker\sft\subseteq\frh\rtimes\frg~,
    \end{align}
    which are isomorphic to \(\frh\) and \(\frg\), respectively.
    The inner semidirect product \(\frh'\rtimes\frg'\) equals the whole Lie algebra \(\frh\rtimes\frg\).  So we can use the primed coordinates to talk about \(\frh\rtimes\frg=\frh'\rtimes\frg'\).
    This amounts to a coordinate transformation (or reparameterisation) of $\frh\rtimes \frg$ to $\frh'\rtimes \frg'$,
    \begin{equation}
       (b,a)\mapsto (b',a')\coloneqq(-b,a+\tilde\sft(b))~,
    \end{equation}
    and it simplifies the differentials in the complex~\eqref{eq:complex_inn_2cm} as follows:
    \begin{equation}\label{eq:inner_algebra_alternative_basis}
        \begin{matrix}
            \ainn(\frh\xrightarrow{\tilde \sft}\frg)&=&(& 
            \frh &\xrightarrow{~\sft~}&\frh' \rtimes \frg'&\xrightarrow{~\sft~}&\frg&)~,\\
            &&&b &\xmapsto{~\sft~} & \big(b,0\big)\\
            &&&&& (b,a) &\xmapsto{~\sft~} & a
        \end{matrix}
    \end{equation}
    The changes to the 2-crossed module structure maps under this reparameterisation are readily derived; we merely note that the semidirect product structure is preserved. Under this coordinate change, the presentations of the chain maps $\upsilon$ and $\pi$ in~\eqref{eq:ses_3-vector} change to
    \begin{equation}
        \begin{tikzcd}
            * \rar["\upsilon_2"] \dar & \frh\rar["\pi_2"] \dar & \frh \dar \\
            \frh\rar["\upsilon_1"] \dar & \frh'\rtimes \frg' \rar["\pi_1"] \dar & \frg \dar\\
            \frg \rar["\upsilon_0"] & \frg \rar["\pi_0"] & * 
        \end{tikzcd}
        ~,\hspace{1cm}
        \begin{aligned}
         \upsilon_2\colon&&*&\mapsto 0~,~~~&\pi_2\colon&&b&\mapsto b~,\\[0.5cm]
         \upsilon_1\colon&&b&\mapsto (-b,\tilde \sft(b))~,~~~&\pi_1\colon&&(b,a)&\mapsto \tilde \sft(b)+a~,\\[0.5cm]
         \upsilon_0\colon&&a&\mapsto a~,~~~&\pi_0\colon&&a&\mapsto *~.
        \end{aligned}
    \end{equation}

    At the finite level, i.e.~the level of the 2-crossed module of Lie groups $\sInn_{\rm \cm}(\CCG)$, we have corresponding Lie closed subgroups
    \begin{align}
        \sH' &\coloneqq \exp\frh' \le \sH \rtimes\sG ~,&
        \sG' &\coloneqq \exp\frg' \le \sH \rtimes\sG ~,
    \end{align}
    and a corresponding reparameterisation of $\sH\rtimes \sG$ as $\sH'\rtimes \sG'$,
    \begin{equation}
        (h,g)\mapsto(h',g')\coloneqq(h^{-1},\tilde \sft(h)g)~,
    \end{equation}
    leading to the normal complex
    \begin{equation}\label{eq:inner_group_alternative_basis}
        \begin{matrix}
            \sInn_{\rm \cm}(\CCG)&=&(&\sH&\xrightarrow{~\sft~}&\sH'\rtimes \sG'&\xrightarrow{~\sft~}&\sG&)~,\\
            &&&h&\xmapsto{~\sft~}&(h,\unit_\sH)\\
            &&& & &(h,g)&\xmapsto{~\sft~}& g
        \end{matrix}
    \end{equation}
    The presentations of the functors in the short exact sequence~\eqref{eq:ses_2-groupoids} change in the obvious manner; in particular,
    \begin{equation}\label{eq:ses_2-groupoids_reparameterisation}
        \Upsilon_1\colon(h,g)\mapsto(h^{-1},g,\tilde \sft(h))\eand \Pi_1\colon(h,g_1,g_2)\mapsto \tilde\sft(h)g_2~.
    \end{equation}

    \subsection{Relation between the Weil algebra and inner derivations}    
    Conceptually, the Weil algebra of a Lie \(n\)-algebra and the inner derivation \(n\)-crossed module of the Lie \(n\)-algebra are similar: both involve doubling the number of generators, with augmented degree, so as to be `topologically (or cohomologically) trivial'. In this section, we show that, under the comparison theorems of \cref{app:Lie3_2_crossed}, the two are in fact precisely the same, for \(n\le2\).
    
    First, we review the case for \(n=1\).
    \begin{thm}\label{thm:Weil_and_inner_Lie1}
        Given a Lie algebra $\frg$, the Chevalley--Eilenberg algebra of the 2-term \(L_\infty\)-algebra corresponding to the crossed module of Lie algebras $\ainn(\frg)$ is isomorphic to $\sW(\frg)$.
    \end{thm}
    \begin{proof}
        The Lie 2-algebra corresponding to the inner derivation crossed module \(\ainn(\frg)\) is $\frg[1]\xrightarrow{~\id~}\frg$
        with binary products
        \begin{equation}
            \mu_2(a_1,a_2)=[a_1,a_2] \eand
            \mu_2(a_1,\hat a_2)=[a_1,\hat a_2]~,
        \end{equation}
        for all $a_1,a_2\in \frg$ and $\hat a_1,\hat a_2\in \frg[1]$. With respect to some basis, its Chevalley--Eilenberg algebra is generated by elements $w^\alpha\in\frg[1]^*$ and $\hat w^\alpha\in\frg[2]^*$ and comes with the differential $Q_{\ainn}$ acting on the generators according to
        \begin{equation}
            \begin{aligned}
                Q_{\ainn}~&\colon~&v^\alpha&\mapsto-\tfrac12 f^\alpha_{\beta\gamma}v^\beta v^\gamma-\hat v^\alpha~,~~~&\hat v^\alpha&\mapsto -f^\alpha_{\beta\gamma}v^\beta\hat v^\gamma~,
            \end{aligned}
        \end{equation}
        where $f^\alpha_{\beta\gamma}$ are the structure constants of $\frg$.
        
        On the other hand, the Weil algebra $\sW(\frg)$ is generated by elements $t^\alpha \in \frg[1]^*$ and $\hat t^\alpha \in \frg[2]^*$ and the differential acts as 
        \begin{equation}
            \begin{aligned}
                Q_{\sW}~&\colon~&t^\alpha&\mapsto-\tfrac12 f^\alpha_{\beta\gamma}t^\beta t^\gamma + \hat t^\alpha~,~~~&\hat t^\alpha&\mapsto -f^\alpha_{\beta\gamma}t^\beta\hat t^\gamma~.
            \end{aligned}
        \end{equation}
        
        Comparing the action of the two differentials, it is obvious that 
        \begin{equation}
            v^\alpha \mapsto t^\alpha~,~~~\hat v^\alpha\mapsto -\hat t^\alpha~
        \end{equation}
        yields an isomorphism (or strict dual quasi-isomorphism) of differential graded algebras.
    \end{proof}
    
    The previous theorem categorifies for Lie 2-algebras.
    \begin{thm}
        Given a crossed module of Lie algebras $(\frh\xrightarrow{~\tilde \sft~}\frg,\tildeacton)$, the Chevalley--Eilenberg algebra of the strict 3-term \(L_\infty\)-algebra obtained as in theorem~\ref{thm:2cm_to_L_infty} from the 2-crossed module of Lie algebras $\ainn(\frh\xrightarrow{~\tilde \sft~}\frg)$ is isomorphic to the Weil algebra of $\frh\xrightarrow{~\tilde \sft~}\frg$.
    \end{thm}
    \begin{proof}
        Theorem~\ref{thm:2cm_to_L_infty} yields the following Lie 3-algebra for  $\ainn(\frh\xrightarrow{~\tilde \sft~}\frg)$:
        \begin{subequations}
        \begin{equation}
            \begin{matrix}
                (&\frh &\xrightarrow{~\mu_1~}& \frh\rtimes \frg&\xrightarrow{~\mu_1~}& \frg&)~, \\
                &b & \xmapsto{~\mu_1~} & (-b,\tilde\sft(b))  \\
                &&& (b,a) & \xmapsto{~\mu_1~} & \tilde\sft(b)+a
            \end{matrix}
        \end{equation}
        with binary products given by
        \begin{equation}
            \begin{aligned}
                \mu_2(a_1,a_2)&=[a_1,a_2]~,~~~&\mu_2\big((b_1,a_1),(b_2,a_2)\big)&=-\big(a_2 \tildeacton b_1+a_1 \tildeacton b_2\big)~,\\
                \mu_2\big(a_1,(b_2,a_2)\big)&=\big(a_1\tildeacton b_2,[a_1,a_2]\big)~,~~~
                &\mu_2(b_1,a_1)&=a_1\tildeacton b_1
            \end{aligned}
        \end{equation}
        \end{subequations}
        for all $a_1,a_2\in \frg$ and $b_1,b_2\in \frh$. Its Chevalley--Eilenberg algebra is generated by elements
        \begin{equation}
            v^\alpha\in\frg[1]^*~,~~~(w^a,\hat v^\alpha)\in(\frh\rtimes\frg)[2]^*~,~~~\hat w^a \in\frh[3]^*
        \end{equation}
        and the differential acts as
        \begin{equation}
            \begin{aligned}
                Q_{\ainn}~&\colon~&v^\alpha&\mapsto-\tfrac12 f^\alpha_{\beta\gamma}v^\beta v^\gamma-f^\alpha_a w^a - \hat v^\alpha~,~~~&\hat v^\alpha&\mapsto -f^\alpha_{\beta\gamma}v^\beta\hat v^\gamma+f_a^\alpha \hat w^a~,\\
                && w^a&\mapsto -f^a_{\alpha b}v^\alpha w^b-\hat w^a~, &\hat w^a &\mapsto -f^a_{\alpha b}v^\alpha \hat w^b+f^a_{\alpha b}\hat v^\alpha w^b~,
            \end{aligned}
        \end{equation}
        where $f^\alpha_a$, $f^\alpha_{\beta\gamma}$, and $f^a_{\alpha b}$ are the structure constants defining $\sft$, the Lie bracket on $\frg$ and the $\frg$-action $\acton$ on $\frh$.
        
        On the other hand, the Weil algebra $\sW(\frh\xrightarrow{~\tilde \sft~}\frg,\tildeacton)$ is generated by elements
        \begin{equation}
            t^\alpha \in \frg[1]^*~,~~~
            r^a \in \frh[2]^*~,~~~
            \hat t^\alpha \in \frg[2]^*~,~~~
            \hat r^a \in \frh[3]^*~,
        \end{equation}
        and the differential acts according to
        \begin{equation}
            \begin{aligned}
                Q_{\sW}~&\colon~&t^\alpha&\mapsto-\tfrac12 f^\alpha_{\beta\gamma}t^\beta t^\gamma-f^\alpha_a r^a+ \hat t^\alpha~,~~~&\hat t^\alpha&\mapsto -f^\alpha_{\beta\gamma}t^\beta\hat t^\gamma+f_a^\alpha \hat r^a~,\\
                && r^a&\mapsto -f^a_{\alpha b}t^\alpha r^b+\hat r^a ~,&\hat r^a &\mapsto -f^a_{\alpha b}t^\alpha \hat r^b+f^a_{\alpha b}\hat t^\alpha r^b~.
            \end{aligned}
        \end{equation}
        
        Comparing the differentials, it is again obvious that 
        \begin{equation}
            v^\alpha \mapsto t^\alpha~,~~~\hat v^\alpha\mapsto -\hat t^\alpha~,~~~w^a\mapsto r^a~,~~~\hat w^a\mapsto -\hat r^a
        \end{equation}
        yields an isomorphism of differential graded algebras.
    \end{proof}
    
    In both theorems, we encountered unfortunate minus signs in the isomorphism, which is a consequence of our being stuck between the hammer of standard conventions for the Weil algebra and the anvil of standard conventions for the semidirect product.
    
    Regardless, the 3-term \(L_\infty\)-algebra encoded in the Weil algebra of a strict Lie 2-algebra is canonically isomorphic as an $L_\infty$-algebra to the 3-term \(L_\infty\)-algebra underlying the inner derivation 2-crossed module of the strict Lie 2-algebra. We stress, however, that the inner derivation 2-crossed module of Lie algebras contains additional data, namely the antisymmetric part of the Peiffer lifting
    \begin{equation}
        \llbracket (b_1,a_1),(b_2,a_2)\rrbracket = \tfrac12\left(a_2 \tildeacton b_1 - a_1 \tildeacton b_2\right)~.
    \end{equation}

    \section{Quasi-isomorphisms and strict 2-group equivalences}\label{app:Equivalence}
    
    Morphisms of $L_\infty$-algebras are most readily understood in their dual formulation: as morphisms of differential graded algebras between the corresponding Chevalley--Eilenberg algebras. Such a morphism descends to a morphism between the $\mu_1$-cohomologies of the $L_\infty$-algebras. A \emph{quasi-isomorphism} between two $L_\infty$-algebras $\frL$ and $\tilde \frL$ is a morphism of $L_\infty$-algebras $\phi\colon\frL\to \tilde \frL$, which descends to an isomorphism $\phi_*\colon\operatorname H^\bullet_{\mu_1}(\frL)\to\operatorname H^\bullet_{\mu_1}(\tilde \frL)$. For more details, see e.g.~\cite{Jurco:2018sby}. Quasi-isomorphisms are indeed the appropriate notion of equivalence for most intents and purposes. For example, quasi-isomorphic gauge $L_\infty$-algebras lead to quasi-isomorphic, and thus physically equivalent, BRST complexes~\cite{Saemann:2019dsl}.
    
    By the {\em minimal model theorem}, any $L_\infty$-algebra $\frL$ is quasi-isomorphic to an $L_\infty$-algebra with underlying graded vector space $H^\bullet_{\mu_1}(\frL)$, which is called a {\em minimal model} for $\frL$.
    
    As an example, we explain the quasi-isomorphism between two strict Lie 2-algebras relevant to our discussion, namely
    \begin{equation}
        \frg=(*\xrightarrow{~~~} \frg)\eand \frg_{\rm lp}\coloneqq (L_0\frg\xhookrightarrow{~~~} P_0\frg)~,
    \end{equation}
    where $L_0\frg$ and $P_0 \frg$ are based loop\footnote{Note that \(L_0\frg\), the ordinary loop space, is \emph{not} the same as \(\hat L_0\frg\), which contains a central extension, used in the loop model of the string 2-algebra.} and path spaces in $\frg$, cf.~\cref{app:path_groups}. Besides the embedding $\mu_1$, the only other non-trivial higher product is in both cases $\mu_2$ given by the obvious commutators. The quasi-isomorphism between these two strict Lie 2-algebras is a truncation of a quasi-isomorphism given in~\cite[Lemma 37]{Baez:2005sn}. We have morphisms of Lie 2-algebras $\phi$ and $\psi$,
    \begin{subequations}\label{eq:quasi-iso-morphs}
    \begin{equation}
        \begin{tikzcd}
            \frg \ar[r,bend left=45, "\phi" pos=0.5] & \frg_{\rm lp}\ar[l,bend left=45, pos=0.45]{}{\psi} 
        \end{tikzcd}
    \end{equation}
    which are given explicitly by the chain maps 
    \begin{equation}
        \begin{tikzcd}
            * \ar[d] \ar[r] & L_0\frg \ar[d,hookrightarrow] \ar[r] & * \ar[d] \\
            \frg \ar[r]{}{\cdot \ell(\tau)}& P_0\frg \ar[r]{}{\dpar} & \frg
        \end{tikzcd}
    \end{equation}
    \end{subequations}
    where $\dpar\colon P_0\frg\to \frg$ is again the endpoint evaluation and $\cdot \ell(\tau)\colon\frg\to P_0\frg$ embeds $\alpha_0\in\frg$ as the line $\alpha(\tau)=\alpha_0 \ell(\tau)$ for some smooth function $\ell\colon [0,1]\to \FR$ with $\ell(0)=0$ and $\ell(1)=1$. Both maps $\phi$ and $\psi$ descend to isomorphisms on the cohomologies 
    \begin{equation}
        \frg\cong\operatorname H^\bullet_{\mu_1}(*\to \frg)\cong\operatorname H^\bullet_{\mu_1}(\frg_{\rm lp})=(*\to \frg)~,
    \end{equation}
    and $(*\to \frg)=\operatorname H^\bullet_{\mu_1}(*\to \frg)$ is thus indeed a minimal model for $\frg_{\rm lp}$.
    
    We can complete the morphisms in~\eqref{eq:quasi-iso-morphs} to a categorical equivalence by adding a contracting homotopy: $(\psi\circ \phi)_0$ is already the identity, and we have a 2-morphisms of Lie 2-algebras $\eta\colon\phi\circ \psi\to \id_{\frg_{\rm lp}}$ encoded in
    \begin{equation}\label{eq:def_chi}
        \eta:P_0\frg\to L_0\frg~,~~~\eta(\gamma)=\gamma-\ell(\tau)\dpar\gamma~.
    \end{equation}

    Strict Lie 2-algebras integrate to particular Lie groupoids, which carry the structure of a 2-group and $\frg_{\rm lp}$ integrates to the 2-group 
    \begin{equation}
        \sG_{\rm lp}\coloneqq(L_0 \sG\rtimes P_0\sG\rightrightarrows P_0\sG)=\CCC(L_0\sG\xrightarrow{~\sft~}P_0\sG)~.
    \end{equation}
    We thus expect $\sG_{\rm lp}$ to be equivalent to the 2-group $*\rightarrow \sG$ in a suitable sense. This is the case as we will show now.
    
    Since both 2-groups are strict, the appropriate notion of morphism is given by butterflies, as defined in~\cref{ssec:BUTTERFLY}, with an equivalence corresponding to a flippable butterfly. In particular, the equivalence of  the strict 2-groups $\sG_{\rm lp}$ and $(*\xrightarrow{~~~}\sG)$ is given by the flippable butterfly
    \begin{equation}
        \begin{tikzcd}
            * \arrow[dd] \arrow[dr]& & L_0\sG \arrow[dl,hookrightarrow] \arrow[dd,hookrightarrow]
            \\
            & P_0\sG \arrow[dl,"\dpar",swap] \arrow[dr,"\sfid"]& 
            \\
            \sG & & P_0\sG
        \end{tikzcd}
    \end{equation}

    \section{Path groupoids}\label{app:path_space}
    
    We need groupoids and higher groupoids of smooth, parameterised paths, but generic such paths with coincident endpoints fail to compose smoothly and associatively. To remedy this, we follow~\cite{Caetano:1993zf,Schreiber:0705.0452} and introduce sitting instants and factor by thin homotopies. This section summarises some of the technical details underlying our path groupoids.
    
    Suppose we are given a manifold \(M\). A \emph{path with sitting instants} is a smooth map $\gamma\colon [0,1]\to M$, regarded as a morphism
    \begin{equation}
        x_1\xleftarrow{~\gamma~}x_0
    \end{equation}
    with \emph{sitting instants} at the endpoints $x_0=\gamma(0)$, $x_1=\gamma(1)$. That is, there is an $\eps>0$ such that for $i\in\{0,1\}$ and all $|t-i|\leq \eps$, the map $\gamma$ is constant: $\gamma(t)=x_i$. We abbreviate this by writing
    \begin{equation}
        t\approx i\in\{0,1\}~~\Rightarrow~~\gamma(t)= x_i~.
    \end{equation}
    This ensures smooth composition of paths.
    
    A \emph{homotopy with sitting instants} between two paths \(\gamma_0,\gamma_1\colon[0,1]\to M\) sharing common endpoints \(x_0,x_1\in M\) is a smooth homotopy
    \begin{equation}
        \sigma\colon [0,1]\times[0,1]\to M~,\hspace{1.5cm}
        \begin{tikzcd}[column sep=1.5cm,row sep=large]
            x_1 & \ar[l, anchor=center,bend left=45, "{\gamma_0}", ""{name=U,inner sep=1pt,above}] \ar[l,anchor=center, bend right=45, "{\gamma_1}", swap, ""{name=D,inner sep=1pt,below}] 
            x_0 \arrow[Rightarrow,from=U, to=D, "{\sigma}",swap]
        \end{tikzcd}
    \end{equation}     
    with sitting instants
    \begin{equation}\label{eq:surface_sitting_instants}
        \begin{aligned}
            s\approx i\in\{0,1\}~~&\Rightarrow~~\sigma(s,t) = \gamma_i(t)~,\\
            t\approx i\in\{0,1\}~~&\Rightarrow~~\sigma(s,t) = x_i~.
        \end{aligned}
    \end{equation}

    A homotopy with sitting instants \(\sigma\) is \emph{thin} if the rank of \(\dd\sigma\) is at most \(1\) everywhere. The \emph{path groupoid} $\CP M$ is the groupoid whose objects are points in \(M\), and whose 1-morphism from \(x_1\in M\) to \(x_2\in M\) is an equivalence class of paths with sitting instants, which we identify any two paths $\gamma_1,\gamma_2\colon x_0\to x_1$, $x_0,x_1\in M$ between which there is a thin homotopy with sitting instants. This ensures that composition of paths is associative.\footnote{The \emph{fundamental groupoid} $\Pi_1(M)$ is finer than $\CP M$, since in that case we do not impose the condition of rank \(\le1\) on the homotopies. A parallel transport functor whose domain is the fundamental groupoid can only describe flat connections.}
    We neglect details of the topology and smooth structure. Such details can be treated rigorously using \emph{diffeological spaces}; see~\cite{Schreiber:0705.0452, Schreiber:0802.0663, Schreiber:2008aa, Waldorf:0911.3212}, as well as~\cite{Stacey:0803.0611} and references therein for further details.
    
    We can also construct the path 2-groupoid \(\CP_{(2)}M\)~\cite{Schreiber:0802.0663} as follows. The objects are points, and the 1-morphisms are equivalence classes of paths (with sitting instants) under thin homotopies (with sitting instants). A 2-morphism \(\sigma_1,\sigma_2\colon x_0\to x_1\) of the path 2-groupoid is, roughly, a \emph{bigon}, i.e.~a surface
bounded by \(\sigma_1\circ\sigma_2^{-1}\). More precisely, 2-morphisms are be equivalence classes of (not necessarily thin!) homotopies (with sitting instants) under thin homotopies of homotopies (with sitting instants), which we now define.
    
    A \emph{homotopy of homotopies with sitting instants} between homotopies \(\sigma_0,\sigma_1\) between the same paths \(\gamma_0,\gamma_1\) between the same endpoints \(x_0,x_1\) is a smooth map
    \begin{equation}
        \rho\colon [0,1]^3 \to M~,\hspace{1.5cm}
        \begin{tikzcd}[column sep=2.5cm,row sep=large]
            x_1 & \ar[l, bend left=60, "\gamma_0", ""{name=U,inner sep=1pt,above}] \ar[l, bend right=60, "\gamma_1", swap, ""{name=D,inner sep=1pt,below}] x_0
            \arrow[Rightarrow, bend left=70,from=U, to=D, "{\sigma_1}",""{name=L,inner sep=1pt,right}]
            \arrow[Rightarrow, bend right=70,from=U, to=D, "{\sigma_0}",""{name=R,inner sep=1pt,left},swap]
            \arrow[triple,from=R,to=L,"{\rho}",swap]
        \end{tikzcd}
    \end{equation}
    with sitting instants 
    \begin{equation}
        \begin{aligned}
            r\approx i\in\{0,1\}~~&\Rightarrow~~\rho(r,s,t) = \sigma_i(t)~,\\
            s\approx i\in\{0,1\}~~&\Rightarrow~~\rho(r,s,t) = \gamma_i(t)~,\\
            t\approx i\in\{0,1\}~~&\Rightarrow~~\rho(r,s,t) = x_i~.
        \end{aligned}
    \end{equation}
    Such a homotopy is called \emph{thin} if $\dd\rho$ has rank $\le2$ everywhere and $\dd\rho$ has rank \(\le1\) at \((r,s,t)\) with \(s\in\{0,1\}\).\footnote{This ensures that domains and codomains are well defined on equivalence classes of homotopies of homotopies.}
    
    We also need the path 3-groupoid \(\CP_{(3)}M\), whose obvious definition we spell out as well. Its objects, 1-morphisms, and 2-morphisms are as before. Its 3-morphisms are equivalence classes of homotopies of homotopies under thin homotopies of homotopies of homotopies, which we define below. A \emph{homotopy of homotopies of homotopies with sitting instants} between homotopies of homotopies \(\rho_0,\rho_1\) between the same homotopies \(\sigma_0,\sigma_1\) between the same paths \(\gamma_0,\gamma_1\) between the same endpoints \(x_0,x_1\) is a smooth map
    \begin{equation}
        \pi\colon [0,1]^4 \to M~,\hspace{1.5cm}
        \begin{tikzcd}[column sep=3.5cm,row sep=large]
            x_1 & \ar[l, bend left=60, "\gamma_0", ""{name=U,inner sep=1pt,above}] \ar[l, bend right=60, "\gamma_1", swap, ""{name=D,inner sep=1pt,below}] x_0
            \arrow[Rightarrow, bend left=70,from=U, to=D, "{\sigma_1}",""{name=L,inner sep=1pt,right}]
            \arrow[Rightarrow, bend right=70,from=U, to=D, "{\sigma_0}",""{name=R,inner sep=1pt,left},swap]
            \arrow[triple,from=R,to=L,bend left=45,"{\rho_1}",""{name=M1,inner sep=1pt,above}]
            \arrow[triple,from=R,to=L,bend right=45,"{\rho_2}",""{name=M2,inner sep=1pt,below},swap]
            \arrow[quad,from=M1,to=M2,"{\pi}",swap]
        \end{tikzcd}
    \end{equation}
    such that, for \(i\in\{0,1\}\),
    with sitting instants 
    \begin{equation}
        \begin{aligned}
            q\approx i\in\{0,1\}~~&\Rightarrow~~\pi(q,r,s,t) = \rho_i(t)~,\\
            r\approx i\in\{0,1\}~~&\Rightarrow~~\pi(q,r,s,t) = \sigma_i(t)~,\\
            s\approx i\in\{0,1\}~~&\Rightarrow~~\pi(q,r,s,t) = \gamma_i(t)~,\\
            t\approx i\in\{0,1\}~~&\Rightarrow~~\pi(q,r,s,t) = x_i~.
        \end{aligned}
    \end{equation}
    Such a homotopy is called \emph{thin} if $\dd\pi$ has rank $\le3$ everywhere, $\dd\pi$ has rank $\le2$ at $(q,r,s,t)$ with $r\in\{0,1\}$ and $\dd\pi$ has rank $\le1$ at $(q,r,s,t)$ with $s\in\{0,1\}$. Thankfully, this is all we need.

    \section{Chen forms}\label{app:chen_forms}
    
    To define path-ordered higher-dimensional integrals, we use the formalism of \emph{Chen forms}. Briefly, the idea is to regard \(n\)-forms as \(1\)-forms on iterated path spaces. The treatment here is not meant to be rigorous, but to give the general flavor of ideas. For technical details the reader should consult~\cite{Baez:2004in,Getzler:1991:339,Hofman:2002ey}.
    
    \paragraph{Surface-ordering.} We want to define a surface-ordered integral of a 2-form, analogous to path-ordered integrals of 1-forms. For this, we must fix an order on the points on a surface $\sigma$, which is evidently not canonical. If $\sigma(s,t)$ is a parameterised surface 
    \begin{equation}
        \sigma\colon [0,1]\times [0,1]\rightarrow M~,
    \end{equation}
    we can define an ordering of points lexicographically: we first sort by \(s\), then by \(t\). This amounts to the following picture.
    
\begin{figure}
    \[
    \begin{tikzpicture}[baseline=-0.1]
    \coordinate (Q0) at (0-1,0);
    \coordinate (Q1) at (1,0);

    \filldraw (Q0) circle (0.05);
    \filldraw (Q1) circle (0.05);
    
    \node [left] at (Q0) {\(x_0\)};
    \node [right] at (Q1) {\(x_1\)};
    
    \begin{scope}[xshift=5cm]
    \coordinate (P0) at (0.7,0.7);
    \coordinate (P1) at (0.7,-0.7);
    \coordinate (P2) at (-0.7,-0.7);
    \coordinate (P3) at (-0.7,0.7);    
    \end{scope}
    
    \begin{scope}[thick,decoration={
        markings,
        mark=at position 0.5 with {\arrow{>}}}
    ]
        \draw [postaction={decorate}] (Q0) .. controls +(1,1) .. (Q1) node[midway, above] {\(\gamma_1\)};
        \draw [postaction={decorate}] (Q0) .. controls +(1,-1) .. (Q1)  node[midway, below] {\(\gamma_0\)};
        \draw (P0) -- (P1) node [right, midway] {\(x_1\)};
        \draw [postaction={decorate}] (P2) -- (P1) node[below, midway] {\(\gamma_0\)};
        \draw (P2) -- (P3) node [left, midway] {\(x_0\)};
        \draw [postaction={decorate}] (P3) -- (P0) node[above, midway] {\(\gamma_1\)};
        
        \begin{scope}[thin]
        \draw [postaction={decorate}] (Q0) -- (Q1);
        \draw [postaction={decorate}] (Q0) .. controls +(1,0.5) .. (Q1);
        \draw [postaction={decorate}] (Q0) .. controls +(1,-0.5) .. (Q1);
        
        \draw [postaction={decorate}] ($(P2)+(0,0.3)$) -- ($(P1)+(0,0.3)$);
        \draw [postaction={decorate}] ($0.5*(P2)+0.5*(P3)$) -- ($0.5*(P1)+0.5*(P0)$);
        \draw [postaction={decorate}] ($(P3)+(0,-0.3)$) -- ($(P0)+(0,-0.3)$);
        \end{scope}
        
    \end{scope}

    \node[above right] at (P0) {\((1,1)\)};
    \node[below right] at (P1) {\((0,1)\)};
    \node[below left] at (P2) {\((s,t)=(0,0)\)};
    \node[above left] at (P3) {\((1,0)\)};
    
    \begin{scope}[dashed]
    \end{scope}
        
    \end{tikzpicture}
    \]
    \caption{A parameterised surface with sitting instants, seen as a parameterised curve on the space of parameterised curves between two fixed points.}\label{fig:parameterised_surface}
    \end{figure}
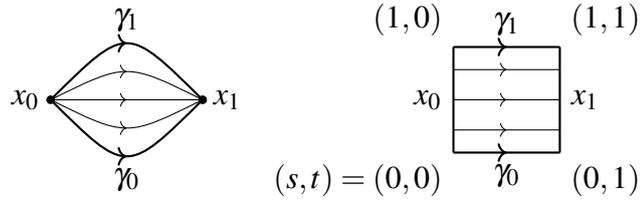

    First, ensure that \(\sigma\) has sitting instants~\eqref{eq:surface_sitting_instants}, reparameterising as necessary; see figure~\ref{fig:parameterised_surface}. Then the parameterised surface \(\sigma\) forms a bigon between the two parameterised curves
    \begin{align}
        \gamma_1(t) &\coloneqq \sigma(0,t)~,&
        \gamma_2(t) &\coloneqq \sigma(1,t)~.
    \end{align}
    We can regard $\sigma$ as a parameterised path \(\check\sigma\) between two points \(\gamma_0,\gamma_1\in P_{x_0}^{x_1}\), where \(P_{x_0}^{x_1}\) is the manifold of parameterised paths\footnote{the manifold of parameterised paths with sitting instants, defined similarly to \(\hom_{\CP M}(x_0,x_1)\), but without quotienting by thin homotopies} between $x_0$ and $x_1$.
    \begin{equation}
        \check\sigma_s(t)\coloneqq\sigma(s,t)~,~~~\check\sigma\in P_{\gamma_0}^{\gamma_1}(P_{x_0}^{x_1}(M))~,~~~\check\sigma_s \in P_{x_0}^{x_1}(M)\text{ for all }s\in[0,1]~.
    \end{equation}
    Note that we represent bigons as paths on path spaces, i.e.~elements of \(P_{\gamma_0}^{\gamma_1}(P_{x_0}^{x_1}(M))\), the space of paths between two points \(\gamma_0\) and \(\gamma_1\) on the path space \(P_{x_0}^{x_1}(M)\).

    An ordinary 2-form on $M$ defines a 1-form $\check B=\chint B$ on the locally convex manifold $P_{x_0}^{x_1}(M)$, also known as a {\em Chen form}. To wit, for each path $\gamma\in P_{x_0}^{x_1}(M)$, we can pull back $B$ along the evaluation map $\operatorname{ev}_t\colon\gamma\mapsto \gamma(t)$. We then contract $\operatorname{ev}^*_t B$ with the vector field tangent $R$ acting as $R(\gamma)=\dot \gamma$, which generates reparameterisations of $\gamma$ and whose pushforward is tangent to $\gamma$:
    \begin{equation}
        \check B=\chint B\coloneqq\int_0^1 \dd t~\iota_R (\operatorname{ev}_t^* B)~.
    \end{equation}
    For details, see again~\cite{Baez:2004in,Getzler:1991:339,Hofman:2002ey}. The 1-form $\check B$ can then be further integrated along the path \(\check\sigma\) in $P_{x_0}^{x_1}(M)$.
    
    \paragraph{Lie-algebra valued 2-forms.} To deal with 2-forms $B\in \Omega^2(M)\otimes\frh$ which transform under a gauge group with connection 1-form $A\in \Omega^1(M)\otimes \frg$, i.e.~equipped with an action of $\frg$ on $\frh$, we extend the picture slightly. The integration to a Chen form is now modified by an underlying parallel transport along the path $\gamma\in P_{x_0}^{x_1}(M)$ described by $A$. The 2-form $B$ is decorated by path-ordered integrals of $A$ along parts of $\gamma$:
    \begin{equation}\label{eq:non-abelian_correction}
        \check B=\chint_AB\coloneqq\int_0^1 \dd t~W^{-1}_t(\iota_R (\operatorname{ev}_t^* B))\ewith W_t=\Pexp\int_{\gamma_t} A~,
    \end{equation}
    where $\gamma_t$ is again the path $\gamma$ truncated at $t$ and reparameterised. For details, see~again~\cite{Baez:2004in,Hofman:2002ey}.
    
    \paragraph{Higher-dimensional generalisations.} The higher-dimensional generalisation on iterated loop spaces is mostly self-evident: we iterate the procedure, producing Chen forms of lower and lower degree. Given a form \(C\in\Omega^k(M)\), we pick two points \(x_0,x_1\in M\), define the space of parameterised paths (with sitting instants) \(P_{x_0}^{x_1}(M)\), and build the Chen form \(\chint C\in \Omega^{k-1}(P_{x_0}^{x_1}(M))\). We then pick two points \(\gamma_0,\gamma_1\in P_{x_0}^{x_1}(M)\) (i.e.~parameterised paths with sitting instants on \(M\)), define the space \(P_{\gamma_0}^{\gamma_1}(P_{x_0}^{x_1}(M))\) of parameterised surfaces (i.e.~parameterised paths with sitting instants on \(P_{x_0}^{x_1}(M)\)), and build the Chen form \(\chichint C\in\Omega^{k-2}(P_{\gamma_0}^{\gamma_1}(P_{x_0}^{x_1}(M))\). We then iterate the process until we obtain a 1-form on an iterated loop space, over which we can then define a path-ordered integral. 
    
    For ``non-abelian'' forms, that is forms $C\in \Omega^{k}(M)\otimes\frv$ taking values in some vector spaces $\frv$ carrying representations of certain Lie algebras, there is also an evident generalisation of~\eqref{eq:non-abelian_correction} by decorating with one-forms on iterated loop spaces.

\chapter{Higher Parallel Transport}

    To simplify the presentation we restrict some parts of our discussion to the example of the loop model of the string Lie 2-group. We expect the result to generalise to arbitrary higher Lie groups admitting an adjustment, since we view the string Lie 2-group is paradigmatic of general adjustable Lie 2-groups. However, it is difficult to write a completely general yet concrete enough discussion since we do not have a general ansatz for every possible adjustment.
    
    \section{Adjusted Weil algebras}\label{ssec:adjusted_Weils}
    The problems outlined in \cref{ssec:limitations} can be eliminated for some gauge $L_\infty$-algebras by deforming their Weil algebras. This deformation was first discussed in the context of the string Lie 2-algebra in~\cite{Sati:2008eg,Sati:2009ic}; see also~\cite{Schmidt:2019pks} and~\cite{Saemann:2019dsl}. 
    
    Given an $L_\infty$-algebra $\frL$, the Weil algebra $\sW(\frL)$ projects onto the Chevalley--Eilenberg algebra $\sCE(\frL)$. We call a deformation $\sW_{\rm adj}(\frL)$ of $\sW(\frL)$ an \emph{adjusted Weil algebra}~\cite{Saemann:2019dsl}, if the underlying graded algebra is isomorphic to $\sW(\frL)$, the projection onto the Chevalley--Eilenberg algebra is not deformed, and the resulting BRST complex is closed. The last condition amounts to closure of the gauge transformations without any restriction on gauge parameters or gauge fields.
    
    This deformation to an adjusted Weil algebra can already be motivated on purely algebraic grounds: the Weil algebra contains the vector space of invariant polynomials, whose definition is only compatible with quasi-isomorphism\footnote{This is the appropriate notion of isomorphism here; see \cref{app:Equivalence}.} after the deformation; see~\cite{Saemann:2019dsl}.
    
    \paragraph{Skeletal model of the string 2-algebra.}
    As a first example, consider the skeletal string Lie 2-algebra 
    \begin{equation}
        \begin{array}{ccccccc}
            \astringsk(\frg)&=&(& \FR &\xrightarrow{~\mu_1~} &\frg&)\\
            &&& r&\xmapsto{~\mu_1~} & 0
        \end{array}
    \end{equation}
    for some metric Lie algebra\footnote{A \emph{metric Lie algebra} is a Lie algebra equipped with a nondegenerate (but not necessarily positive-definite) bilinear form \(\langle-,-\rangle\) that is invariant under the adjoint action.} $\frg$ with
    \begin{subequations}
        \begin{align}
            &\mu_2\colon\frg\wedge \frg \to \frg~, &\mu_2(a_1,a_2)&=[a_1,a_2]~,\\
            &\mu_3\colon\frg\wedge \frg\wedge \frg \to \FR~, &\mu_3(a_1,a_2,a_3)&=\big\langle a_1,[a_2,a_3]\big\rangle~,
        \end{align}
    \end{subequations}
    where $\langle-,-\rangle \colon\frg\times\frg\to\mathbb R$ denotes the invariant metric on $\frg$. Let $e_\alpha$ be a basis of $\frg$ and $f^\alpha_{\beta\gamma}$ and $\kappa_{\alpha\beta}$ be the structure constants and the components of the metric, respectively. The unadjusted Weil algebra is generated by coordinate functions $t^\alpha, r$ of degrees~1 and~2 respectively on $\frL[1]$ as well as their shifted copies $\hat t^\alpha=\sigma t^\alpha$ and $\hat r=\sigma r$ of degrees~2 and~3 respectively. The differential acts according to 
    \begin{equation}
        \begin{aligned}
            Q_\sW~&:~&t^\alpha &\mapsto -\tfrac12 f^{\alpha}_{\beta\gamma} t^\beta  t^\gamma + \hat t^\alpha~,~~~&r &\mapsto \tfrac1{3!} f_{\alpha\beta\gamma} t^\alpha  t^\beta   t^\gamma + \hat r~,\\
            &&\hat t^\alpha &\mapsto -f^\alpha_{\beta\gamma} t^\beta   \hat t^\gamma~,~~~&\hat r &\mapsto -\tfrac12 f_{\alpha\beta\gamma} t^\alpha  t^\beta   \hat t^\gamma~,
        \end{aligned}
    \end{equation}
    where $f_{\alpha\beta\gamma}\coloneqq \kappa_{\alpha\delta}f^\delta_{\beta\gamma}$. An adjusted form of this Weil algebra which we shall denote by $\sW_{\rm adj}(\astringsk(\frg))$ has (by definition) the same generators, but the differential $Q_{\sW_{\rm adj}}$ acts as
    \begin{equation}
        \begin{aligned}
            Q_{\sW_{\rm adj}}~&:~&t^\alpha &\mapsto -\tfrac12 f^\alpha_{\beta\gamma} t^\beta  t^\gamma + \hat t^\alpha~,~~~&r &\mapsto \tfrac{1}{3!} f_{\alpha\beta\gamma} t^\alpha  t^\beta  t^\gamma -\kappa_{\alpha\beta}t^\alpha\hat t^\beta+ \hat r~,\\
            &&\hat t^\alpha &\mapsto -f^\alpha_{\beta\gamma} t^\beta  \hat t^\gamma~,~~~&\hat r&\mapsto \kappa_{\alpha\beta}\hat t^\alpha\hat t^\beta~.
        \end{aligned}
    \end{equation}
    Now, following adjustment, the kinematical data for a gauge theory on a local patch $U$ has changed from that given in \eqref{eq:unadjusted_fields} to 
    \begin{subequations}
        \begin{align}
            A&\in\Omega^1(U)\otimes\frg~,\\
            B&\in\Omega^2(U)~,\\
            F&\coloneqq \dd A+\tfrac12 [A,A]\in \Omega^2(U)\otimes \frg~,&
            \dd F+[A,F]&=0~,\\
            H&\coloneqq \dd B-\tfrac{1}{3!}\mu_3(A,A,A)+\langle A,F\rangle\in \Omega^3(U)~,&
            \dd H-\langle F,F\rangle&=0
        \end{align}
    \end{subequations}
    while gauge transformations have changed from \eqref{eq:unadjusted_gauge_transf} to
    \begin{subequations}
        \begin{align}
            \delta A&=\dd \alpha+\mu_2(A,\alpha)~,\\
            \delta B&=\dd \Lambda+\langle\alpha,F\rangle-\tfrac12 \mu_3(A,A,\alpha)~,\\
            \delta F&=-\mu_2(F,\alpha)~,\\
            \delta H&=0~,
        \end{align}
    \end{subequations}
    where \(\alpha\in\Omega^0(U)\otimes\frg\) and \(\Lambda\in\Omega^1(U)\) parameterise infinitesimal gauge transformations. The commutator of two gauge transformations now closes as expected, and the BRST complex of these fields is indeed closed; see~\cite{Saemann:2019dsl}. Moreover, writing down covariant field equations for $H$ has become easier.
    
    Such connections arise naturally in the context of heterotic supergravity, as well as in non-abelian self-dual strings and six-dimensional superconformal field theories~\cite{Saemann:2017rjm,Saemann:2017zpd,Saemann:2019dsl}. For references to the original literature on string structures and a detailed explanation of their relevance, see also~\cite{Saemann:2017zpd}.
    
    \paragraph{Loop model of the string 2-algebra.}
    Since we wish to discuss parallel transport, we need finite descriptions of gauge transformations and their actions. The skeletal model is not a strict \(L_\infty\)-algebra, and hence not well suited for integration. It is more convenient to work with the loop model, which is quasi-isomorphic\footnote{Quasi-isomorphism implies that the two models define physically equivalent kinematical data; see \cref{app:Equivalence} for definitions.} to the skeletal model:
    \begin{equation}
        \begin{array}{ccccccc}
            \astringl(\frg)&=&(& \hat L_0\frg &\xrightarrow{~\mu_1~} &P_0\frg&)\\
            &&& (\lambda,r)&\xmapsto{~\mu_1~} & \lambda
        \end{array}
    \end{equation}
    where $P_0\frg$ and $L_0\frg$ are based path and loop spaces, respectively, of $\frg$ and $\hat L_0\frg=L_0\frg\oplus\FR$ is the vector space underlying the Lie algebra obtained by the Kac--Moody extension; for technical details see \cref{app:path_groups}. The loop model is a strict 2-term \(L_\infty\)-algebra (i.e.~\(\mu_i=0\) for \(i\ge3\)); the unary product \(\mu_1\) was given above, and the binary product \(\mu_2\) is as follows:
    \begin{subequations}
        \begin{align}
            P_0\frg\wedge  P_0\frg&\to P_0\frg~,&(\gamma_1,\gamma_2)&\mapsto[\gamma_1,\gamma_2]~,\\
            P_0\frg\otimes\hat L_0 \frg[1]&\to \hat L_0 \frg[1]~,~~~&\big(\gamma,(\lambda,r)\big)&\mapsto\left([\gamma,\lambda]\; ,\; -2\int_0^1 \dd\tau \big\langle\gamma(\tau),\dot \lambda(\tau)\big\rangle\right)~,
        \end{align}
    \end{subequations}
    where $\dot{-}$ labels the derivative with respect to the path or loop parameter.
    
    The corresponding Weil algebra is generated by coordinate functions $t^{\alpha \tau}$, $r^{\alpha \tau}$, $r_0$ as well as their shifted counterparts. The differential \(Q_\sW\) acts as
    \begin{equation}\label{eq:Weil_string_Lie_2_loop}
        \begin{aligned}
            t^{\alpha\tau} &\mapsto -\tfrac12 f^{\alpha}_{\beta\gamma} t^{\beta\tau} t^{\gamma\tau} - r^{\alpha\tau} +\hat{t}^{\alpha\tau}~,~~~ 
            &\hat{t}^{\alpha\tau}&\mapsto -f^\alpha_{\beta\gamma}t^{\beta\tau}\hat{t}^{\gamma\tau}+\hat{r}^{\alpha\tau}~,\\
            r^{\alpha\tau} &\mapsto -f^\alpha_{\beta\gamma} t^{\beta\tau}r^{\gamma\tau} + \hat{r}^{\alpha\tau}~,~~~
            &\hat{r}^{\alpha\tau}&\mapsto-f^\alpha_{\beta\gamma} t^{\beta\tau}\hat{r}^{\gamma\tau} + f^\alpha_{\beta\gamma} \hat{t}^{\beta\tau}r^{\gamma\tau} ~,\\
            r_0 &\mapsto 2\int_0^1\dd \tau\, \kappa_{\alpha\beta}t^{\alpha\tau} \dot{r}^{\beta\tau}+\hat{r}_0 ~,~~~&\hat{r}_0&\mapsto 2\int_0^1\dd\tau\,\kappa_{\alpha\beta}\left(t^{\alpha\tau} \smash{\dot{\hat{r}}}^{\beta\tau}-\hat{t}^{\alpha\tau}\dot{r}^{\beta\tau}\right)~.
        \end{aligned}
    \end{equation}
    As we saw in Chapter~2, the unadjusted Weil algebra is not suitable for physics. Instead, although it is not obvious, we will see below that the following constitutes a suitable adjustment:
    \begin{equation}\label{eq:Weil_string_Lie_2_loop_adj}
        \begin{aligned}
            Q_{\sW_{\rm adj}}:~t^{\alpha\tau} &\mapsto -\tfrac12 f^{\alpha}_{\beta\gamma} t^{\beta\tau} t^{\gamma\tau} - r^{\alpha\tau} +\hat{t}^{\alpha\tau}~,~~~&\hat{t}^{\alpha\tau}&\mapsto -f^\alpha_{\beta\gamma}t^{\beta\tau}\hat{t}^{\gamma\tau}+\chi^{\alpha\tau}(t,\hat{t})+\hat{r}^{\alpha\tau}~,\\
            r^{\alpha\tau} &\mapsto -f^\alpha_{\beta\gamma} t^{\beta\tau}r^{\gamma\tau} +\chi^{\alpha\tau}(t,\hat t)+ \hat{r}^{\alpha\tau}~,~
            &\hat{r}^{\alpha\tau}&\mapsto0~,\\
            r_0 &\mapsto 2\int_0^1\dd\tau\,\kappa_{\alpha\beta}t^{\alpha\tau} \dot r^{\beta\tau}+\chi(\dot t,\hat t)+\hat{r}_0&\hat{r}_0&\mapsto -\chi(\dot{\hat t},\hat t)~,
        \end{aligned}
    \end{equation}
    where we introduced a function $\chi$ with components
    \begin{subequations}
    \begin{align}
        \chi^{\alpha\tau}(t,\hat t)&\coloneqq f^\alpha_{\beta\gamma}(t^{\beta\tau}\hat t^{\gamma\tau}-\ell(\tau)t^{\beta 1}\hat t^{\gamma 1})~,\\
        \chi(\dot t,\hat{t})&\coloneqq 2\int_0^1\dd\tau\,\kappa_{\alpha\beta} \dot t^{\alpha\tau} \hat t^{\beta\tau}~,\\
        \chi(\dot{\hat t},\hat{t})&\coloneqq 2\int_0^1\dd\tau\,\kappa_{\alpha\beta} \dot{\hat t}^{\alpha\tau} \hat t^{\beta\tau}~,
    \end{align}
    \end{subequations}
    and where $\ell(\tau)$ is an arbitrary smooth function $\ell\colon[0,1]\to [0,1]$ with $\ell(0)=0$ and $\ell(1)=1$. The kinematical data encoded in a morphism $\sW_{\rm adj}(\astringl(\frg))\to \sW(U)$ is then
    \begin{subequations}\label{eq:adjusted_loop_fields}
        \begin{align}
            A&\in \Omega^1(U)\otimes P_0\frg~,\\
            B&\in\Omega^2(U)\otimes \hat L_0\frg~,\\
            F&\coloneqq \dd A+\tfrac12 [A,A]+\mu_1(B)~, &
            \dd F+[A,F]-\mu_1(\chi(A,F))&=\mu_1(H)~,\label{eq:fake_curvature_bianchi_identity}\\
            H&\coloneqq \dd B+\mu_2(A,B)-\chi(A,F)~, &
            \dd H+\chi(F,F)&=0\label{eq:H_bianchi_identity}
        \end{align}
    \end{subequations}
    with gauge transformations
    \begin{subequations}
        \begin{align}
            \delta A&=\dd \alpha+\mu_2(A,\alpha)+\mu_1(\Lambda)~,\\
            \delta B&=\dd \Lambda+\mu_2(A,\Lambda)+\mu_2(\alpha,B)-\chi(\alpha,F)~,\\
            \delta F&=-\mu_1(\chi(\alpha,F))-\mu_2(F,\alpha)~,\\
            \delta H&=0~,
        \end{align}
    \end{subequations}
    where the gauge transformations are parameterised by elements
    \begin{equation}
        \alpha\in\Omega^0(U)\otimes P_0\frg\eand
        \Lambda\in\Omega^1(U)\otimes \hat L_0 \frg
    \end{equation}
    and where $\chi$ is here the function
    \begin{equation}\label{eq:ext-kappa-definition}
        \begin{aligned}
            \chi & \colon &P_0\frg \times P_0\frg &\to \hat L_0\frg \\
            &&(\gamma_1,\gamma_2)&\mapsto \left([\gamma_1,\gamma_2] - \ell(\tau)\partial([\gamma_1,\gamma_2]), 2\int_0^1\dd\tau \langle\dot\gamma_1,\gamma_2\rangle \right).
        \end{aligned}
    \end{equation}
    
    If we now look at just the transformations parameterised by $\alpha$ and trivial $\Lambda$, the various fields transform under different \(P_0\sG\)-representations, as a result of the adjustment. For example, \(H\), which before adjustment transformed under the adjoint representation of \(\sG\), is now invariant. Similarly, the fake curvature \(F\) now transforms differently, and the covariant derivative acts on it as
    \begin{equation}
        \mathrm d_AF \coloneqq \dd F+[A,F] - \mu_1(\chi(A,F))~,
    \end{equation}
    which can be seen from~\eqref{eq:fake_curvature_bianchi_identity}. The 2-form potential \(B\), which used to transform on its own, now forms a multiplet with \(F\), unlike in the unadjusted case.\footnote{After adjustment, the fake curvature \(F\) still transforms as a representation on its own, but \(B\) only forms a representation together with \(F\).} This reflects the fact that the adjustment of the Weil algebra requires an adjustment of the 2-crossed module (in which the parallel transport functor takes value) encoding the representations.
    
    The advantage of the crossed module of Lie algebras $\astring_{\rm lp}(\frg)$ over the 2-term $L_\infty$-algebra $\astring_{\rm sk}(\frg)$ is now that it readily integrates to the crossed module of Lie groups
    \begin{equation}
        \sString_{\rm lp,\cm}(\sG)=(L_0\sG\rightarrow P_0\sG)~.
    \end{equation}
    The integration of $\astring_{\rm sk}(\frg)$ is much harder; see~\cite{Schommer-Pries:0911.2483,Demessie:2016ieh}.

    \paragraph{The loop model of a Lie algebra.}
    There is an interesting truncation in the loop model of the string Lie 2-algebra, namely the 2-term \(L_\infty\)-algebra
    \begin{equation}\label{eq:loop_model_of_Lie_algebra}
        \frg_{\rm lp}\coloneqq  (L_0\frg\hooklongrightarrow P_0\frg)~.
    \end{equation}
    This 2-term \(L_\infty\)-algebra is quasi-isomorphic to the Lie algebra $\frg$; see \cref{app:Equivalence}. Together with the adjustment, it also allows us to interpret the connection on an ordinary principal fiber bundle as a connection on a principal 2-bundle~\cite{Saemann:2017rjm}.
    We can construct a $\frg_{\rm lp}$-valued connection \((A_\text{lp},B_\text{lp})\) from a $\frg$-connection \(A\in\Omega^1(M)\otimes\frg\) as
    \begin{equation}
        A_{\rm lp}=A\ell(\tau)~,~~~B_{\rm lp}=\tfrac12[A,A](\ell(\tau)-\ell^2(\tau))~.
    \end{equation}
    Note that 
    \begin{equation}
        F_{\rm lp}=\dd A_{\rm lp}+\tfrac12[A_{\rm lp},A_{\rm lp}]+\sft(B_{\rm lp})=\ell(\tau)F_{\rm sk}~. 
    \end{equation}
    Infinitesimal gauge transformations translate according to
    \begin{equation}
        \alpha_{\rm lp}=\alpha_{\rm sk}\ell(\tau)\eand \Lambda_{\rm lp}=[\alpha_{\rm sk},A_{\rm sk}](\ell(\tau)-\ell^2(\tau))~.
    \end{equation}
    Thus, gauge transformations are mapped to gauge transformations and gauge orbits are mapped to gauge orbits. The inverse map is the endpoint evaluation map $\dpar\colon P_0\frg\to \frg$:
    \begin{equation}
        A_{\rm sk}=\dpar A_{\rm lp}\eand \alpha_{\rm sk}=\dpar \alpha_{\rm lp}~.
    \end{equation}
    
    We use both 2-term \(L_\infty\)-algebras $\astringl(\frg)$ and $\frg_{\rm lp}$ as examples for our further discussion leading to an adjusted parallel transport.

    \section{Adjusted Weil algebras and inner derivations}
    Recall from \cref{app:inner_weil} that the Weil algebra can be interpreted as the inner derivation 2-crossed module of Lie algebras, and the exponentials of potentials and curvatures take values in the Lie 2-group corresponding to this 2-crossed module. After we adjust the Weil algebra, we need to construct the corresponding adjusted 2-crossed module and the Lie 2-group. This is a prerequisite to discussing the parallel transport functor, which takes values in this 2-group. 
    
    \subsection{Example: loop model of a Lie algebra}\label{ssec:loop_model_of_Lie_algebra}
    
    Before treating the adjusted Weil algebra of the string Lie 2-algebra, we first consider the simpler example of the adjusted and unadjusted Weil algebras of the 2-term \(L_\infty\)-algebra $\frg_{\rm lp}\coloneqq L_0\frg\overset\sft\to P_0\frg$, which is quasi-isomorphic to the Lie (1-)algebra $\frg$; see~\eqref{eq:quasi-iso-morphs}.
    
    The Weil algebra $\sW(\frg_{\rm lp})$ is generated by coordinate functions $(t^{\alpha \tau},r^{\alpha \tau},\hat t^{\alpha \tau},\hat r^{\alpha \tau})$, cf.~the similar parameterisation of $\sW(\astring_{\rm lp}(\frg))$ in~\eqref{eq:Weil_string_Lie_2_loop}. We first perform the reparameterisation explained in the previous section, which amounts to the coordinate change 
    \begin{equation}\label{eq:coord_change}
        (t^{\alpha \tau},r^{\alpha \tau},\hat t^{\alpha \tau},\hat r^{\alpha \tau})\rightarrow(t^{\alpha \tau},r^{\alpha \tau},\tilde t^{\alpha \tau},\hat r^{\alpha \tau})\ewith\tilde t^{\alpha \tau}=\hat t^{\alpha \tau}-r^{\alpha \tau}~.
    \end{equation}
    This simplifies the differential of the unadjusted Weil algebra to
    \begin{equation}
        \begin{aligned}
            Q_\sW~&\colon~&t^{\alpha\tau} &\mapsto -\tfrac12 f^{\alpha}_{\beta\gamma} t^{\beta\tau} t^{\gamma\tau} +\tilde{t}^{\alpha\tau}~,~~~ 
            &\tilde{t}^{\alpha\tau}&\mapsto -f^\alpha_{\beta\gamma}t^{\beta\tau}\tilde{t}^{\gamma\tau}~,\\
            &&r^{\alpha\tau} &\mapsto -f^\alpha_{\beta\gamma} t^{\beta\tau}r^{\gamma\tau} + \hat{r}^{\alpha\tau}~,~~~
            &\hat{r}^{\alpha\tau}&\mapsto-f^\alpha_{\beta\gamma} t^{\beta\tau}\hat{r}^{\gamma\tau} + f^\alpha_{\beta\gamma} \tilde{t}^{\beta\tau}r^{\gamma\tau} ~.
        \end{aligned}
    \end{equation}
    The differential of the adjusted Weil algebra also simplifies to
    \begin{equation}
        \begin{aligned}
            Q_{\sW_{\rm adj}}\colon~t^{\alpha\tau} &\mapsto -\tfrac12 f^{\alpha}_{\beta\gamma} t^{\beta\tau} t^{\gamma\tau} + \tilde t^{\alpha \tau}~,~~~&\tilde{t}^{\alpha\tau}&\mapsto -f^\alpha_{\beta\gamma}t^{\beta\tau}\tilde {t}^{\gamma\tau}~,\\
            r^{\alpha\tau} &\mapsto f^\alpha_{\beta\gamma}(t^{\beta\tau}\tilde t^{\gamma\tau}-\ell(\tau)t^{\beta 1}\tilde t^{\gamma 1})+ \hat{r}^{\alpha\tau}~,~
            &\hat{r}^{\alpha\tau}&\mapsto0~.
        \end{aligned}
    \end{equation}
    
    We now focus on the adjusted Weil algebra, as the unadjusted case is trivially constructed following the discussions in section~\ref{ssec:inner_derivations} and \cref{app:inner_weil}. Dualisation to a 3-term $L_\infty$-algebra yields the complex of Lie algebras 
    \begin{equation}
        \begin{array}{cccccccc}
            \sW_\text{adj}(L_0\frg \to P_0\frg) &=(&L_0 \frg &\xrightarrow{~\mu_1~} & L_0\frg'\rtimes P_0\frg' & \xrightarrow{~\mu_1~} & P_0\frg&)\\
            &&\lambda &\xrightarrow{~\mu_1~} &(\lambda,0)\\
            &&&&(\lambda,\gamma)&\xrightarrow{~\mu_1~}&\gamma
        \end{array}
    \end{equation}
    endowed with binary products
    \begin{subequations}
        \begin{align}
            \mu_2&\colon&P_0\frg\wedge P_0\frg&\to P_0 \frg~, &
            (\gamma_1,\gamma_2)&\mapsto [\gamma_1,\gamma_2]~,
            \\
            &&P_0\frg\wedge(L_0\frg'\rtimes P_0\frg')&\to L_0\frg'\rtimes P_0\frg'~, &
            (\gamma_1,(\lambda_2,\gamma_2))&\mapsto\big(-\chi(\gamma_1,\gamma_2),[\gamma_1,\gamma_2]\big)~,\\
            &&P_0\frg\wedge L_0\frg &\to L_0\frg~, &
            (\gamma_1,\lambda_2)&\mapsto0~,
            \\
            &&(L_0\frg'\rtimes P_0'\frg)^{\times2}&\to L_0 \frg~, &
            \big((\lambda_1,\gamma_1),(\lambda_2,\gamma_2)\big)&\mapsto0~,
        \end{align}
    \end{subequations}
    where
    \begin{align}\label{eq:chi_loop_model_of_Lie_algebra}
        \chi&\colon & P_0\mathfrak g \times P_0\mathfrak g &\to L_0\mathfrak g~, & (\gamma_1,\gamma_2) & \mapsto [\gamma_1,\gamma_2] - \ell \cdot \partial[\gamma_1,\gamma_2]
    \end{align}
    is the projection of the Lie bracket of two paths to based loops.
    
    Just as in the unadjusted case (see~\cref{app:inner_weil}), the adjusted Weil algebra admits a lift to a 2-crossed module. There are, in fact, two possible 2-crossed modules of Lie algebras. Both options have the same underlying complex of graded vector spaces,
    \begin{equation}
        L_0 \frg \quad \overset\sft\longrightarrow \quad L_0\frg\oplus P_0\frg \quad\overset\sft\longrightarrow\quad P_0\frg~,
    \end{equation}
    but their Lie brackets, induced by a choice of the antisymmetric parts of the Peiffer liftings \(\llbracket-,-\rrbracket\), differ.
    
    The first option is fixed by imposing the ordinary Lie brackets on \(L_0\frg\) and \(L_0\frg\rtimes P_0\frg\). This determines the Peiffer brackets uniquely by~\eqref{eq:h_Lie_bracket} and~\eqref{eq:l_Lie_bracket}, as \(\mu_1\colon L_0\frg\to L_0\frg \rtimes P_0\frg\) is injective. All other compatibility relations hold, and the required Peiffer bracket is
    \begin{equation}
        \{(\lambda_1,\gamma_1),(\lambda_2,\gamma_2)\}
        =
        \llbracket(\lambda_1,\gamma_1),(\lambda_2,\gamma_2)\rrbracket
        =
        \chi(\lambda_1+\gamma_1,\lambda_2+\gamma_2)~,
    \end{equation}
    leading to the 2-crossed module
    \begin{equation}
        \ainn_{\rm adj}(\frg_{\rm lp})\coloneqq (L_0 \frg \quad \overset\sft\longrightarrow \quad L_0\frg\rtimes P_0\frg \quad\overset\sft\longrightarrow\quad P_0\frg)~.
    \end{equation}
    The two 2-crossed modules of Lie algebras $\ainn(\frg_{\rm lp})$ and $\ainn_{\rm adj}(\frg_{\rm lp})$ are not isomorphic as 2-crossed modules, but their underlying complexes of Lie algebras agree, including the Lie brackets. They differ in the Peiffer lifting and the actions of $P_0\frg$ on $L_0\frg\rtimes P_0\frg$ and $L_0\frg$.

    The second option arises from setting 
    \begin{equation}
        \{-,-\}=\llbracket-,-\rrbracket=0~,
    \end{equation}
    which is possible according to~corollary~\ref{cor:trivial_peiffer_lifting}, since \(\mu_2\colon (L_0\frg \rtimes P_0\frg)^{\wedge2}\to L_0\frg\) vanishes. This case is simpler, but it changes the Lie brackets of the components considerably. Let \(\overset\circ\frg\) denote the abelian Lie algebra over the vector space \(\frg\). Then this case corresponds to the 2-crossed module of Lie algebras
    \begin{equation}
        L_0 \overset\circ\frg \quad \overset\sft\longrightarrow \quad L_0\overset\circ\frg\rtimes P_0\frg \quad\overset\sft\longrightarrow\quad P_0\frg~.
    \end{equation}
    The Lie bracket on $L_0\frg$ vanishes by~\eqref{eq:l_Lie_bracket}, and the Lie bracket on $L_0\frg\oplus P_0\frg$ also becomes “more abelian” by~\eqref{eq:h_Lie_bracket}. 
    
    While both options are mathematically consistent, the second option “forgets” the natural structure of the path and loop spaces, and deviates too far from our original $L_\infty$-algebra. More importantly, only the first option seems possible after we extend \(L_0\frg\) to \(\hat L_0\frg\) for the string 2-algebra; see section~\ref{ssec:adjusted_inner_derivation}.
    Finally, it seems very significant that for the “correct” option, the antisymmetric part of the Peiffer lifting is \emph{precisely} the map \(\chi\), required for adjusting the Weil algebra, that also appears during the lifting of 3-term \(L_\infty\)-algebras to 2-crossed modules. This fact hints at a deeper connection between \(\llbracket-,-\rrbracket\) and $\chi$.
    
    We now integrate the 2-crossed module obtained from the first option,\footnote{The second option can also be straightforwardly integrated; this produces the 2-crossed module of Lie groups \((L_0\sG\to L_0\overset\circ\frg\rtimes P_0\sG\to P_0\sG)\), where the vector space \(L_0\overset\circ\frg\) is now interpreted as an abelian Fréchet--Lie group.} which is essentially straightforward\footnote{While general 3-term $L_\infty$-algebras are very hard to integrate, there is no difficulty or obstruction to integrating 2-crossed modules of Lie algebras~\cite[Theorem~10]{Martins:2009aa}. The fact that we deal with 2-crossed modules of Fréchet–Lie algebras is not a problem, since all of the components, being path or loop algebras on the Lie algebra \(\frg\), admit obvious integrations to path or loop algebras on the Lie group \(\sG\). See also~\cite{Baez:2005sn} for more details on Fr\'echet--Lie algebras and groups. The only possible ambiguity is the usual one involving the center/fundamental group, which amounts to consistently using the same integration \(\sG\) of the Lie algebra \(\frg\).}: : we simply have to integrate the Lie algebras in each component of the crossed module. The integration of the actions is then automatically compatible. A verification of the successful integration is then the straightforward differentiation. 
    
    For example, the crossed module of Lie algebras $\frg_{\rm lp}$ integrates to the crossed module of Lie groups $\sG_{\rm lp,\rm \cm}=(L_0\sG\xrightarrow{~\sft~} P_0\sG)$, with pointwise multiplication, pointwise action of $P_0\sG$ on $L_0\sG$ and $\sft$ being the embedding. Differentiation (by applying the tangent functor) directly recovers $\frg_{\rm lp}$. Correspondingly the 2-crossed module of Lie groups resulting from the integration of $\ainn_{\rm adj}(\frg_{\rm lp})$ is
    \begin{equation}
        \sInn_{\rm adj,\cm}(\sG_{\rm lp})\coloneqq(L_0 \sG\xrightarrow{~\sft~}L_0\sG\rtimes P_0\sG\xrightarrow{~\sft~}P_0\sG)
    \end{equation}
    with the given product structure and the evident pointwise $P_0\sG$-actions. The Peiffer lifting is fixed by the relation
    \begin{equation}
     \sft(\{h_1,h_2\})=h_1 h_2 h_1^{-1}(\sft(h_1)\acton h_2^{-1})~,
    \end{equation}
    cf.~\eqref{eq:global_peiffer_lifting_identity}, because $\sft$ is injective. The fact that $\sInn_{\rm adj,\cm}(\sG_{\rm lp})$ integrates $\ainn_{\rm adj}(\frg_{\rm lp})$ follows from straightforward differentiation.
    
    Without adjustment, we would have arrived at the 2-crossed module of Lie groups $\sInn_{\rm \cm}(\sG_{\rm lp})$. The difference between the latter and $\sInn_{\rm adj,\cm}(\sG_{\rm lp})$ is seen from the difference of the corresponding 2-crossed modules of Lie algebras: While the underlying normal complexes and the products in each degree agree, the Peiffer lifting and the action of $P_0\sG$ on $L_0 \sG\rtimes P_0 \sG$ and $L_0\sG$ are different. Since $\sft\colon L_0\sG\rightarrow L_0\sG\rtimes P_0\sG$ is injective, the \(P_0\sG\)-actions fix the Peiffer lifting. At the level of the corresponding monoidal 2-categories encoding the Gray groups $\sInn_{\rm \cm}(\sG_{\rm lp})$ and $\sInn_{\rm adj,\cm}(\sG_{\rm lp})$, we thus encounter the same globular structure. Also, there is no modification to the short exact sequence~\eqref{eq:ses_2-groupoids}.

    \subsection{Adjusted inner derivations of the string Lie 2-algebra}\label{ssec:adjusted_inner_derivation}

    We now readily construct the main example: the adjusted Weil algebras of the string Lie 2-algebra $\astringl(\frg)$ defined by the differential~\eqref{eq:Weil_string_Lie_2_loop_adj}. The coordinate change~\eqref{eq:coord_change} leads to the differential graded algebra
    \begin{equation}\label{eq:Weil_string_Lie_2_loop_adj_cc}
        \begin{aligned}
            Q_{\sW_{\rm adj}}\colon~t^{\alpha\tau} &\mapsto -\tfrac12 f^{\alpha}_{\beta\gamma} t^{\beta\tau} t^{\gamma\tau} + \tilde t^{\alpha \tau}~,~~~&\tilde{t}^{\alpha\tau}&\mapsto -f^\alpha_{\beta\gamma}t^{\beta\tau}\tilde {t}^{\gamma\tau}~,\\
            r^{\alpha\tau} &\mapsto f^\alpha_{\beta\gamma}(t^{\beta\tau}\tilde t^{\gamma\tau}-\ell(\tau)t^{\beta 1}\tilde t^{\gamma 1})+ \hat{r}^{\alpha\tau}~,~
            &\hat{r}^{\alpha\tau}&\mapsto0~,\\
            r_0 &\mapsto 2\int_0^1\dd\tau\,\kappa_{\alpha\beta} \dot t^{\alpha\tau} \tilde t^{\beta\tau}+\hat{r}_0~,&\hat{r}_0&\mapsto -2\int_0^1\dd\tau\,\kappa_{\alpha\beta} \dot{\tilde t}^{\alpha\tau}\tilde t^{\beta\tau}~.
        \end{aligned}
    \end{equation}
    Dually, we have the 3-term $L_\infty$-algebras $\sW^*_{\mathrm{adj}}(\astringl(\frg))$ with cochain complex
    \begin{equation}
        \begin{matrix}
            (&
            \hat L_0 \frg & \xrightarrow{~\mu_1~} & \hat L_0\frg'\rtimes P_0\frg' & \xrightarrow{~\mu_1~} & P_0\frg &)\\
            &\lambda+r & \xmapsto{~\mu_1~} &(\lambda+r,0)\\
            &&& (\lambda+r,\gamma) & \xmapsto{~\mu_1~} &\gamma
        \end{matrix} 
    \end{equation}
    which is endowed with the binary products \(\mu_2\)
    \begin{subequations}
        \begin{align}
            P_0\frg\wedge P_0\frg&\to P_0 \frg~, &
            (\gamma_1,\gamma_2)&\mapsto[\gamma_1,\gamma_2]~, \\
            P_0\frg\wedge(\hat L_0\frg'\rtimes P_0\frg')&\to \hat L_0\frg'\rtimes P_0\frg'~, &
            \big(\gamma_1,(\lambda_2+r_2,\gamma_2)\big)&\mapsto\left(-\chi(\gamma_1,\gamma_2),[\gamma_1,\gamma_2]\right)~, \\
            P_0\frg\wedge \hat L_0\frg &\to \hat L_0\frg~, &
            (\gamma_1,\lambda_2+r_2)&\mapsto0~, \\
            (\hat L_0\frg'\rtimes P_0\frg')^{\times2}&\to \hat L_0 \frg~, &
            \!\!\!\!\!\!\!\!\!\!\!\!\!\!
            \big((\lambda_1+r_1,\gamma_1),(\lambda_2+r_2,\gamma_2)\big)&\mapsto
            -\chi(\gamma_1,\gamma_2)-\chi(\gamma_2,\gamma_1)~,
        \end{align}
    \end{subequations}
    where \(\chi\) was defined in~\eqref{eq:ext-kappa-definition}.\footnote{This \(\chi\) is analogous to, but naturally different from, the \(\chi\) in~\eqref{eq:chi_loop_model_of_Lie_algebra} used in section~\ref{ssec:loop_model_of_Lie_algebra}.} The extension to a 2-crossed module of Lie algebras 
    \[
    \ainn_\text{adj}(\hat L_0\frg\to P_0\frg)=\big(\ainn_\text{adj}(\hat L_0\frg\to P_0\frg),\sft,\acton,\{-,-\}\big)
    \]
    has underlying cochain complex of Lie algebras
    \begin{equation}
        \begin{matrix}
            \ainn_\text{adj}(\hat L_0\frg\to P_0\frg) &=& (&
            \hat L_0\frg & \overset\sft\longrightarrow & \hat L_0\frg'\rtimes P_0\frg' & \overset\sft\longrightarrow & P_0\frg &)\\
            &&&\lambda+r & \xmapsto{~\sft~} & (\lambda+r,0) \\
            &&&&& (\lambda+r,\gamma) & \xmapsto{~\sft~} & \gamma
        \end{matrix} 
    \end{equation}
    with
    \begin{subequations}
        \begin{align}
            [\gamma_1,\gamma_2]_{P_0\frg}&=\mu_2(\gamma_1,\gamma_2)~,\\
            [(\lambda_1+r_1,\gamma_1),(\lambda_2+r_2,\gamma_2)]_{\hat L_0\frg\rtimes P_0\frg}&=
            \big([\lambda_1,\lambda_2]+[\gamma_1,\lambda_2]+[\lambda_1,\gamma_2]
            ,~[\gamma_1,\gamma_2]\big)\notag\\
            &=\big(\chi(\lambda_1+\gamma_1,\lambda_2+\gamma_2)-\chi(\gamma_1,\gamma_2)
            ,~[\gamma_1,\gamma_2]\big)~,\\
            [\lambda_1+r_1,\lambda_2+r_2]_{\hat L_0\frg}&=\chi(\lambda_1,\lambda_2)~,\\
            \gamma_1\acton(\lambda_2+r_2,\gamma_2)&=\mu_2\big(\gamma_1,(\lambda_2+r_2,\gamma_2)\big)~,\\
            \gamma_1\acton(\lambda_2+r_2)&=0~,\\
            \{(\lambda_1+r_1,\gamma_1),(\lambda_2+r_2,\gamma_2)\}&=\chi(\lambda_1+\gamma_1,\lambda_2+\gamma_2)~.
        \end{align}        
    \end{subequations}
    The Peiffer bracket is again precisely the function $\chi$ encoding the adjustment of the Weil algebra. Unlike the case of \(\frg_\text{lp}\) in section~\ref{ssec:loop_model_of_Lie_algebra}, here $\chi$ (and thus the Peiffer lifting $\{-,-\}$) is no longer purely antisymmetric, due to a boundary term. The symmetric part of the Peiffer bracket corresponds to the non-vanishing higher product \(\mu_2\colon(\hat L_0\frg\rtimes P_0\frg)^{\times2}\to\hat L\frg\). The antisymmetric part of the Peiffer bracket is the additional structure map $\llbracket-,-\rrbracket$ of the 2-crossed module of Lie algebras.
    
    Integrating $\ainn_{\rm adj}(\astring_{\rm lp}(\frg))$, we arrive at the 2-crossed module of Lie groups 
    \begin{equation}
        \begin{matrix}
            \sInn_{\text{adj},\cm}(\sString_{\rm lp}(\sG)) &=& (&
            \hat L_0 \sG & \overset\sft\longrightarrow & \hat L_0\sG'\rtimes P_0\sG' & \overset\sft\longrightarrow & P_0\sG  &)\\
            &&&(l,r) & \xmapsto{~\sft~} &\big((l,r),\unit_{P_0\sG}\big)\\
            &&&&& ((l,r),p) & \xmapsto{~\sft~} & p
        \end{matrix} 
    \end{equation}
    The $P_0\sG$-actions on the bases $L_0\sG$ and $L_0\sG\rtimes P_0\sG$ of the principal \(\sU(1)\)-bundles $\hat L_0 \sG$ and $\hat L_0\sG\rtimes P_0\sG$ are the same as in $\sInn_{\rm adj,\cm}(\sG_{\rm lp})$. The $P_0\sG$-action on the $\sU(1)$-fibers are the canonical ones as in the loop model of the string Lie 2-group. Explicit expressions are best constructed indirectly, after trivialising the circle bundles;\footnote{that is, the technique of constructing a nontrivial \(\sU(1)\)-bundle on a Fréchet–Lie group \(\sG\) as a quotient of the trivial \(\sU(1)\)-bundle on the path group \(P_0\sG\)} see~\cite{Baez:2005sn}.
    The Peiffer lifting is fixed by~\eqref{eq:global_peiffer_lifting_identity}:
    \begin{multline}
        \Big\{\big((l_1,r_1),p_1\big),\big((l_2,r_2),p_2\big)\Big\}=\\
        \big((l_1,r_1),p_1\big)\big((l_2,r_2),p_2\big)
        \big((l_1,r_1),p_1\big)^{-1}
        \left(p_1\acton\big((p_2,r_2),p_2\big)^{-1}\right)~,
    \end{multline}
    where all products are taken in the semidirect product \(\hat L_0\sG'\rtimes P_0\sG'\).
    
    The 3-group constructed from $\sInn_{\text{adj},\cm}(\sString_{\rm lp}(\sG))$ as in~\eqref{eq:mon_cat_from_3_group} is then 
    \begin{subequations}
    \begin{equation}
        \sInn_\text{adj}(\sString_{\rm lp}(\sG))=\left( 
            \hat L_0\sG\rtimes\big((\hat L_0'\sG\rtimes P_0\sG')\rtimes P_0\sG\big)   \rightrightarrows   (\hat L_0\sG'\rtimes P_0\sG')\rtimes P_0\sG)  \rightrightarrows P_0\sG \right)
    \end{equation}
    with the following globular structure.
    \begin{equation}
        \begin{tikzcd}[column sep=4.0cm,row sep=large]
            p_1p_2 & \ar[l, bend left=45, "{\big(\ell_2,p_1,p_2\big)}", ""{name=U,inner sep=1pt,above}] \ar[l, bend right=45, "{\big(\ell_1\ell_2,p_1,p_2\big)}", swap, ""{name=D,inner sep=1pt,below}] p_2
            \arrow[Rightarrow,from=U, to=D, "{\big(\ell_1,\ell_2,p_1,p_2\big)}",swap]
        \end{tikzcd}
    \end{equation}
    \end{subequations}
    
    We have a short exact sequence of 2-groupoids,
    \begin{equation}\label{eq:ses_groupoids_adjusted}
        *\xrightarrow{~~~}\sString_{\rm lp}(\sG)\xrightarrow{~\Upsilon~} \sInn_{\rm adj}(\sString_{\rm lp}(\sG)) \xrightarrow{~\Pi~}\sB\sString_{\rm lp}(\sG) \xrightarrow{~~~} *~,
    \end{equation}
    where the functors $\Upsilon$ and $\Pi$ are again given by the obvious embedding and projection functors. This is the adjusted analogue of~\eqref{eq:ses_2-groupoids}. As complexes of globular sets, this complex is identical to that in~\eqref{eq:ses_2-groupoids};\footnote{Of course, as 2-functors between 2-groupoids, the 2-functors \(\Upsilon\) and \(\Pi\) in~\eqref{eq:ses_groupoids_adjusted} are different from the 2-functors \(\Upsilon\) and \(\Pi\) in~\eqref{eq:ses_2-groupoids}, simply because the (co)domains are inequivalent 2-groupoids. As we are not much concerned with 2-groupoids beyond their globular structure, however, we abuse notation and do not notate the two pairs differently.} in particular the presentation~\eqref{eq:3-group_short_exact_sequence_components}, as well as the presentation~\eqref{eq:ses_2-groupoids_reparameterisation} with the reparameterisation~\eqref{eq:inner_algebra_alternative_basis}, continue to be valid after adjustment.
    
    \section{Parallel transport}\label{sec:parallel_transport}
    
    We now discuss the main topic of this chapter: the consistent definition of a higher, truly non-abelian parallel transport. The key features are already visible over local patches, and gluing the construction to a global one is, in principle, a mere technicality; see~e.g.~\cite{Wang:1512.08680}. For clarity of our discussion, we always work on local patches or, equivalently, a contractible manifold $U$.

    \subsection{Ordinary parallel transport and connections}\label{ssec:parallel}
    
    The fact that the holonomies around all smooth loops encode a connection has been known in the literature since at least the 1950s~\cite{Kobayashi:1954aa}. The picture we use is inspired by the treatment of loops in~\cite{Barrett:1991aj} (see also~\cite{Gambini:1980yz}), and generalised to paths in~\cite{Caetano:1993zf} (see also~\cite{Schreiber:0705.0452}).
    
    Let \(\sG\) be a Lie group. Parallel transport encoded in a connection on a principal $\sG$-bundle $P$ over the contractible manifold $U$ amounts to an assignment of a group element $g\in \sG$ to each path $\gamma\colon[0,1]\rightarrow U$ in the base manifold. Composition of paths translates to multiplication of the corresponding group elements.
    The paths and points of the base manifold naturally combine to the path groupoid $\CP U$, as defined in \cref{app:path_space}. Regarding $\sG$ as the one-object groupoid $\sB\sG=(\sG\rightrightarrows *)$, we see that parallel transport is precisely a functor
    \begin{equation}
        \Phi\colon \CP U\longrightarrow \sB\sG~,\hspace{2cm}
        \begin{tikzcd}
            \mbox{paths} \arrow[d,shift left] \arrow[d,shift right]\arrow[r]{}{\Phi_1} & \sG \arrow[d,shift left] \arrow[d,shift right]\\
            U \arrow[r]{}{\Phi_0} & * 
        \end{tikzcd}
    \end{equation}
    
    Given a connection in terms of a $\frg\coloneqq \sLie(\sG)$-valued 1-form $A$ on $U$, we can construct the parallel transport functor as
    \begin{equation}
        \Phi_1(\gamma)=\Pexp\int_\gamma A\in \sG~,
    \end{equation}
    where $\Pexp(\dots)$ is the path-ordered exponential well-known in physics. Mathematically, $\Phi_1(\gamma)=g(1)$, where $g$ is the (unique) solution $g(t)$ to the differential equation\footnote{given here for clarity for matrix Lie groups, the abstract analogue being evident}
    \begin{equation}\label{eq:path_ordered_ode}
        \left(\frac{\dd}{\dd t} g(t)\right)g(t)^{-1}=-\iota_{\dot \gamma(t)}A(\gamma(t)) ~,~~~g(0)=\unit_\sG~,
    \end{equation}
    where $\iota_{\dot \gamma(t)}$ denotes the contraction with the tangent vector to $\gamma$ at $\gamma(t)$. Conversely, given a functor $\Phi$, the corresponding connection \(A\in\Omega^1(U)\otimes\frg\) is obtained as follows. Let $x\in U$ be a point and $v\in T_xU$ a tangent vector at $x$. We choose a path $\gamma\colon[0,1]\rightarrow U$ such that $\gamma(\tfrac12)=x$ and $\dot \gamma(\tfrac12)=v$. Just as any general path, $\gamma$ gives rise to a function $g(t)=\Phi(\gamma_t)\colon[0,1]\rightarrow \sG$, where $\gamma_t$ is the truncation of $\gamma$ at $\gamma(t)$ (with appropriate reparameterisation). We can then use equation~\eqref{eq:path_ordered_ode} and define
    \begin{equation}
        -\iota_v A(x)=\left.\left(\frac{\dd}{\dd t} g(t)\right)g(t)^{-1}\right|_{t=\tfrac12}~,
    \end{equation}
    where $A$ is independent of the choice of $\gamma$ and the reparameterisation in the truncation of $\gamma$ to $\gamma_t$. Thus the parallel transport functor \(\Phi\) contains exactly the same information as the connection \(A\).    
    
    Since connections correspond to functors, it is rather obvious that gauge transformations correspond to natural transformations.\footnote{In general, for functors between general categories, one distinguishes between \emph{natural transformations} and \emph{natural isomorphisms}, where the latter is a natural transformation whose components are all isomorphisms. For functors between groupoids, as in our case, all natural transformations are natural isomorphisms.}\footnote{Gauge transformations can be thought of in two different but equivalent perspectives: the “physicist’s”, where the gauge fields and Wilson lines are objects defined on the manifold that are acted upon by gauge transformations; and the “mathematician’s”, where the gauge fields are invariant objects defined on the total space of principal bundles, where the apparent gauge transformations correspond to different local trivialisations of the principal bundle. In this chapter, we work with an already locally trivialised bundle, so that the formulae appear as actions of the gauge transformations; but they can be equally well interpreted as the result of the gauge transformations’ changing the choice of local trivialisation.} A natural transformation $\eta\colon\Phi\Rightarrow\tilde \Phi$ between two functors of Lie groupoids $\Phi,~\tilde\Phi\colon\CP U\to \sB\sG$ is encoded in a function $\eta\colon U\to \sG$ such that 
    \begin{equation}
        \tilde \Phi_1(\gamma)=\eta(\gamma(1))^{-1}\Phi_1(\gamma)\eta(\gamma(0))
    \end{equation}
    for each path $\gamma$. This is precisely the gauge transformation law for a Wilson line. Let $A$ and $\tilde A$ be the connection 1-forms associated with $\Phi$ and $\tilde \Phi$, respectively. The functions $g(t)$ and $\tilde g(t)$ appearing in equation~\eqref{eq:path_ordered_ode}, are related by
    \begin{equation}
        \tilde g(t)=\eta(\gamma(t))^{-1} g(t)\eta(\gamma(0))~,
    \end{equation}
    and equation~\eqref{eq:path_ordered_ode} for $\tilde A$ induces then the usual gauge transformations,
    \begin{equation}
        \tilde A(x)=\eta(x)^{-1} A(x) \eta(x)+\eta(x)^{-1} \dd \eta(x)~.
    \end{equation}
    
    Altogether, the parallel transport functor is kinematically omniscient: it contains all information about gauge configurations and gauge orbits.

    \subsection{Ordinary parallel transport and the derivative parallel transport functor}\label{ssec:derived_parallel_ordinary}
    
    To see the curvature 2-form $F=\dd A+\tfrac12[A,A]$ of $A$ arise from parallel transport, we trivially extend $\Phi$ to a strict 2-functor $\varPhi$ as follows. First, we extend the path groupoid $\CP U$ to a path 2-groupoid $\CP_{(2)} U$, whose objects are the points of $U$, whose 1-morphisms are the paths, and whose 2-morphisms between two paths $\gamma_1,\gamma_2\colon x\to y$ are bigons, as defined in \cref{app:path_space}.
    Similarly, we extend $\sB\sG$ to 
    \begin{equation}
        \sB\sInn(\sG) = (\sG \rtimes \sG \rightrightarrows \sG \rightrightarrows *)~,
    \end{equation}
    which is a 2-groupoid with one object $*$, over which we have the morphism 2-group $\sInn(\sG)$. As explained in section~\ref{ssec:inner_group}, this is the action groupoid for the action of $\sG$ onto itself by left-multiplication with morphisms
    \begin{equation}
        g_1g_2 \xleftarrow{~~(g_1,g_2)~~} g_2~.
    \end{equation}
    Then we can construct the \emph{derivative parallel transport 2-functor}\footnote{not to be confused with the unrelated concept of derived functors in homological algebra}~\cite{Schreiber:0802.0663}, which is a strict 2-functor
    \begin{equation}
        \varPhi\colon \CP_{(2)} U\longrightarrow \sB\sInn(\sG)~,\hspace{2cm}
        \begin{tikzcd}
            \text{surfaces} \rar["\varPhi_2"] \dar[shift left] \dar[shift right] & \sG \rtimes \sG \dar[shift left] \dar[shift right] \\
            \text{paths} \arrow[d,shift left] \arrow[d,shift right]\arrow[r]{}{\varPhi_1} & \sG \arrow[d,shift left] \arrow[d,shift right]\\
            U \arrow[r]{}{\varPhi_0} & * 
        \end{tikzcd}
    \end{equation}
    It assigns to each path $\gamma$ an element $\varPhi_1(\gamma)=g_\gamma$ in $\sG$ and to each surface $\sigma$ an element $\varPhi(\sigma)=(g^1_\sigma,g^2_\sigma)$ in $\sG\rtimes \sG$, as follows: 
    \begin{equation}
        \begin{tikzcd}[column sep=2.5cm,row sep=large]
            x_2 & \ar[l, bend left=45, "\gamma_1", ""{name=U,inner sep=1pt,above}] \ar[l, bend right=45, "\gamma_2", swap, ""{name=D,inner sep=1pt,below}] x_1
            \arrow[Rightarrow,from=U, to=D, "\sigma",swap]
        \end{tikzcd}~~~\overset\varPhi\longmapsto~~~
        \begin{tikzcd}[column sep=2.5cm,row sep=large]
            * & \ar[l, bend left=45, "g_{\gamma_1}", ""{name=U,inner sep=1pt,above}] \ar[l, bend right=45, "g_{\gamma_2}", swap, ""{name=D,inner sep=1pt,below}] *
            \arrow[Rightarrow,from=U, to=D, "{(g^1_\sigma,g^2_\sigma)}",swap]
        \end{tikzcd}
    \end{equation}
    Compatibility with the domain and codomain maps \(\dom,\codom\) implies that
    \begin{equation}
        g_{\gamma_1}=\dom(\varPhi(\sigma))=g^2_\sigma\eand g_{\gamma_2}=\codom(\varPhi(\sigma))=g^1_\sigma g^2_\sigma~.
    \end{equation}
    Thus $\varPhi(\sigma)$ is fully fixed by the $g_\gamma$, and the strict 2-functor $\varPhi$ is determined by the \mbox{(1-)functor} $\Phi$.
    
    At an infinitesimal level, the additional data for surfaces encodes the curvature, and $\varPhi$ being determined by $\Phi$ amounts to a non-abelian version of Stokes' theorem~\cite[Section~3.2]{Schreiber:0802.0663}. In terms of the component fields \(A\in\Omega^1(U)\otimes\frg\) and its curvature \(F\in\Omega^2(U)\otimes\frg=\dd A+\tfrac12[A,A]\), we can write $g_\gamma$ and $g_\sigma$ as 
    \begin{equation}
        g_\gamma = \Pexp\int_\gamma A\eand g_\sigma = \Pexp\ichint_{A}(-F)~.
    \end{equation}
    The additional minus sign in front of the curvature $F$ arises from the globular identities and is explained below. The second integral is a path-ordered integral over a path in path space, and $\chint_A(-F)$ is a \emph{Chen form} as described in \cref{app:chen_forms}. Briefly, we view \(\sigma\) as a path \(\check\sigma\) on the space of paths between two points on the boundary of $\sigma$, \(x_0\) and \(x_1\), and the 2-form \(F\) as a 1-form \(-\check F = \chint_A(-F)\) on the space of paths between \(x_0\) and \(x_1\). Then
    \[
    \Pexp\ichint_{A} (-F) \coloneqq \Pexp\int_{\check\sigma}(-\check F)~.
    \]
    This is now of course equivalent to a differential equation on the path space.
    
    Given a bigon \(\sigma\colon\gamma_1\to\gamma_2\), since \(\dpar\sigma = \gamma_1\cup\bar\gamma_2\), the globular identity
    \begin{equation}
        g_{\gamma_1}g_{\gamma_2}^{-1} = g_\sigma^{-1}
    \end{equation}
    becomes
    \begin{equation}
        \Pexp\oint_{\dpar\sigma}A=\Pexp\ichint_{A} F~,
    \end{equation}
    where \(F=\dd A+\tfrac12[A,A]\) is the ordinary curvature and
    \begin{equation}
        g_\sigma = \Pexp\ichint_{A}(-F)~.
    \end{equation}
    
    Conversely, we can recover the fields and curvatures from the derivative parallel transport 2-functor \(\varPhi\). We have already explained how to recover \(A\) as above. As for \(F\), since \(\varPhi\) assigns elements of \(\sG\) to parameterised surfaces \(\sigma\), i.e.~paths \(\check\sigma\) in the space of paths between \(x_1\) and \(x_2\), we can do the same procedure as for \(A\) to recover the corresponding 1-form \(\check F\) on path space, and translate it to a 2-form \(F\) on \(U\).
    
    We now discuss gauge transformations. Just as for the plain parallel transport functor, we should identify gauge transformations with natural transformations. The general notion of 2-natural transformations between 2-functors between 2-groupoids is that of \emph{pseudonatural transformations}.\footnote{For 2-natural transformations between 2-functors between general 2-categories, one distinguishes between \emph{lax natural 2-transformations}, whose component 2-cells need not be invertible, and \emph{weak natural 2-transformations} or {\em pseudonatural transformations}, whose component 2-morphisms must be invertible (but not necessarily trivial) by definition; see e.g.~\cite{Jurco:2014mva}. However, for 2-groupoids, all 2-morphisms are invertible, and the two classes coincide.} A pseudonatural transformation $\eta\colon\varPhi\rightarrow \tilde \varPhi$ between two strict 2-functors $\varPhi,~\tilde \varPhi\colon \CP_{(2)}U\rightarrow \sB\sInn(\sG)$ is encoded in maps
    \begin{equation}
        \eta_1\colon \CP_{(2)}U_0=U\mapsto \sG\eand \eta_2\colon \CP_{(2)}U_1\mapsto \sG\rtimes \sG~,
    \end{equation}
    where $\CP_{(2)}U_1$ are the paths or 1-morphisms in $\CP_{(2)}U$, such that for each path $x_1\xleftarrow{~~\gamma~~}x_0$, we have the commuting diagram
    \begin{equation}
        \begin{tikzcd}[column sep=2.5cm,row sep=large]
            * \ar[d,swap]{}{\eta_1(x_1)} & * \ar[l,swap]{}{\varPhi(\gamma)}\ar[d]{}{\eta_1(x_0)}\\
            * & \ar[l]{}{\tilde\varPhi(\gamma)}* \ar[Rightarrow,ul,swap]{}{\eta_2(\gamma)}
        \end{tikzcd}
    \end{equation}
    implying that 
    \begin{equation}
        \eta_2(\gamma)=\big(\eta_1(x_1)\varPhi(\gamma)\eta_1^{-1}(x_0)\tilde \varPhi^{-1}(\gamma) ,\tilde \varPhi(\gamma)\eta_1(x_0)\big)\in \sG\rtimes \sG~.
    \end{equation}
    The coherence axioms for a pseudonatural transformation are then automatically satisfied. 
    
    The additional freedom in the gauge transformations allows for a pseudonatural transformation $\eta$ between any strict 2-functor $\varPhi$ and the trivial strict 2-functor $\unit$
    \begin{equation}
        \unit(x)=*~,~~~\unit(\gamma)=\unit_\sG~,~~~\unit(\sigma)=(\unit_\sG,\unit_\sG)
    \end{equation}
    for all $x\in \CP_{(2)}U_0$, $\gamma\in \CP_{(2)}U_1$ and $\sigma\in \CP_{(2)}U_2$. Explicitly, $\eta$ is given by 
    \begin{equation}
        \eta_1(x)=\unit_\sG\eand\eta_2(\gamma)=(\varPhi(\gamma),\unit_\sG)~.
    \end{equation}
    This transformation reflects the fact that $\sInn(\sG)$ is equivalent to the trivial 2-group and that $\sB\sInn(\sG)$ is equivalent to the trivial 3-groupoid.
    
    We thus need to restrict the allowed gauge transformations in an obvious way. The short exact sequence~\eqref{ses:groupoids_Lie} leads to the following commutative diagram:
    \begin{equation}
        \begin{tikzcd}
            \CP U\ar[d,"\Phi",swap] \ar[r,hookrightarrow] & \CP_{(2)}U  \ar[d,"\varPhi"] \\
            \sB \sG \ar[r,hookrightarrow] & \sB \sInn(\sG)
        \end{tikzcd}
    \end{equation}    Furthermore, if we fix endpoints \(x_0,x_1\in U\), we can decategorify\footnote{in the sense of taking hom-categories, thus shifting 1-morphisms to be objects and 2-morphisms to be 1-morphisms} the above diagram, and add a new functor \(\varPhi_\mathrm{curv}(x_0,x_1)\), which is a truncation of \(\varPhi(x_0,x_1)\) to surfaces only:
    \begin{equation}
        \begin{tikzcd}
            \CP U(x_0,x_1)\ar[d,"{\Phi(x_0,x_1)}",swap] \ar[r,hookrightarrow] & \CP_{(2)}U(x_0,x_1)  \ar[d,"{\varPhi(x_0,x_1)}"']  \ar[dr,"{\varPhi_\mathrm{curv}(x_0,x_1)}"] \\
            \sG \ar[r,hookrightarrow] & \sInn(\sG) \ar[r,twoheadrightarrow,"\Pi"] & \sB\sG
        \end{tikzcd}
    \end{equation}
    Here, $\sG$ is regarded as the discrete category $\sG\rightrightarrows\sG$. Note that the decategorification is necessary because \(\sB\sB\sG\) does not make sense as a 2-category in general: there is no compatible monoidal product for non-abelian $\sG$ due to the Eckmann--Hilton argument. The functor \(\varPhi_\text{curv}(x_0,x_1)\) therefore does \emph{not} extend to a 2-functor \(\varPhi\colon\CP_{(2)}U\to\sB\sB\sG\). 
    
    Clearly, we are only interested in transformations of $\varPhi$ that originate from transformations of $\Phi$ and which become trivial\footnote{This does {\em not} imply that the curvatures do not transform under gauge transformations.} on $\varPhi_{\rm curv}$. That is, for any two derivative parallel transport 2-functors $\varPhi,\tilde \varPhi$ connected by such a transformation, we have
    \begin{equation}
        \varPhi_{\rm curv}(x_0,x_1)=\Pi\circ \varPhi(x_0,x_1)=\Pi\circ \tilde \varPhi(x_0,x_1)=\tilde \varPhi_{\rm curv}(x_0,x_1)~.
    \end{equation}
    Equivalently, these natural transformations are rendered trivial by the whiskering\footnote{{\em Whiskering} is the horizontal composition of a trivial 2-morphism, here $\id_\Pi\colon\Pi\Rightarrow \Pi$ in the higher category of 2-groupoids, 2-functors and 2-natural isomorphisms, with another 2-morphism, here $\eta$.}
    \begin{equation}
        \begin{tikzcd}
            \sB\sG & \sInn(\sG) \lar["\Pi",swap]  &  \CP_{(2)} U(x_0,x_1) \lar[bend left=50, "{\varPhi(x_0,x_1)}", ""'{name=A}] \lar[bend right=50, "{\tilde \varPhi(x_0,x_1)}"', ""{name=B}] \ar[from=A, to=B, Rightarrow,"\eta"]
        \end{tikzcd}
    \end{equation}
    This is simply achieved by demanding that $\eta_2$ be trivial:
    \begin{equation}
        \eta_2(\gamma)=\big(\unit,\tilde \Phi(\gamma)\eta_1(x_0)\big)~.
    \end{equation}
    Such natural transformations are known as {\em strict 2-natural transformations}.

    \subsection{Unadjusted higher parallel transport and connections}\label{ssec:derived_parallel_higher}
    
    Higher-dimensional generalisations of parallel transport have been studied since the 1990s. First discussions for higher principal bundles are found in~\cite{0817647309}; appropriate higher path spaces where discussed in~\cite{caetano1998family}. The higher-dimensional parallel transport for abelian higher principal bundles was then fully developed in~\cite{Gajer:1997:155-207,Gajer:1999:195-235,Mackaay:2000ac}. The non-abelian extension was discussed in~\cite{Chepelev:2001mg},~\cite{Baez:2002jn}~\cite{Girelli:2003ev},~\cite{Baez:2004in} and further, in great detail, in the papers~\cite{Schreiber:0705.0452,Schreiber:0802.0663,Schreiber:2008aa}; see also~\cite{Alvarez:1997ma} for earlier considerations and~\cite{Li:2019nqu} for a recent discussion. We also need the structures underlying the higher parallel transport along volumes, discussed in~\cite{Martins:2009aa}.
    
    Let $\CCG$ be a strict Lie 2-group with underlying monoidal category $(\sH\rtimes \sG\rightrightarrows \sG)$ with morphisms 
    \begin{equation}
        \sft(h)g\xleftarrow{~~(h,g)~~}g~;
    \end{equation}
    the corresponding crossed module of Lie groups is $\CCG_{\rm \cm}=(\sH\xrightarrow{~\sft~}\sG,\acton)$, cf.~\cref{app:hypercrossed_modules}. Parallel transport over a local patch $U$ with gauge 2-group $\CCG$ is then described by strict 2-functors from the path 2-groupoid $\CP_{(2)} U$ to $\sB\CCG$,
    \begin{equation}
        \Phi\colon \CP_{(2)}U \to \sB\CCG~,\hspace{2cm}
        \begin{tikzcd}
            \text{surfaces} \rar["\Phi_2"] \dar[shift left] \dar[shift right] & \sH\rtimes\sG \dar[shift left] \dar[shift right] \\
            \text{paths} \rar["\Phi_1"] \dar[shift left] \dar[shift right] & \sG \dar[shift left] \dar[shift right] \\
            U \rar["\Phi_0"] & *
        \end{tikzcd}
    \end{equation}
    which assign to each path $\gamma$ a group element $g_\gamma\in \sG$ and to each surface $\sigma$ a group element $\Phi(\sigma)=(h_\sigma,g_\sigma)\in \sH\rtimes \sG$:
    \begin{equation}
        \begin{tikzcd}[column sep=2.5cm,row sep=large]
            x_2 & \ar[l, bend left=45, "\gamma_1", ""{name=U,inner sep=1pt,above}] \ar[l, bend right=45, "\gamma_2", swap, ""{name=D,inner sep=1pt,below}] x_1
            \arrow[Rightarrow,from=U, to=D, "\sigma",swap]
        \end{tikzcd}~~~\overset\Phi\longmapsto~~~
        \begin{tikzcd}[column sep=2.5cm,row sep=large]
            * & \ar[l, bend left=45, "g_{\gamma_1}", ""{name=U,inner sep=1pt,above}] \ar[l, bend right=45, "g_{\gamma_2}", swap, ""{name=D,inner sep=1pt,below}] *
            \arrow[Rightarrow,from=U, to=D, "{(h_\sigma,g_\sigma)}",swap]
        \end{tikzcd}
    \end{equation}
    Compatibility with domain and codomain maps in the morphism categories amounts to
    \begin{equation}
        g_{\gamma_1}=g_\sigma\eand g_{\gamma_2}=\sft(h_\sigma)g_\sigma~.
    \end{equation}
    Let \(\frg\) and \(\frh\) be the Lie algebras of \(\sG\) and \(\sH\), respectively. Then, the kinematical data consists of fields
    \begin{equation}
        A\in\Omega^1(U)\otimes\frg\eand B\in\Omega^2(U)\otimes\frh
    \end{equation}
    and their relation to the parallel transport functor is given by
    \begin{equation}
        g_\gamma=\Pexp\int_\gamma A\eand
        h_\sigma = \Pexp\ichint_A B~,
    \end{equation}
    where $\check B=\chint_A B$ is again a Chen form; see~\cref{app:chen_forms}. Part of the data defining this Chen form is the \(P_0\sG\)-representation of \(B\) (which is part of the data of the crossed module \(\hat L_0\sG\xrightarrow\sft P_0\sG\)) as well as the \(P_0\sG\)-connection \(A\).
    
    In general, the globular structure of the codomain of the 2-functor (in this case, the crossed module of Lie groups $\CCG$) translate to (possibly non-abelian) Stokes’ theorems on the curvatures. In this case, the globular structure requires that the condition known as \emph{fake flatness} holds, namely
    \begin{equation}
        F = \mathrm dA+\tfrac12[A,A]+\mu_1(B) = 0~. 
    \end{equation}
    
    To derive this, one needs some technical setup. The crux of the argument, however, is simple to describe. The identity
    \begin{equation}
        g_{\gamma_1}g_{\gamma_2}^{-1} = \sft(h_\sigma^{-1})
    \end{equation}
    for $\dpar\sigma=\gamma_1\cup\bar\gamma_2$ translates to
    \begin{equation}
        \Pexp\int_{\dpar \sigma}A = \sft\left(\Pexp\ichint_A (-B)\right).
    \end{equation}
    By the non-abelian Stokes' theorem,
    \begin{equation}
        \Pexp\int_{\dpar\sigma}A = \Pexp\ichint_A(\mathrm dA+\tfrac12[A,A])~.
    \end{equation} 
    Since our closed surface was arbitrary, we get
    \begin{equation}\label{eq:fake_flatness}
        \mathrm dA+\tfrac12[A,A] = -\sft(B)~,
    \end{equation}
    and thus \(F \coloneqq \mathrm dA+\tfrac12[A,A] + \sft(B) = 0\). This sketch can be made rigorous~\cite{Baez:2004in} (see also~\cite{Schreiber:2008aa}) using Chen forms; see \cref{app:chen_forms} for details.
    
    In other words, the globular structure of the crossed module means that the parallel transport 2-functor induces a Stokes’ theorem that, unfortunately, renders all physical theories based on it essentially abelian, as reviewed in section~\ref{ssec:limitations}.
    
    Gauge transformations between two strict 2-functors $\Phi,~\tilde \Phi\colon \CP_{(2)}U\rightarrow \sB\CCG$ are again given by appropriate natural transformations, which are here the general pseudonatural transformations $\eta\colon\Phi\rightarrow \tilde \Phi$. These are encoded in maps
    \begin{equation}
        \eta_1\colon \CP_{(2)}U_0=U\mapsto \sG\eand \eta_2=(\eta_2^1,\eta_2^2)\colon \CP_{(2)}U_1\mapsto \sH\rtimes \sG~,
    \end{equation}
    where $\CP_{(2)}U_1$ are the paths or 1-morphisms in $\CP_{(2)}U$, such that for each path $x_1\xleftarrow{~~\gamma~~}x_0$, we have the commutative diagram
    \begin{equation}
        \begin{tikzcd}[column sep=2.5cm,row sep=large]
            * \ar[d,swap]{}{\eta_1(x_1)} & * \ar[l,swap]{}{\Phi(\gamma)}\ar[d]{}{\eta_1(x_0)}\\
            * & \ar[l]{}{\tilde\Phi(\gamma)}* \ar[Rightarrow,ul,swap]{}{\eta_2(\gamma)}
        \end{tikzcd}~~~\Rightarrow\hspace{1.5cm}
        \begin{aligned}
            \eta_2^2(\gamma)&=\tilde \Phi(\gamma)\eta_1(x_0)~,\\
            \sft(\eta_2^1(\gamma))\tilde \Phi(\gamma)\eta_1(x_0)&=\eta_1(x_1)\Phi(\gamma)~.
        \end{aligned}
    \end{equation}
    
    We also have higher-order natural transformations (sometimes called \emph{modifications}) between the pseudonatural transformations \(\eta,\tilde\eta\colon\Phi\Rightarrow\tilde\Phi\); these correspond to the fact that the gauge parameters themselves gauge-transform. 
    
    \subsection{Unadjusted higher derivative parallel transport}    

    To make the curvatures visible, we can again categorify once more and consider a strict 3-functor $\varPhi$ from the path 3-groupoid $\CP_{(3)} U$ to $\sB\sInn(\CCG)$. The path 3-groupoid $\CP_{(3)} U$ is the evident extension of the path 2-groupoid $\CP_{(2)}U$ by adding 3-morphisms consisting of 3-dimensional homotopies between pairs of bigons; for details see \cref{app:path_space}. The 3-groupoid $\sB\sInn(\CCG)$ has one object and its morphism 2-category is $\sInn(\CCG)$, as defined in section~\ref{ssec:inner_derivations}.
    \begin{equation}
        \varPhi\colon \CP_{(3)}U \to \sB\sInn(\CCG)~,\qquad
        \begin{tikzcd}
            \text{volumes} \rar["\varPhi_3"] \dar[shift left] \dar[shift right] & \sH \rtimes \big((\sH \rtimes \sG) \rtimes \sG\big) \dar[shift left] \dar[shift right] \\
            \text{surfaces} \rar["\varPhi_2"] \dar[shift left] \dar[shift right] & (\sH \rtimes \sG) \rtimes \sG \dar[shift left] \dar[shift right]\\
            \text{paths} \rar["\varPhi_1"] \dar[shift left] \dar[shift right] & \sG \dar[shift left] \dar[shift right]\\
            U \rar["\varPhi_0"] & *
        \end{tikzcd}
    \end{equation}
    
    Explicitly, the strict 3-functor $\varPhi$ therefore amounts to assignments
    \tikzset{Rightarrow/.style={double equal sign distance,>={Implies},->},triple/.style={-,preaction={draw,Rightarrow}},quad/.style={preaction={draw,Rightarrow,shorten >=0pt},shorten >=1pt,-,double,double distance=0.2pt}}
    \begin{equation}
        \begin{tikzcd}[column sep=3cm,row sep=large]
            x_2 & \ar[l, bend left=60, "\gamma_1", ""{name=U,inner sep=1pt,above}] \ar[l, bend right=60, "\gamma_2", swap, ""{name=D,inner sep=1pt,below}] x_1
            \arrow[Rightarrow, bend left=60,from=U, to=D, "\sigma_2",""{name=L,inner sep=1pt,right}]
            \arrow[Rightarrow, bend right=60,from=U, to=D, "\sigma_1",""{name=R,inner sep=1pt,left},swap]
            \arrow[triple,from=R,to=L,"\rho",swap]
        \end{tikzcd}~~~\overset{\varPhi}\longmapsto~~~
        \begin{tikzcd}[column sep=5cm,row sep=large]
            * & \ar[l, bend left=60, "\varPhi(\gamma_1)", ""{name=U,inner sep=1pt,above}] \ar[l, bend right=60, "\varPhi(\gamma_2)", swap, ""{name=D,inner sep=1pt,below}] *
            \arrow[Rightarrow, bend left=70,from=U, to=D, "{\varPhi(\sigma_2)}",""{name=L,inner sep=1pt,right}]
            \arrow[Rightarrow, bend right=70,from=U, to=D, "{\varPhi(\sigma_1)}",""{name=R,inner sep=1pt,left},swap]
            \arrow[triple,from=R,to=L,"{\varPhi(\rho)}",swap]
        \end{tikzcd}
    \end{equation}
    where, using the reparameterisation introduced in section~\ref{ssec:simplification},
    \begin{equation}
        \begin{gathered}
            \varPhi(\gamma)=g_\gamma\in \sG~,~~~\varPhi(\sigma)=(h_\sigma,g^1_\sigma,g^2_\sigma)\in (\sH'\rtimes \sG')\rtimes \sG~,\\
            \varPhi(\rho)=(h^1_\rho,h^2_\rho,g^1_\rho,g^2_\rho)\in \sH\rtimes \big((\sH' \rtimes\sG')\rtimes \sG\big)
        \end{gathered}
    \end{equation}
    with
    \begin{subequations}
        \begin{align}
            g_{\gamma_1}&=g^2_{\sigma_1}=g^2_{\sigma_2}~,\\
            g_{\gamma_2}&=g^1_{\sigma_1}g_{\gamma_1}=g^1_{\sigma_2}g_{\gamma_1}~,\\
            \big(h_{\sigma_1},g^1_{\sigma_1},g^2_{\sigma_1}\big)&=\big(h^2_\rho,g^1_\rho,g^2_\rho\big)~,\\
            \big(h_{\sigma_2},g^1_{\sigma_2},g^2_{\sigma_2}\big)&=\big(h^1_\rho h^2_\rho,g^1_\rho,g^2_\rho\big)~.
        \end{align}
    \end{subequations}
    Now, $h_\sigma$ fixes $h^1_\rho$ and $h^2_\rho$, and $g_\gamma$ fixes \(g^1_\sigma\) and $g^2_\sigma$, which in turn fix \(g^1_\rho\) and \(g^2_\rho\). Altogether, the strict 3-functor $\varPhi\colon\CP_{(3)}U\to \sB\sInn(\CCG)$ is fully determined by the strict 2-functor $\Phi\colon\CP_{(2)}U\to \sB\CCG$.
    
    In terms of the gauge potential and curvature forms~\eqref{eq:unadjusted_fields}, the 3-functor \(\varPhi\) can be parameterised according to
    \begin{subequations}
        \begin{align}
            &&g_\gamma &= \Pexp\int_\gamma A~,&& \\
            &&h_\sigma = \Pexp\ichint_{A}&B~,~~~
            g_\sigma = \Pexp\ichint_{A}(-\tilde F)~, \\
            &&h_\rho &= \Pexp\ichichint_{A,B} (-H)~,\label{eq:443c}
        \end{align}
    \end{subequations}
    where \(\tilde F \coloneqq F - \mu_1(B) = \dd A+\tfrac12[A,A]\) is the ordinary Yang--Mills curvature. This assignment is fixed by the mapping between the Weil algebra and the inner automorphism 2-crossed module. In the iterated integral \eqref{eq:443c},
    $\chint_A (-H)$ is defined as in~\eqref{eq:non-abelian_correction}, which is consistent as $H$ takes values in $\frh$, which carries a representation of the Lie group $\sG$ integrating $\frg$. Moreover, the second integral is again defined as in~\eqref{eq:non-abelian_correction}, but now on the path space $P_{x_0}^{x_1}(M)$ with the 1-form
    \begin{equation}
        \check B\coloneqq\chint_A B\in\Omega^1(P_{x_0}^{x_1}(M))\otimes\frh~.
    \end{equation}
    Again, $\frh$ clearly carries a representation of the Lie group $\sH$ integrating $\frh$, and thus~\eqref{eq:443c} is indeed well-defined.
    
    The Chen form (see \cref{app:chen_forms}) relating $H$ to $h_{\rho}$ is obtained by lifting $H$ first to a 2-form $\chint H$ on path space using the $\sG$-connection $A$ and then, further to a 1-form $\chichint H$ on surface space. The last step requires that $H$ form an $\sH$-representation, which is only the case if $F=0$, according to equation~\eqref{eq:unadjusted_H_gauge_transformation}. Under a \(\sH\)-gauge transformation parameterised by \(\Lambda\), \(H\) mixes with \(F\), and cannot form an \(\sH\)-representation by itself. Fake flatness enters the picture yet again. 
    
    Similarly, in defining \(\chint_A B\), we must require \(\sB\) to form a \(\sG\)-representation, which is only the case if \(\mu_3(A,A,-)=0\) according to equation~\eqref{eq:unadjusted_B_gauge_transformation}.
        
    The globular structure of $\sB\sInn(\CCG)$ now induces Stokes’ theorems as follows. Given a 1-morphism \(\sigma\colon\gamma_1\to\gamma_2\) and a 2-morphism \(\rho\colon\sigma_1\to\sigma_2\), we have the globular identities
    \begin{equation}\label{eq:derivative_globular_identities}
            g_{\gamma_1}g_{\gamma_2}^{-1} = g_\sigma^{-1}~,~~~h_{\sigma_1}h_{\sigma_2}^{-1} = h_\rho^{-1}~,~~~
            g_{\sigma_1}g_{\sigma_2}^{-1} = \unit~.        
    \end{equation}
    The first identity fixes
    \begin{subequations}\label{eq:derivative_stokes_theorems}
        \begin{align}
            \dd A+\tfrac12[A,A] &= \tilde F \coloneqq F-\mu_1(B) ~.\label{eq:derivative_stokes_theorem_1}
            \intertext{The second and third translate to the identities}
            \dd_AB &= H~, \label{eq:derivative_stokes_theorem_2}\\
            \dd_A\tilde F &= 0~. \label{eq:derivative_stokes_theorem_3}
        \end{align}
    \end{subequations}
    Equations~\eqref{eq:derivative_stokes_theorem_1} and~\eqref{eq:derivative_stokes_theorem_3} hold automatically; equation~\eqref{eq:derivative_stokes_theorem_2}, however, only holds if \(\tfrac1{3!}\mu_3(A,A,A) = 0\), according to equation~\eqref{eq:unadjusted_fields_H}.

    The derivative parallel transport 3-functor now fits into the following commutative diagram:
    \begin{equation}\label{diag:commut_functors}
        \begin{tikzcd}
            \CP_{(2)} U\ar[d,"\Phi",swap] \ar[r,hookrightarrow] & \CP_{(3)}U  \ar[d,"\varPhi"] \\
            \sB \CCG \ar[r,hookrightarrow] & \sB \sInn(\CCG)
        \end{tikzcd}
    \end{equation}
    which makes it clear how gauge transformations should be defined. As in the case of ordinary gauge theory, we can fix endpoints \(x_0,x_1\in U\) and decategorify, considering the hom 2-categories. Then we can add the 2-functor \(\varPhi_\text{curv}\), which is a truncation of \(\varPhi\) to integrals of field strengths only:
    \begin{equation}
        \begin{tikzcd}
            \CP_{(2)} U(x_0,x_1)\ar[d,"{\Phi(x_0,x_1)}",swap] \ar[r,hookrightarrow] & \CP_{(3)}U(x_0,x_1)  \ar[d,"{\varPhi(x_0,x_1)}"']  \ar[dr,"{\varPhi_\mathrm{curv}(x_0,x_1)}"] & \\
            \CCG \ar[r,hookrightarrow] & \sInn(\CCG) \ar[r,twoheadrightarrow,"\Pi"] & \sB\CCG
        \end{tikzcd}
    \end{equation}
    Gauge transformations are then 3-natural transformations \(\varPhi\Rightarrow\tilde\varPhi\), which are general enough to include the pseudonatural transformations of 2-categories, and whose induced 2-natural transformations become trivial on the induced curvature 2-functors:
    \begin{equation}
        \varPhi_{\rm curv}(x_0,x_1)=\Pi\circ \varPhi(x_0,x_1)=\Pi\circ \tilde \varPhi(x_0,x_1)=\tilde \varPhi_{\rm curv}(x_0,x_1)~.
    \end{equation}

    \subsection{Adjusted higher parallel transport}\label{ssec:adjusted_parallel_transport}
    
    Above, we saw that we have two equivalent definitions of parallel transport. For an ordinary parallel transport based on a Lie group $\sG$ over a contractible manifold $U$, we can use either a functor $\Phi\colon\CP U\to \sB\sG$ or a strict 2-functor $\varPhi\colon\CP_{(2)}U\to \sB\sInn(\sG)$ with a restricted set of (higher) natural isomorphisms. This picture clearly generalises to higher categorifications\footnote{One should use simplicial models for the higher categories in order to avoid the technicalities arising from higher coherence conditions.}.
    
    In the case of a strict gauge 2-group $\CCG=(\sH\rtimes\sG\rightrightarrows \sG)$, the globular structure of the 2-crossed module $\sInn(\CCG)$ induces fake flatness~\eqref{eq:fake_flatness}, which renders the theory essentially abelian. We have seen before that an adjustment of the Weil algebra, if it exists,\footnote{At the time of writing, a characterisation of necessary or sufficient conditions for the existence of adjustments is an open problem, although the firmly adjusted Weil algebras of Chapter~5 offer a partial solution to this important problem.} can remove the necessity for fake flatness~\eqref{eq:fake_flatness}. The same is true in the case of higher parallel transport: the adjusted Weil algebra leads to an adjusted 2-crossed module of Lie groups, whose adjusted globular structure obviates the need for fake flatness.
    
    Since we need an adjustment, we must start from a gauge 2-group that admits one. Adjusted parallel transport for an adjustable crossed module of Lie groups $\CCG$ is then defined as a 3-functor
    \begin{equation}\label{eq:adj_parallel_transport_functor}
        \varPhi^\text{adj}\colon\CP_{(3)}U\to \sB\sInn_{\rm adj}(\CCG)~,
    \end{equation}
    which is the analogue of the derivative parallel transport 3-functors. There is no analogue of the 2-functor 
    \begin{equation}
        \Phi\colon\CP_{(2)}U\rightarrow \sB\scG
    \end{equation}
    for $\varPhi^{\rm adj}$, unlike the other cases discussed so far in this section. This is as expected: adjustment is crucial to the existence of a well-defined notion of non-abelian higher parallel transport, and this is only visible at the level of the Weil algebra \(\sW_\text{adj}(\sLie(\CCG))\) or, correspondingly, the inner automorphism 2-group \(\sInn_\text{adj}(\CCG)\). It \emph{is}, however, possible to truncate the 3-functor to a 2-functor sensitive only to the curvatures, and for every pair of endpoints \(x_0,x_1\in U\) we have a commutative diagram
    \begin{equation}
        \begin{tikzcd}
            & \CP_{(3)}U(x_0,x_1)  \ar[d,"{\varPhi^\mathrm{adj}(x_0,x_1)}"']  \ar[dr,"{\varPhi^\mathrm{adj}_\mathrm{curv}(x_0,x_1)}"] & \\
            \CCG \ar[r,hookrightarrow] & \sInn_{\rm adj}(\CCG) \ar[r,twoheadrightarrow,"\Pi"] & \sB\CCG
        \end{tikzcd}
    \end{equation}
    where the bottom row is the short exact sequence~\eqref{eq:ses_groupoids_adjusted}. This diagram is the adjusted analogue of diagram~\eqref{diag:commut_functors}, without the nonexistent 2-functor $\Phi$. Similarly to the previous cases, the admissible gauge transformations are those natural transformations \(\eta\colon\varPhi\to\varPhi\) that are rendered trivial by the following whiskering.
    \begin{equation}
        \begin{tikzcd}
            \sB\CCG & \sInn(\CCG) \lar["\Pi",swap]  &  \CP_{(3)} U(x_0,x_1)\lar[bend left=50, "{\varPhi(x_0,x_1)}", ""'{name=A}] \lar[bend right=50, "{\tilde \varPhi(x_0,x_1)}"', ""{name=B}] \ar[from=A, to=B, Rightarrow,"\eta"]
        \end{tikzcd}
    \end{equation}
    
    To explain the 3-functor in more detail, let us focus on the archetypical example: (the generalisation of) the loop model of the string group,
    \begin{equation}
        \sString_{\rm lp}(\sG)=\big(\hat L_0\sG\rtimes P_0\sG\rightrightarrows P_0\sG\big)\ewith \sLie(\CCG)=\big(\hat L_0 \frg\rightarrow P_0\frg\big)~,
    \end{equation}
    where $\sG$ is a finite-dimensional Lie group whose Lie algebra $\frg$ is metric.\footnote{Thus, $\sG$ admits a bi-invariant (pseudo-)Riemannian metric.} The Weil algebra of $\sLie(\CCG)$ admits an adjustment as discussed in section~\ref{ssec:adjusted_Weils}, and thus $\CCG$ admits an adjusted 3-group of inner automorphisms as explained in section~\ref{ssec:adjusted_inner_derivation}. Other examples of 2-groups admitting an adjustment can be treated similarly; in particular the discussion for the group $\sG_{\rm lp}=(L_0\sG\rtimes P_0\sG\rightrightarrows P_0\sG)$ discussed in section~\ref{ssec:loop_model_of_Lie_algebra} follows by truncation. 
    
    The 3-functor~\eqref{eq:adj_parallel_transport_functor} is then of the following form:
    \begin{subequations}
        \begin{equation}
            \varPhi^\text{adj}\colon\CP_{(3)}U\to \sB\sInn_{\rm adj}(\sString_{\rm lp}(\sG))
        \end{equation}
        with components consisting of the following maps:
        \begin{equation}
            \begin{tikzcd}
                \text{volumes} \dar[shift left] \dar[shift right] \rar["\varPhi^\text{adj}_3"] & \hat L_0\sG \rtimes \big((\hat L_0\sG \rtimes P_0\sG) \rtimes P_0\sG \big) \dar[shift left] \dar[shift right] \\
                \text{surfaces} \dar[shift left] \dar[shift right] \rar["\varPhi^\text{adj}_2"] & (\hat L_0\sG \rtimes P_0\sG) \rtimes P_0\sG \dar[shift left] \dar[shift right] \\
                \text{paths}  \dar[shift left] \dar[shift right] \rar["\varPhi^\text{adj}_1"] & P_0\sG \dar[shift left] \dar[shift right] \\
                U \rar["\varPhi^\text{adj}_0"] & *
            \end{tikzcd}
        \end{equation}
    \end{subequations}
    
    In terms of the fields~\eqref{eq:adjusted_loop_fields} taking values in the adjusted Weil algebra, all components of the 3-functor \(\varPhi\) can be covariantly defined:
    \begin{subequations}\label{eq:adj_hol_3_functor}
        \begin{align}
            g_\gamma &= \Pexp\int_\gamma A~, \\
            (h_\sigma,g_\sigma) &= \Pexp\ichint_{A}\binom B{-\tilde F}~, \\
            h_\rho^{-1} &= \Pexp\chichint_{A,B} (-H)~,
        \end{align}
    \end{subequations}
    where \(\tilde F\) is the ordinary Yang--Mills field strength
    \begin{equation}
        \tilde F = F - \mu_1(B) = \dd A+\tfrac12[A,A]~.
    \end{equation}
    Notice that the field \(B\) does not form a \(P_0\sG\)-representation by itself, which is similar to the problem with \(H\) in the unadjusted case. Happily, in the 2-crossed module \(B\) occurs together with \(\tilde F\), and \((B,\tilde F)\) does form a \(P_0\sG\)-representation, which can be exponentiated. Also, now \(H\) is gauge-\emph{in}variant, so that there is no problem defining it.
    We do not have any freedom to choose how to define the components of the parallel transport 3-functor; this is determined by the mapping between the adjusted Weil algebra and the adjusted inner derivation 2-crossed module.
    
    It remains to check that the required Stokes’ theorems hold. The globular identities~\eqref{eq:derivative_globular_identities} are unchanged from the unadjusted case and these correspond to the same Stokes’ theorems~\eqref{eq:derivative_stokes_theorems}, which we rewrite for clarity:
    \begin{subequations}
        \begin{align}
            \dd A+\tfrac12[A,A] &= \tilde F \coloneqq F-\mu_1(B)~,\\
            \dd_A\binom B{-\tilde F} &= \binom H0~.
        \end{align}
    \end{subequations}
    We write it thus to emphasise that \(B\) only forms a \(P_0\sG\)-representation together with \(\tilde F\). The first is the non-abelian Stokes’ theorem as before, and one can easily check that the second equation corresponds to the correct Bianchi identities~\eqref{eq:fake_curvature_bianchi_identity} and~\eqref{eq:H_bianchi_identity} for the adjusted Weil algebra.

    We make a few final remarks. The assignment~\eqref{eq:adj_hol_3_functor} indeed defines a strict 3-functor; verifying functoriality mostly consists of drawing elaborate diagrams, meditating on them, and concluding that they are trivial, especially since this 3-functor is strict. We leave this to the vigilant reader with free time (much as Cervantes dedicates \emph{Don Quijote} to the \emph{desocupado lector}).
    
    Technically, our path 3-groupoids are equivalence class of paths, surfaces, and volumes under \emph{thin homotopy}, which are homotopies of “zero volume” (see \cref{app:path_space}). Once we grant that the 3-functors are well-defined without this quotienting, a transformation by thin homotopy corresponds to a parallel transport along a zero-volume homotopy, which are given by the integral of the relevant curvatures, but this vanishes because the volume is zero. (In the case of the top curvature \(H\), one uses the Bianchi identity~\eqref{eq:H_bianchi_identity} for it.)

    In retrospect, the assertion of~\cite{Baez:2004in} that fake flatness is required for thin homotopy invariance was but an avatar of the fact that, without adjustment, gauge transformations generically only close if fake curvature vanishes. In the adjusted case, this defect is absent.

\chapter{Tensor Hierarchies as Higher Gauge Algebras}
\chaptermark{\(E_2L_\infty\)-algebras}

    \section{Review of tensor hierarchies}
    
    Before analyzing the algebraic structure underlying tensor hierarchies in more detail, let us briefly review the physical context. Consider the Lagrangian of ungauged Einstein--Maxwell-scalar theory in $d$ dimensions,
    \begin{equation}\label{Lth}
        \mathcal{L}_{\text{ungauged}} = \star R +\tfrac{1}{2} g_{xy} \rmd \varphi^x \wedge \star \rmd \varphi^y  -\tfrac{1}{2} a_{ij} F^{i} \wedge \star F^{j} +\cdots
    \end{equation}
    with scalars $\varphi^x$ mapping spacetime to a scalar manifold $\caM$ and 1-form abelian gauge potentials $A^i$ with field strengths $F^{i}=\rmd A^{i}$. Here, $g_{xy}(\varphi)$ and $a_{ij}(\varphi)$ are symmetric and positive-definite on the entire scalar manifold $\caM$. The ellipsis denotes possible deformations, such as a scalar potential $\mathcal{V}(\varphi)$ or topological terms such as, e.g., ${d}_{ijk}F^{i} \wedge  F^{j}\wedge A^{k}$ for $d=5$, as familiar from supergravity. The set of constant tensors controlling these deformation terms, which includes those appearing in the tensor hierarchies that do not enter~\eqref{Lth}, will be referred to as {\em deformation tensors}. 
    
    We assume that there is a global symmetry group $\sfG$ acting on scalars and 1-form potentials such that the undeformed action~\eqref{Lth} is invariant under this action. In particular, the total gauge potential one-form $A$ takes value in a representation $V_{-1}$ of $\sfG$, which is isomorphic to the fibers of the tangent space of the scalar manifold. In the presence of deformations, we assume that there is a non-abelian subgroup $\sfK\subseteq \sfG$ leaving the full action invariant.    
    
    Infinitesimally, the corresponding Lie algebra actions of $\frg=\sfLie(\sfG)$ on the scalars and the gauge potential are given by 
    \begin{equation}    
        \delta_\lambda \varphi^x = \lambda^{\alpha} k_{\alpha}{}^{x}(\varphi)~,~~~ \delta_\lambda A^{i} =  \lambda^{\alpha} t_{\alpha}{}^{i}{}_{j} A^{j}~,
    \end{equation}
    where  $t_{\alpha}{}^{i}{}_{j}, \alpha=1,2,\ldots,\dim \frg$, are the generators of $\frg$ in the representation $V_{-1}$ with respect to some bases $e_\alpha$ of $\frg$ and $e_i$ of $V_{-1}$. Invariance under $\sfG$ requires that $k(\phi)$ be  Killing vectors of the scalar manifold and $\mathcal{L}_{k_\alpha}a_{ij}=-2\ t_{\alpha}{}^{k}{}_{(i} a_{j)k}$. 
    
    In order to gauge\footnote{That is, we promote a global symmetry $\sfH$ to a local one by adding a principal $\sfH$-bundle on our spacetime and consider (a part of) the one-form potential as a connection on this bundle.} a subgroup $\sfH\subseteq \sfK\subseteq \sfG$ with Lie algebra $\frh=\sfLie(\sfH)$, we first note that we can trivially regard the pair $(V_{-1},\frg$) as a differential graded Lie algebra
    \begin{equation}\label{eq:trivialdgLA}
        V=(~V_{-1}~\xrightarrow{~0~}~V_{0}=\frg~)
    \end{equation}
    with evident Lie bracket on $V_0$ and the Lie bracket $[-,-]:V_{0}\times V_{-1}\rightarrow V_{-1}$ given by the action of $\frg$ on $V_{-1}$. Because the gauge potential takes values in $V_{-1}$, it does not make sense to gauge a Lie subalgebra of $\frg$ which is larger than $V_{-1}$. Therefore, we can identify the subalgebra $\frh$ with the image of a linear map 
    \begin{equation}
        \begin{aligned}
            \Theta: V_{-1}&\twoheadrightarrow \frh~~~~~~\subset \frg~,
            \\
            e_i&\mapsto \Theta_i{}^\alpha e_\alpha~.
        \end{aligned}
    \end{equation}
    The $(\Theta_{i}{}^{\alpha} e_\alpha)$ then form a spanning set\footnote{but not necessarily a basis} of the Lie algebra $\frh$. Moreover, we have an induced action of $\frh$ on $V_{-1}$, given by 
    \begin{equation}
        (\Theta_{i}{}^{\alpha} e_\alpha) \acton e_j=\Theta_{i}{}^{\alpha} t_{\alpha j}{}^k e_k\eqqcolon X_{ij}{}^k e_k
    \end{equation}
    with 
    \begin{equation}
        t_{\alpha j}{}^k=-t_{\alpha}{}^k{}_j\eand t_{\alpha i}{}^jt_{\beta j}{}^k-t_{\beta i}{}^jt_{\alpha j}{}^k=-f_{\alpha\beta}{}^\gamma t_{\gamma i}{}^k~.
    \end{equation}
    
    In order to guarantee closure of the Lie bracket on $\frh$ and consistency of the action, we can assume that we can incorporate $\Theta$ into~\eqref{eq:trivialdgLA} such that 
    \begin{equation}\label{eq:ThetadgLA}
        V_\Theta=(~V_{-1}~\xrightarrow{~\Theta~}~V_{0}=\frg~)
    \end{equation}
    is again a differential graded Lie algebra. To jump ahead of the story, note that this guarantees the existence of a higher gauge algebra via~\cref{prop:dgLA_to_L_infty}, which we anticipate as part of the construction of a higher gauge theory. The fact that $\Theta$ is a derivation for the graded Lie bracket then implies the quadratic {\em closure constraint}
    \begin{equation}\label{eq:quad_closure}
        f_{\alpha\beta}{}^{\gamma}\Theta^{\alpha}_{i}\Theta^{\beta}_{j}=\Theta^{\gamma}_{k}X_{ij}{}^{k}~~~\Leftrightarrow~~X_{im}{}^\ell X_{j\ell}{}^n-X_{jm}{}^\ell X_{i\ell}{}^n=-X_{ij}{}^\ell X_{\ell m}{}^n~. 
    \end{equation}
    It is well known that this quadratic closure constraint implies that the $X_{ij}{}^k$ form the structure constants of a Leibniz algebra on $V_{-1}$,
    \begin{equation}\label{eq:action_V_1_V_1}
        e_i\circ e_j\coloneqq X_{ij}{}^k e_k~,
    \end{equation}
    cf.~e.g.~\cite{Lavau:2017tvi,Hohm:2018ybo,Kotov:2018vcz}. This is unsatisfactory given that the 1-form gauge potentials $A$ will take values in $V_{-1}$ and $V_{-1}$ should therefore have some Lie structure. As noted in~\cite{Saemann:2019dsl}, and as evident from \cref{prop:Leib_is_hemistrict_ELinfty}, this Leibniz algebra can be promoted to an $\ophLie$-algebra. Moreover, the fact that we have the differential graded algebra~\eqref{eq:ThetadgLA} guarantees that we will have an $\ophLie$-algebra via \cref{thm:antisym_hLie} (or, if preferred, the corresponding $L_\infty$-algebra obtained from \cref{thm:antisym_hLie}). This will turn out to be indeed the higher gauge algebra underlying the tensor hierarchies.
    
    But let us continue with the tensor hierarchy from the physicists' perspective. The quadratic closure constraint~\eqref{eq:quad_closure} allows us to introduce a consistent combination of a covariant derivative on the scalar fields and local transformations parameterised by $\Lambda_{(0)}\in C^\infty(M)\otimes V_{-1}$:
    \begin{equation}\label{eq:first_gauge}
        \begin{gathered}
            \nabla\varphi^i\coloneqq  \rmd \varphi^i + \Theta^{\alpha}_{j} A^{j} k_{\alpha}{}^{i}(\varphi)~,
            \\
            \delta_{\Lambda_{(0)}} \varphi^i\coloneqq  {\Lambda_{(0)}}^{i}\Theta^{\alpha}{}_{j}  k_{\alpha}{}^{j}(\varphi)~,~~~
            \delta_{\Lambda_{(0)}} A^i\coloneqq \rmd {\Lambda_{(0)}}^i+X_{jk}{}^iA^j{\Lambda_{(0)}}^k~.            
        \end{gathered}
    \end{equation}
    Note that the action~\eqref{eq:action_V_1_V_1} of $V_{-1}$ on $V_{-1}$ is usually not faithful, and the parameterisation by $\Lambda^i$ is thus usually highly degenerate. 
    
    In light of our above discussion of the higher Lie algebra arising from the Leibniz algebra~\eqref{eq:action_V_1_V_1}, it is not surprising that the naive gauge transformation~\eqref{eq:first_gauge} of the gauge potential $A$ does not render the naive definition of curvature $\rmd A^{i}+\tfrac12 X_{jk}{}^{i}A^{j}\wedge A^{k}$ covariant. This is remedied by introducing a second $\sfG$-module $V_{-2}\subset \text{Sym}^2(V_{-1})$, where $r=1,2\ldots\dim V_{-2}$ for some basis $(e_{r})$ together with a map
    \begin{equation}
        \begin{aligned}
            Z: V_{-2}&\rightarrow V_{-1}~,
            \\
            e_r&\mapsto Z^i{}_r e_i~.
        \end{aligned}
    \end{equation}
    This allows us to introduce a $V_{-2}$-valued 2-form potential $B$ and a $V_{-2}$-valued 1-form gauge parameter $\Lambda_{(1)}$ to generalised gauge transformations and curvatures as usual in higher gauge theory:
    \begin{equation}
        \begin{gathered}
            \delta A^{i} = \rmd \Lambda^i_{(0)}+X^i_{jk}A^j\Lambda^k_{(0)} +Z^{i}{}_{r}\Lambda^{r}_{(1)}~,~~~
            \delta B^r= \nabla\Lambda^{r}_{(1)}+\ldots~,
            \\
            F^{i}=\rmd A^{i}+\tfrac12 X^{i}{}_{jk}A^{j}\wedge A^{k}+Z^{i}{}_{r}B^{r}~,~~~H^r=\nabla B^r+\ldots~,
        \end{gathered}
    \end{equation}
    where here $\nabla$ is the covariant derivative given by the natural action of $\frh$ on $V_{-2}$ and the ellipses refer to covariantizing terms that are needed to complete the kinematical data to that of an adjusted higher gauge theory. In particular, the latter will include terms involving the various deformation tensors. This process is then iterated in a reasonably obvious fashion until the full kinematical data of an adjusted higher gauge theory is obtained\footnote{The fact that this iteration terminates is guaranteed because spacetime is finite-dimensional.}.
    
    In the gauged supergravity literature there is also often a  linear \emph{representation constraint}
    \begin{equation}    
        P_\Theta \Theta =\Theta~,
    \end{equation}
    where $P_\Theta$ is the projector onto the representation contained in $V_{-1}^*\otimes \mathfrak{g}$  carried by $\Theta$, which  will be denoted $\rho_\Theta$. This can be understood as a requirement of supersymmetry~\cite{deWit:2002vt,deWit:2005hv}, the mutual locality of the action~\cite{deWit:2005ub} or anomaly cancellation~\cite{DeRydt:2008hw}. 
    
    A final important ingredient is now that the electromagnetic duality contained in U-duality needs to link potential $p$-forms to potential $d-p-2$-forms, and correspondingly the $\sfG$-modules $V_{-p}$ and $V_{p+2-d}$ have to be dual to each other in the lowest degrees that involve physical gauge potentials. 
    
    The above constraints restrict severely the choices of representations $V_{-2}$, $V_{-3}$. In \cref{Mcharges} we listed some important concrete examples of maximal supergravities, in which $\sfK=\sfG$. In this case, there is a tensor hierarchy dgLa determined by the U-duality group~\cite{Palmkvist:2011vz, Palmkvist:2013vya}, with graded vector space described in \autoref{Mcharges} and derivation given by the action of $\Theta$. Also, the electromagnetic duality is visibly reflected in the duality of representations in the cases $d=5,6,7$. 
    \begin{table}[!ht]\small
        \begin{tabular*}{\textwidth}{cccccccc}
            \hline
            \hline
            $d$ &$ \sfG                           $&$ V_{-1}            $&$ V_{-2}           $&$ V_{-3}           $&$ V_{-4}            $&$ V_{-5}       $&$ V_{-6}       $   \\
            \hline
            7   &$ \sfSL(5,\IR )                  $&$ \rep{10}_c    $&$ \rep{5}     $&$ \rep{5}_c    $&$ \rep{10} $&$ \rep{24} $&$ \rep{15}_c\oplus\rep{40} $ \\
            6   &$ \sfSO(5,5)                $&$ \rep{16}_c     $&$ \rep{10}    $&$ \rep{16} $&$ \rep{45} $&$ \rep{144} $&$ \rep{10\oplus126\oplus320} $\\
            5   &$ \sfE_{6(6)}              $&$ \rep{27}_c     $&$ \rep{27} $&$ \rep{78} $&$ \rep{351}_c $&$ \rep{27\oplus1728} $\\
            4   &$ \sfE_{7(7)}              $&$ \rep{56} $&$ \rep{133} $&$ \rep{912} $&$ \rep{133\oplus8645} $\\
            3   &$ \sfE_{8(8)}              $&$ \rep{248} $&$ \rep{1\oplus 3875} $&$ \rep{3875\oplus147250} $\\
            \hline
            \hline
        \end{tabular*}
        \caption{Global symmetry groups $\sfG$ of maximal supergravity in $3\leq d\leq 7$ spacetime dimensions (ignoring discrete factors). The $\sfG$ representations $V_{-p}$ are carried by  $p$-forms in the tensor hierarchy. The scalars ($0$-forms) are valued in $\caM\coloneqq \sfG/\sfG_0$,  where $\sfG_0\subset \sfG$ is the maximal compact subgroup. } \label{Mcharges}
    \end{table}
    
    We note that in the presence of generic deformations, the differential graded Lie algebra constructed in the maximally supersymmetric case is actually insufficient and needs to be extended further by at least one step in both directions. We shall explain this below, when discussing the example $d=5$.

    \section{\texorpdfstring{$\ophLie$}{hLie}-algebras and \texorpdfstring{$\opEilh$}{Eilh}-algebras}
    
    We start with the definition of the two Koszul-dual concepts of $\ophLie$-algebras and $\opEilh$-algebras that underlie our definition of $E_2L_\infty$-algebras.
    Koszul duality was briefly reviewed in \cref{app:higher_Lie_algebras}; we stress, however, that more than an intuitive understanding of Koszul duality is not required for our discussion.
    
    \subsection{\texorpdfstring{$\ophLie$}{hLie}-algebras}
    
    In this section, we define a generalisation of differential graded Lie algebras that contains hemistrict Lie 2-algebras as defined in~\cite{Roytenberg:0712.3461}. Besides a differential, Lie 2-algebras contain a binary operator of degree~$0$ and an alternator, i.e.~a binary operator of degree~$-1$. In the case of Lie 3-algebras, we also obtain a higher alternator of degree~$-2$. It is thus natural to allow for higher alternators of arbitrary degree, and to minimally extend the algebraic relations known from Lie 2- and Lie 3-algebras in such a way that they are compatible with the differential. This leads to the following definition.
    \begin{definition}
        An $\ophLie$-algebra is a graded vector space $\frE$ together with a differential $\eps_1$ and a collection of binary products $\eps_2^i$ of degree~$i$ for $i\in\{0,1\}$,
    \begin{equation}
        \begin{aligned}
            \eps_1&: \frE\rightarrow \frE~,~~~|\eps_1|=1~,
            \\
            \eps^i_2&: \frE\otimes \frE\rightarrow \frE~,~~~|\eps^i_2|=-i~,
        \end{aligned}
    \end{equation}
    that satisfy the following relations: 
        \begin{equation}\label{eq:hLie-relations}
            \begin{aligned}
                \eps_1(\eps_1(x_1))&=0~,
                \\
                \eps_1(\eps^i_2(x_1,x_2))&=(-1)^i\big(\eps^i_2(\eps_1(x_1),x_2)+(-1)^{|x_1|}\eps^i_2(x_1,\eps_1(x_2))\big)
                \\
                &\hspace{1cm}+\eps^{i-1}_2(x_1,x_2)-(-1)^{i+|x_1|\,|x_2|}\eps^{i-1}_2(x_2,x_1)~,
                \\
                \eps^i_2(\eps_2^i(x_1,x_2),x_3)&=(-1)^{i(|x_1|+1)}\eps^i_2(x_1,\eps^i_2(x_2,x_3))-(-1)^{(|x_1|+i)|x_2|}\eps^i_2(x_2,\eps_2^i(x_1,x_3))
                \\
                &\hspace{1cm}-(-1)^{(|x_2|+|x_3|)|x_1|+(i-1)|x_2|}\eps^{i+1}_2(x_2,\eps^{i-1}_2(x_3,x_1))~,
                \\
                \eps^{j}_2(\eps^{i}_2(x_1,x_2),x_3)&=
                (-1)^{1+j(i+1)+|x_1|(|x_2|+|x_3|)+(j-1)|x_2|}\eps_2^{i+1}(x_2,\eps_2^{j-1}(x_3,x_1))~,
                \\
                \eps^{i}_2(\eps^{j}_2(x_1,x_2),x_3)&=
                (-1)^{i(j+|x_1|)}\eps_2^{j}(x_1,\eps_2^{i}(x_2,x_3))-(-1)^{(|x_1|+j)|x_2|}\eps_2^i(x_2,\eps_2^j(x_1,x_3))
                \\
                &\hspace{1cm}-(-1)^{j+|x_3|(j+|x_2|-1)+|x_1|(|x_2|+|x_3|)}\eps_2^{i+1}(x_3,\eps_2^{j-1}(x_2,x_1))
            \end{aligned}
        \end{equation}
        for all $i,j\in \IN$ and $j<i$, where we regard $\eps_2^i=0$ for $i\not\in\{0,1\}$, and for all $x_1,x_2,x_3\in \frE$.
    \end{definition}
    
    We note that the first three relations in~\eqref{eq:hLie-relations} for $i=0$ amount to the relations for a differential graded Leibniz algebra with differential $\eps_1$ and Leibniz product $\eps_2^0$. If $\eps_2^1$ vanishes, then $\eps_2^0$ is graded antisymmetric, and the Leibniz bracket becomes Lie. If we restrict to the case $\eps_2^i=0$ for $i\neq 0$, we simply recover differential graded Lie algebras. If we restrict ourselves to 2-term $\ophLie$-algebras, i.e.~$\ophLie$-algebras concentrated in degrees~$-1$ and $0$, then only $\eps_2^0$ and $\eps_2^1$ can be non-trivial for degree reasons, and we obtain the hemistrict Lie 2-algebras of~\cite{Roytenberg:0712.3461} with a graded symmetric $\eps_2^1$. The latter map is a chain homotopy sometimes called the \uline{alternator}, capturing the failure of $\eps_2^0(x_1,x_2)$ to be antisymmetric:
    \begin{equation}
        \begin{aligned}
            &\eps^0_2(x_1,x_2)+(-1)^{|x_1|\,|x_2|}\eps^0_2(x_2,x_1)
            \\
            &\hspace{2cm}=\eps_1(\eps^1_2(x_1,x_2))
            +\eps^1_2(\eps_1(x_1),x_2)+(-1)^{|x_1|}\eps^1_2(x_1,\eps_1(x_2))~.
        \end{aligned}
    \end{equation}
    
    We close this section with two results on $\ophLie$-algebras that allow us to construct new $\ophLie$-algebras from existing ones. First, we can create a larger $\ophLie$-algebra by tensoring $\ophLie$-algebras with differential graded commutative algebras, just as in the case of Lie algebras, 
    \begin{proposition}\label{prop:tensor_product_dgca_hLie}
        The tensor product of a differential graded commutative algebra and an $\ophLie$-algebra carries a natural $\ophLie$-algebra structure.
    \end{proposition}
    \begin{proof}
        On the tensor product of a differential graded commutative algebra $\frA$ and an $\ophLie$-algebra $\frE$, we define 
        \begin{equation}
            \begin{aligned}
                \hat \frE\coloneqq \frA\otimes \frE=\oplus_{k\in \IN}(\frA\otimes \frE)_k~,~~~(\frA\otimes \frE)_k=\sum_{i+j=k} \frA_i\otimes \frE_j~,
                \\
                \hat\eps_1(a_1\otimes x_1)=(\rmd a_1)\otimes x_1+(-1)^{|a_1|}a_1\otimes \eps_1(x_1)~,
                \\
                \hat\eps^i_2(a_1\otimes x_1,a_2\otimes x_2)=(-1)^{i(|a_1|+|a_2|)}(a_1a_2)\otimes \eps^i_2(x_1,x_2)~,
            \end{aligned}
        \end{equation}
        One then readily verifies the axioms~\eqref{eq:hLie-relations}.
    \end{proof}
    
    Second, we note that the axioms~\eqref{eq:hLie-relations} of $\ophLie$-algebras have some translation invariance built in, which allows us to construct new $\ophLie$-algebras by shifting degrees.
    \begin{proposition}
        Given an $\ophLie$-algebra $(\frE,\eps_1,\eps_2^i)$, there is an $\ophLie$-algebra structure on the grade-shifted complex $\tilde \frE=s\frE$ with
        \begin{equation}\label{eq:shifted_hLie_relations}
            \begin{aligned}
                \tilde \eps_1(sa)&=-s\eps_1(a)~,
                \\
                \tilde \eps^i_2(sa,sb)&=(-1)^{|a|+i}s\eps^{i-1}_2(a,b)~.
            \end{aligned}
        \end{equation}
        Here, $s$ is the shift isomorphism $s:\frE\rightarrow \frE[-1]$ with that $|sa|=|a|+1$.
    \end{proposition}
    \begin{proof}
        The proof is a straightforward check of the relations~\eqref{eq:hLie-relations} for $\tilde \frE$. The fact that $\tilde \eps_1$ is a differential is evident. The remaining relations follow rather directly by replacing all arguments $x_i$ in~\eqref{eq:hLie-relations} by $sx_i$, shifting all $|x_i|$ by $1$, and pulling out factors of $s$ using the rules~\eqref{eq:shifted_hLie_relations}. The resulting equations are then satisfied due to~\eqref{eq:hLie-relations}.
    \end{proof}
    
    \subsection{\texorpdfstring{$\opEilh$}{Eilh}-algebras}\label{ssec:Eilh-algebras}
    
    For our constructions, it will be convenient to have the Koszul-dual concept to that of $\ophLie$-algebras at our disposal. 
    \begin{definition}
        An $\opEilh$-algebra is a differential graded vector space endowed with binary operations $\oslash_i$ of degree\footnote{Note that $\oslash_i$ should really be regarded as a binary function of degree~$i$ in order to identify the correct Koszul signs. For example,
            \begin{equation}
                t_1 \oslash_i t_2\coloneqq \oslash_i(t_1,t_2)~,~~~(-\oslash_i (-\oslash_j-))(t_1,t_2,t_3)=(-1)^{j|t_1|}t_1\oslash_i (t_2\oslash_j t_3)~.
            \end{equation}} $i\in \{0,1\}$ that satisfy the following quadratic identities:
        \begin{subequations}
        \begin{equation}\label{eq:Eilh-relations}
            \begin{aligned}
                a\oslash_0(b\oslash_0 c)&=(a\oslash_0 b+(-1)^{|a|\,|b|} b\oslash_0 a)\oslash_0 c~,
                \\
                -(-1)^{|a|}  a\oslash_1(b\oslash_1 c)&=(a\oslash_1 b+(-1)^{|a|\,|b|}b\oslash_1 a) \oslash_1 c~,
                \\
                a\oslash_0 (b\oslash_1 c)&=(-1)^{|a|}(a \oslash_0 b)\oslash_1 c~,
                \\
                a\oslash_1(b\oslash_0 c)&=
                (-1)^{|a|\,|b|}(b \oslash_0 a)\oslash_1 c~,
            \end{aligned}
        \end{equation} 
        and the differential satisfies the evident deformed Leibniz rule given by
        \begin{equation}\label{eq:def_Leibniz}
            \begin{aligned}
                Q(a\oslash_0 b)&=(Qa)\oslash_0 b+(-1)^{|a|}a\oslash_0 Qb+ a \oslash_1 b-(-1)^{|a|\,|b|}b\oslash_1a~,
                \\
                Q(a\oslash_1 b)&=-\big((Qa)\oslash_1 b+(-1)^{|a|}a\oslash_1 Qb\big)~.
            \end{aligned}
        \end{equation}
    \end{subequations}
    \end{definition}
    The relations~\eqref{eq:Eilh-relations} and~\eqref{eq:def_Leibniz} are equivalent (up to some signs) to those defining the operad (an abstraction of the axioms of an algebraic strutcure) $\caR^!$ introduced in Squires~\cite{Squires:2011aa} to capture 2-term $E_2L_\infty$-algebras. The free algebra over $\caR^!$ is dual to our notion of 2-term $E_2L_\infty$-algebra with a graded symmetric alternator $\sfalt(x,y)=(-1)^{|x|\,|y|}\sfalt(y,x)$.
    
    The relation between $\opEilh$-algebras and  $\ophLie$-algebras is equivalent to that between Lie algebras and commutative algebras: it provides another example of Koszul duality, cf.~\cref{app:higher_Lie_algebras}; see also~\cite[Section 7.6.4]{Loday:2012aa} for details. Instead of going into further details, let us construct the Chevalley--Eilenberg algebra of an $\ophLie$-algebra by constructing the corresponding differential semifree $\opEilh$-algebra.
    
    We start from an $\ophLie$-algebra $\frE$ which we assume for clarity of the discussion to admit a nice basis $(\tau_\alpha)$ and whose underlying graded vector space can be dualised degreewise. We will make this assumption for all the graded vector spaces from here on, mostly for pedagogical reasons: it allows us to give very explicit formulas. More generally, one may want to work with (graded) pseudocompact vector spaces, cf.~the discussion in~\cite[\S 1.1]{Guan:1909.11399}.
    
    Consider the graded vector space\footnote{Our convention for degree shift is the common one, $V[k]\coloneqq \bigoplus_i V[k]_i$ with $V[k]_i\coloneqq V_{k+i}$, implying that $V[k]$ is the graded vector space $V$ shifted by $-k$.} $V\coloneqq \frE[1]^*$ and together with its free, non-associative tensor algebra $\bigoslash^\bullet_\bullet V$ with respect to all the products $\oslash_i$ simultaneously. The quadratic identities~\eqref{eq:Eilh-relations} allow us to rearrange all tensor products such that they are nested from left to right. We thus define 
    \begin{equation}
        \begin{aligned}
            {\caE(V)}&\coloneqq \IR~\oplus~V~\oplus~\bigoplus_{i\in \IN}V\oslash_i V~\oplus~\bigoplus_{i,j\in \IN}(V\oslash_i V)\oslash_j V~\oplus~\dotsb
            \\
            &\eqqcolon\caE^{(0)}(V)~\oplus~\caE^{(1)}(V)~\oplus~\caE^{(2)}(V)~\oplus~\caE^{(3)}(V)~\oplus~\dotsb~.
        \end{aligned}
    \end{equation}
    We also define the reduced tensor algebra 
    \begin{equation}\label{eq:red_ten_algebra}
        \overline{\caE(V)}\coloneqq V~\oplus~\bigoplus_{i\in \IN}V\oslash_i V~\oplus~\bigoplus_{i,j\in \IN}(V\oslash_i V)\oslash_j V~\oplus~\dotsb~,
    \end{equation}    
    which is, in fact, sufficient for the description of $\ophLie$-algebras. 
    \begin{definition}
        We call an $\opEilh$-algebra of the form $(\caE(V),Q)$ for some graded vector space $V$ without any restrictions on the products $\oslash_i$ beyond~\eqref{eq:Eilh-relations} \uline{semifree}.
    \end{definition}
    
    Semifree $\opEilh$-algebras yield the Chevalley--Eilenberg description of $\ophLie$-algebras:
    \begin{definition}
        The \uline{Chevalley--Eilenberg algebra} $\sfCE(\frE)$ of an $\ophLie$-algebra $\frE$ whose differential and binary products are given by 
        \begin{equation}
            \begin{aligned}
                \eps_1&: \frE\rightarrow \frE~,~~~&\tau_\alpha&\mapsto m^\beta_\alpha \tau_\beta~,~~~&&|\eps_1|=1~,
                \\
                \eps^i_2&: \frE\otimes \frE\rightarrow \frE~,~~~&\tau_\alpha\otimes \tau_\beta&\mapsto m^{i,\gamma}_{\alpha\beta}\tau_\gamma~,~~~&&|\eps^i_2|=i
            \end{aligned}
        \end{equation}
        for some $m^\beta_\alpha$ and $m^{i,\gamma}_{\alpha\beta}$ taking values in the underlying ground field is the $\opEilh$-algebra $\caE(V)$ with $V=\frE[1]^*$ and the differential
        \begin{equation}\label{eq:Q_hLie}
            Q t^\alpha=-(-1)^{|\beta|}m^\alpha_\beta t^\beta-(-1)^{i(|\beta|+|\gamma|)+|\gamma|(|\beta|-1)}\,m^{i,\alpha}_{\beta\gamma}\,t^\beta \oslash_i t^\gamma~.
        \end{equation}
        Here, $|\beta|$ is shorthand for $|t^\beta|$, the degree of $t^\beta$ in $V$.
    \end{definition}
    In the case of Lie algebras and $L_\infty$-algebras, the (homotopy) Jacobi relations are equivalent to a nilquadratic differential in the corresponding Chevalley--Eilenberg algebra, and this is also the case here:
    \begin{proposition}
        The equation $Q^2=0$ for the differential of the Chevalley--Eilenberg algebra of an $\opEilh$-algebra that is of the form~\eqref{eq:Q_hLie} is equivalent to the relations~\eqref{eq:hLie-relations}.
    \end{proposition}
    \begin{proof}
        The proof is again a straightforward but tedious verification of the axioms, which is better left to a computer algebra program.
    \end{proof}

    Note, however, that we cannot simply put the products $\oslash_i=0$ for $i\notin I$ if $I$ does not contain $0$. The deformed Leibniz rule~\eqref{eq:def_Leibniz} would render the lowest non-vanishing product graded symmetric or antisymmetric and the quadratic relations~\eqref{eq:Eilh-relations} would then imply severe restrictions on the products at cubic order. Restrictions of $\caE(V)$ to $\caE_I(V)$ that are evidently sensible are of the form $I=\{0,\ldots,n\}$.

    \subsection{Cohomology of semifree \texorpdfstring{$\opEilh$}{Eilh}-algebras}\label{ssec:cohomology_Eilh}
    
    An important tool in studying Lie algebras is Lie algebra cohomology, and we consider the generalisation to $\ophLie$-algebras in the following. As we saw above, this amounts to the cohomology of semifree $\opEilh$-algebras. Due to the deformation of the usual Leibniz rule to~\eqref{eq:def_Leibniz}, a subtle and important point arises. For ordinary differential graded algebras, the cohomology again carries the structure of a differential graded algebra of the same type, with product induced from the product on the original algebra. In particular, the product of two cocyles is again a cocycle. The deformation of the Leibniz rules can now evidently break this connection.   
    
    To start, let us consider the cohomology of the semifree $\opEilh$-algebra $(\caE(V),Q_0)$ with the most trivial differential $Q_0$ satisfying
    \begin{equation}
        \begin{aligned}
            Q_0(v)&=0~,
            \\
            Q_0(a\oslash_i b)&=(-1)^i\big((Q_0a)\oslash_i b+(-1)^{|a|}a\oslash_i Q_0b\big)
            \\
            &\hspace{1cm}+(-1)^i (a \oslash_{i+1}b)-(-1)^{|a|\,|b|} (b\oslash_{i+1} a)~.
        \end{aligned}
    \end{equation}
    for all $v\in V$ and $a,b\in \caE(V)$. 
    \begin{proposition}\label{prop:Q0_cohomology}
        The $Q_0$-cohomology of $(\caE(V),Q_0)$ is the vector space\footnote{Here, $\odot$ denotes the symmetrised tensor product.} $\sfE_0(\bigodot^\bullet V)$, with the embedding map $\sfE_0:\bigodot^\bullet V\mapsto \caE(V)$ defined by 
        \begin{equation}
            \sfE_0(v_1\odot \dotsb \odot v_n)\coloneqq \sum_{\sigma\in S_n}(\dotsb(v_{\sigma(1)}\oslash_0 v_{\sigma(2)})\oslash_0\dotsb )\oslash_0 v_{\sigma(n)}~.
        \end{equation}
        This vector space carries the evident structure of a differential graded commutative algebra.
    \end{proposition}
    \begin{proof}
        It is clear that the image of $\sfE_0$ forms the kernel of $Q_0$ inside $\caE_{\{0\}}(V)$ and that $\caE_{\{0\}}(V)$ cannot contain any coboundaries.  Moreover, it is easy to show that all cocycles are coboundaries in the case of elements of $V\oslash_\bullet V$: here, the kernel of $Q_0$ consists of elements of the form 
        \begin{equation}
            v_1\oslash_i v_2+(-1)^{i+|v_1|\,|v_2|}v_2\oslash_i v_1=(-1)^{i-1}Q(v_1\oslash_{i-1}v_2)~.
        \end{equation}
        For elements that are cubic and of higher order in $V$, the proof is subtle and lengthy, and we only sketch it here.
        We first restrict to elements $a\in\caE(V)$ of a particular degree and order in $V$, as both the deformed Leibniz rule~\eqref{eq:def_Leibniz} and the algebra relations~\eqref{eq:Eilh-relations} preserve these. We then split $Q a\in \caE(V)$ into polynomials of the same monomial type, i.e.~the same order and type of products $\oslash_i$. These terms have to vanish separately. The condition $Q a=0$ requires the application of the relations~\eqref{eq:Eilh-relations} $d-2$ times, where $d$ is the degree of $a$ in $V$. This application enforces a particular symmetry structure on $a$ which forces it to lie in the image of $V$.
    \end{proof}
    
    \Cref{prop:Q0_cohomology} together with the usual arguments underlying a general Hodge--Ko\-daira decomposition\footnote{see e.g.~\cite{Weibel:1994aa}} then yield the following theorem:
    \begin{theorem}
        Consider the trivial semifree $\opEilh$-algebra from \cref{prop:Q0_cohomology}. Then we have the following contracting homotopy between $\caE(V)$ and the differential graded commutative algebra $(\bigodot^\bullet V,0)$:
        \begin{subequations}
            \begin{equation}\label{eq:contracting_hom_Q0}
                \begin{tikzcd}
                    \ar[loop,out=194,in= 166,distance=20,"\sfH_0"](\caE(V),Q_0)\arrow[r,shift left]{}{\sfP_0} & (\bigodot^\bullet V,0) \arrow[l,shift left]{}{\sfE_0}~,
                \end{tikzcd}
            \end{equation}
            with the projection 
            \begin{equation}
                \begin{aligned}
                    \sfP_0(v_1)&=v_1
                    \\
                    \sfP_0\big(((v_0\oslash_{i_1} v_2)\oslash_{i_2} v_3)\dotsb \oslash_{i_n} v_n\big)&=\begin{cases}
                        \tfrac{1}{(n+1)!}v_0\odot v_1\odot\dotsb\odot v_n& i_1=\dotsb=i_n=0~,
                        \\
                        0 & \mbox{else}
                    \end{cases}
                \end{aligned}
            \end{equation}
            for all $v_0,\ldots, v_n\in V$, such that
            \begin{equation}\label{eq:homotopy_transfer_relations}
                \begin{gathered}
                    \sfH_0\circ \sfH_0= \sfH_0\circ  \sfE_0=0~,~~~ \sfP_0\circ  \sfH_0=0~,
                    \\
                    \sfid_{\caE(V)}- \sfE_0\circ  \sfP_0= \sfH_0\circ  Q_0+ Q_0\circ  \sfH_0~.
                \end{gathered}
            \end{equation}
        \end{subequations}
    \end{theorem}
    
    As usual, the map $\sfH_0$ is not unique. 
    
    We note that such a contracting homotopy for ordinary differential graded algebras often induces an algebra morphism $\Phi\coloneqq \sfE_0\circ \sfP_0$. This is not the case here, as clearly $\Phi(a)\oslash_i\Phi(b)\neq \Phi(a\oslash_i b)$ in general. We shall return to this point in~\cref{ssec:EL_infty_and_L_infty}.
    
    The contracting homotopy~\eqref{eq:contracting_hom_Q0} has a number of important generalisations and applications. Here, we note that it evidently extends to differentials $Q_{\rm lin}=Q_0+\rmd$, where $\rmd:V\rightarrow V$ and $\rmd$ satisfies the ordinary Leibniz rules on $\caE(V)$ and $\bigodot^\bullet V$:
    \begin{equation}\label{eq:homotopy_for_transfer_EL_to_L}
        \begin{tikzcd}
            \ar[loop,out=194,in= 166,distance=20,"\sfH_0"](\caE(V),Q_0+\rmd)\arrow[r,shift left]{}{\sfP_0} & (\bigodot\nolimits^\bullet V,\rmd) \arrow[l,shift left]{}{\sfE_0}~.
        \end{tikzcd}
    \end{equation}
    Moreover, if we have a differential $\rmd:\bigodot^\bullet V\rightarrow \bigodot^\bullet V$ non-linear in $\odot$, we still have a corresponding contracting homotopy
    \begin{equation}
        \begin{tikzcd}
            \ar[loop,out=194,in= 166,distance=20,"\sfH_0"](\caE(V),Q_0+Q_1)\arrow[r,shift left]{}{\sfP_0} & (\bigodot\nolimits^\bullet V,\rmd) \arrow[l,shift left]{}{\sfE_0}
        \end{tikzcd}
    \end{equation}
    with
    \begin{equation}
        Q_1=\sfE_0\circ \rmd\circ \sfP_0~.
    \end{equation}
    We therefore arrive at the following statement:
    \begin{theorem}\label{thm:lift_dgca_to_Eilh}
        Any semifree differential graded commutative algebra $(\bigodot^\bullet V,\rmd)$ gives rise to the semifree $\opEilh$-algebra $(\caE(V),Q_0+\sfE_0\circ \rmd\circ \sfP_0)$. 
    \end{theorem}
    
    We will pick up this thread of our discussion again later.
    
    \section{\texorpdfstring{$E_2L_\infty$}{EL-infinity}-algebras}\label{sec:EL_infty}
    
    \subsection{\texorpdfstring{$E_2L_\infty$}{EL-infinity}-algebras and their morphisms}
    
    We define $E_2L_\infty$-algebras to be the homotopy version of $\ophLie$-algebras.
    \begin{definition}
        An \(E_2L_\infty\)-algebra is a graded vector space \(\frE\) equipped with a nilpotent differential on the semi-free \(\opEilh\)-algebra $\sfCE(\frE)\coloneqq \caE(V)$ for $V=\frE[1]^*$.
    \end{definition}
    
    
    The differential $Q$ is fully specified by its action on $V$. With respect to a basis $(t^\alpha)$ on $V$, we encode this action in structure constants $m$ taking values in the ground field as follows:
    \begin{equation}\label{eq:structure_constants_ELinfty}
        Q t^\alpha=\pm m^\alpha\pm m^\alpha_\beta t^\beta~\pm~m^{i_1,\alpha}_{\beta_1\beta_2}~t^{\beta_1}\oslash_{i_1} t^{\beta_2}~\pm~m^{i_1i_2,\alpha}_{\beta_1\beta_2\beta_3}~(t^{\beta_1}\oslash_{i_1} t^\gamma)\oslash_{i_2} t^\delta+\ldots~.
    \end{equation}
    Here the choice of signs $\pm$ is a convention\footnote{For $L_\infty$-algebras, we follow the conventions of~\cite{Jurco:2018sby} for the structure constants and the differential in the Chevalley--Eilenberg algebra.} and we shall be more explicit below, cf.\linebreak also~\eqref{eq:Q_hLie}. The structure constants $m$ define \uline{higher products},
    \begin{equation}
        \begin{gathered}
            \eps^I_n:\frE^{\otimes n}\rightarrow \frE~,
            \\
            \eps_0=m^\alpha \tau_\alpha~,~~~
            \eps_1(\tau_\alpha)=m^\beta_\alpha \tau_\beta~,~~~\eps_2^i(\tau_\alpha,\tau_\beta)=m^{i,\gamma}_{\alpha\beta}\tau_\gamma~,~~~\ldots
            \\
            \eps^I_n(\tau_{\alpha_1},\ldots,\tau_{\alpha_n})=m_{\alpha_1\ldots\alpha_n}^{I,\beta} \tau_\beta~,
        \end{gathered}
    \end{equation}
    where $I$ is a multi-index consisting of $n-1$ indices $i_1,i_2,\ldots,\in \{0,1\}$ and we define $|I|\coloneqq i_1+i_2+\ldots$. The products $\eps^I_n$ have degree~$-|I|$.
    
    It is useful to identify the following special cases:
    \begin{definition}\label{def:special_EL_infty}
        Let $(\frE,\eps_k^I)$ be an $E_2L_\infty$-algebra. If $\eps_0\neq 0$, we call $\frE$ \uline{curved}, otherwise $\frE$ is \uline{uncurved}. An uncurved $E_2L_\infty$-algebra is called \uline{hemistrict}, if $\eps_k^I=0$ for $k\geq 3$. It is called \uline{strict} if it is hemistrict and $\eps_2^i=0$ for $i>0$. Finally, $\frE$ is called \uline{semistrict} if $\epsilon_n^I=0$ for $I\neq (0,\ldots,0)$.
    \end{definition}
    In the following, all our $E_2L_\infty$-algebras will be uncurved.
    Clearly, hemistrict $E_2L_\infty$-algebras are simply $\ophLie$-algebras, and therefore $E_2L_\infty$-algebras subsume differential graded Lie algebras.
    
    As an example, let us consider a family of weak models of the string Lie 2-algebra which we will use to exemplify many of our constructions in the following. We consider the graded vector space $V=(\frg\oplus \IR[1])[1]^*$, where $\frg$ is a finite-dimensional quadratic (i.e.~metric) Lie algebra with structure constants $f^\alpha_{\beta\gamma}$ and Cartan--Killing form $\kappa_{\alpha\beta}$ with respect to a basis $(\tau_\alpha)$. On $V$, we introduce basis vectors $t^\alpha\in \frg[1]^*$ and $r\in \IR[2]^*$ of degrees $1$ and $2$, respectively. The differential is given by
    \begin{equation}
        \begin{aligned}
            Q t^\alpha&=-f^\alpha_{\beta\gamma}\,t^\beta \oslash_0 t^\gamma+(1-\vartheta)\kappa_{\alpha\delta} f^\delta_{\beta\gamma}\,(t^\beta \oslash_0 t^\gamma) \oslash_0 t^\delta-2\vartheta \kappa_{\beta\gamma}\,t^\beta \oslash_0 t^\gamma~,
            \\
            Q r &=0
        \end{aligned}
    \end{equation}
    with $\vartheta\in \IR$, and a direct calculation verifies $Q^2=0$. This defines the family of $E_2L_\infty$-algebras $\frstring^{\rmwk,\vartheta}_\rmsk(\frg)$ with the following underlying graded vector space and higher products:
    \begin{equation}\label{eq:skeltal_model_string_Lie_2_algebra}
        \begin{gathered}
            \frstring^{\rmwk,\vartheta}_\rmsk(\frg)~\coloneqq ~(\IR[1]\xrightarrow{~0~}\frg)~,
            \\
            \eps_2(x_1,x_2)=[x_1,x_2]~,~~~\eps_2^1(x_1,x_2)=2\vartheta(x_1,x_2)
            \\
            \eps^{00}_3(x_1,x_2,x_3)=(1-\vartheta)(x_1,[x_2,x_3])~,
        \end{gathered}
    \end{equation}
    for $x\in \frg$ and $y\in \IR$. All other higher products vanish. We notice that $\frstring^{\rmwk,\vartheta}_\rmsk(\frg)$ is an uncurved $E_2L_\infty$-algebra for each $\vartheta\in\IR$. It becomes semistrict for $\vartheta=0$ and hemistrict for $\vartheta=1$.
    
    
    Another very general and useful example is the $E_2L_\infty$-algebra of inner derivations $\frinn(\frE)$ of another $E_2L_\infty$-algebra $\frE$. This is obtained as a straightforward generalisation of the definition of the (unadjusted) Weil algebra of an $L_\infty$-algebra.
    \begin{definition}\label{def:Weil-EL-infty}
        The \uline{(unadjusted) Weil algebra} of an $E_2L_\infty$-algebra $\frE$ is the $\opEilh$-algebra
        \begin{equation}
            \sfW(\frE)\coloneqq \Big(~\bigoslash\nolimits^\bullet~(\frE[1]^*\oplus \frE[2]^*)~,~Q_\sfW~\Big)~,
        \end{equation}
        where the Weil differential is defined by the relations
        \begin{equation}
            Q_\sfW=Q_\sfCE+\sigma~,~~~Q_\sfW\sigma=-\sigma Q_\sfW~,
        \end{equation}
        in which \(Q_\sfCE\) is the differential on \(\sfCE(\frE)\subset\sfW(\frE)\) and
        $\sigma\colon\frE[1]^*\rightarrow \frE[2]^*$ is the shift isomorphism, trivially extended to $\sfW(\frE)$ by the (undeformed) Leibniz rule, i.e.
        \begin{equation}
            \sigma(a\oslash_i b)=(-1)^i\big(\sigma a\oslash_i b+(-1)^{|a|}a \oslash_i \sigma b\big)~.
        \end{equation}
        The $E_2L_\infty$-algebra dual to $\sfW(\frE)$ is the \uline{inner derivation $E_2L_\infty$-algebra} $\frinn(\frE)$ of $\frE$.
    \end{definition}    
    
    In terms of semifree $\opEilh$-algebras, the notion of a morphism of $E_2L_\infty$-algebras becomes evident. 
    \begin{definition}\label{def:morphisms}
        A \uline{morphism of $E_2L_\infty$-algebras} $\phi:\frE\rightarrow \tilde \frE$ is a morphism dual to the corresponding morphism $\Phi:\sfCE(\frE)\rightarrow \sfCE(\tilde \frE)$ of $\opEilh$-algebras. For $\sfCE(\frE)=({\caE(V)}, Q)$ and $\sfCE(\tilde \frE)=({\caE(\tilde V)},\tilde Q)$, such a morphism is compatible with the differential,
        \begin{equation}
            Q\circ \Phi=\Phi\circ \tilde Q~,
        \end{equation}
        and the product structure,
        \begin{equation}
            \Phi(x\oslash_i y)=\Phi(x)\oslash_i\Phi(y)
        \end{equation}
        for all $x,y\in \sfCE(\frE)$ and $i\in\{0,1\}$.
    \end{definition}
    
    Recall that the appropriate notion of equivalence for homotopy algebras is that of a quasi-isomorphism, which we make explicit in the following definition.
    \begin{definition}
        Two $E_2L_\infty$-algebras $\frE$ and $\tilde \frE$ are called \emph{quasi-isomorphic} if there is a morphism of $E_2L_\infty$-algebras $\phi:\frE\rightarrow \tilde \frE$ such that the contained chain map $\phi_1$, the dual to the linear component of the dual morphism of $\opEilh$-algebras $\Phi$, descends to an isomorphism between the cohomologies of $\frE$ and $\tilde \frE$.
    \end{definition}
    We note that the cohomology $H^\bullet_{\eps_1}(\frE)$ of an $E_2L_\infty$-algebra $\frE$ is dual to the cohomology with respect to the linear part of the Chevalley--Eilenberg differential $Q$. In the case of the Weil algebra, the included shift isomorphism $\sigma$ renders the cohomology $H^\bullet_{\eps_1}(\frinn(\frE))$ trivial. We therefore obtain a quasi-isomorphism between $\frinn(\frE)$ and the trivial $E_2L_\infty$-algebra, extending the situation for $L_\infty$-algebras.
    
    In the future, it may be useful to have an inner product structure on an $E_2L_\infty$-algebra. The appropriate notion here is a simple generalisation of cyclic structures on $L_\infty$-algebras.
    \begin{definition}
        An $E_2L_\infty$-algebra $\frE$ is called \uline{cyclic} if it is equipped with a non-degenerate graded-symmetric bilinear form $\langle-,-\rangle:\frE\times \frE\rightarrow K$, where $K$ is the ground field, such that
        \begin{equation}\label{eq:cyclicity}
            \langle e_1,\eps_i(e_2,\ldots,e_{i+1})\rangle=(-1)^{i+i(|e_1|+|e_{i+1}|)+|e_{i+1}|\sum_{j=1}^{i}|e_j|}\langle e_{i+1},\eps_i(e_1,\ldots,e_{i})\rangle
        \end{equation}
        for all $e_i\in \frE$.
    \end{definition}
    
    \subsection{Homotopy transfer and minimal model theorem}\label{ssec:homotopy_transfer}
    
    A minimal characteristic of homotopy algebras is that they provide a nice notion of homotopy transfer. The latter will be important for all our future constructions, and we therefore develop a form of the homological perturbation lemma below. 
    
    We start from deformation retract between two differential graded complexes $(\frE,\eps_1)$ and $(\tilde \frE,\tilde \eps_1)$. That is, we have chain maps $\sfp$ and $\sfe$, together with a map $\sfh$ of degree~$-1$, which fit into the diagram
    \begin{subequations}\label{eq:HE_data_diff_complex}
        \begin{equation}
            \begin{tikzcd}
                \ar[loop,out=194,in= 166,distance=20,"\sfh"](\frE,\eps_1)\arrow[r,shift left]{}{\sfp} & (\tilde \frE,\tilde \eps_1) \arrow[l,shift left]{}{\sfe}~,
            \end{tikzcd}
        \end{equation}
        and satisfy the relations
        \begin{equation}\label{eq:homotopy_transfer_relations_dual}
            \begin{gathered}
                \sfp\circ \sfe=\sfid_{\tilde \frE}~,~~~\sfid_{\frE}-\sfe\circ \sfp=\sfh\circ \eps_1+\eps_1\circ \sfh~,
                \\
                \sfh\circ\sfh=0~,~~~\sfh\circ \sfe=0~,~~~\sfp\circ \sfh=0~.
            \end{gathered}
        \end{equation}
    \end{subequations}
    We now want to consider the homological perturbation lemma for the semifree $\opEilh$-algebras on ${\caE(V)}$ and ${\caE(\tilde V)}$ with differentials $Q$ and $\tilde Q$, respectively, defined by
    \begin{equation}
        V=\frE[1]^*~,~~~Q=\eps_1^*\eand \tilde V=\tilde \frE[1]^*~,~~~\tilde Q=\tilde \eps_1^*~.
    \end{equation}
    
    \begin{theorem}
        The deformation retract~\eqref{eq:HE_data_diff_complex} lifts to the deformation retract
        \begin{subequations}\label{eq:HE_data_Eilh}
            \begin{equation}
                \begin{tikzcd}
                    \ar[loop,out=194,in= 166,distance=20,"\sfH_0"]({\caE(V)},Q)\arrow[r,shift left]{}{\sfP_0} & ({\caE(\tilde V)},\tilde Q) \arrow[l,shift left]{}{\sfE_0}~,
                \end{tikzcd}
            \end{equation}
            with\footnote{The notation is chosen to match more closely the formulas of \cref{ssec:cohomology_Eilh}.}
            \begin{equation}
                \begin{aligned}
                    \sfE_0(v)&=\sfp^*(v)~,~~~&\sfE_0(x\oslash_i y)&=\sfE_0(x)\oslash_i\sfE_0(y)~,
                    \\
                    \sfP_0(v)&=\sfe^*(v)~,~~~&\sfP_0(x\oslash_i y)&=\sfP_0(x)\oslash_i\sfP_0(y)~,
                \end{aligned}
            \end{equation}
            for all $v\in V$ and $x,y\in {\caE(V)}$. The dual to the contracting homotopy is continued by a modification of the tensor trick, 
            \begin{equation}
                \begin{aligned}
                    \sfH_0(v)&=\sfh^*(v)~,
                    \\
                    \sfH_0(x\oslash_i y)&=(-1)^i\sfH_0(x)\oslash_i\sfE_0(\sfP_0(y))+(-1)^{i+|x|}x\oslash_i\sfH_0(y)
                    \\
                    &\hspace{4cm}+(-1)^{|y|+|x|\,|y|}\sfH_0(x)\oslash_{i+1}\sfH_0(y)~.
                \end{aligned}
            \end{equation}
            These maps satisfy the relations
            \begin{equation}\label{eq:homotopy_transfer_relations_0}
                \begin{gathered}
                    \sfP_0\circ\sfE_0=\sfid_{\caE(\tilde V)}~,~~~\sfid_{\caE(V)}-\sfE_0\circ \sfP_0=\sfH_0\circ Q+Q\circ \sfH_0~,
                    \\
                    \sfH_0\circ\sfH_0=0~,~~~\sfH_0\circ \sfE_0=0~,~~~\sfP_0\circ \sfH_0=0~.
                \end{gathered}
            \end{equation}
        \end{subequations}
    \end{theorem}
    \begin{proof}
        The proof of~\eqref{eq:homotopy_transfer_relations_0} is a straightforward computation for elements in $\oslash^2_\bullet V$. The general case follows then by iterating the algebra relations and applying~\eqref{eq:homotopy_transfer_relations_dual}.
    \end{proof}
    
    The higher products $\eps_i$ with $i>1$ on $\frE$ can now be regarded as a perturbation of the differential. Dually, we have a perturbation of $Q$,
    \begin{equation}
        Q\rightarrow \hat Q\coloneqq Q+Q_\delta~.
    \end{equation}
    We can then use the homological perturbation lemma~\cite{Gugenheim1989:aa,gugenheim1991perturbation,Crainic:0403266} to transfer these structures over to higher products $\tilde \eps_i$ on $\tilde \frE$, or, dually, to a perturbed differential $\tilde Q$ on $\caE(\tilde V)$. The formulas for this are the usual ones, cf.~\cite{Crainic:0403266}.
    
    \begin{lemma}\label{prop:HPL}
        Starting from the deformation retract~\eqref{eq:HE_data_Eilh}, we can construct another deformation retract
        \begin{subequations}\label{eq:HE_data_Eilh2}
            \begin{equation}
                \begin{tikzcd}
                    \ar[loop,out=194,in= 166,distance=20,"\sfH"]({\caE(V)},\hat Q)\arrow[r,shift left]{}{\sfP} & ({\caE(\tilde V)},\hat{\tilde Q}) \arrow[l,shift left]{}{\sfE}~,
                \end{tikzcd}
            \end{equation}
            with
            \begin{equation}
                \begin{gathered}
                    \hat Q\coloneqq Q+Q_\delta~,~~~\hat{\tilde Q}=\tilde Q+ \sfP\circ Q_\delta \circ \sfE_0~,
                    \\
                    \sfP\coloneqq \sfP_0\circ(\sfid+Q_\delta\circ\sfH_0)^{-1}~,~~~\sfE\coloneqq \sfE_0-\sfH\circ Q_\delta\circ \sfE_0~,~~~\sfH=\sfH_0\circ(\sfid+Q_\delta\circ \sfH_0)^{-1}~,
                \end{gathered}
            \end{equation}
            where inverse operators are defined via the evident geometric series, such that 
            \begin{equation}\label{eq:homotopy_transfer_relations2}
                \begin{gathered}
                    \sfP\circ \sfE=\sfid_{\caE(\tilde V)}~,~~~
                    \sfid_{\caE(V)}-\sfE\circ \sfP=\sfH\circ \hat Q+\hat Q\circ \sfH~,
                    \\
                    \sfH_0\circ\sfH_0=\sfH_0\circ \sfE=0~,~~~\sfP\circ \sfH=0~,
                \end{gathered}
            \end{equation}
        \end{subequations}
    \end{lemma}
    \begin{proof}
        The lemma follows from the usual perturbation lemma, cf.~\cite{Crainic:0403266}, with the specialisation that $\sfE$ and $\sfP$ are here morphisms of $\opEilh$-algebras. To see this, we note that $Q_\delta$ acts as a derivation,
        \begin{equation}
            Q_\delta(x\oslash_i y)=(-1)^i\big((Q_\delta x)\oslash_i y+(-1)^{|x|}x\oslash_i Q_\delta y\big)~.
        \end{equation}
        Moreover, in the non-vanishing terms of
        \begin{equation}
            \begin{aligned}
                \sfP(x\oslash_i y)&=(\sfP_0\circ(\sfid-Q_\delta \circ \sfH_0+Q_\delta\circ \sfH_0\circ Q_\delta\circ \sfH_0-\ldots)\big)(x\oslash_i y)
                \\
                &=\sfP_0(x)\oslash_i\sfP_0(y)-\sfP_0(Q_\delta(\sfH_0(x))\oslash_i\sfP_0(y)-\sfP_0(x)\oslash_i\sfP_0(Q_\delta(\sfH_0(x)))+\dotsb~,
            \end{aligned}
        \end{equation}
        all the $\sfH_0$ are precomposed by a $Q_\delta$, as otherwise the map $\sfP_0$, which is precomposed to all summands, would annihilate the term due to $\sfP_0\circ \sfH_0=0$. The relation
        \begin{equation}
            \sfP(x\oslash_i y)=\sfP(x)\oslash_i\sfP(y)
        \end{equation}
        follows then by a direct computation. The same holds for $\sfE$.
    \end{proof}
    We note that for small perturbations $Q_\delta$, the homological perturbation \cref{prop:HPL} implies that
    \begin{equation}\label{eq:HPL_geom_series}
        \hat{\tilde Q}=\tilde Q+\sfP_0\circ Q_\delta\circ \sfE_0-\sfP_0\circ Q_\delta\circ \sfH_0\circ Q_\delta \circ \sfE_0+\sfP_0\circ Q_\delta\circ \sfH_0\circ Q_\delta\circ \sfH_0\circ Q_\delta \circ \sfE_0-\dotsb~.
    \end{equation}
    
    A direct consequence of homotopy transfer is the existence of minimal models for homotopy algebras. Consider the deformation retract~\eqref{eq:HE_data_diff_complex} with $(\tilde \frE,\tilde \eps_1=0)=H^\bullet_{\eps_1}(\frE)$ the cohomology of $(\frE,\eps_1)$ as well as $\sfp$ and $\sfe$ the projection onto the cohomology and a choice of embedding, respectively. Then the homotopy transfer yields the structure of a quasi-isomorphic $E_2L_\infty$-algebra on the cohomology of $(\frE,\eps_1)$. This implies the minimal model theorem.
    \begin{theorem}
        Any $E_2L_\infty$-algebra $\frE$ comes with a quasi-isomorphic $E_2L_\infty$-algebra structure on its cohomology $H^\bullet_{\eps_1}(\frE)$. We call this a \uline{minimal model} of $\frE$.
    \end{theorem}
    
    \subsection{\texorpdfstring{$L_\infty$}{L-infinity}-algebras as \texorpdfstring{$E_2L_\infty$}{EL-infinity}-algebras}\label{ssec:L_infty_as_EL_infty}
    
    As expected, $L_\infty$-algebras are special cases of $E_2L_\infty$-algebras.
    \begin{theorem}\label{thm:el_infty_contain_l_infty}
        A semistrict $E_2L_\infty$-algebra $\frE$ is an $L_\infty$-algebra. Conversely, any $L_\infty$-algebra is a (semistrict) $E_2L_\infty$-algebra. Dually, the data contained in a differential $Q$ in a semifree $\opEilh$-algebra $({\caE(V)},Q)$ with $\rmIm(Q|_V)\subset \bigoslash_0^\bullet V$ is equivalent to the data contained in a differential $\tilde Q$ on the semifree differential graded commutative algebra $(\bigodot^\bullet V,\tilde Q)$.
    \end{theorem}
    \begin{proof}
        It suffices to show the dual statement, which is a direct consequence of \cref{thm:lift_dgca_to_Eilh}.
    \end{proof}
    
    Concretely, given an $L_\infty$-algebra $\frL$ with higher products $\mu_k$ this yields a semistrict $E_2L_\infty$-algebra with higher products
    \begin{equation}
        \eps^I_k=\begin{cases}
            \mu_k & \mbox{for~$|I|=0$}~,
            \\
            0 & \mbox{else}~.
        \end{cases}
    \end{equation}
    Dually, we have agreement in the structure constants in the corresponding Chevalley--Eilenberg differentials $Q_{\rm L}$ and $Q_{\rm EL}$ for the $L_\infty$-algebra and its trivial lift to $E_2L_\infty$-algebra up to combinatorial factors:
    \begin{equation}
        \begin{aligned}
            Q_{\rm L} t^\alpha&= q^\alpha_\beta t^\beta+\tfrac12 q^\alpha_{\beta\gamma}t^\beta t^\gamma+\tfrac1{3!}q^\alpha_{\beta\gamma\delta}t^\beta t^\gamma t^\delta+\ldots~,
            \\
            Q_{\rm EL} t^\alpha&= q^\alpha_\beta t^\beta+q^\alpha_{\beta\gamma}t^\beta \oslash_0t^\gamma+q^\alpha_{\beta\gamma\delta}(t^\beta\oslash_0 t^\gamma)\oslash_0 t^\delta+\ldots~.
        \end{aligned}
    \end{equation}
    
    Conversely, any semistrict $E_2L_\infty$-algebra $\frE$ is an $L_\infty$-algebra with higher products $\mu_k=\eps^0_k$. As an example, consider the $\vartheta=0$ case of the family of weak string Lie 2-algebra models~\eqref{eq:skeltal_model_string_Lie_2_algebra}. This is a semistrict $E_2L_\infty$-algebra and therefore an $L_\infty$-algebra.
    
    For consistency, we obviously expect the following.
    \begin{theorem}\label{thm:lift_L_infty_morphism}
        Any $L_\infty$-algebra morphism $\phi:\frL\rightarrow \tilde \frL$ lifts to an $E_2L_\infty$-algebra morphism $\hat \phi:\hat \frL\rightarrow \hat{\tilde \frL}$, where $\hat \frL$ and $\hat{\tilde \frL}$ are the $L_\infty$-algebras $\frL$ and $\tilde \frL$, regarded as $E_2L_\infty$-algebras.
    \end{theorem}
    \begin{proof}
        We prove this statement again from the dual perspective. Let $(\bigodot^\bullet V,Q)$ and $(\bigodot^\bullet \tilde V,\tilde Q)$ be the Chevalley--Eilenberg algebras of $\frL$ and $\tilde \frL$, respectively. The Chevalley--Eilenberg algebras of $\hat \frL$ and $\hat{\tilde \frL}$ are then
        \begin{equation}
            (\caE(V),\hat Q=Q_0+\sfE\circ Q\circ \sfP)
            \eand 
            (\caE(\tilde V),\hat{\tilde Q}=Q_0+\sfE\circ \tilde Q\circ \sfP)~,
        \end{equation}
        cf.~\cref{thm:lift_dgca_to_Eilh}. The dual of the morphism $\phi$,
        \begin{equation}
            \Phi:\sfCE(\frL)\rightarrow \sfCE(\tilde \frL)~,
        \end{equation}
        trivially lifts to the following dual of an $E_2L_\infty$-algebra morphism 
        \begin{equation}
            \hat \Phi:\sfCE(\hat\frL)\rightarrow \sfCE(\hat{\tilde \frL})\ewith\hat \Phi(v)\coloneqq \sfE(\Phi(v))~,
        \end{equation}
        and we note that $\hat \Phi \circ \sfE=\sfE\circ \Phi$. It then follows that 
        \begin{equation}
            \begin{aligned}
                (\hat \Phi\circ \sfE\circ Q)(v)&=(\sfE\circ \Phi\circ Q)(v)
                \\
                \hat \Phi(\hat Qv)&=(\sfE\circ \tilde Q\circ \sfP\circ \sfE\circ\Phi)(v)~,
                \\
                \hat \Phi(\hat Qv)&=\hat{\tilde Q}\hat \Phi(v)~,
            \end{aligned}
        \end{equation}
        and $\hat\Phi$ is the dual to the desired morphism of $E_2L_\infty$-algebras $\hat \phi$.
    \end{proof}
    
    \subsection{2-term \texorpdfstring{$E_2L_\infty$}{EL-infinity}-algebras}\label{ssec:royten}
    
    Having identified $L_\infty$-algebras within $E_2L_\infty$-algebras, let us also make contact with the 2-term $E_2L_\infty$-algebras of~\cite{Roytenberg:0712.3461}.
    
    In~\cite{Baez:2003aa}, Baez--Crans introduced semistrict Lie 2-algebras, which are linear categories equipped with a strictly antisymmetric  bilinear functor,  the categorified Lie bracket, that is only required to satisfy the Jacobi identity up to a coherent trilinear natural transformation, the \emph{Jacobiator}.  In~\cite{Roytenberg:0712.3461}  semistrict Lie 2-algebras were fully categorified  to \emph{weak} Lie 2-algebras by also relaxing  antisymmetry  to hold only up to a coherent natural transformation, the \emph{alternator}.  Of course, a weak Lie 2-algebra with trivial alternator is a semistrict Lie 2-algebra. Similarly,  a weak Lie 2-algebra  with trivial Jacobiator is referred to as a  \emph{hemi}-strict Lie 2-algebra. If both the alternator and Jacobiator are trivial, it is a \emph{strict} Lie 2-algebra. 
    
    By passing to its normalised chain complex, we transition from the categorical description containing many redundancies to a more convenient description in terms of differential graded algebras. In particular, a semistrict Lie 2-algebra is seen to be equivalent to a 2-term $L_\infty$-algebra~\cite{Baez:2003aa}, i.e.~an $L_\infty$-algebra with underlying graded vector space concentrated in degrees $-1$ and $0$. Analogously,   by    passing to its normalised chain complex, any weak Lie 2-algebra is seen to be equivalent to a  2-term $E_2L_\infty$-algebra in the sense of~\cite{Roytenberg:0712.3461}, where the letter $E$ was added to indicate that `everything' is relaxed up to homotopy. 
    
    \begin{theorem}\label{thm:2-term}
        An $E_2L_\infty$-algebra structure on a two-term complex concentrated in degrees $-1$ and $0$, $\frE:\frE_{-1}\xrightarrow{~\eps_1~}\frE_0$ has only three non-trivial higher products,
        \begin{subequations}\label{eq:rels_2-term-EL_infty}
            \begin{equation}
                \begin{aligned}
                    \eps_2\coloneqq \eps_2^0~&:~\frE_i\otimes \frE_j\rightarrow \frE_{i+j}~,
                    \\
                    \eps_3\coloneqq \eps_3^{00}~&:~\frE_0\otimes \frE_0\otimes \frE_0\rightarrow \frE_{-1}~,
                    \\
                    \sfalt\coloneqq \eps_2^1~&:~\frE_0\otimes \frE_0\rightarrow \frE_{-1}~.
                \end{aligned}
            \end{equation}
            The map $\eps_2$ is a chain map, and the maps $\sfalt$ and $\eps_3$ are chain homotopies\footnote{Here, $\sigma_{12}$ denotes the obvious permutation.}
            \begin{equation}
                \begin{aligned}
                    \sfalt&\colon\eps_2(-,-)+\eps_2(-,-)\circ \sigma_{12} \rightarrow 0~,\\
                    \eps_3&\colon\eps_2(-,\eps_2(-,-))-\eps_2(\eps_2(-,-),-)-\eps_2(-,\eps_2(-,-))\circ \sigma_{12}\rightarrow 0~.
                \end{aligned}
            \end{equation} 
            In addition, the higher products satisfy the relations
            \begin{equation}\label{eq:EL:axioms}
                \begin{aligned}
                    &\sfalt(x_1,x_2)=\sfalt(x_2,x_1)~,
                    \\
                    &\eps_3(x_1,x_2,x_3)+\eps_3(x_2,x_1,x_3)=\eps_2(\sfalt(x_1,x_2),x_3)~,
                    \\
                    &\eps_3(x_1,x_2,x_3)+\eps_3(x_1,x_3,x_2)=\sfalt(\eps_2(x_1,x_2),x_3)+\sfalt(x_2,\eps_2(x_1,x_3))\\
                    &\hspace{8cm}-\eps_2(x_1,\sfalt(x_2,x_3))~,
                    \\
                    &\eps_2(x_1,\eps_3(x_2,x_3,x_4))+\eps_3(x_1,\eps_2(x_2,x_3),x_4)+\eps_3(x_1,x_3,\eps_2(x_2,x_4))+\\
                    &\hspace{1cm}+\eps_2(\eps_3(x_1,x_2,x_3),x_4)+\eps_2(x_3,\eps_3(x_1,x_2,x_4))\\
                    &\hspace{0.2cm}=\eps_3(x_1,x_2,\eps_2(x_3,x_4))+\eps_3(\eps_2(x_1,x_2),x_3,x_4)+\eps_2(x_2,\eps_3(x_1,x_3,x_4))\\
                    &\hspace{1cm}+\eps_3(x_2,\eps_2(x_1,x_3),x_4)+\eps_3(x_2,x_3,\eps_2(x_1,x_4))
                \end{aligned}
            \end{equation}
        \end{subequations}
        for all $x_i\in \frE_0$.
    \end{theorem}
    \begin{proof}
        Perhaps the easiest way of proving the above relations is to consider the corresponding Chevalley--Eilenberg algebra (and we again assume, for simplicity, that the degree-wise duals of $\frE$ are nice, cf.~\cref{ssec:Eilh-algebras}). That is, we consider the tensor algebra $\caE(V)$ for $V=\frE[1]^*$. The differential $Q$ is determined by its action on the basis elements $t^\alpha$. Since the latter can be of degree~$1$ or~$2$, it follows that $Qt^\alpha$ is of degree~$2$ or~$3$, and therefore it has to be of the form
        \begin{equation}\label{eq:form_EL_infty_Q}
            \begin{aligned}
                Q t^\alpha&=-(-1)^{|\beta|}m^\alpha_\beta t^\beta-(-1)^{|\gamma|(|\beta|-1)}\,m^{0,\alpha}_{\beta\gamma}\,t^\beta \oslash_0 t^\gamma
                \\
                &~~~-(-1)^{|\beta|+|\gamma|+|\delta|+|\gamma|(|\beta|-1)+|\delta|(|\beta|+|\gamma|-2)}\,m^{00,\alpha}_{\beta\gamma\delta}\,(t^\beta \oslash_0 t^\gamma) \oslash_0 t^\delta
                \\
                &~~~-(-1)^{|\beta|+|\gamma|+|\gamma|(|\beta|-1)}\,m^{1,\alpha}_{\beta\gamma}\,t^\beta \oslash_1 t^\gamma~,
            \end{aligned}
        \end{equation}
        where the $(t^\alpha)$ form a basis on $V$. The above formula can be reduced further when we split $(t^\alpha)=(r^a,s^i)$, where $|r^a|=1$ and $|s^i|=2$:
        \begin{equation}\label{eq:form_2-term-EL_infty_Q}
            \begin{aligned}
                Q r^a&=-m^a_a s^a-m^a_{bc}\,r^b \oslash_0 r^c~,
                \\
                Q s^i&=-m^i_{a j}\,r^a\oslash_0 s^j+m^a_{ja}\,s^j\oslash_0 r^a\\
                &~~~+m^i_{abc}\,(r^a\oslash_0 r^b) \oslash_0 r^c-n^i_{ab}\,r^a\oslash_1 r^b~.
            \end{aligned}
        \end{equation}
        Defining 
        \begin{equation}
            \begin{gathered}
                \eps_1(\tau_\alpha)=m^\beta_\alpha\tau_\beta,~~\eps_2(\tau_\alpha,\tau_\beta)=m^{0,\gamma}_{\alpha\beta}\tau_\gamma,~~\eps_3(\tau_\alpha,\tau_\beta,\tau_\gamma)=m^{00,\delta}_{\alpha\beta\gamma}\tau_\delta,~~\sfalt(\tau_\alpha,\tau_\beta)=m^{1,\gamma}_{\alpha\beta}\tau_\gamma
            \end{gathered}
        \end{equation}
        with respect to the basis $(\tau_\alpha)$ of $\frE$, which is shifted-dual to the basis $(t^\alpha)$ of $V=\frE[1]^*$, we readily verify that $Q^2=0$ corresponds to the equations~\eqref{eq:rels_2-term-EL_infty}.
    \end{proof}
    
    The properties~\eqref{eq:rels_2-term-EL_infty}, with a slightly weaker condition on $\sfalt$, were given as the defining axioms in the definition of a 2-term $E_2L_\infty$-algebra in the sense of~\cite{Roytenberg:0712.3461}.
    \begin{corollary}
        An $E_2L_\infty$-algebra of the form considered in \cref{thm:2-term} is a 2-term $E_2L_\infty$-algebra in the sense of~\cite{Roytenberg:0712.3461} with a graded symmetric alternator.
    \end{corollary}
    Our additional condition of graded symmetric alternator is, in fact, very natural. It guarantees a rectification to semistrict Lie 2-algebras, and we shall return to this point in \cref{ssec:EL_infty_and_L_infty}.

    \subsection{\texorpdfstring{$E_2L_\infty$}{EL-infinity}-algebras as \texorpdfstring{$L_\infty$}{L-infinity}-algebras and the rectification theorem}\label{ssec:EL_infty_and_L_infty}
    
    In \cref{ssec:L_infty_as_EL_infty}, we identified $L_\infty$-algebras with semistrict $E_2L_\infty$-algebras, which suggests that there should be a rectification of $E_2L_\infty$-algebras to $L_\infty$-algebras. This is indeed the case, as we show below.
    
    We start with a projection of $E_2L_\infty$-algebras onto $L_\infty$-algebras, extending results of~\cite{Roytenberg:0712.3461,Dehling:1710.11104}.
    \begin{theorem}\label{thm:antisym}
        Any $E_2L_\infty$-algebra $(\frE,\eps_i)$ induces an $L_\infty$-algebra structure on the graded vector space $\frE$. This $L_\infty$-algebra structure is induced by homotopy transfer using the homotopy~\eqref{eq:homotopy_for_transfer_EL_to_L}.
    \end{theorem}
    \begin{proof}
        The proof is readily obtained by applying the homological perturbation lemma to the contracting homotopy 
        \begin{equation}
            \begin{tikzcd}
                \ar[loop,out=204,in= 156,distance=70,"\sfH_0"](\caE(V),Q_0+Q_1+Q_\delta)\arrow[r,shift left]{}{\sfP_0} & (\odot^\bullet V,Q_L) \arrow[l,shift left]{}{\sfE_0}~,
            \end{tikzcd}
        \end{equation}
        cf.~\eqref{eq:homotopy_for_transfer_EL_to_L}. Consider the Chevalley--Eilenberg algebra $\sfCE(\frE)$ of $\frE$, and split the differential $Q=Q_0+Q_1+Q_\delta$ into $Q_0$, a linear part $Q_1$, and a perturbation $Q_\delta$. Then homotopy transfer yields a differential
        \begin{equation}\label{eq:HPL_formula_Q_EL_to_L}
            Q_L=Q_1+\sfP_0\circ Q_\delta\circ\sfE_0-\sfP_0\circ Q_\delta\circ \sfH_0\circ Q_\delta\circ \sfE_0+\sfP_0\circ Q_\delta\circ \sfH_0\circ Q_\delta\circ \sfH_0\circ Q_\delta\circ \sfE_0+\ldots
        \end{equation}
        on $\odot^\bullet(\frE[1]^*)$. By construction, $Q_L^2=0$.
        Moreover, $Q_L$ satisfies the Leibniz rule on $\odot^\bullet(\frE[1]^*)$: the deformation terms in the Leibniz rule~\eqref{eq:def_Leibniz} are graded antisymmetric, and this graded antisymmetry is preserved by subsequent applications of $\sfH$ and $Q_\delta$. The final projector $\sfP_0$ then eliminates these terms.
    \end{proof}
    As an example, we can compute the antisymmetrisation of a 2-term $E_2L_\infty$-algebra and reproduce\footnote{Due to different conventions, there is a relative minus sign between our $\mu_3$ and that of~\cite{Roytenberg:0712.3461}.}~\cite[Proposition 3.1]{Roytenberg:0712.3461}.
    \begin{corollary}
        Given a 2-term $E_2L_\infty$-algebra $\frE$, there is an $L_\infty$-algebra structure on the graded vector space $\frE$ with higher products
        \begin{equation}\label{eq:antisymmetrisation_2-term}
            \begin{aligned}
                \mu_1(y)&\coloneqq \eps_1(y)~,\\
                \mu_2(x_1,x_2)&\coloneqq \tfrac12(\eps_2(x_1,x_2)-\eps_2(x_2,x_1))~,\\
                \mu_2(x_1,y)=-\mu_2(y,x_1)&\coloneqq \tfrac12(\eps_2(x_1,y)-\eps_2(y,x_1))~,\\
                \mu_3(x_1,x_2,x_3)&\coloneqq \tfrac{1}{3!}\sum_{\sigma\in S_3}\chi(\sigma;x_1,x_2,x_3)\Big(\eps_3(x_{\sigma(1)},x_{\sigma(2)},x_{\sigma(3)})\\
                &\hspace{4.5cm}+\tfrac12 \sfalt(\eps_2(x_{\sigma(1)},x_{\sigma(2)}),x_{\sigma(3)})\Big)
            \end{aligned}
        \end{equation}
        for all $x\in \frE_0$ and $y\in \frE_{-1}$.
    \end{corollary}
    \begin{proof}
        We start from the Chevalley--Eilenberg algebra $\sfCE(\frE)$. As always, we assume for convenience that there is a basis, explicitly given by elements $r^\alpha$, $s^a$ of degrees $1$ and $2$ of $\frE[1]^*$. The differential then reads as
        \begin{equation}\label{eq:CE_differential_2-term}
            \begin{aligned}
                Q r^a&=-m^a_a s^a-m^a_{bc}\,r^b \oslash_0 r^c~,
                \\
                Q s^i&=-m^i_{a j}\,r^a\oslash_0 s^j+m^a_{ja}\,s^j\oslash_0 r^a\\
                &~~~+m^i_{abc}\,(r^a\oslash_0 r^b) \oslash_0 r^c-n^i_{ab}\,r^a\oslash_1 r^b~,
            \end{aligned}
        \end{equation}
        cf.~\eqref{eq:form_2-term-EL_infty_Q}. Evaluate formula~\eqref{eq:HPL_formula_Q_EL_to_L} yields
        \begin{equation}
            \begin{aligned}
                Q_L r^a &=-m^a_i s^i-\tfrac12 m^a_{bc}\,r^b \odot r^c~,
                \\
                Q_L s^i&=-(m^i_{a j}+m^i_{ja})r^a \odot s^j+\tfrac1{3!}(m^i_{abc}+\tfrac12 n^i_{dc}m^d_{ab})r^a\odot r^b\odot r^c~.
            \end{aligned}
        \end{equation}
        This is the differential for the Chevalley--Eilenberg algebra of $(\frE,\mu_i)$ with the higher products~\eqref{eq:antisymmetrisation_2-term}.
    \end{proof}
    
    As observed already in~\cite{Roytenberg:0712.3461,Dehling:1710.11104}, the antisymmetrisation map is functorial for 2-term $E_2L_\infty$-algebras, and any morphism of $E_2L_\infty$-algebra induces a morphism between the corresponding antisymmetrised $L_\infty$-algebras. The antisymmetrisation map fails, however, to be functorial for 3-term $E_2L_\infty$-algebras. It is natural to conjecture that the antisymmetrisation is $n$-functorial for an $n$-term $E_2L_\infty$-algebra. We will not need this result and refrain from going further into these technicalities.
    
    A new result is that this antisymmetrisation map lifts indeed to a quasi-isomorphism of $E_2L_\infty$-algebras. 
    \begin{theorem}\label{thm:quasi-iso_of_antisym}
        Let $\frE$ be an $E_2L_\infty$-algebra, and let $\frE'$ be the $L_\infty$-algebra induced by \cref{thm:antisym}, regarded as an $E_2L_\infty$-algebra. Then there is a quasi-isomorphism $\phi:\frE\rightarrow \frE'$.
    \end{theorem}
    \begin{proof}
        We prove this statement again using the Chevalley--Eilenberg algebras $\sfCE(\frE)$ and $\sfCE(\frE')$. Note that as graded vector spaces, $\frE[1]^*=\frE'[1]^*$. If $Q=Q_0+Q_1+Q_\delta$ is the differential on $\sfCE(\frE)$, then the differential on $\sfCE(\frE')$ reads as
        \begin{equation}\label{eq:lifted_differential}
            Q'v=Q_1v+\sfE_0\circ \sfP_0\circ (\sfid+Q_\delta\circ \sfH_0)^{-1}\circ Q_\delta v
        \end{equation}
        for all $v\in \frE[1]^*$. We must construct an invertible $\opEilh$-algebra morphism $\Phi:\caE(\frE[1]^*)\rightarrow \caE(\frE[1]^*)$ satisfying $Q\Phi=\Phi Q'$. The desired morphism on $\frE[1]^*$ is\footnote{This morphism implements a coordinate transformation such that the image of $Q$ on $\tilde v=\Phi(v)$ has no component in the subspace $Q_0\sfH_0\caE(\frE[1]^*)$. This then implies that it has no component in $\sfH_0 Q_0\caE(\frE[1]^*)$ either. The only remaining component is in $\sfE_0\sfP_0\caE(\frE[1]^*)$, which implies that $Q$ is the Chevalley--Eilenberg differential of an $L_\infty$-algebra, trivially regarded as an $E_2L_\infty$-algebra.}
        \begin{equation}
            \Phi(v)=(\sfid-\sfH_0\circ Q_\delta+\sfH_0\circ Q_\delta\circ \sfH_0\circ Q_\delta-\ldots)(v)=(\sfid+\sfH_0\circ Q_\delta)^{-1}(v)~,
        \end{equation}
        and using 
        \begin{equation}
            \begin{aligned}
                Q_0\circ \sfH_0&=\sfid-\sfE_0\circ \sfP_0-\sfH_0\circ Q_0~,
                \\
                Q_0Q_\delta&=-Q_\delta^2-Q_\delta Q_0~,
            \end{aligned}
        \end{equation}
        one readily verifies that $Q\Phi v=\Phi Q'v$ for all $v\in \frE[1]^*$, which is sufficient. Since the morphism is clearly invertible, this is a quasi-isomorphism.
    \end{proof}
    
    We can now combine \cref{thm:quasi-iso_of_antisym}, \cref{thm:lift_L_infty_morphism} as well as the strictification theorem for $L_\infty$-algebras to obtain the following.
    \begin{corollary}\label{thm:strictification}
        Any $E_2L_\infty$-algebra is quasi-isomorphic to a differential graded Lie algebra, trivially regarded as an $E_2L_\infty$-algebra.
    \end{corollary}
    More directly, this follows from the strictification theorem for generic homotopy algebras, see e.g.~\cite[Proposition 11.4.9]{Loday:2012aa}.
    
    \Cref{thm:quasi-iso_of_antisym} also shows that the choice of~\cite{Roytenberg:0712.3461} not to symmetrise the alternator was perhaps not the best. It leads to a classification of 2-term $E_2L_\infty$-algebras which is generally larger than that of $L_\infty$-algebras~\cite[Theorem 4.5]{Roytenberg:0712.3461}, contradicting the rectification theorem expected in line with the situation for $L_\infty$-algebras.
    
    A consequence of the strictification theorem and homotopy transfer is the following. Just as for $\ophLie$-algebras, we can also tensor an $E_2L_\infty$-algebra by a differential graded commutative algebra\footnote{One may be tempted to replace the differential graded commutative algebra with an $\opEilh$-algebra, but already the product between an $\opEilh$-algebra and an $\ophLie$-algebra does {\em not} carry a natural $\ophLie$-algebra structure.}:
    \begin{theorem}
        The tensor product of an $E_2L_\infty$-algebra and a differential graded commutative algebra carries a natural $E_2L_\infty$-algebra structure.
    \end{theorem}
    \begin{proof}
        We can invoke the argument presented in~\cite{Borsten:2021hua} for the existence of general tensor products between certain homotopy algebras. That is, by \cref{thm:strictification}, $\frE$ is quasi-isomorphic to a hemistrict $E_2L_\infty$-algebra $\frE^{\rm hst}$, and the chain complexes $\frA\otimes \frE^{\rm hst}$ and $\frA\otimes \frE$ are quasi-isomorphic. By \cref{prop:tensor_product_dgca_hLie}, $\frA\otimes \frE^{\rm hst}$ carries an $\ophLie$-algebra structure, and the homological perturbation lemma allows us to perform a homotopy transfer from $\frA\otimes \frE^{\rm hst}$ to $\frA\otimes \frE$, leading to the desired $E_2L_\infty$-algebra structure.
    \end{proof}
    Instead of using the above elegant but abstract argument, we can also perform a direct computation in the dual Chevalley--Eilenberg picture. This leads to the following explicit formulas for the tensor product $\hat \frE=\frA\otimes \frE$ of a differential graded commutative algebra $\frA$ and an $E_2L_\infty$-algebra $\frE$:
    \begin{equation}
        \begin{gathered}
            \hat \frE\coloneqq \frA\otimes \frE=\oplus_{k\in \IZ}(\frA\otimes \frE)_k~,~~~(\frA\otimes \frE)_k=\sum_{i+j=k} \frA_i\otimes \frE_j~,
            \\
            \hat\eps_1(a_1\otimes x_1)=(\rmd a_1)\otimes x_1+(-1)^{|a_1|}a_1\otimes \eps_1(x_1)~,
            \\
            \hat\eps^I_k(a_1\otimes x_1,\ldots,a_k\otimes x_k)=(-1)^{|I|(|a_1|+\ldots+|a_k|)}(a_1\ldots a_k)\otimes \eps^I_k(x_1,\ldots,x_k)~.
        \end{gathered}
    \end{equation}
    
    \subsection{Examples: String Lie algebra models}\label{ssec:string_models}
    
    Let us illustrate the above structure theorems using the important and archetypal examples of 2-term $E_2L_\infty$-algebra models for the string Lie algebra. We have already encountered the $E_2L_\infty$-algebras $\frstring^{\rmwk,\vartheta}_\rmsk(\frg)$ in~\eqref{eq:skeltal_model_string_Lie_2_algebra}. A short computation using formulas~\eqref{eq:antisymmetrisation_2-term} shows that these all antisymmetrise to the skeletal model of the string 2-term $L_\infty$-algebra (previously discussed in \cref{ssec:adjusted_Weils}):
    \begin{equation}\label{eq:skeletal_string}
        \begin{gathered}
            \frstring_\rmsk(\frg)~\coloneqq ~(\IR\xrightarrow{~0~}\frg)~,
            \\
            \mu_2(x_1,x_2)=[x_1,x_2]~,~~~\mu_3(x_1,x_2,x_3)=(x_1,[x_2,x_3])~.
        \end{gathered}
    \end{equation}
    
    Recall from \cref{ssec:adjusted_Weils} that this $L_\infty$-algebra (which is a minimal model for its quasi-isomorphism class) is quasi-isomorphic to the loop model,
    \begin{equation}
        \begin{gathered}
            \frstring_\rmlp(\frg)~\coloneqq ~(L_0\frg\oplus\IR\frg\xrightarrow{~\mu_1~}P_0\frg)~,
            \\
            \mu_1(\lambda,r)=\lambda~,
            \\
            \mu_2(\gamma_1,\gamma_2)=[\gamma_1,\gamma_2]~,~~~\mu_2(\gamma_1,(\lambda,r))=([\gamma_1,\lambda],2\int_0^1\rmd \tau~(\dot \gamma_1,\lambda)~,
            \\
            \mu_3(\gamma_1,\gamma_2,\gamma_3)=0~,
        \end{gathered}
    \end{equation}
    where $L_0\frg$ and $P_0\frg$ are the based path and based loop spaces of $\frg$ defined in \cref{app:path_groups}. There are two quasi-isomorphisms,
    \begin{equation}
        \begin{tikzcd}
            \frstring_\rmsk(\frg) \arrow[r,bend left=10]{}{\psi} & \frstring_\rmlp(\frg) \arrow[l,bend left=10]{}{\phi}
        \end{tikzcd}~,
    \end{equation}
    and their explicit forms are found e.g.~in~\cite{Saemann:2019dsl}. This implies that there is a quasi-isomorphic family of $E_2L_\infty$-algebras that antisymmetrise to $\frstring_\rmlp(\frg)$, which is readily found:
    \begin{equation}
        \begin{gathered}
            \frstring^{\rmwk,\vartheta}_\rmlp(\frg)~\coloneqq ~(L_0\frg\oplus \IR\xrightarrow{~\eps_1~}P_0\frg)~,
            \\
            \eps_1(\lambda,r)=\lambda~,
            \\
            \eps^0_2(\gamma_1,\gamma_2)=[\gamma_1,\gamma_2]~,~~~\eps^0_2(\gamma_1,(\lambda,r))=\left([\gamma_1,\lambda],2\int_0^1\rmd \tau~(\dot \gamma_1,\lambda)\right)~,
            \\
            \eps_2^1(\gamma_1,\gamma_2)=\big(0,2\vartheta(\dpar\gamma_1,\dpar\gamma_2)\big)
            \\
            \eps^{00}_3(\gamma_1,\gamma_2,\gamma_3)=\vartheta(\dpar \gamma_1,[\dpar \gamma_2,\dpar \gamma_3])~.
        \end{gathered}
    \end{equation}
    
    Altogether, we can summarise the situation in the following commutative diagram:
    \begin{equation}
        \begin{tikzcd}
            \frstring_\rmsk^{\rmwk,\vartheta}(\frg) \arrow[r,bend left=10]{}{\hat\psi} \arrow[d,"\rm asym"] & \frstring_\rmlp^{\rmwk,\vartheta}(\frg) \arrow[d,"\rm asym"] \arrow[l,bend left=10]{}{\hat\phi}
            \\
            \frstring_\rmsk(\frg) \arrow[r,bend left=10]{}{\psi} & \frstring_\rmlp(\frg) \arrow[l,bend left=10]{}{\phi}
        \end{tikzcd}
    \end{equation}
    The morphisms ${\rm asym}$ are special cases of the antisymmetrisation map~\eqref{eq:antisymmetrisation_2-term}, and the morphisms $\hat\phi$ and $\hat\psi$ are formed by lifts of the morphisms $\phi$ and $\psi$ as given by~\cref{thm:lift_L_infty_morphism}.
    
    Generically, on top of every $L_\infty$-algebra, there is a family of $E_2L_\infty$-algebras that antisymmetrise to it. The additional structure constants contained in the alternators of the $E_2L_\infty$-algebra will turn out to be crucial in the construction of higher gauge theories.
    
    \section{Relations to other algebras}
    
    In the following, we explain the relation between $E_2L_\infty$-algebras and homotopy Leibniz algebras and, in particular, to differential graded Lie algebras. The latter prepares our interpretation of tensor hierarchies.    
    
    \subsection{Relation to homotopy Leibniz algebras}
    
    Just as Lie algebras are Leibniz algebras that happen to have an antisymmetric Leibniz bracket, $E_2L_\infty$-algebras are $\opLeib_{\infty}$-algebras whose higher Leibniz brackets are antisymmetric up to homotopies. Homotopy Leibniz algebras were defined in~\cite{Ammar:0809.4328,Khudaverdyan:2013cta}, and they can be defined as differential free Zinbiel algebras.~\cite{JSTOR:24491899,Ginzburg:0709.1228} which, as suggested by the name\footnote{This nomenclature is a successful joke suggested by J.~M.~Lemaire. Zinbiel algebras are also known as {\em (commutative) shuffle algebras}, and the free Zinbiel algebra over a vector space is the shuffle algebra on its tensor algebra.}
    
    Explicitly, consider the semifree non-associative tensor algebra $\oslash_0^\bullet V$ for a graded vector space $V$ with only the first relation of~\eqref{eq:Eilh-relations} imposed. A (nilquadratic) differential $Q$ on this algebra which satisfies the ordinary Leibniz rule then defines a homotopy Leibniz algebra. All the additional structure in $\opEilh$ (as well as the resulting additional structure in $E_2L_\infty$-algebras) capture the appropriate notion of symmetry up to homotopy of the higher Leibniz brackets.
    
    Ordinary Leibniz algebras form an interesting source of 2-term $\ophLie$-algebras, which had been observed before:
    \begin{proposition}[\cite{Roytenberg:0712.3461}]\label{prop:Leib_is_hemistrict_ELinfty}
        Any Leibniz algebra induces canonically a hemistrict 2-term $E_2L_\infty$-algebra concentrated in degrees $-1$ and $0$.
    \end{proposition}   
    \noindent Explicitly, let $\frg$ be a Leibniz algebra, and write $\frg^{\rm ann}=[\frg,\frg]$. Then 
    \begin{equation}
        \frE(\frg)=\big(\frE(\frg)_{-1}\xrightarrow{~\eps_1~}\frE(\frg)_{0}\big)\coloneqq \big(\frg^{\rm ann} \xhookrightarrow{~~~~} \frg\big)
    \end{equation}
    is a differential graded Leibniz algebra, and we promote it to a 2-term $E_2L_\infty$-algebra by 
    \begin{equation}
        \sfalt(e_1,e_2)\coloneqq [e_1,e_2]+[e_2,e_1]\in \frg^{\rm ann}
    \end{equation}
    for all $e_1,e_2\in \frg$.\footnote{We note that this result, together with \cref{thm:antisym}, immediately implies that any Leibniz algebra gives rise to a 2-term $L_\infty$-algebra as shown separately in~\cite{Sheng:2015:1-5}.}

    \subsection{\texorpdfstring{$\ophLie$}{hLie}-algebras from differential graded Lie algebras and derived brackets}\label{ssec:antisym_hLie}
    
    Given a differential graded Lie algebra $\frg=\bigoplus_{k\in \IZ}\frg_k$, one can construct an associated $L_\infty$-algebra on the grade-shifted partial complex $\frL=\bigoplus_{k\leq 0}\frg[1]$. As explained in~\cite{Getzler:1010.5859}, this is a corollary to the result of~\cite{Fiorenza:0601312} that the mapping cone of a morphism between two differential graded Lie algebras carries a natural $L_\infty$-algebra structure. In this section, we present a refinement of this associated $L_\infty$-algebra to an $\ophLie$-algebra. The existence of the $L_\infty$-algebra is then a corollary to the antisymmetrisation \cref{thm:antisym}. Our construction extends the construction of $\opLeib_\infty$-algebras from $\opLeib$-algebras in~\cite{Uchino:0902.0044} as well as the construction of 2-term $E_2L_\infty$-algebras from 3-term differential graded Lie algebras in~\cite{Roytenberg:0712.3461}. 
    
    Given a differential graded Lie algebra, we readily construct a grade-shifted $\ophLie$-algebra.
    \begin{theorem}\label{thm:ophLie_from_dgLA}
        Given a differential graded Lie algebra $(\frg,\rmd_\frg,\{-,-\})$ with $\frg=\bigoplus_{k\in \IZ}\frg_k$, we have an associated $\ophLie$-algebra 
        \begin{equation}
            \frE=\bigoplus_{k\leq 0} \frE_k~,~~~\frE_k=\frg_{k-1}
        \end{equation}
        with higher products
        \begin{equation}
            \begin{aligned}
                \eps_1(x_1)&\coloneqq \begin{cases}
                    \rmd_\frg x~&\mbox{for}~|x|_\frE<0~,
                    \\
                    0 &\mbox{else}~,
                \end{cases}
                \\
                \eps_2^i(x_1,x_2)&\coloneqq \begin{cases}
                    \{\delta x_1,x_2\} & \mbox{for}~i=0~,
                    \\
                    (-1)^{|x_1|_\frE}\{x_1,x_2\} & \mbox{for}~i=1~,
                    \\
                    0& \mbox{else}
                \end{cases}
            \end{aligned}
        \end{equation}
        for all $x_1,x_2\in \frE$. Here, $\delta\coloneqq \rmd_{\frg}|_{\frg_{-1}}$ and $|x_1|_\frE$ denotes the degree of $x_1$ in $\frE$.
    \end{theorem}
    \begin{proof}
        The proof consists of a straightforward verification of the axioms of an $\ophLie$-algebra~\eqref{eq:hLie-relations}, which is most conveniently done again with a computer algebra program.
    \end{proof}
    
    Let us discuss the explicit form of the antisymmetrisation in some more detail. We assume, as usual, that $\frE$ admits a nice basis $(\tau_\alpha)$, so that $\frE[1]^*$ has a dual basis $(t^\alpha)$. the Chevalley--Eilenberg differential then reads as
    \begin{equation}\label{eq:Q_hLie2}
        Q t^\alpha=-(-1)^{|\beta|}m^\alpha_\beta t^\beta-(-1)^{i(|\beta|+|\gamma|)+|\gamma|(|\beta|-1)}\,m^{i,\alpha}_{\beta\gamma}\,t^\beta \oslash_i t^\gamma~,
    \end{equation}
    and we have the following theorem.
    \begin{theorem}\label{thm:antisym_hLie}
        For each $\ophLie$-algebra $(\frE,\eps^i_j)$ (with the above mentioned restrictions), there is an $L_\infty$-algebra $(\frE,\mu_i)$ with first four higher products reading as
        \begin{equation}\label{eq:antisymmetrisation_hLie}
            \begin{aligned}
                \mu_1(x_1)&\coloneqq \eps_1(x_1)~,\\
                \mu_2(x_1,x_2)&\coloneqq \tfrac12(\eps^0_2(x_1,x_2)-\eps^0_2(x_2,x_1))~,\\
                \mu_3(x_1,x_2,x_3)&\coloneqq \tfrac{1}{3!}\sum_{\sigma\in S_3}\chi(\sigma;x_1,x_2,x_3)\Big(\eps^0_3(x_{\sigma(1)},x_{\sigma(2)},x_{\sigma(3)})\\
                &\hspace{1.7cm}+\tfrac14\big( \eps^1_2(\eps^0_2(x_{\sigma(1)},x_{\sigma(2)}),x_{\sigma(3)})
                +\eps^1_2(x_{\sigma(1)},\eps^0_2(x_{\sigma(2)},x_{\sigma(3)}))\big)\Big),
                \\
                \mu_4(x_1,x_2,x_3,x_4)&\coloneqq 0~,
            \end{aligned}
        \end{equation}
        for all $x_i\in \frE$.
    \end{theorem}
    \begin{proof}   
        We use again \cref{thm:antisym} and determine the Chevalley--Eilenberg differential~\eqref{eq:HPL_formula_Q_EL_to_L} of $(\frE,\mu_i)$,
        which allows us to compute $Q_L$ up to quartic order. This produces the higher products~\eqref{eq:antisymmetrisation_hLie}.
    \end{proof}
    
    
    We can now compose the map from differential graded Lie algebras to $\ophLie$-algebras with the antisymmetrisation~\cref{thm:antisym}. This reproduces the following propostion of~\cite{Getzler:1010.5859}, which in turn is a specialisation of~\cite{Fiorenza:0601312}:
    \begin{proposition}\label{prop:dgLA_to_L_infty}
        Given a differential graded Lie algebra $(\frg,\rmd,[-,-])$, we have an $L_\infty$-algebra structure on the truncated complex
        \begin{equation}
            \frg_{\leq 0}=\big(~~
            \ldots~\xrightarrow{~\rmd~}~\frg_{-2}~\xrightarrow{~\rmd~}~\frg_{-1}~\xrightarrow{~\rmd~}~\frg_{0}~\xrightarrow{~0~}~*~\xrightarrow{~0~}~\ldots~~)
        \end{equation}
        with
        \begin{equation}
            \begin{aligned}
                \mu_1(x_1)&=\begin{cases}
                    \rmd x_1 & \mbox{for $|x_1|<0$}~,
                    \\
                    0 & \mbox{for $|x_1|=0$}
                \end{cases}
                \\
                \mu_k(x_1,\ldots,x_k)&=\frac{(-1)^k}{(k-1)!}B_{k-1}\sum_{\sigma\in S_{k}}\chi(\sigma;x_1,\ldots,x_k)[[\ldots[[\delta x_{\sigma(1)},x_{\sigma(2)}],\ldots],x_{\sigma(k)}]~,
            \end{aligned}
        \end{equation}
        where
        \begin{equation}
            \delta(x_1)=\begin{cases}
                \rmd x_1 & \mbox{for $|x_1|=0$}~,
                \\
                0 & \mbox{else}
            \end{cases}
        \end{equation}
        for all $x_i\in \frg_{\leq 0}$. Here, $B_k$ are the Bernoulli numbers\footnote{i.e.\ $B_0,B_1,\ldots=1,-\tfrac12,\tfrac{1}{6},0,-\tfrac{1}{30},0,\tfrac{1}{42},\ldots$}.
    \end{proposition}
    
    Altogether, our above constructions suggest the following picture:
    \begin{equation}\label{eq:diagram_algebras}
        \begin{tikzcd}
            \mbox{dg Lie algebra} \arrow[r,"\text{\Cref{thm:ophLie_from_dgLA}}", bend left=30] \arrow[rr,"\text{\Cref{prop:dgLA_to_L_infty}}",swap, bend right=20]& \mbox{$\ophLie$-algebra} \arrow[r,"\text{\Cref{thm:antisym_hLie}}", bend left=30] & \mbox{$L_\infty$-algebra}
        \end{tikzcd}
    \end{equation}
    Our formulas~\eqref{eq:HPL_formula_Q_EL_to_L} show that this picture is true for differential graded Lie algebras concentrated in degrees $d\geq -3$. For more general differential graded Lie algebras, this picture is still very plausible from the expression for~\eqref{eq:HPL_formula_Q_EL_to_L}. A complete proof, however, would require an explicit expression of the homotopy $\sfH_0$ to all orders, which has not been completed at the time of writing of this thesis.
    
    From \cref{prop:dgLA_to_L_infty} it is also clear that $\mu_4$ in~\eqref{eq:antisymmetrisation_hLie} vanishes because $B_3=0$. Similarly, all even higher brackets $\mu_{2i}$ with $i\geq 1$ vanish, as the odd Bernoulli numbers $B_k$ for $k\geq 3$ vanish.
    
    As a simple example, consider a quadratic Lie algebra $\frg$, and construct the differential graded Lie algebra
    \begin{equation}
        \frG=(~\ldots\xrightarrow{~0~}*\xrightarrow{~0~}\underbrace{\IR}_{\frG_{-2}}\xrightarrow{~0~}\underbrace{\frg}_{\frG_{-1}}\xrightarrow{~\sfid~}\underbrace{\frg}_{\frG_{0}}\xrightarrow{~0~}*\xrightarrow{~0~}\ldots~)~,
    \end{equation}
    concentrated in degrees $-2$, $-1$, $0$ with differential and Lie brackets
    \begin{equation}
        [x_1,x_2]_\frG=2[x_1,x_2]~,~~~[y_1,x_1]_\frG=-[x_1,y_1]_\frG~,~~~[y_1,y_2]_\frG=(y_1,y_2)
    \end{equation}
    for all $x_1,x_2\in \frG_0\cong\frg$ and $y_1,y_2\in \frG_{-1}\cong \frg$, where $[-,-]$ and $(-,-)$ are the Lie bracket and the Cartan--Killing form on $\frg$. Then the associated $\ophLie$-algebra is 
    \begin{equation}
        \begin{gathered}
            \frE=(~\IR\xrightarrow{~0~}\frg~)~,
            \\
            \eps_1(r)\coloneqq 0~,
            \\
            \eps^0_2(x_1,x_2)=[x_1,x_2]~,~~~\eps^1_2(x_1,x_2)=2(x_1,x_2)~.
        \end{gathered}
    \end{equation}
    We thus recover the hemistrict $E_2L_\infty$-algebra $\frstring^{\rm wk,1}_{\rm sk}(\frg)$ introduced in \cref{ssec:string_models}. The antisymmetrisation of this $\ophLie$-algebra then yields the skeletal string Lie 2-algebra model $\frstring_{\rm sk}(\frg)$. Interestingly, a quick consideration of the case leads to the conclusion that there is no differential graded Lie algebra that reproduces the strict string Lie 2-algebra model $\frstring_\rmlp(\frg)=\frstring^{\rmwk,0}_\rmlp(\frg)$. This points towards a possible extension of \cref{thm:ophLie_from_dgLA} producing $E_2L_\infty$-algebras from certain $L_\infty$-algebras.
        
    \section{Higher gauge theory with \texorpdfstring{$E_2L_\infty$}{EL-infinity}-algebras}
    
    In this section, we develop and explore the generalities of higher gauge theory using $E_2L_\infty$-algebras as higher gauge algebras.    
    
    \subsection{Homotopy Maurer--Cartan theory for \texorpdfstring{$E_2L_\infty$}{EL-infinity}-algebras}
    
    Recall that given an $L_\infty$-algebra $\frL$ with higher products $\mu_i$, there is a functor $\sfMC(\frL,-)$ taking a differential graded commutative algebra $\fra$ to Maurer--Cartan elements with values in $\fra$, cf.~e.g.~\cite{Chuang:0912.1215}. This functor is represented by the Chevalley--Eilenberg algebra $\sfCE(\frL)$ of the $L_\infty$-algebra. 
    
    What we usually call Maurer--Cartan elements in $\frL$ are Maurer--Cartan elements with values in $\IR$, where the latter is regarded as a trivial differential graded algebra $\IR_\fra$ with underlying vector space $\IR$, spanned by a generator $w$ subject to the relation $w^2=w$, and trivial differential. 
    
    For concreteness' sake, let us assume that $\frL$ is degree-wise finite, and let $(t^A)$ be the generators of $\frL[1]^*$ dual to some basis $(\tau_A)$ of $\frL$. A Maurer--Cartan element is encoded in a morphism of differential graded commutative algebras $a:\sfCE(\frL)\rightarrow \IR_\fra$, which is fully determined by the image of the generators $(t^\alpha)$ of degree~$0$,
    \begin{equation}\label{eq:dga_morphism}
        a: \sfCE(\frL)\rightarrow \IR~,~~~t^\alpha \mapsto a^\alpha w
    \end{equation}
    for $a^\alpha\in \IR$. Dually, we have an element $a\coloneqq a^\alpha\tau_\alpha\in \frL_1$, the {\em gauge potential}. Compatibility with the differential requires the {\em curvature}
    \begin{equation}
        f\coloneqq \mu_1(a)+\tfrac12 \mu_2(a,a)+\tfrac1{3!}\mu_3(a,a,a)+\dotsb~~\in \frL_2
    \end{equation}
    to vanish, and the equation $f=0$ is called the homotopy Maurer--Cartan equation. This curvature satisfies the {\em Bianchi identity}
    \begin{equation}
        \sum_{k\geq0}\frac{1}{k!}\mu_{k+1}(a,\ldots,a,f)=0~.
    \end{equation}
    Infinitesimal gauge transformations are obtained from infinitesimal homotopies between morphisms from $\sfCE(\frL)$ to $\IR$. They are 
    parameterised by elements $c\in \frg_0$ and act according to
    \begin{equation}\label{eq:GaugeTrafo}
        \delta_{c} a=\sum_{k\geq 0} \frac{1}{k!}\mu_{k+1}(a,\ldots,a, c)~.
    \end{equation}
    Higher homotopies yield higher gauge transformations.
    
    Similarly, one defines Maurer--Cartan elements of an $A_\infty$-algebra with values in a differential graded algebra.
    
    In the case of $E_2L_\infty$-algebras, we can still consider tensor products of a base $E_2L_\infty$-algebra $\frE$ and a differential graded commutative algebra $\frA$. However, the Chevalley--Eilenberg algebra $\sfCE(\frE)$ is an $\opEilh$-algebra and not a differential graded commutative algebra. Therefore the homotopy Maurer--Cartan functor cannot be represented by it directly.
    
    There are two loopholes to this obstruction. First, we can lift the differential graded commutative algebra $\frA$, if it is semifree, to an $\opEilh$-algebra $\hat \frA$ as explained in \cref{thm:lift_dgca_to_Eilh}. We can then consider $\opEilh$-algebra morphisms 
    \begin{equation}
        a:\sfCE(\frE)\rightarrow \hat \frA~.
    \end{equation}
    Second, we can project $\sfCE(\frE)$ to the Chevalley--Eilenberg algebra of the $L_\infty$-algebra $\frL$ induced by $\frE$ and consider the usual morphisms 
    \begin{equation}
        a:\sfCE(\frL)\rightarrow \frA~.
    \end{equation}
    A third approach is simply to consider general morphisms of $\opEilh$-algebras. In particular, one may want to replace differential forms with more general objects, cf.~also~\cite{Ritter:2015zur}.
    
    We note that, in general, the three different types of morphism will give rise to different sets of Maurer--Cartan elements with the first one encompassing the second one. In all the applications we are aware of, however, the second approach is the appropriate one. While the difference between an $E_2L_\infty$-algebra and the corresponding $L_\infty$-algebra obtained by antisymmetrisation is then invisible at the level of homotopy Maurer--Cartan theory, the additional algebraic structure in an $E_2L_\infty$-algebra is important in adjusting non-flat higher gauge theories.
    
    \subsection{Firmly adjusted Weil algebras from \texorpdfstring{$\ophLie$}{hLie}-algebras}\label{ssec:firmly_adjusted}
    Special cases of adjusted Weil algebras whose corresponding morphisms~\eqref{eq:naive_dga_morphism} into differential forms yield adjusted higher gauge theories with closed BRST complex are the following ones:
    \begin{definition}
        A \uline{firmly adjusted Weil algebra} of an $L_\infty$-algebra $\frL$ is a differential graded commutative algebra obtained from the Weil algebra $\sfW(\frL)$ by a coordinate change
        \begin{equation}\label{eq:firm_rotation}
            \hat t^A\mapsto \hat t'^A\coloneqq \hat t^A+p^A_{B_1B_2\dotso B_mC_1C_2\dotso C_n}\hat t^{B_1}\dotsm\hat t^{B_m}t^{C_1}\dotsm t^{C_n}~,
        \end{equation}
        where $t^A\in \frL[1]^*$, $\hat t^A\in \frL[2]^*$, $m\ge1$, and $n\ge0$, such that the image of the resulting differential $Q_{\rm fadj}$ on generators in $\frL[2]^*$ contains no generator in $\frL[1]^*$ except for at most one of degree~$1$.
    \end{definition}
    We note that putting the generators $(\sigma t^A)$ to zero still recovers the Chevalley--Eilenberg algebra $\sfCE(\frL)$ of $\frL$. In this sense, the coordinate change has not changed the underlying $L_\infty$-algebra. Moreover, note that any Weil algebra is fully contractible in the sense that the cohomology of its linearised differential is trivial. Dually, it is the Chevalley--Eilenberg algebra of an $L_\infty$-algebra which is quasi-isomorphic to the trivial $L_\infty$-algebra. The non-trivial information contained in the Weil algebra is the relation between the generators $(t^A)$ and $(\sigma t^A)$, which translates under the morphism~\eqref{eq:naive_dga_morphism} into the relation between gauge potentials and their curvatures. Our coordinate change thus changes the definition of the curvatures and, as partially flat homotopies describe gauge transformations, also the gauge transformations. Firmly adjusted Weil algebras ensure that the corresponding BRST complex closes: the restricted terms govern the Bianchi identities, which fix the gauge transformations of the curvatures. Closure of the latter is what induces the fake curvature conditions. Thus, firmly adjusted Weil algebras are adjusted Weil algebras.   
    
    As an example, consider the following firmly adjusted Weil algebra of the string Lie 2-algebra~\eqref{eq:skeletal_string}:
    \begin{equation}\label{eq:ext_twt_string_sk_differential}
        \begin{aligned}
            Q_{\rm fadj}~&:~&t^\alpha &\mapsto -\tfrac12 f^\alpha_{\beta\gamma} t^\beta  t^\gamma + \hat t^\alpha~,~&r &\mapsto \tfrac{1}{3!} f_{\alpha\beta\gamma} t^\alpha  t^\beta  t^\gamma -\kappa_{\alpha\beta}t^\alpha\hat t^\beta+ \hat r'~,
            \\
            &&\hat t^\alpha &\mapsto -f^\alpha_{\beta\gamma} t^\beta  \hat t^\gamma~,~
            &\hat r'&\mapsto \kappa_{\alpha\beta}\hat t^\alpha\hat t^\beta~,
        \end{aligned}
    \end{equation}
    which is obtained from the coordinate transformation $\hat r\mapsto \hat r'=\hat r+\kappa_{\alpha\beta} \hat t^\alpha t^\beta$. Here, $t\in \frg[1]^*$, $r\in \IR[2]^*$ and $\hat t=\sigma t$, $\hat r=\sigma r$. Under the morphism~\eqref{eq:naive_dga_morphism}, this firmly adjusted Weil algebra gives rise to the usual string connections
    \begin{equation}
        \begin{aligned}
            a&=A+B\in \Omega^1(M,\frg)\oplus \Omega^2(M,\IR)~,
            \\
            f&=F+H\in \Omega^2(M,\frg)\oplus \Omega^3(M,\IR)~,
            \\
            F&=\rmd A+\tfrac12[A,A]~,
            \\
            H&=\rmd B-\tfrac1{3!}(A,[A,A])+(A,F)=\rmd B+{\rm cs}(A)~.
        \end{aligned}
    \end{equation}
    
    More generally, consider an $L_\infty$-algebra obtained from an $\ophLie$-algebra by antisymmetrisation. For simplicity, we also assume that the $L_\infty$-algebra has maximally ternary brackets. Its Weil algebra then reads as
    \begin{equation}
        \begin{aligned}
            Q_\sfW t^\alpha&=-(-1)^{|\beta|}m^\alpha_\beta t^\beta-(-1)^{|\gamma|(|\beta|-1)}\tfrac12\,m^{\alpha}_{\beta\gamma}\,t^\beta t^\gamma
            \\
            &~~~~~-(-1)^{|\beta|(|\gamma|+1)+|\delta|(|\beta|+|\gamma|+1)}\tfrac{1}{3!}\,m^{\alpha}_{\beta\gamma\delta}\,t^\beta t^\gamma t^\delta +\hat t^\alpha~,
            \\
            Q_\sfW \hat t^\alpha&=(-1)^{|\beta|}m^\alpha_\beta \hat t^\beta+(-1)^{|\gamma|(|\beta|-1)}\,m^{\alpha}_{\beta\gamma}\,\hat t^\beta t^\gamma
            \\
            &~~~~~+(-1)^{|\beta|(|\gamma|+1)+|\delta|(|\beta|+|\gamma|+1)}\tfrac{1}{2}\,m^{\alpha}_{\beta\gamma\delta}\,\hat t^\beta t^\gamma t^\delta~.
        \end{aligned}
    \end{equation}
    In general, this Weil algebra is clearly not firmly adjusted because of the explicit form of $Q_\sfW\hat t^\alpha$. Let us therefore perform the coordinate change
    \begin{equation}
        \hat t^\alpha\mapsto \hat t'^\alpha\coloneqq \hat t^\alpha+s^\alpha_{\beta\gamma} \hat t^\beta t^\gamma~.
    \end{equation}
    The new Weil differential then reads as follows.
    \begin{equation}\label{eq:fadjust_Qp}
        \begin{aligned}
            Q'_\sfW \hat t'^\alpha&=(-1)^{|\beta|}m^\alpha_\beta \hat t'^\beta
            +(-1)^{1+|\beta|}s^\alpha_{\beta\gamma}\hat t'^\beta \hat t'^\gamma
            +(-1)^{|\gamma|(|\beta|-1)}\,m^{\alpha}_{\beta\gamma}\,\hat t'^\beta t^\gamma
            \\
            &~~~~~+\big(-(-1)^{|\beta|} m^\alpha_\beta s^\beta_{\gamma\delta}+(-1)^{|\gamma|} s^\alpha_{\beta\delta}m^\beta_\gamma
            +(-1)^{|\gamma|+|\delta|} s^\alpha_{\gamma\beta}m^\beta_\delta\big)\hat t'^\gamma t^\delta+\cdots~,
        \end{aligned}
    \end{equation}
    where the ellipsis denotes cubic and higher terms. Let us now further restrict to $\ophLie$-algebras obtained from a differential graded algebra via \cref{thm:antisym_hLie} with differential $\Theta^\alpha_\beta$ and structure constants $f^\alpha_{\beta\gamma}$. In this case, we have
    \begin{subequations}
        \begin{equation}
            m^\alpha_\beta=\Theta^\alpha_\beta
            ~,~~~
            m^\alpha_{\beta\gamma}=\begin{cases}
                \tfrac12 f^\alpha_{\delta\gamma}\Theta^\delta_\beta & \mbox{if $|\beta|=1$}~,
                \\
                0 & \mbox{else}~;
            \end{cases}
        \end{equation}
        we also put
        \begin{equation}
            s^\alpha_{\beta\gamma}=\tfrac12(-1)^{|\beta|(|\gamma|+1)} f^\alpha_{\beta\gamma}~.
        \end{equation}
    \end{subequations}
    In the above formulas, $|\alpha|,|\beta|,|\gamma|\geq 1$, and $|\delta|=0$. Together with the Jacobi identity for the $f^\alpha_{\beta\gamma}$, one can then easily verify that $Q'$ becomes a firmly adjusted Weil differential,
    \begin{equation}\label{eq:fadjust_Q}
        Q_{\rm fadj} \hat t'^\alpha=(-1)^{|\beta|}m^\alpha_\beta \hat t'^\beta
        +(-1)^{1+|\beta|\,|\gamma|}\tfrac12f^\alpha_{\beta\gamma}\hat t'^\beta \hat t'^\gamma~.
    \end{equation}
    We conclude with the following theorem.
    \begin{theorem}\label{thm:firmly_adjust}
        Given an $L_\infty$-algebra with maximally ternary brackets that is obtained from the antisymmetrisation of a differential graded Lie algebra by~\cref{prop:dgLA_to_L_infty}, then there is a corresponding firmly adjusted Weil algebra. The data necessary for an adjustment arises from the alternators in the corresponding $\ophLie$-algebra.
    \end{theorem}
    
    Below, we shall give examples motivated from higher gauge theory. We stress, however, that the definition of an adjustment is also interesting for purely algebraic considerations, as it allows for the definition of a differential graded algebra of invariant polynomials for an $L_\infty$-algebra which is compatible with quasi-isomorphisms of this $L_\infty$-algebra, cf.~the discussion in~\cite{Saemann:2019dsl}. 
    
    We also note that our construction highlights the features needed for obtaining a firmly adjusted Weil algebra. In particular, it is not necessary that the $\ophLie$-algebra was obtained from a differential graded Lie algebra; it was sufficient that there be a relation between the parameters $s^\alpha_{\beta\gamma}$ of the coordinate change and the structure constants $f^\alpha_{\beta\gamma}$ of the Lie algebra to ensure that~\eqref{eq:fadjust_Qp} reduces to~\eqref{eq:fadjust_Q}. This is the case, for example, in the tensor hierarchies in non-maximally supersymmetric gauged supergravity.
    
    \subsection{Example: (1,0)-gauge structures}\label{ssec:10-structures}
    
    As a first more involved example of $E_2L_\infty$-algebras arising in the context of higher gauge theory, let us consider the higher gauge algebra defined in~\cite{Saemann:2017rjm}, see also~\cite{Saemann:2017zpd,Saemann:2019dsl,Rist:2020uaa}. This algebra is a specialisation of the general non-abelian algebraic structure identified in~\cite{Samtleben:2011fj} and can be derived from tensor hierarchies, to which we shall return shortly. The latter had received an interpretation as an $L_\infty$-algebra with some ``extra structure'' before, cf.~\cite{Palmer:2013pka} as well as~\cite{Lavau:2014iva}. Here, we show that it is, in fact an $E_2L_\infty$-algebra.
    
    The higher gauge algebra $\hat\frg^\omega$ for $\frg$ a quadratic Lie algebra has underlying graded complex
    \begin{equation}\label{eq:ghsk-complex}
        \hat\frg^\omega=\left(
        \begin{tikzcd}[row sep=0cm,column sep=2cm]
            \frg^*_v\arrow[r]{}{\mu_1=\sfid} & \frg^*_u & \IR^*_s \arrow[r]{}{\mu_1=\sfid} & \IR_p^*
            \\
            & \oplus & \oplus & \oplus
            \\
            \underbrace{\phantom{\frg^*_v}}_{\hat \frg^\omega_{\rm sk,-3}} & \underbrace{\IR_q}_{\hat \frg^\omega_{\rm sk,-2}} \arrow[r]{}{\mu_1=\sfid} & \underbrace{\IR_r}_{\hat \frg^\omega_{\rm sk,-1}} & \underbrace{\frg_t}_{\hat \frg^\omega_{\rm sk,0}}
        \end{tikzcd}\right)~,
    \end{equation}
    where the subscripts merely help to distinguish between isomorphic subspaces. In~\cite{Saemann:2017zpd}, this differential complex was extended to an $L_\infty$-algebra $\hat\frg^\omega$ with higher products
    \begin{equation}\label{eq:ghsk-brackets}
        \begin{aligned}
            \mu_2(t_1,t_2)&=[t_1,t_2]\in\frg_t~,\\
            \mu_2(t,u)&=u\big([-,t]\big)\in\frg^*_u~,~~~&
            \mu_2(t,v)&=v\big([-,t]\big)\in\frg_v^*~,\\
            \mu_3(t_1,t_2,t_3)&=(t_1,[t_2,t_3])\in\IR_r~,~~~&
            \mu_3(t_1,t_2,s)&= s\big(\,(-,[t_1,t_2])\,\big)\in\frg^*_u~,
        \end{aligned}
    \end{equation}
    where $t\in \frg_t$, etc. Moreover, $[-,-]$ and $(-,-)$ denote the Lie bracket and the quadratic form in $\frg$, respectively. When constructing gauge field strengths based on this $L_\infty$-algebra, the following, additional maps feature as part of the adjustment data:
    \begin{equation}
        \begin{aligned}
            \nu_2(t_1,t_2)&=-2(t_1,t_2)\in \IR_r~,~~~&\nu_2(t,s)&=2s(-,t)\in\frg_u^*~,
            \\
            \nu_2(t_1,u_1)&=u_1\big([-,t_1]\big)\in \frg_v^*~.
        \end{aligned}
    \end{equation}
    
    As motivated in more detail later, it is useful to first perform a quasi-isomorphism on $\hat\frg^\omega$ leading to the higher brackets
    \begin{equation}\label{eq:ghsk-brackets2}
        \begin{aligned}
            \mu_2(t_1,t_2)&=[t_1,t_2]\in\frg_t~,\\
            \mu_2(t,u)&=\tfrac12u\big([-,t]\big)\in\frg^*_u~,~~~&
            \mu_2(t,v)&=\tfrac12v\big([-,t]\big)\in\frg_v^*~,\\
            \mu_3(t_1,t_2,t_3)&=(t_1,[t_2,t_3])\in\IR_r~,~~~&
            \mu_3(t_1,t_2,s)&= s\big(\,(-,[t_1,t_2])\,\big)\in\frg^*_u~,
            \\
            \mu_3(t_1,t_2,u)&= \tfrac14v\big(\,(-,[t_1,t_2])\,\big)\in\frg^*_v~.
        \end{aligned}
    \end{equation}
    This is the $L_\infty$-algebra obtained by \cref{thm:antisym_hLie} from the $\ophLie$-algebra $\frE$ with differential complex~\eqref{eq:ghsk-complex} with $\eps_1=\mu_1$ and the additional binary products
    \begin{equation}
        \begin{aligned}
            \eps_1&=\mu_1~,
            \\
            \eps^0_2(t_1,t_2)&=-\eps^0_2(t_2,t_1)=[t_1,t_2]\in \frg_t~,
            \\
            \eps^0_2(t,u)&=u\big([-,t]\big)\in \frg^*_u~,~~~&
            \eps^0_2(t,v)&=v\big([-,t]\big)\in \frg_v^*~,
            \\
            \eps^1_2(t_1,t_2)&=\eps^1_2(t_2,t_1)=2(t_1,t_2)\in \IR_r~,
            \\
            \eps^1_2(t,s)&=3!s\big(-,t\big)\in\frg_u^*~,~~~
            &\eps^1_2(s,t)&=\eps^1_2(t,s)=3!s\big(-,t\big)\in \frg_u^*~,
            \\
            \eps^1_2(t,u)&=u\big([-,t]\big)\in \frg_v^*~,~~~&\eps^1_2(u,t)&=\eps^1_2(t,u)=u\big([-,t]\big)\in \frg_v^*~,
        \end{aligned}
    \end{equation}
    as one verifies by direct computation. This $\ophLie$-algebra is obtained from a differential graded Lie algebra $\frG$ by \cref{thm:ophLie_from_dgLA}, and we have
    \begin{equation}\label{eq:ghsk-complex2}
        \frG=\left(
        \begin{tikzcd}[row sep=0cm,column sep=1.6cm]
            \frg^*_v\arrow[r]{}{\mu_1=\sfid} & \frg^*_u & \IR^*_s \arrow[r]{}{\mu_1=\sfid} & \IR_p^*
            \\
            & \oplus & \oplus & \oplus
            \\
            \underbrace{\phantom{\frg^*_v}}_{\frG_{-4}} & \underbrace{\IR_q}_{\frG_{-3}} \arrow[r]{}{\mu_1=\sfid} & \underbrace{\IR_r}_{\frG_{-2}} & \underbrace{\frg_t}_{\frG_{-1}} \arrow[r]{}{\mu_1=\sfid} & \underbrace{\frg_{\hat t}}_{\frG_0}
        \end{tikzcd}\right)
    \end{equation}
    with the non-trivial Lie brackets $[-,-]_\frG$ fixed by
    \begin{equation}
        \begin{aligned}
            [\hat t_1,\hat t_2]_\frG&\coloneqq [\hat t_1,\hat t_2]\in\frg_{\hat t}~,
            ~~~&[\hat t_1,t_2]_\frG&\coloneqq [\hat t_1,t_2]\in\frg_{t}~,
            \\
            [\hat t_1,u]_\frG&\coloneqq u([-,t])\in\frg_u^*~,
            ~~~&[\hat t_1,v]_\frG&\coloneqq v([-,t])\in\frg_u^*~,
            \\
            [t_1,t_2]_\frG&\coloneqq (t_1,t_2)\in\IR_t~,
            ~~~&[t_1,s]_\frG&\coloneqq \alpha_2 s(-,t_1)\in\frg_u^*~.
        \end{aligned}
    \end{equation}
    This is an extension of the example presented at the end of~\cref{ssec:antisym_hLie}. 
    
    We thus see that we have the following sequence that leads to a construction of $\hat\frg^\omega$:
    \begin{equation}
        \mbox{dg Lie algebra}~\frG~\xrightarrow{\text{\Cref{thm:ophLie_from_dgLA}}}~~\mbox{$\ophLie$-algebra}~\frE~\xrightarrow{\text{\Cref{thm:antisym_hLie}}}~~\mbox{$L_\infty$-algebra}~\hat\frg^\omega,
    \end{equation}
    specializing the picture~\eqref{eq:diagram_algebras}. The additional information (i.e.~structure constants) contained in the $E_2L_\infty$-algebra are vital for constructing the adjusted form of the curvatures. 
    
    A corresponding adjusted Weil algebra was found in~\cite{Saemann:2019dsl}, and it agrees with the one obtained from our construction of a firmly adjusted Weil algebra from \cref{ssec:firmly_adjusted}:
    \begin{equation}
        \def\arraystretch{1.5}
        \begin{aligned}
            Q_{\rm fadj}~&:~&t^\alpha &\mapsto  -\tfrac12 f^\alpha_{\beta\gamma} t^\beta t^\gamma+\hat t^\alpha~,~~~& p&\mapsto -s+\hat p~,\\
            &&\hat t^\alpha &\mapsto  -f^\alpha_{\beta\gamma} t^\beta \hat t^\gamma~,~~~& \hat p&\mapsto \hat s~,\\
            &&r&\mapsto \tfrac{1}{3!} f_{\alpha\beta\gamma} t^\alpha t^\beta t^\gamma-\kappa_{\alpha\beta}t^\alpha \hat t^\beta+q+\hat r~,~~~& s&\mapsto \hat s~,\\
            &&\hat r&\mapsto \kappa_{\alpha\beta}\hat t^\alpha \hat t^\beta-\hat q~,~~~& \hat s&\mapsto 0~,\\
            &&u_\alpha &\mapsto -f^\gamma_{\alpha\beta}t^\beta u_\gamma-\tfrac12 f_{\alpha\beta\gamma}t^\beta t^\gamma s-v_\alpha+\hat u_\alpha~,~~~&
            q &\mapsto \hat q~,\\
            &&\hat u_\alpha &\mapsto -f^\gamma_{\alpha\beta}t^\beta \hat u_\gamma+\hat v_\alpha~,~~~&
            \hat q &\mapsto 0~,\\
            &&v_\alpha&\mapsto -f^\gamma_{\alpha\beta}t^\beta v_\gamma - f^\gamma_{\alpha\beta}\hat t^\beta u_\gamma +f_{\alpha\beta\gamma}t^\beta\hat t^\gamma s-\tfrac12 f_{\alpha\beta\gamma} t^\beta t^\gamma \hat s+\hat v_\alpha~,\\
            &&\hat v_\alpha&\mapsto -f^\gamma_{\alpha\beta}t^\beta \hat v_\gamma+f^\gamma_{\alpha\beta}\hat t^\beta \hat u_\gamma~.
        \end{aligned}
    \end{equation}

    \section{Tensor hierarchies in terms of \(E_2L_\infty\)-algebras}\label{sec:tensor_hierachies}
    \label{ssec:gen_tensor_hierarchies}
    
    Let us ignore the link between tensor hierarchies and gauged supergravity for a moment; clearly, the resulting kinematical data is potentially of interest in higher gauge theory in a much wider context. 
    
    The construction prescription is rather straightforward. We consider a Lie algebra $\frg$, which we enlarge to a differential graded Lie algebra 
    \begin{equation}
        \begin{aligned}
            V=\Big(~\dotsb \xrightarrow{~\rmd~} V_{-2} \xrightarrow{~\rmd~} V_{-1} \xrightarrow{~\rmd~} V_0=\frg \xrightarrow{~\rmd~} V_1 \xrightarrow{~\rmd~} \dotsb~\Big)~,
        \end{aligned}
    \end{equation}
    where we allowed for additional vector spaces $V_i$ with $i>0$. All vector spaces $V_i$ are $\frg$-modules, and the Lie bracket on $V_0$ as well as the Lie brackets on $V_0\otimes V_i$ are given. Further Lie brackets $[-,-]:V_i\otimes V_j\rightarrow V_{i+j}$ can be introduced, but due to the Jacobi identity, the underlying structure constants have to be invariant tensors of $\frg$ (as we shall also see below in an example). The differentials do not have to satisfy this restriction. As an additional constraint, we can also impose the condition that $V_{-p}^*=V_{p+2-d}$ as required by the U-duality condition from supergravity. This can be useful in the construction of action principles.
    
    To illustrate the above, let us construct a generic example in $d=5$. Let $\frg$ be a Lie algebra and $V_{-1}$ any representation. Imposing the duality constraint and allowing for an extension in one degree on either side leads to the differential complex 
    \begin{equation}
        \begin{aligned}
            V=\Big(~V_{-4}\cong{\rm coker}(\Theta)^* \xrightarrow{~\rmd~} V_{-3}\cong\frg^* &\xrightarrow{~\rmd~} V_{-2}\cong V_{-1}^* 
            \\
            &\xrightarrow{~\rmd~} V_{-1} \xrightarrow{~\Theta~} V_0=\frg \xrightarrow{~\rmd~} V_1\cong{\rm coker}(\Theta)~\Big)~.
        \end{aligned}
    \end{equation}
    Let us now switch to the Chevalley--Eilenberg description $\sfCE(V)$ of the differential graded Lie algebra $V$ we want to construct, which is generated by coordinates $r^\mu$, $r^\alpha$, $r^a$, $r_a$, $r_\alpha$, $r_\mu$ of degrees $0,1,2,3,4,5$, respectively. We note that we have a natural symplectic form on $V[1]^*$ of degree 5,
    \begin{equation}
        \omega=\rmd r^\alpha \wedge \rmd r_\alpha+\rmd r^a\wedge \rmd r_a+\rmd r^\mu \wedge \rmd r_\mu~.
    \end{equation}
    Compatibility of the Lie algebra action with the duality pairing amounts to the fact that the Chevalley--Eilenberg differential $Q$ is Hamiltonian for the Poisson bracket of degree~$-5$,
    \begin{equation}\label{eq:Poisson_5d}
        \begin{aligned}
            \{f,g\}\coloneqq &-\parder[f]{r_\alpha}~\parder[g]{r^\alpha}+(-1)^{|f|+1}\parder[f]{r^\alpha}~\parder[g]{r_\alpha}-\parder[f]{r_a}~\parder[g]{r^a}+(-1)^{|g|+1}\parder[f]{r^a}~\parder[g]{r_a}
            \\
            &-\parder[f]{r_\mu}~\parder[g]{r^\mu}+(-1)^{|g|+1}\parder[f]{r^\mu}~\parder[g]{r_\mu}
        \end{aligned}
    \end{equation}
    induced by $\omega$. That is,
    \begin{equation}
        Q=\{\caQ,-\}~,~~~|\caQ|=6~.
    \end{equation}
    The most generic Hamiltonian $\caQ$ of degree~$6$ that is at most cubic in the generators\footnote{This restriction is required to obtain a differential graded Lie algebra, as opposed to an $L_\infty$-algebra} is
    \begin{equation}
        \begin{aligned}
            \caQ&=\tfrac12 f_{\beta\gamma}{}^\alpha r_\alpha r^\beta r^\gamma+t_{\alpha a}{}^b r^\alpha r^a r_b+\tfrac1{3!}d_{abc} r^ar^br^c+\tfrac12 Z^{ab}r_ar_b+\Theta_{a}{}^{\alpha}r^ar_\alpha
            \\
            &~~~~+g^{\mu}_{1\alpha}r_\mu r^\alpha+g^{\mu}_{2\alpha\nu}r_\mu r^\alpha r^\nu+g_{3\mu a}^\alpha r^\mu r^ar_\alpha+g_{4\mu}^{ab} r^\mu r_ar_b~,
        \end{aligned}
    \end{equation}
    where besides the structure constants $f_{\beta\gamma}{}^\alpha$ and the embedding tensor $\Theta_{a}{}^{\alpha}$ we have the deformation tensors $d_{abc}$ and $Z^{ab}$, which are totally symmetric and antisymmetric, respectively, due to the grading of the generators. The remaining structure constants will be called {\em auxiliary}. For $\caQ$ to give rise to a Chevalley--Eilenberg differential, we have to impose
    \begin{equation}
        Q^2=0~~~\Leftrightarrow~~~\{\caQ,\caQ\}=0~.
    \end{equation}
    This equation imposes conditions on the structure constants. For example, we have
    \begin{equation}\label{eq:cond_1}
        \Theta_{a}{}^{\gamma} f_{\beta\gamma}{}^{\alpha}+t_{\beta a}{}^b\Theta_{b}{}^{\alpha}-g^\mu_{1\beta}g_{3\mu a}^\alpha=0~.
    \end{equation}
    For $g_{1}=g_{3}=0$, this implies that the embedding tensor is an invariant tensor, which is clearly too strong a condition. We can make a non-canonical choice of an embedding
    \begin{equation}
        i:{\rm coker}(\Theta)\hookrightarrow\frg~,
    \end{equation}
    which is given by structure constants $i^\alpha_\mu$ such that
    \begin{equation}
        i^\alpha_\mu g_{1\alpha}^\nu=\delta_\mu^\nu~.
    \end{equation}
    With this choice, we can split the condition~\eqref{eq:cond_1} into
    \begin{equation}
        \begin{aligned}
            \Theta_c{}^\beta\Theta_{a}{}^{\gamma} f_{\beta\gamma}{}^{\alpha}+X_{ca}^b\Theta_{b}{}^{\alpha}&=0~,
            \\
            i^\beta_\mu(\Theta_{a}{}^{\gamma} f_{\beta\gamma}{}^{\alpha}+t_{\beta a}{}^b\Theta_{b}{}^{\alpha})&=g_{3\mu a}^\alpha~,
        \end{aligned}
    \end{equation}
    and the first condition is the usual one encountered in the $d=5$ tensor hierarchy, while the second condition fixes one of the auxiliary structure constants. Besides the above condition and the fact that $f_{\beta\gamma}{}^{\alpha}$ and $t_{\alpha a}{}^b$ are the structure constants of the Lie algebra $\frg$ and a representation of $\frg$, we also have
    \begin{equation}
        \begin{aligned}
            Z^{ab}\Theta_{b}{}^{\alpha} &=0
            ~,~~~&
            \Theta_{a}{}^{\alpha} g_{1\alpha}^\mu&=0~,
            \\
            Z^{ab}d_{acd}-2X^{a}_{(cd)}&=0
            ~,~~~&
            Z^{a[b} t_{\alpha a}{}^{c]}+2g_{1\alpha}^\mu g_{4\mu}^{bc}&=0~,
            \\
            t_{\alpha (a}{}^d d_{bc)d}&=0~,
        \end{aligned}
    \end{equation}
    as well as a number of conditions for the auxiliary structure constants. As expected, the tensor $d_{abc}$ capturing the Lie bracket $V_{-1}\otimes V_{-1}\rightarrow V_{-2}$ has to be an invariant tensor.
    
    The kinematical data of a generic tensor hierarchy can then be constructed from the firmly adjusted Weil algebra of the corresponding $L_\infty$-algebra as described in detail in \cref{ssec:firmly_adjusted}.
    
    We note that the condition that $d_{abc}$ be an invariant tensor is too strong a constraint, e.g.~for the non-maximally supersymmetric case. From the formulas of the curvatures, it is clear that there is no differential graded Lie algebra underlying this case, if the higher gauge algebra is constructed using the formulas of \cref{thm:ophLie_from_dgLA}. This observation strongly suggests that there are  generalisations of these derived bracket constructions, but this is beyond the scope of this thesis.
    
    \subsection{Example: \texorpdfstring{$d=5$}{d=5} maximal supergravity}
    
    Let us give a concrete and complete picture of the interpretation of a tensor hierarchy using $\ophLie$-algebras, including the construction of curvatures. We choose the case $d=5$, which allows us to recycle observations made in \cref{ssec:gen_tensor_hierarchies}. For a detailed discussion of this theory, see~\cite{deWit:2004nw}.
    
    Maximal supergravity in $d=5$ dimensions has the non-compact global symmetry group $\sfE_{6(6)}(\IR)$~\cite{Cremmer:1980gs}. When dimensionally reducing from $d=11$, in order to make manifest the $\fre_{6(6)}$ structure of the scalar sector in $d=5$, one must first dualise the 3-form potential, as described in detail in~\cite{Cremmer:1997ct}. This gives a total of 42~scalars parameterizing
    $\sfE_{6(6)}(\IR)/\sfUSp(8)$.
    
    The fully dualised bosonic Lagrangian with manifest $\sfE_{6(6)}(\IR)$-invariance can be written as
    \begin{equation}\label{D5L}
        \mathcal{L}_5 = R\star 1 +\tfrac{1}{2} g_{xy} \rmd \varphi^x \wedge \star \rmd \varphi^y  -\tfrac{1}{2} a_{ab} F^{a} \wedge \star F^{b} -\tfrac{1}{6} d_{abc}F^{a}_{\2} \wedge F^{b}_{\2} \wedge A^{c}_{\1}~.
    \end{equation}
  The 1-form potentials transform linearly in the ${\rep{27 } }_c$ of  $\mathfrak{e}_{6(6)}$, and $a,b,c \in \{ 1,\ldots, 27\}$. In addition to the singlet $\rep{1}\in{\rep{27 } }_c\otimes \rep{27}$, used to construct the 1-form kinetic term, there is a singlet in the totally symmetric 3-fold tensor product $\rep{1}\in\bigodot^3(\mathbf{{27} }_c)$, which is used to construct the topological cubic term.
    
    For the construction of the tensor hierarchy, we shall need the following $\sfE_6$-invariant tensors: 
    \begin{equation}        
        \begin{aligned}
            f_{\alpha\beta\gamma} & \in \bigwedge\nolimits^3~\mathbf{78}
            ~,~~~&
            t_{\alpha}{}_{a}{}^{b}& \in \mathbf{78}\otimes \mathbf{{27}}\otimes\mathbf{{27} }_c~,
            \\
            d^{abc}& \in \bigodot\nolimits^3~\mathbf{27}
            ~,~~~&
            d_{abc}& \in \bigodot\nolimits^3~\mathbf{{27} }_c~.
        \end{aligned}
    \end{equation}
    To optimise our notation, we also introduce the following tensors:
    \begin{equation}        
        \begin{aligned}
            X_{ab}{}^{c}& = \Theta_{a}{}^{\alpha} t_{\alpha}{}_{b}{}^{c}~,~~~&Y_{a}{}_{\alpha}{}^{\beta}& =  \Theta_{a}{}^{\gamma} f_{\gamma\alpha}{}^{\beta}+ t_{\alpha}{}_{a}{}^{b}\Theta_{b}{}^{\beta}\equiv\delta_\alpha \Theta_{b}{}^{\beta}~,
            \\
            X_{a\alpha}{}^{\beta}& =\Theta_{a}{}^{\gamma} f_{\gamma\alpha}{}^{\beta}~,~~~
            &Z^{ab}& = \Theta_{c}{}^{\alpha} t_{\alpha}{}_{d}{}^{a}d^{bcd}= X_{cd}{}^{a}d^{bcd}=Z^{[ab]}~,
        \end{aligned}
    \end{equation}
    The above tensors satisfy the following identities~\cite{deWit:2004nw}:
    \begin{subequations}\label{eq:def_tens_identities}
        \begin{equation}
            d_{a cd}d^{b cd} = \delta_{a}{}^{b}~,~~~
            X_{(ab)}{}^{c} = d_{ab d}Z^{cd}~,~~~X_{[ab]}{}^{c} = 10 d_{adf}d_{beg}d^{c de}Z^{fg}~,
        \end{equation}
        and in addition, we have the following three equivalent forms of the closure constraints:
        \begin{equation}        
            2X_{[a|c|}{}^{d}X_{b]d}{}^{e}+X_{[ab]}{}^{d}X_{dc}{}^{e}=0~,~~~
            Z^{ab}X_{bc}{}^{d}=0~,~~~
            X_{dc}{}^{[a}Z^{b]c}=0~.
        \end{equation}
    \end{subequations}
    Using these, we can now apply the formalism of \cref{ssec:gen_tensor_hierarchies} and construct the differential graded Lie algebra. It helps to broaden the perspective a bit and derive the latter from a graded Lie algebra $V$, with underlying vector space consisting of $\mathfrak{e}_{6(6)}$-modules:
    \begin{equation}    
        \begin{array}{ccccccccccccccccccccc}
            V_{\mathfrak{e}_{6(6)}}& =& V_{-5}&\oplus& V_{-4} &\oplus& V_{-3}&\oplus& V_{-2}&\oplus& V_{-1}&\oplus& V_{0} &\oplus& V_{1}\\[5pt]
            \rho_{\sst{(k)}}&& \mathbf{27\oplus 1728}&& \mathbf{351}_c && \mathbf{78} && \mathbf{27} && \mathbf{27}_c && \mathbf{78} && \mathbf{351} \\[5pt]
            e_{\sst{(k)}}&& (e^{a}, e_{ab}{}^{\alpha}) && e_{a}{}^{\alpha} && e^{\alpha} &&e^{a} && e_a &&  e_\alpha  &&  e_{\alpha}{}^{a}
        \end{array}
    \end{equation}
    We have indicated the $\mathfrak{e}_{6(6)}$-representations $\rho_{\sst{(k)}}$ carried by each $V_{\mathfrak{e}_{6(6)}}$-degree~$k$ summand, $V_k$, and their corresponding basis elements $e_{\sst{(k)}}$, e.g.~$(e_\0)_\alpha=e_\alpha$, where $(e_\alpha)$ is some basis for the exceptional Lie algebra $\mathfrak{e}_{6(6)}$. Note that the embedding tensor $\Theta=\Theta_{a}{}^{\alpha}e_{\alpha}{}^{a}$ is an element of $V_1$ and $e_{\alpha}{}^{a}= P_{\mathbf{351}}e_{\alpha}\otimes e^{a}$.  
    
    The graded Lie bracket on $V$ is now given mostly by the obvious projections of the graded tensor products,
    \begin{equation}\label{D5_gla_com}
        [e_\alpha,e_\beta]=f_{\alpha\beta}{}^\gamma e_\gamma~,~~~
        [e_\alpha, e_{\sst{(k)}}] = \rho_{\sst{(k)}}(e_\alpha) e_{\sst{(k)}}~,~~~
        [e_{\sst{(k)}}, e_{\sst{(l)}}] = T_{k,l}e_{\sst{(k+l)}}~.
    \end{equation}
    Here, $T_{k,l}$ are the intertwiners dual to the projectors  $P_{k,l}:V_k\wedge V_l\rightarrow V_{k+l}$. For example, 
    \begin{equation}\label{eq:e6_gla_comm_def}
        [e_a, e_b] = 2d_{abc}e^c~,~~~[e_a, e^b] = (t_{\alpha})_{a}{}^{b} e^\alpha~,
    \end{equation}
    where
    \begin{equation}
    e^a \coloneqq \tfrac12 d^{abc}[e_b, e_c], \qquad  e^\alpha \coloneqq (t_{\alpha})_{b}{}^{a} [e_a, e^b]~. 
        \end{equation}
The adjoint indices are raised/lowered with $\eta_{\alpha\beta}= \tr (t_\alpha t_\beta)$, which is proportional to the Cartan--Killing form. 
    
    Selecting an element $\Theta=\Theta_{a}{}^{\alpha}e_{\alpha}{}^{a}\in V_1$ now defines a differential 
    \begin{equation}    
        \rmd v \coloneqq [\Theta, v] 
    \end{equation}
    for $v\in V$, and we note that $[\Theta,\Theta]=0$ for degree reasons. The explicit action of $\rmd e_{\sst{(k)}} \coloneqq [\Theta, e_{\sst{(k)}} ] $ can be determined using the graded Jacobi identity  from the initial condition
    \begin{equation}    
        [ e_{\alpha}{}^{a} , e_b] = P_{\rep{351}}\delta_{b}{}^{a}e_\alpha~, 
    \end{equation}
    where $P_{\rep{351}}$ is the projector $P_{\mathbf{351}}: \rep{78}\otimes \rep{27}_c \rightarrow \rep{351}$,
    \begin{equation}    
        (P_{\rep{351}})_{\alpha}{}^{a}{}_{b}{}^{\beta} = -\tfrac65 t_{\alpha}{}_{b}{}^{c}t^{\beta}{}_{c}{}^{a}+\tfrac{3}{10} t_{\alpha}{}_{c}{}^{a}t^{\beta}{}_{b}{}^{c}+\tfrac15 \delta_{\alpha}{}^{\beta}\delta^{a}{}_{b}~.
    \end{equation}
    We thus obtain a differential graded Lie algebra, and this is a special case of the algebra called dgLie (THA$'$) in \cref{ssec:tensor_hierarchy_algebras}.
    
    Let us now construct the $\ophLie$-algebra $\frE$ of this differential graded Lie algebra using \cref{thm:ophLie_from_dgLA}. We arrive at the graded vector space 
    \begin{equation}    
        \begin{array}{ccccccccccccccccccccc}
            \frE_{\mathfrak{e}_{6(6)}}=&  \frE_{-4} &\oplus& \frE_{-3}&\oplus& \frE_{-2}&\oplus& \frE_{-1}&\oplus& \frE_{0} \\
            & \mathbf{27\oplus 1728} && \mathbf{351}_c && \mathbf{78} && \mathbf{27} && \mathbf{27}_c \\
            & (e^{a}, e_{ab}{}^{\alpha}) && e_{a}{}^{\alpha} &&e^{\alpha} &&  e^{a}&&  e_a
        \end{array}
    \end{equation}
    with non-trivial products
    \begin{equation}
        \eps_1(x)\coloneqq [\Theta, x]
        ~,~~~
        \eps^0_2(x,y)\coloneqq     [[\Theta, x] , y]
        ~,~~~ 
        \eps^{1}_{2}(x,y)\coloneqq     (-1)^{|x|}[x , y]~.
    \end{equation}
    Explicitly, we have the differentials
    \begin{subequations}
        \begin{equation}
            \begin{split}
                \eps_1(e^a)&=  \Theta_{b}{}^{\alpha} t_{\alpha}{}_{c}{}^{d}d^{bca}e_d = X_{bc}{}^{d}d^{bca}e_d=-Z^{ad}e_d~,
                \\
                \eps_1(e^\alpha)&=   \Theta_{b}{}^{\alpha}e^{b}~,\\
                \eps_1(e_{a}{}^{\alpha})&=  - \delta_{\beta}\Theta_{a}{}^{\alpha}e^{\beta}=-Y_{a}{}_{\beta}{}^{\alpha}e^{\beta}~,         
            \end{split}
        \end{equation}
        the Leibniz-like products
        \begin{equation}
            \begin{aligned}
                \eps^0_2(e_a, e_b)&=    [\Theta_{a}{}^{\alpha} e_\alpha  , e_b] =\Theta_{a}{}^{\alpha} t_{\alpha}{}_{b}{}^{c}e_{c}  =X_{ab}{}^{c}e_{c}~,
                \\
                \eps^0_2(e_a, e^b)& =    [\Theta_{a}{}^{\alpha} e_\alpha  , e^b] =-\Theta_{a}{}^{\alpha} t_{\alpha}{}_{c}{}^{b}e^{c}  =-X_{ac}{}^{b}e^{c}~,
                \\
                \eps^0_2(e_a, e^\beta)& =    [\Theta_{a}{}^{\alpha} e_\alpha  , e^\beta] =-\Theta_{a}{}^{\alpha}f_{\alpha\gamma}{}^\beta e^\gamma =-X_{a\gamma}{}^\beta e^\gamma~,
                \\
                \eps^0_2(e_a, e_{b}{}^{\beta})& =    [\Theta_{a}{}^{\alpha} e_\alpha  , e_{b}{}^{\beta}] =-\Theta_{a}{}^{\alpha}f_{\alpha\gamma}{}^\beta e_{b}{}^{\gamma}+\Theta_{a}{}^{\alpha}t_{\alpha b}{}^{c} e_{c}{}^{\beta} =-X_{a\gamma}{}^\alpha   e_{b}{}^{\gamma}+X_{ab}{}^{c}e_{c}{}^{\beta}~,
            \end{aligned}
        \end{equation}
        as well as the alternator-type products
        \begin{equation}
            \begin{aligned}
                \eps^{1}_{2}(e_a, e_b)&=  2 d_{abc}e^{c}~,
                \\
                \eps^{1}_{2}(e_a, e^b)&=  t_{\alpha a}{}^{b}e^{\alpha}   =\eps^{1}_{2}(e^b, e_a)~,
                \\
                \eps^{1}_{2}(e_a, e^\alpha)&= e_{a}{}^{\alpha}  = P_{\rep{351}_c}e_{b}{}^{\beta} =-\eps^{1}_{2}(e^\alpha, e_a)~,
                \\
                \eps^{1}_{2}(e^a, e^b)&= -e^{[ab]}  = t_{\alpha c}{}^{[a}d^{b]cd}e_{d}{}^{\alpha}~,
            \end{aligned}
        \end{equation}
        where we used that  $t_{\alpha c}{}^{[a}d^{b]cd}$ is the intertwiner between the $\mathbf{351 }_c\in \mathbf{27 }_c \otimes \mathbf{78}$ and $\mathbf{351 }_c\cong  \bigwedge^2\,\mathbf{27}$. 
    \end{subequations}
    
    We can now construct the corresponding curvatures. We start from the Chevalley--Eilenberg algebra of $\frE_{\mathfrak{e}_{6(6)}}$ with the following generators $(r^A)$ spanning $\frE_{\mathfrak{e}_{6(6)}}[1]^*$:
    \begin{equation}    
        \begin{array}{ccccccccccccccccccccc}
            \text{degree} & 1&& 2&&3&&4&&5\\[5pt]
            \frE_{\mathfrak{e}_{6(6)}}[1]^*= & \mathbf{27} &\oplus& \mathbf{27 }_c &\oplus& \mathbf{78} &\oplus& \mathbf{351} &\oplus& \mathbf{27 }_c\oplus\rep{1728 }_c \\[5pt]
            & r^a &&r_{a} &&r_{\alpha} &&  r_{\alpha}{}^{a} &&  (r_{a}, r_{\alpha}{}^{ab})
        \end{array}
    \end{equation}
    Consider now the Weil algebra $\sfW(\frE_{\mathfrak{e}_{6(6)}})$, cf.~\cref{def:Weil-EL-infty}. Here, we introduce a second copy of shifted generators $(\hat r^A)$ spanning $\frE_{\mathfrak{e}_{6(6)}}[2]^*$ with $|\hat r^A|=|r^A|+1$. The usual Weil differential up to degree~$3$ elements, dual to scalars in $d=5$, then reads as 
    \begin{equation}\label{N=8D5QW}
        \begin{aligned}
            Q_{\sf W} r^{a} &=-Z^{ab}r_{b} - X_{bc}{}^{a}    r^{b} \oslash_0 r^{c}+\hat r^a~,
            \\
            Q_{\sf W} r_{a} &= 
            \Theta_{a}{}^{\alpha}r_{\alpha}
            +X_{ba}{}^{c}r^b \oslash_0 r_c 
            - d_{abc} r^{b} \mathbin{\hat \oslash_1}   r^{c}+ \hat r_a~,
            \\
            Q_{\sf W}r_{\alpha} &=  
            Y_{a\alpha}{}^{\beta}r_{\beta}{}^{a}
            +X_{a\alpha}{}^{\beta}r^{a} \oslash_0 r_{\beta}
            + t_{\alpha}{}_{a}{}^{b}  r^{a} \mathbin{\hat \oslash_1} r_{b} 
            +  \hat r_\alpha~,
            \\ 
            Q_{\sf W} \hat r^{a} &=
            Z^{ab} \hat r_{b} + X_{bc}{}^{a}  \hat  r^{b} \oslash_0 r^{c}- X_{bc}{}^{a}    r^{b} \oslash_0 \hat r^{c}~,
            \\
            Q_{\sf W} \hat r_{a} &= 
            - \Theta_{a}{}^{\alpha}\hat r_{\alpha}
            - X_{ba}{}^{c}\hat r^b \oslash_0 r_c 
            + X_{ba}{}^{c} r^b \oslash_0 \hat r_c 
            +2 d_{abc} \hat r    ^{b} \mathbin{\hat \oslash_1} r    ^{c}~,
            \\
            Q_{\sf W} \hat r_{\alpha} &=-Y_{a\alpha}{}^{\beta}\hat r_{\beta}{}^{a}
            -X_{a\alpha}{}^{\beta} \hat r^{a} \oslash_0 r_{\beta}
            + X_{a\alpha}{}^{\beta}  r^{a} \oslash_0 \hat r_{\beta}
            - t_{\alpha}{}_{a}{}^{b} \hat r^{a} \mathbin{\hat \oslash_1} r_{b} 
            + t_{\alpha}{}_{a}{}^{b}  \hat r_{b} \mathbin{\hat \oslash_1} r^{a}~,
        \end{aligned}
    \end{equation}
    where we have introduced the notation
    \begin{equation}
        \begin{split}
            a \mathbin{\hat \oslash_i} b& =  a  \oslash_i b +(-1)^{i+|a|\,|b|} b  \oslash_i a~, \\
            a \mathbin{\check \oslash_i} b& = a  \oslash_i b -(-1)^{i+|a|\,|b|} b  \oslash_i a~.\\
        \end{split}
    \end{equation} 
    The deformed Leibniz rule~\eqref{eq:def_Leibniz}, together with the remaining $\opEilh$-relations~\eqref{eq:Eilh-relations} and the identities~\eqref{eq:def_tens_identities}, then imply $Q_\sfW^2=0$ as one can check by direct computation. 
    
    In order to define the curvatures of the tensor hierarchy, we symmetrise to an $L_\infty$-algebra using \cref{thm:antisym_hLie}. We can then use the formalism of \cref{ssec:firmly_adjusted} to construct an adjusted Weil algebra in the sense of~\cite{Saemann:2019dsl}, ensuring closure of the gauge algebra without any further constraints on the field strengths.
    
    To illustrate in more detail the procedure and what it achieves, we can perform the coordinate change already at the level of the Weil algebra of the $\ophLie$-algebra. This coordinate change yields a symmetrised and firmly adjusted Weil algebra through an  evident  coordinate change, $r^A\mapsto \tilde r^A$, which removes all appearances of $\oslash_1$ in $Q_{\sf W}\tilde r^A$ via the deformed Leibniz rule~\eqref{eq:def_Leibniz}. Hence,  by \cref{thm:el_infty_contain_l_infty} we are left with an $L_\infty$-algebra. Explicitly, the following  coordinate change manifestly removes all appearances of $\oslash_1$:
    \begin{equation}    
        \begin{aligned}\label{eq:can_rotation}
            r^a&\mapsto a^a \coloneqq  r^a~,
            \\
            r_a&\mapsto b_a \coloneqq  r_a  +\tfrac12  d_{abc} r^b \mathbin{\check \oslash_0} r^c~,
            \\
            r_{\alpha}&\mapsto c_{\alpha}\coloneqq  r_{\alpha}- \tfrac12   t_{\alpha}{}_{a}{}^{b}  r^{a} \mathbin{\check  \oslash_0} r_{b}~,
            \\
            r_{\alpha}{}^{a}&\mapsto d_{\alpha}{}^{a} \coloneqq  r_{\alpha}{}^{a}+\tfrac12 P_{\rep{351}_c} r^{b} \mathbin{\check  \oslash_0} r_{\beta}+\tfrac12 t_{\alpha c}{}^{[b}d^{c]ad}r_{b} \mathbin{\check  \oslash_0} r_{c}~,
        \end{aligned}
    \end{equation}
    where $ d_{\alpha}{}^{a}$ is included as it is needed for $Q_{\sf W}\tilde r_{\alpha}$. The corresponding coordinate change on $\hat r^A$ is firmly adjusted by simply first ordering the occurrences of  $\hat r^B$ in  $\hat{\tilde{r}}^A$ to the left (which is permitted by the appearance of only $\check \oslash_0$ in $\hat{\tilde{r}}^A$) and then  sending $\check \oslash_i$ to $\check \oslash_i+\hat \oslash_i=2\oslash_i$. The choice of left ordering follows from the choice of left Leibniz rule, which is a matter of convention. Applied to \eqref{eq:can_rotation} this yields
    \begin{equation}    
        \begin{aligned}
            \hat   r^a&\mapsto f^a \coloneqq  \hat r^a~,
            \\
            \hat    r_a&\mapsto h_a \coloneqq \hat  r_a  + 2d_{abc}\hat r^b  \oslash_0  r^c~,
            \\
            \hat    r_{\alpha}&\mapsto g_{\alpha} \coloneqq  \hat r_{\alpha}-  t_{\alpha}{}_{a}{}^{b} (\hat  r^{a}   \oslash_0  r_{b} -  \hat r_{b}  \oslash_0  r^{a})~, 
            \\
            \hat  r_{\alpha}{}^{a}&\mapsto k_{\alpha}{}^{a} \coloneqq  \hat r_{\alpha}{}^{a}+ P_{\rep{351}_c} (\hat r^{b}   \oslash_0  r_{\beta} +\hat r_{\beta}  \oslash_0  r^{b})+2 t_{\alpha c}{}^{[b}d^{c]ad} \hat r_{b}   \oslash_0  r_{c}~.
        \end{aligned}
    \end{equation}
    Note, this is a special case of the transformation \eqref{eq:firm_rotation} for a firm adjustment.

    The result of this coordinate change is the differential graded commutative algebra $\sfW_{\rm adj}(\frE_{\mathfrak{e}_{6(6)}})$ generated by $\frE_{\mathfrak{e}_{6(6)}}[1]^*\oplus\frE_{\mathfrak{e}_{6(6)}}[2]^*$ and differential
    \begin{equation}\label{symN=8D5QW}
		\begin{aligned}
            Q_{\sfW_{\rm adj}} a^{a} &=-Z^{ab}b_{b} - \tfrac12 X_{bc}{}^{a}    a^{b}   a^{c}+f^a~,
            \\
            Q_{\sfW_{\rm adj}} b_{a} &= 
            \Theta_{a}{}^{\alpha}c_{\alpha}
            + \tfrac12 X_{ba}{}^{c}a^b   b_c + \tfrac16 d_{abc} X_{de}{}^{b}a^c   a^d   a^e 
            - d_{abc} f^{b}   a^{c}+ h_a~,
            \\
            Q_{\sfW_{\rm adj}}c_{\alpha} &=  
            Y_{a\alpha}{}^{\beta}d_{\beta}{}^{a}
            +\tfrac12 X_{a\alpha}{}^{\beta}a^{a}   c_{\beta}
            +(\tfrac14 X_{a\alpha}{}^{\beta}t_{\beta b}{}^{c}+\tfrac13t_{\alpha a}{}^{d} X_{(db)}{}^{c})a^a  a^b b_c
            \\ 
            &~~~~~+\tfrac12 t_{\alpha}{}_{a}{}^{b} f^a   b_b -\tfrac12 t_{\alpha}{}_{a}{}^{b} h_b a^a    -\tfrac{1}{6} t_{\alpha}{}_{a}{}^{b} d_{bcd}a^{a}  a^{c}  f^{d}+ g_\alpha~,\\
            Q_{\sfW_{\rm adj}} f^{a} &=
            Z^{ab} h_{b} + X_{bc}{}^{a} a^{b}    f^{c} ~,
            \\
            Q_{\sfW_{\rm adj}} h_{a} &= 
            -\Theta_{a}{}^{\alpha}g_{\alpha}
            + X_{ab}{}^{c} a^b   h_c 
            + d_{abc}f^b   f^c~,
            \\
            Q_{\sfW_{\rm adj}} g_{\alpha} &=-Y_{a\alpha}{}^{\beta}k_{\beta}{}^{a}
            +X_{a\alpha}{}^{\beta}  a^{a}   g_{\beta}
            - t_{\alpha}{}_{a}{}^{b}   h_{b}    f^{a}~.
		\end{aligned}
	\end{equation}
    
    We can now define the corresponding curvatures in the adjusted higher gauge theory as usual as a morphism of differential graded algebras
    \begin{equation}\label{eq:dga_morphism_5d}
        (\caA,\caF): \sfW_{\rm adj}(\frE_{\mathfrak{e}_{6(6)}})~\longrightarrow~\Omega^\bullet(M)~,
    \end{equation}
    where\footnote{The additional signs here follow from the choice of sign convention in \eqref{eq:e6_gla_comm_def}.}
    \begin{equation}
        \begin{split}
            (a^a,b_a,c_\alpha,  d_{\alpha}{}^{a}) &\mapsto  (A^a, B_a, -C_\alpha,  -D_{\alpha}{}^{a})~,\\
            (f^a,h_a,g_\alpha,  k_{\alpha}{}^{a}) &\mapsto  (F^a, H_a, -G_\alpha, -K_{\alpha}{}^{a})~.
        \end{split}
    \end{equation}
    This indeed yields  the gauge potentials and curvatures of the $d=5$ tensor hierarchy:
    \begin{subequations}\label{eq:field_strengths}
        \begin{align} 
            F^a &= \rmd A^a +\tfrac{1}{2} X_{bc}{}^{a} A^b\wedge A^c  + Z^{ab}B_b~,
            \\
            H_a 
            &= \rmd B_a -\tfrac 12 X_{ba}{}^{c} A^b\wedge B_c   - \tfrac{1}{6} d_{abc} X_{de}{}^{b} A^c\wedge A^d \wedge A^e  +d_{abc} A^b\wedge F^c + \Theta_{a}{}^{\alpha}  C_\alpha~,
            \\
            G_\alpha  &= \rmd  C_\alpha-\tfrac12 X_{a\alpha}{}^{\beta}A^a\wedge  C_\gamma+(\tfrac14 X_{a\alpha}{}^{\beta}t_{\beta b}{}^{c}+\tfrac13t_{\alpha a}{}^{d} X_{(db)}{}^{c})A^a\wedge  A^b \wedge B_c \\
            &\phantom{=} +\tfrac12 t_{\alpha}{}_{a}{}^{b} F^a \wedge B_b -\tfrac12 t_{\alpha}{}_{a}{}^{b}H_b  \wedge A^a  -\tfrac{1}{6} t_{\alpha}{}_{a}{}^{b} d_{bcd}A^{a}\wedge A^{c}\wedge F^{d}- Y_{a\alpha}{}^{\beta} D_{\beta}{}^{a}~,
        \end{align}
    \end{subequations}
    along with the corresponding Bianchi identities,
    \begin{subequations}\label{D5Bianchi}
        \begin{align} 
            0&=\rmd F^{a} 
            - X_{bc}{}^{a} A^{b} \wedge   F^{c} -Z^{ab} H_{b}  ~,
            \\
            0 &= \rmd H_{a}
            - X_{ab}{}^{c} A^b  \wedge H_c 
            - d_{abc}F^b \wedge  F^c-\Theta_{a}{}^{\alpha}G_{\alpha}~,
            \\
            0&=  \rmd G_{\alpha}- X_{a\alpha}{}^{\beta}  A^{a} \wedge  G_{\beta}- t_{\alpha}{}_{a}{}^{b}   H_{b} \wedge   F^{a} +Y_{a\alpha}{}^{\beta}K_{\beta}{}^{a}~.
        \end{align}
    \end{subequations}
    We note that the full kinematical data is determined in this way: the Bianchi identities are implied by compatibility of the morphism~\eqref{eq:dga_morphism_5d} with the differential, and the gauge transformations are constructed as infinitesimal partially flat homotopies, cf.~e.g.~\cite{Saemann:2019dsl} for details.

    To make contact with the expressions in the supergravity literature, cf.~\cite{deWit:2004nw, Hartong:2009vc}, one must make the field redefinitions 
    \begin{equation}
        \begin{aligned}
            C_{\alpha}&\mapsto C_\alpha+\tfrac12t_{\alpha a}{}^b A^a\wedge B_b~,
            \\
            D_{\alpha}{}^{a} &\mapsto D_{\alpha}{}^{a} - \tfrac12 P_{\rep{351 }_c}  A^a\wedge  C_\alpha~.
        \end{aligned}
    \end{equation}
    Similar field redefinitions were also used in~\cite{Greitz:2013pua} to link another elegant derivation of the curvature forms (in which, however, the link to higher gauge algebras also is somewhat obscured) to the supergravity literature. We stress that from the higher gauge algebra point of view, the form~\eqref{eq:field_strengths} is special in the sense that all exterior derivatives of gauge potentials in non-linear terms have been absorbed in field strengths. This makes~\eqref{eq:field_strengths} particularly useful, as it exposes cleanly the separation of unadjusted curvature and adjustment. From the former, one can straightforwardly identify the higher Lie algebra of the structure group of the underlying higher principal bundle. Moreover, gauge transformations are readily derived from partially flat homotopies, as mentioned above. As a side effect, it is interesting to note that the arising higher products are at most ternary.
    
    An interesting aspect of~\eqref{eq:field_strengths} is the fact that the covariantisations of the differentials $\rmd B$ and $\rmd C$ contain a perhaps unexpected factor of $\tfrac12$. This factor is a clear indication that
    the origin of the gauge $L_\infty$-algebra is indeed an $\ophLie$-algebra: the action $\acton$ of $A$ on $B$ and $C$ is encoded in an $\ophLie$-algebra with
    \begin{equation}
        \eps_2^0(A,B)\coloneqq A\acton B\eand \eps_2^0(A,C)\coloneqq A\acton C~,
    \end{equation}
    which is then antisymmetrised by \cref{thm:antisym_hLie} to 
    \begin{equation}
        \mu_2(A,B)\coloneqq \tfrac12 \eps^0_2(A,B)\eand \mu_2(A,C)\coloneqq \tfrac12 \eps^0_2(A,C)~,
    \end{equation}
    at the cost of introducing non-trivial higher products $\mu_3$, cf.~\eqref{eq:antisymmetrisation_hLie}.
    
    \section{Comparison to the literature}\label{sec:comparison}
    
    We compare our results with algebraic structures previously introduced in the literature to capture the gauge structure underlying the higher gauge theories obtained in the tensor hierarchies of gauged supergravity. We shall focus on the particularities of the gauge algebraic structures of the tensor hierarchies; for other work linking the tensor hierarchy to ordinary $L_\infty$-algebras, see also~\cite{Cagnacci:2018buk}.
    
    \subsection{Enhanced Leibniz algebras}
    
    A notion of enhanced Leibniz algebras was introduced in~\cite{Strobl:2016aph,Strobl:2019hha} to capture the parts of the higher gauge algebraic structures appearing in the tensor hierarchy. See also~\cite{Kotov:2018vcz} for a discussion of the higher gauge theory employing these enhanced Leibniz algebras and the link to the tensor hierarchy.
    
    \begin{definition}[\cite{Strobl:2019hha}]
        An \uline{enhanced Leibniz algebra} is a Leibniz algebra $(\sfV,[-,-])$ together with a vector space $\sfW$ and a linear map $t:\sfW\rightarrow \sfV$ as well as a binary operation $\circ:\sfV\otimes\sfV\to\sfW$ such that
        \begin{equation}
            \begin{aligned}
                [t(w),v]&=0~~~&u\stackrel{s}{\circ}[v,v]&=v\stackrel{s}{\circ}[u,v]
                \\
                t(w)\circ t(w)&=0~,~~~&[v,v]&=t(v\circ v)
            \end{aligned}
        \end{equation}
        for all $u,v\in \sfV$ and $w\in \sfW$, where $u\stackrel{s}{\circ}v$ denotes the symmetric part of $u\circ v$.
        
        A \uline{symmetric enhanced Leibniz algebra} additionally satisfies the condition that
        \begin{equation}
            u\circ v=v\circ u
        \end{equation}
        for all $u,v\in \sfV$.
    \end{definition}
    
    A symmetric enhanced Leibniz algebra is an $\ophLie$-algebra concentrated in degrees $-1$ and $0$ with a few axioms missing. We can identify the structure maps as follows.
    \begin{equation}
        \begin{gathered}
            \frE=(\frE_{-1}\xrightarrow{~\eps_1~}\frE_0)~=~(\sfW\xrightarrow{~\sft~}\sfV)~,
            \\
            \eps_2(v_1,v_2)=[v_1,v_2]~,~~~\eps_2(v,w)=0~,~~~\sfalt(v_1,v_2)=v_1\circ v_2~,
        \end{gathered}
    \end{equation}
    for $v,v_1,v_2\in \sfV$ and $w\in\sfW$. The $\ophLie$-algebra relations~\eqref{eq:hLie-relations} are trivially satisfied since $\eps_2$ is a Leibniz bracket. Moreover, $\eps_1$ is trivially a differential and a derivation of $\eps_2$. The relation $\eps_2(v_1,v_2)+\eps_2(v_2,v_1)=\eps_1(\sfalt(v_1,v_2))$ is the polarisation of $[v,v]=t(v\circ v)$. The relation $u\stackrel{s}{\circ}[v,v]=v\stackrel{s}{\circ}[u,v]$ fails to accurately reproduce the relation between $\eps_2$ and the alternator, $\sfalt(v_1,\eps_2(v_2,v_3))=\sfalt(\eps_2(v_2,v_3),v_1)$. Moreover, the relation $t(w)\circ t(w)=0$ fails to reproduce the appropriate relation for the alternator, $\sfalt(v_1,\sft(w_1))=\sfalt(\sft(w_1),v_1)=0$.
    
    The original definition in~\cite{Strobl:2019hha} of a (not necessarily symmetric) ``enhanced Leibniz algebra'' is slightly more general, allowing for the operation $\circ$ to be not symmetric. However, this is not very natural, as discussed in \cref{ssec:royten,ssec:EL_infty_and_L_infty}. Moreover, the algebraic structure underlying the tensor hierarchy is an $\ophLie$-algebra, so enhanced Leibniz algebras require axiomatic completion.
    
    \subsection{\texorpdfstring{$\infty$}{Infinity}-Enhanced Leibniz algebras}
    
    A similar notion of extended Leibniz algebras was formulated in~\cite{Bonezzi:2019ygf}, see also~\cite{Bonezzi:2019bek} as well as the previous work on Leibniz algebra gauge theories~\cite{Hohm:2018ybo}.
    \begin{definition}[\cite{Bonezzi:2019ygf}]
        An \uline{\(\infty\)-enhanced Leibniz algebra} is an $\IN$-graded differential complex $(X=\oplus_{i\in\IN} X_i,\rmd)$ with differential of degree~$-1$, endowed with two binary operations 
        \begin{subequations}
            \begin{equation}
                \begin{aligned}
                    \circ&:X_0\otimes X_0\rightarrow X_0~,
                    \\
                    \bullet&:X_i\otimes X_j\rightarrow X_{i+j+1}~,
                \end{aligned}
            \end{equation}
            satisfying the following relations:
            \begin{align}
                (x\circ y)\circ z&=x \circ(y\circ z)-y\circ (x\circ z)~,\label{eq:eL_Leibniz}
                \\
                a\bullet b&=(-1)^{|a|\,|b|}(b\bullet a)~,\label{eq:eL_sym_bullet}
                \\
                (\rmd w)\circ x &= 0~, \label{eq:ielax1}
                \\ 
                \rmd(x\bullet y)&= x\circ y+y\circ x~,\label{eq:ielax2}
                \\
                \rmd(u\bullet v) &= - (\rmd u)\bullet v + (-1)^{|u|+1}u\bullet\rmd v~, \label{eq:ielax5} 
                \\
                (a\bullet b)\bullet c&= 
                (-1)^{|a|+1}a\bullet(b\bullet c)-(-1)^{(|a|+1)|b|} b\bullet(a\bullet c)~, \label{eq:ielax6} 
                \\ 
                \rmd(x\bullet(y\bullet z)) &= (x\circ y)\bullet z + (x\circ z)\bullet y - (y\circ z+z\circ y)\bullet x~, \label{eq:ielax3} 
                \\
                \left[\rmd(x\bullet(y\bullet z))\right]_{x\leftrightarrow y}&=\left[(x\circ y)\bullet u-2x\bullet\rmd(y\bullet u)-x\bullet(y\bullet \rmd u)
                \right]_{x\leftrightarrow y}~, \label{eq:ielax4} 
            \end{align}
        \end{subequations}
        where \(x,y,z\) range over degree~$0$ elements, \(w\) ranges over degree \(1\) elements, \(u,v\) range over positive degree elements, and \(a,b,c\) over arbitrary elements of homogeneous degrees, and where \(
        [\dotsb]_{x\leftrightarrow y}
        \)
        signifies that the enclosed expression is antisymmetrised with respect to the permutation between \(x\) and \(y\).
    \end{definition}
    
    An $\infty$-enhanced Leibniz algebra is a particular type of $\ophLie$-algebra with some axioms missing. Clearly, to compare the axioms, we have to invert the sign of the degree. We thus consider an $\ophLie$-algebra $\frE$ concentrated in non-positive degrees with $\eps_2^0=\circ$ non-trivial only on elements of degree~$0$. Moreover, we are led to identify $\eps_2^1$ with $\bullet$; all other $\eps_2^i$ are trivial. Then we have the following relations between the axioms of an $\infty$-enhanced Leibniz algebra and an $\ophLie$-algebra:
    \begin{itemize}
        \item[\eqref{eq:eL_Leibniz}] is simply the Leibniz identity and follows from the quadratic relation for $\eps_2^0$.
        \item[\eqref{eq:eL_sym_bullet}] amounts to $\eps_2^1$ being graded symmetric and follows from the modified Leibniz rule, as do~\eqref{eq:ielax1}--\eqref{eq:ielax5}.
        \item[\eqref{eq:ielax6}] follows from the $\ophLie$-axiom for $\eps_2^1\circ \eps_2^1$.
        \item[\eqref{eq:ielax3}] follows from the modified Leibniz rule together with the $\ophLie$-axioms for $\eps_2^1\circ \eps_2^0$ and $\eps_2^0\circ \eps_2^1$:
        \begin{equation}
            \begin{aligned}
                \eps_1(\eps_2^1(x,\eps_2^1(y,z))) &=
                \eps_2^0(x,\eps_2^1(y,z))
                +  \eps_2^0(\eps_2^1(y,z),x)
                - \eps_2^1(x,\eps_2^0(y,z)+\eps_2^0(z,y)) \\
                &= \eps_2^1(\eps_2^0(x,y),z) + \eps_2^1(y,\eps_2^0(x,z))
                - \eps_2^1(x,\eps_2^0(y,z)+\eps_2^0(z,y))~,
            \end{aligned}
        \end{equation}
        as does~\eqref{eq:ielax4}: we have:
        \begin{equation}
            \begin{aligned}
                \eps_2^0(\eps_2^1(-,-),-)&=0~,
                \\
                \eps_1(\eps_2^1(x,\eps_2^1(y,u))) &= \eps_2^1(x,\eps_2^1(y,\eps_1(u)))
                + \eps_2^0(x,\eps_2^1(y,u)) - \eps_2^1(x,\eps_2^0(y,u)+\eps_2^0(u,y)) 
                \\
                &= \eps_2^1(x,\eps_2^1(y,\eps_1(u)))
                +\eps_2^1(\eps_2^0(x,y),u) + \eps_2^1(y,\eps_2^0(x,u))
                \\
                &\qquad - \eps_2^1(x,\eps_2^0(y,u)) -\eps_2^1(x,\eps_2^0(u,y))~,
                \\
                \eps_2^1(x,\eps_1(\eps_2^1(y,u))) &= -\eps_2^1(x,\eps_2^1(y,\eps_1(u))) + \eps_2^1(x,\eps_2^0(y,u)+\eps_2^0(u,y))~,
            \end{aligned}
        \end{equation}
        and putting this together, we obtain 
        \begin{equation}
            \begin{aligned}
                &\left[\rmd(\eps_2^1(x,\eps_2^1(y,z)))
                +2\eps_2^1(x,\rmd(\eps_2^1(y,u))
                \right]_{x\leftrightarrow y}
                \\
                &\hspace{3cm}=[
                -\eps_2^1(x,\eps_2^1(y,\rmd u))
                +\eps_2^1(\eps_2^0(x,y),u)
                +\eps_2^1(x,\eps_2^0(u,y))
                ]_{x\leftrightarrow y}~.
            \end{aligned}
        \end{equation}
    \end{itemize}
    Note, however, that while the $\ophLie$-algebra axioms imply the axioms of an $\infty$-enhanced Leibniz algebra, the reverse statement is not true, even for $\infty$-enhanced Leibniz algebras concentrated in degrees $0$ and $1$. The latter essentially implies that $\infty$-enhanced Leibniz algebras are an incomplete abstraction of homotopy Lie algebras and thus do not give the full picture. Altogether, we arrive at the same conclusion as for enhanced Leibniz algebras. 
    
    As a side remark, we note that in the outlook of~\cite{Bonezzi:2019ygf}, the authors mentioned the desire for the interpretation of $\infty$-enhanced Leibniz algebras as the homotopy algebras of some simpler algebraic structure. Our discussion suggests that this is not possible; instead, the axiomatic completion of $\infty$-enhanced Leibniz algebras yields $\ophLie$-algebras whose homotopy algebras form $E_2L_\infty$-algebras, a much weaker version of $L_\infty$-algebras.
    
    \subsection{Algebras producing the tensor hierarchies}\label{ssec:tensor_hierarchy_algebras}
    
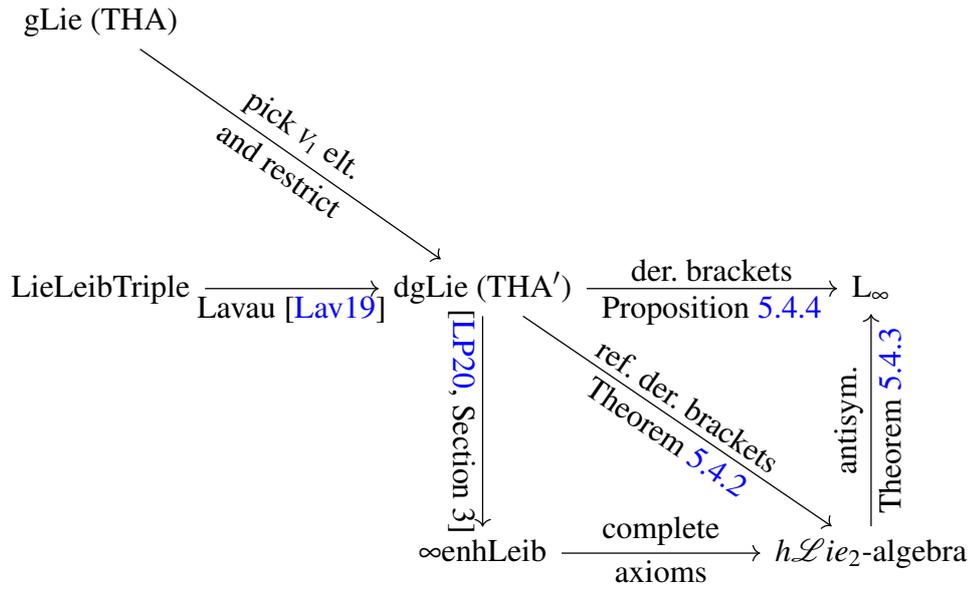
\begin{figure}[h]
        \begin{center}
            \begin{tikzcd}[row sep=2.8cm,column sep=2.3cm]
                \mbox{gLie (THA)} \arrow{rd}[sloped]{\mbox{pick $V_1$ elt.}}[sloped,swap]{\mbox{and restrict}}& 
                \\
                \mbox{LieLeibTriple} \rar[swap,"\mbox{Lavau~\cite{Lavau:2017tvi}}"] & \mbox{dgLie (THA\('\))} \arrow{r}{ \mbox{der.~brackets}}[swap]{\mbox{\Cref{prop:dgLA_to_L_infty}}}
                \arrow{d}[sloped,swap]{\mbox{\cite[Section 3]{Lavau:2019oja}}}
                \arrow{rd}[sloped]{\mbox{ref.~der.~brackets}}[swap,sloped]{\mbox{\Cref{thm:ophLie_from_dgLA}}}
                & \mbox{L\(_\infty\)}\\
                & \mbox{$\infty$enhLeib} \arrow{r}{\mbox{complete}}[swap]{\mbox{axioms}}& \mbox{$\ophLie$-algebra} \arrow{u}[sloped]{ \mbox{antisym.}}[swap,sloped]{\mbox{\Cref{thm:antisym_hLie}}}
            \end{tikzcd}
        \end{center}
        \caption{The relation between the various algebraic structures in the literature and how $\ophLie$-algebras fit into the picture.}\label{fig:alg_diagram}
    \end{figure}

    We now come to larger picture of algebras that lead to the gauge structures visible in the tensor hierarchies, see \cref{fig:alg_diagram}. Note that this picture has only been applied in the context of the tensor hierarchy for maximal supersymmetry. We shall be less detailed in the following.
    
    In~\cite{Palmkvist:2013vya}, Palmkvist constructs an infinite-dimensional $\IZ/2$-graded Lie algebra, which he calls the {\em tensor hierarchy algebra}, ``gLie (THA)'' in \cref{fig:alg_diagram}. For further work on the tensor hierarchy algebra, see also~\cite{Cederwall:2019qnw,Cederwall:2019bai,Cederwall:2021ymp}. As observed in~\cite{Greitz:2013pua}, see also~\cite{Palmkvist:2013vya}, this $\IZ/2$-grading can be naturally refined into a $\IZ$-grading, and picking an element of degree~1 and subsequent restriction induces the structure of a differential graded Lie algebra, ``dgLie (THA\('\))'' in~\cref{fig:alg_diagram}. In~\cite{Lavau:2017tvi}, Lavau called this differential graded Lie algebra the ``tensor hierarchy algebra'' (not to be confused with Palmkvist's larger graded Lie algebra), and derived it from a further algebraic structure called {\em Lie--Leibniz triples}, ``LieLeibTriple'' in~\cref{fig:alg_diagram}. This differential graded Lie algebra then naturally gives rise to $\infty$-enhanced Leibniz algebras, as described in~\cite[Section 3]{Lavau:2019oja}. As explained above, the $\infty$-enhanced Leibniz algebra were an incomplete ``guess'' of the axioms of an $\ophLie$-algebra with $\eps_2^i=0$ for $i\geq 2$. Thus, from our perspective, $\infty$-enhanced Leibniz algebras are appropriately replaced by these, and we then have the construction of the gauge $L_\infty$-algebra via the picture~\eqref{eq:diagram_algebras}, which is refined in \cref{fig:alg_diagram}. We note that the composition of the arrows ``complete axioms'' and ``antisym.'', which produces an $L_\infty$-algebra from an $\infty$-enhanced Leibniz algebra, is found in~\cite[Appendix B]{Bonezzi:2019ygf}. As indicated in~\cref{fig:alg_diagram}, the direct construction of an $L_\infty$-algebra from a differential graded Lie algebra is the Fiorenza--Manetti--Getzler construction~\Cref{prop:dgLA_to_L_infty}, as pointed out in~\cite{Lavau:2019oja}, where Getzler's formulas were specialised to the tensor hierarchy differential graded Lie algebra.
    
    For prior relations amongst tensor hierarchies, the embedding tensor formalism and (homotopy) algebras see also~\cite{Hohm:2018ybo, Kotov:2018vcz}. We again stress that from our point of view, it is not natural to consider gauge theories with infinitesimal symmetries that are not (weaker forms of) Lie algebras. Axiomatically completing the various forms of Leibniz algebras to $\ophLie$-algebras solves this issue.
    
    As a side remark, let us note that the fact that Leibniz algebras naturally produce $L_\infty$-algebras has been pointed out in~\cite{Lavau:2020pwa}. This is \cref{prop:Leib_is_hemistrict_ELinfty} stating that any Leibniz algebra naturally extends to an $\ophLie$-algebra combined with \cref{thm:antisym_hLie} antisymmetrising this $\ophLie$-algebra to an $L_\infty$-algebra.

\chapter{T-duality via Higher Gauge Theory}
\chaptermark{T-duality}
    \section{Introduction}
        Throughout this chapter, we work with affine torus bundles as defined in~\cite{Baraglia:1105.0290}, extending the discussion of~\cite{Waldorf:2022lzs} for principal torus bundles even in the topological case. We argue that this forces us to work with the larger T-duality group\footnote{Here and throughout this chapter, $\IZ_2$ refers to the additive group of integers modulo $2$, not to the 2-adic integers.} $\sfGO(n,n;\IZ)\coloneqq \sfO(n,n;\IZ)\ltimes \IZ_2$. This group appears in the context of T-duality~\cite{Mathai:2004qc,Mathai:2005fd}, and it is a natural part of the automorphism 2-group of $\sfTD_n$~\cite{Nikolaus:2018qop,Waldorf:2022lzs}. 
    
    We begin with an explicit construction of a 2-group of automorphisms $\scGO(n,n;\IZ)$ of $\sfTD_n$ and the corresponding semidirect product 2-group $\scGO(n,n;\IZ)\ltimes \sfTD_n$ in \cref{sec:higher_groups}. The 2-group $\scGO(n,n;\IZ)$ is equivalent as a 2-group to the full automorphism 2-group of $\sfTD_n$ constructed in~\cite{Waldorf:2022lzs}. 
    
    We then give the explicit description of geometric T-dualities in terms of principal 2-bundles in \cref{sec:geometric_t-duality}, extending the topological picture of~\cite{Nikolaus:2018qop} to general torus bundle and providing a differential refinement. We explicitly show how to treat the well-known case of the three-dimensional nilmanifold and how to recover the individual T-dual geometries from the principal 2-bundle data.
    
    This picture is extended in \cref{sec:T-folds} to the case of T-folds. We review the arguments that T-duality is closely related to Kaluza--Klein theory and show how the group $\sfTD_n$ arises naturally from this perspective. We then construct the appropriate Lie 2-groupoid $\scTD_n$ that governs T-dualities between T-folds. An explicit description of the cocycles of principal $\scTD_n$-bundles is given, and we discuss an explicit example of a T-fold in this context. We also show how the half-geometric T-dualities of~\cite{Nikolaus:2018qop} are subsumed in our construction.
    
    The final extension to general T-dualities involving $R$-spaces is then made in \cref{sec:non_geometric}. To complete the picture, we recall that non-geometric fluxes are related to the embedding tensor in supergravity. That allows us to identify the correct representation of the $R$-fluxes, which we then adjoin as $(-1)$-simplices to the simplicial form of the Lie 2-groupoid $\scTD_n$. The result is the augmented Lie 2-quasi-groupoid $\scTD^\text{aug}_n$, and it is not hard to write down the explicit cocycles for the principal $\scTD^\text{aug}_n$-bundles describing general T-dualities. We then use these cocycles to classify the branes arising in toroidal compactifications of string theory. Finally, we also comment on explicit examples of $R$-spaces from this perspective.
    
    We note that principal $\scTD^\text{aug}_n$-bundles naturally truncate to principal $\scTD_n$-bundles describing T-dualities with T-folds and principal $\sfTD_n$-bundles describing geometric T-dualities. Moreover, all constructions are manifestly $\sfGO(n,n;\IZ)$-covariant: the action of the T-duality group is always explicit. In this sense, our approach is similar in spirit to double field theory.
    
    \section{Lightning review: T-duality}
    
    In the following, we collect some basic results about T-duality from the literature; helpful reviews for further reading include~\cite{Giveon:1994fu,Plauschinn:2018wbo}.
    
    \subsection{Topological T-duality}
    
    We start with a brief review of topological T-duality~\cite{Bouwknegt:2003vb,Bouwknegt:2003zg} with an emphasis on the T-correspondences of~\cite{Bunke:2005sn,Bunke:2005um}. 
    
    \paragraph{T-backgrounds.} The low-energy sector of a geometric string theory background, or an $\caN=0$ supergravity background, is given by a smooth Riemannian manifold $(M,g)$ that carries an abelian gerbe $\scG$, whose connective structure provides the Kalb--Ramond field $B$~\cite{Gawedzki:1987ak,Freed:1999vc}.\footnote{Technically, the $\caN=0$ supergravity background also includes the dilaton $\phi$. Its T-duality transformation, however, is trivial: the rescaled combination $\exp(-2\phi)\sqrt{|\det g|}$ remains invariant. We therefore neglect it in this work. It will, however, become important in the extension to U-duality \cite{Borsten:2022ab}.} Recall that abelian gerbes can be described in a geometrically appealing fashion as bundle gerbes~\cite{Murray:9407015,Murray:2007ps} or as central groupoid extensions; here, however, we will be using the equivalent but simpler Hitchin--Chatterjee gerbes~\cite{Hitchin:1999fh,Chatterjee:1998}. For us, a topological abelian gerbe is thus simply a cocycle in Čech cohomology $h_{\rm top}\in \rmH^3(M,\IZ)$. It becomes differentially refined, i.e.~equipped with a connection, if this Čech cocycle is extended to a cocycle in Deligne cohomology $h_{\rmD}\in \rmH^3_{\rmD}(M,\IZ)$. The cohomology class of $h_{\rm top}$ is called the Dixmier--Douady class of the gerbe; if the gerbe carries a connection with 2-form potential $B$, then the image of $[h_D]$ in de~Rham cohomology is the cohomology class of the 3-form curvature $H=\rmd B\in \Omega^3(M,\IZ)$ of the gerbe.
    
    Most commonly, T-duality is defined for string theory backgrounds with a circle or, more generally, a torus action\footnote{or, even more generally, a $\sfGL(n;\IZ)\ltimes\IT^n$-action on certain infinite ($|\sfGL(n;\IZ)|$-fold) families of possibly unoriented $n$-torus bundles, associated to the 2-group $\sfTD^\ltimes_n$; see \cref{sec:geometric_t-duality}} that preserves the metric and the curvature 3-form of the gerbe. We therefore focus on backgrounds containing a number of 1-cycles that are fibered as a torus bundle $M=P$ over a base manifold $X$. Recall that principal torus bundles are always oriented; we want to explicitly permit unoriented affine torus bundles as considered in~\cite{Baraglia:1105.0290}. As an additional geometric datum, there is an abelian gerbe $\scG$ on the total space of this bundle. We call the triple $(X,P,\scG)$ a \emph{topological, geometric (toric) T-background}; if both $P$ and $\scG$ carry connections and $X$ carries a Riemannian structure, we speak of a \emph{differentially refined, geometric (toric) T-background}, cf.~\cite{Bunke:2005um,Nikolaus:2018qop}. Note that a T-background is not necessarily a consistent background of supergravity or string theory.
    
    \paragraph{Classification.} There is now a useful classification of toric T-backgrounds. The Serre spectral sequence associated to the fibration $\pi\colon P\rightarrow X$ defines a filtration
    \[
    \pi^*\rmH^k(X)\eqqcolon F^k \subset F^{k-1} \subset \dotsb \subset F^0 \coloneqq\rmH^k(P)
    \]
    relating the cohomologies of the base $X$ and the total space $P$, cf.~\cite{Bunke:2005um,Baraglia:1105.0290}.
    In particular, the Dixmier--Douady class $h\in\rmH^3(P,\IZ)$ of a gerbe lies within the filtration $F^3\subset F^2\subset F^1\subset F^0$. If the gerbe carries a connection with curvature $H$, then $h$ belongs to $F^i$ if some contractions of $H$ with $3-i$ vector fields along the fiber directions are non-trivial. We say that a toric T-background is \underline{(of type) $F^i$} or simply an \underline{$F^i$-background} if its Čech cocycle lies in $F^i$ but not in $F^{i+1}$.
    
    This classification now allows us to make clear statements about the image of a toric T-background $(X,\check P,\check \scG)$ under T-duality along fiber directions.
    \begin{itemize}\itemsep-2pt
        \item[$F^3$:] The gerbe $\check\scG$ is the pullback of a gerbe on $X$ along $\check \pi\colon\check P\rightarrow X$, and T-duality maps the toric T-background $(X,\check P,\check \scG)$ to itself.
        \item[$F^2$:] As shown in~\cite{Bunke:2005um}, this is the minimum requirement for having a \emph{geometric T-dual}. A \emph{geometric T-duality} relates a geometric toric T-background to another geometric toric T-background, $(X,\check P,\check \scG)\mapsto (X,\hat P,\hat \scG)$, preserving the total dimension of $\check P$ but generically not the topologies of $\check P$ or $\check \scG$, cf.~e.g.~\cite{Bouwknegt:2003vb}. In particular, if $\check P$ is a principal $\IT^n\cong \sfU(1)^n$-bundle, then so is $\hat P$.
        \item[$F^1$:] T-duality along the fibers maps such a toric T-background to a ``non-geometric'' background~\cite{Mathai:2004qc,Mathai:2005fd,Hull:2006qs}. Locally, an $F^1$-background is always $F^2$, and it has local T-duals, which can then be glued together into a T-fold. This was in particular the perspective adopted in~\cite{Nikolaus:2018qop}. Another possible interpretation is to regard certain T-folds as bundles of non-commutative tori, see e.g.~\cite{Mathai:2004qq}.
        \item[$F^0$:] The T-dual of an $F^0$-background is not even locally geometric, and the image is sometimes called an $R$-space and interpreted in terms of non-associative geometry, see~\cite{Bouwknegt:2004ap,Ellwood:2006my,Blumenhagen:2011ph} as well as~\cite{Szabo:2018hhh,Plauschinn:2018wbo} for helpful reviews.
    \end{itemize}
    
    \paragraph{Example: trivial fibration.}
    Let us briefly consider the simple case of a trivial torus bundle $P=X\times \IT^n$ together with a gerbe $\scG$ with curvature 3-form $H$. In this case, the Dixmier--Douady 3-class $h$ and, correspondingly, the curvature $H$ naturally decomposes into four parts:
    \begin{equation}\label{eq:direct_product_cohomology_decomposition}
        \begin{aligned}
            h\in\rmH^3(X\times\IT^n)&\cong\underbrace{\rmH^3(X)\oplus\underbrace{(\rmH^2(X))^{\oplus n}\oplus\underbrace{(\rmH^1(X))^{\oplus\binom n2}\oplus\underbrace{(\rmH^0(X))^{\binom n3}}_{F^3}}_{F^2}}_{F^1}}_{F^0}~,
            \\
            H&=~H^{(3)}~\,+~\,\sum_{i=1}^nH^{(2)}_i~\,\,+~\,\sum_{i,j=1}^nH^{(1)}_{ij}~~\,+~\sum_{i,j,k=1}^nH^{(0)}_{ijk}~.
        \end{aligned}
    \end{equation}
    In this decomposition, the 2-classes $H^{(2)}_1,\dotsc,H^{(2)}_n\in\rmH^2(X)$ dualise, under T-duality, to the Chern classes of a non-trivial torus bundle on $X$, thus to a geometric background.
    The 1-classes $H^{(1)}_{ij}$ correspond under T-duality to $Q$-fluxes. That is, the T-dual is formed by starting with a geometric universal cover and taking a possibly non-geometric quotient given by $\sfO(n,n;\IZ)$-transformations encoded by $H^{(1)}_{ij}$ to form T-folds.
    Finally, the 0-classes $H^{(0)}_{ijk}$ correspond to $R$-fluxes, which encode the degree to which even local geometry fails to exist.
    
    \paragraph{Topological geometric T-duality.} Let us consider the case of geometric T-duality in more detail and focus on the purely topological aspect. Topological T-duality~\cite{Bouwknegt:2003vb,Bouwknegt:2003zg} is based on the existence of the \emph{Gysin sequence}~\cite{Gysin:1941:61-122}, see also~\cite[Prop.~14.33]{Bott:1982aa}. Given a principal \(\sfU(1)\)-bundle $\:\check P\rightarrow X$ with first Chern class $\check F\in \rmH^2(X,\IZ)$, the following sequence is exact:
    \begin{equation}\label{eq:Gysin_sequence}
        \ldots~\xrightarrow{~~~}~\rmH^k(X,\IZ)~\xrightarrow{~\check \pi^*~}~\rmH^k(\check P,\IZ)~\xrightarrow{~\check \pi_*~}~\rmH^{k-1}(X,\IZ)~\xrightarrow{~\check F\,\smile~}~\rmH^{k+1}(X,\IZ)~\xrightarrow{~~~}~\ldots
    \end{equation}
    For topological T-duality, we are interested in this segment for $k=3$. Any 3-form $\check H\in \rmH^3(P,\IZ)$ comes with an associated element $\hat F\coloneqq \check\pi_* \check H\in \rmH^2(X,\IZ)$ with $\check F\smile \hat F=0$ in $\rmH^4(X,\IZ)$. We can now consider a second torus bundle $\hat \pi\colon\hat P\rightarrow X$ with first Chern class $\hat F$. Because $\check F\smile \hat F=\hat F\smile \check F=0$, exactness of the Gysin sequence with hatted maps now shows that there is an $\hat H$ such that $\hat\pi_*\hat H=\check F$. \emph{Topological geometric T-duality} is then the transition from $(\check P,\check H)$ to $(\hat P,\hat H)$. As shown in~\cite{Bouwknegt:2003vb}, this construction matches the various expectations from string theory considerations. It also extends to the case of affine torus bundles, and there is a corresponding Gysin sequence~\cite{Baraglia:1105.0290}.
    
    \paragraph{T-duality correspondence.} We can arrive at a more geometric picture if we include the correspondence space $\check P\times_X \hat P$ and regard $\check H$ and $\hat H$ as the Dixmier--Douady classes of some bundle gerbes $\check \scG$ and $\hat \scG$, respectively. This then leads to the commutative diagram
    \begin{equation}\label{eq:T-duality_correspondence}
        \begin{tikzcd}[column sep=1cm, row sep=0.8cm]
            & & \scG_\rmC=\check\sfp^*\check\scG\otimes \hat \sfp^*\hat \scG^{-1} \arrow[d]& & \\
            & & \arrow[ld,"\check \sfp",swap] \check P\times_X\hat P \arrow[rd,"\hat \sfp"]& & \\
            \check\scG \arrow[r] & \check P \arrow[rd,"\check \pi"] & & \hat P \arrow[ld,"\hat \pi",swap] & \hat \scG \arrow[l]\\
            & & X & &             
        \end{tikzcd}
    \end{equation}
    which appears crucially in the definition of topological T-duality in terms of T-duality triples~\cite{Bunke:2005um}. Such a T-duality triple is given by the data $((\check P,\check H),(\hat P,\hat H), u)$, where $u$ is a trivialisation of the gerbe~$\scG_\rmC$, relating it to the Poincar\'e bundle (or the higher-dimensional generalisation thereof) over the correspondence space, cf.~also~\cite[Rem.~6.3]{Fiorenza:2016oki} for a string theoretic interpretation. 
    
    \subsection{Differential refinement of topological T-duality}\label{ssec:review_diff_refine}
    
    In order to describe a geometric T-background $(X,P,\scG)$ completely, we need to provide a Riemannian metric on $X$ and connections on $P$ and $\scG$.
    
    \paragraph{Principal $G$-connection and Kaluza--Klein metric.} We can describe the connection on the principal $\sfU(1)^n$-bundle $P$ as a principal $G$-connection $\theta$. Recall that such a connection is a $\frt^n\coloneqq \sfLie(\sfU(1)^n)\cong \IR^n$-valued $1$-form $\theta\in\Omega^1(P,\frt^n)$ such that, for any fundamental vector field $X_\xi\in \Gamma(TX)$ of $\xi\in \frt^n$, the $1$-form $\theta$ is equivariant, $\caL_{X_\xi}\theta=0$, and reproduces $\xi$ in the sense that $\iota_{X_\xi}\theta=\xi$.
    
    Together with a Riemannian metric $g$ on $X$, the connection $\theta$ induces the \emph{Kaluza--Klein metric} $\tilde g$ on $P$ defined by
    \begin{equation}\label{eq:KK_metric}
        \tilde g\coloneqq \pi^*g+\theta^i\otimes \theta^i~.
    \end{equation}
    In the case of an affine torus bundle, we use the connection on the corresponding principal $(\sfGL(n;\IZ)\ltimes \sfU(1)^n)$-bundle, which corresponds to locally defined $\fru(1)^n$-valued vector fields defined up to invertible integer linear transformations. This ambiguity, however, drops out of~\eqref{eq:KK_metric}.
    
    \paragraph{The group $\sfO(n,n;\IZ)$.} T-duality is often presented as an involution given by a $\IZ_2$-action. On a string background, this action maps the radius of the involved circle direction $R$ to the inverse radius\footnote{We put $\alpha'=1$.} $\frac{1}{R}$ and interchanges the momentum and the winding modes of the string. There is an additional freedom of reversing the sign in the latter interchange so that the full T-duality group for T-duality along a circle direction should be identified with $\IZ_2\times \IZ_2\cong\sfO(1,1;\IZ)$.
    
    For an $n$-torus $\IT^n$, this group is enlarged to the group $\sfO(n,n;\IZ)$, see~\cite{Giveon:1988tt,Shapere:1988zv}. Elements $g$ of $\sfO(n,n;\IZ)$ are $2n\times2n$ integer matrices that leave the form 
    \begin{equation}\label{eq:Onn-metric}
        \eta\coloneqq \begin{pmatrix} 0 & \unit_n \\ \unit_n & 0\end{pmatrix}
    \end{equation}
    invariant in the sense that $g^\rmT \eta g=\eta$, which in components becomes 
    \begin{equation}\label{eq:g-parameterisation}
        \begin{gathered}
            g=\begin{pmatrix} A & B \\ C & D \end{pmatrix}~,~~~A,B,C,D\in \sfMat(n;\IZ)~,
            \\
            A^\rmT C+C^\rmT A=B^\rmT D+D^\rmT B=0~,~~~A^\rmT D+C^\rmT B=\unit_n~.
        \end{gathered}
    \end{equation}
    
    The group $\sfO(n,n;\IZ)$ is a subgroup of the larger group $\sfG\sfO(n,n;\IZ)\coloneqq \sfO(n,n;\IZ)\rtimes \IZ_2$ originally defined in~\cite{Mathai:2004qc,Mathai:2005fd}, which becomes relevant for T-duality with general torus bundles. For $n>0$, this group can be identified with the $2n\times2n$ integer matrices that leave $\eta$ invariant up to sign in the sense that $g^\rmT \eta g=\pm\eta$, which in components becomes
    \begin{equation}\label{eq:def_GO}
        A^\rmT C+C^\rmT A=B^\rmT D+D^\rmT B=0~,~~~A^\rmT D+C^\rmT B=\pm\unit_n~.
    \end{equation}
    For $n=0$, we have $\sfG\sfO(0,0;\IZ)\cong\IZ_2$. For future convenience, we introduce the indicator function $|-|\colon\sfG\sfO(n,n;\IZ)\rightarrow \{0,1\}$, which is simply the projection onto the $\IZ_2$ component. In particular, $(-1)^{|g|}=+1$ for all $g\in \sfO(n,n;\IZ)$.
    
    \paragraph{Subgroups of $\sfGO(n,n;\IZ)$.} It is convenient to introduce the following subgroups of the T-duality group $\sfGO(n,n;\IZ)$, which together generate the entirety of $\sfGO(n,n;\IZ)$, cf.~\cite{Giveon:1994fu}:
    \begin{itemize}
        \item[$A$)] The subgroup $\sfGL(n;\IZ)\subset \sfO(n,n;\IZ)$ of \emph{$A$-transformations} consists of group elements
        \begin{equation}\label{eq:trafos_A}
            g_A=\begin{pmatrix} A & 0 \\ 0 & (A^\rmT )^{-1} \end{pmatrix}~~~\mbox{with}~~~A\in \sfGL(n;\IZ)~.
        \end{equation}
        These transformations are simply the automorphism $\sfAut(\IT^n)\cong \sfGL(n;\IZ)$ of the $n$-dimensional torus $\IT^n$ forming the fibers of the torus bundle, and it is therefore also sometimes called the \emph{geometric (sub)group}.
        \item[$B$)] The abelian torsion-free subgroup $\fro(n;\IZ)\subset \sfO(n,n;\IZ)$ of \emph{$B$-transformations} consists of group elements
        \begin{equation}\label{eq:trafos_B}
            g_B=\begin{pmatrix} \unit_n & B  \\ 0 & \unit_n \end{pmatrix}~~~\mbox{with}~~~B\in\{A\in \sfMat(n;\IZ)~|~A^\rmT =-A\}~.
        \end{equation}
        We note that if we tensor this subgroup with functions along the $n$-torus, then certain $B$-transformations are naturally identified with the 2-form $\rmd \Lambda$ for a 1-form $\Lambda$ along the torus direction. The corresponding $B$-transformations then describe gauge transformations, as familiar from the Courant algebroid description.
        \item[$\beta$)] The abelian torsion-free subgroup $\fro(n;\IZ)\subset \sfO(n,n;\IZ)$ of \emph{$\beta$-transformations} consists of group elements
        \begin{equation}\label{eq:trafos_beta}
            g_B=\begin{pmatrix} \unit_n & 0 \\ \beta & \unit_n \end{pmatrix}~~~\mbox{with}~~~\beta\in\{A\in \sfMat(n;\IZ)~|~A^\rmT =-A\}~.
        \end{equation}
        \item[$T_k$)] The abelian torsion subgroup $\sfO(1,1,\IZ)^{n}\cong (\IZ_2)^{2n}\subset \sfO(n,n;\IZ)$ of the \emph{factorised dualities} is generated by group elements
        \begin{equation}\label{eq:trafos_fac_dual}
            g^\pm_{T_k}=\begin{pmatrix} \unit_n-1_k & \pm 1_k \\ \pm 1_k & \unit_n-1_k \end{pmatrix}~~~\mbox{with}~~~1_k=\rmdiag(\underbrace{0,\ldots,0}_{k-1},1,\underbrace{0,\ldots,0}_{n-k-1})~.
        \end{equation}
        These transformations can be identified with the involutions that are T-dualities along the $k$th circle direction. Clearly, $(g^\pm_{T_k})^2=\unit_{2n}$.
        \item[$G$)] The abelian torsion subgroup $\IZ_2\times\IZ_2\subset\sfGO(n,n;\IZ)$ consists of group elements
        \begin{equation}
            g_G^{s_1,s_2}=\begin{pmatrix}
                s_1\unit_n & 0 \\
                0 & s_2\unit_n
            \end{pmatrix}~~~\mbox{with}~~~s_1,s_2\in\{\pm1\}~.
        \end{equation}
        Note that $(-1)^{|g_G^{s_1,s_2}|}=s_1s_2$. In other words, when $s_1s_2=-1$, then $g_G^{s_1,s_2}\not\in\sfO(n,n;\IZ)$.
    \end{itemize}
    
    \paragraph{$\sfGO(n,n;\IZ)$ versus $\sfO(n,n;\IZ)$.} Throughout this chapter, we will work with the larger T-duality group $\sfGO(n,n;\IZ)$ instead of $\sfO(n,n;\IZ)$. The difference between the two groups is that the former contains the additional generator $g_G^{-+}=\rmdiag(-\unit_n,\unit_n)$. We will see in our later discussion that this group element flips the sign of the Kalb--Ramond field along the fiber directions of $\check P\times_X\hat P\rightarrow X$. Identifying gerbes of opposite orientation becomes a necessity because working with general torus bundles implies that we also identify principal bundles with opposite orientation. As an example, consider a principal $\sfU(1)$-bundle $P$ regarded as a general circle bundle, i.e.~a principal $\sfO(2)$-bundle. A constant coboundary equal to the additional $\IZ_2$-factor in $\sfO(2)$ over $\sfU(1)\cong \sfSO(2)$ now flips the orientation of $P$, rendering $P$ and its dual isomorphic. It is well-known that T-duality can interchange the topological invariant of the torus bundle with the topological invariant of the gerbe. Thus, working with general torus bundles implies that we have to enlarge the T-duality group from $\sfO(n,n;\IZ)$ to $\sfGO(n,n;\IZ)$. For further discussion, see also~\cite{Mathai:2004qc,Mathai:2005fd}.
    
    \paragraph{Non-geometric backgrounds.} If local descriptions of a T-background are glued together with elements of the geometric subgroup of the T-duality group, then we have a geometric T-background. T-folds\footnote{also called monodrofolds in~\cite{Hellerman:2002ax}; see also references therein for earlier suggestions for extending the monodromy group}~\cite{Hull:2004in,Hull:2006va,Belov:2007qj} are T-backgrounds, most importantly torus fibrations with $B$-field, that are locally geometric, but whose local descriptions are glued together by general elements of the T-duality group, i.e.~$\sfGO(n,n;\IZ)$ for torus fibrations. Therefore, T-folds always have a global double geometry~\cite{Hull:2006va}. Because some T-folds arise as T-duals of geometric T-backgrounds, it is clear that string theory is well-defined on these and, in particular, that they have to be included in the space of possible string backgrounds. There is, in fact, a constrained sigma model description of certain T-fold backgrounds~\cite{Hull:2004in}. We also note that a higher geometric local model of T-folds was given in~\cite{Fiorenza:2016oki}.
    
    A further generalisation are the so-called $R$-spaces \cite{Shelton:2005cf,Wecht:2007wu} which do not even locally admit a geometric description.
    
    \paragraph{Vanishing Dixmier--Douady class.} In the case in which the Dixmier--Douady class of the gerbe vanishes, and, as a consequence, the Kalb--Ramond $B$-field is globally defined, we can combine it with the metric $g$ into the tensor $\caE\coloneqq g+B$; under an element $g\in \sfGO(n,n;\IZ)$ of the form~\eqref{eq:g-parameterisation}, $\caE$ transforms in the Möbius-like, non-linear fashion
    \begin{equation}
        \tilde \caE=g\acton \caE\coloneqq \frac{A \caE+B}{C \caE+D}~.
    \end{equation}
    For the factorised dualities $T_k$, this transformation reproduces the Buscher rules~\cite{Buscher:1987sk,Buscher:1987qj} for the transformations of the metric and $B$-field along a circle direction.
    
    In order to render the above transformation linear, we can switch to the \emph{generalised metric}~\cite{Shapere:1988zv,Giveon:1988tt,Maharana:1992my,Gualtieri:2003dx}
    \begin{equation}
        \caH\coloneqq \begin{pmatrix}
            g-Bg^{-1} B & B g^{-1}
            \\
            -g^{-1} B & g^{-1}
        \end{pmatrix}~,
    \end{equation}
    satisfying $\caH^{-1}=\eta \caH \eta$. We then have a simple adjoint action of $\sfGO(n,n;\IZ)$ on $\caH$,
    \begin{equation}
        \tilde \caH=g\acton \caH\coloneqq g\caH g^\rmT ~.
    \end{equation}
    In particular, the generalised metric is obtained as follows from $B$-transformations:
    \begin{equation}
        \caH=\begin{pmatrix} \unit_n & B  \\ 0 & \unit_n \end{pmatrix}\begin{pmatrix} g & 0  \\ 0 & g^{-1}\end{pmatrix}\begin{pmatrix} \unit_n & B  \\ 0 & \unit_n \end{pmatrix}^\rmT ~.
    \end{equation}
    
    \paragraph{Non-trivial Dixmier--Douady class.} In the case in which the $B$-field is not globally defined, we still have the group $\sfGO(n,n;\IZ)$ as the relevant group of T-dualities. Its action on the data making up the differential refinement, however, is more complicated. In particular, the combination $\caE\coloneqq g+B$ can exist only locally (for T-folds) or not exist at all (for $R$-spaces). For further details, see e.g.~\cite{Hull:2006qs,Belov:2007qj}.
    
    \section{Higher geometric groups for T-duality}\label{sec:higher_groups}
    
    \subsection{The Lie 2-group \texorpdfstring{$\underline{\sfTD}_n$}{TDn}}\label{ssec:Onn_subgroup_and_actions}
    
    \paragraph{Generalities.} Recall that T-duality correspondences exist for toric T-backgrounds of type $F^2$. As shown in~\cite{Nikolaus:2018qop}, there is a strict Lie 2-group $\sfTB^\text{F2}_n$ that represents such backgrounds. In other words, we can regard a toric T-background as a principal 2-bundle, i.e.~a higher or categorified principal bundle with $\sfTB^\text{F2}_n$ as its structure 2-group. The interesting observation of~\cite{Nikolaus:2018qop} is then that not only the T-backgrounds but also the correspondence space and the gerbe $\scG_\rmC$ over it can be replaced by a principal 2-bundle. A T-duality correspondence then amounts to a double fibration or span of principal 2-bundles which is induced by an underlying span of Lie 2-groups. It is clear that the principal 2-bundle taking over the role of the correspondence space should describe the bundle $\check P\times_X \hat P\rightarrow X$, so the structure group should contain the abelian group $\sfU(1)^{2n}$. As observed in~\cite{Nikolaus:2018qop}, this group needs to be extended to a categorical torus, see~\cite{Ganter:2014zoa}, denoted by $\sfTD_n$.
    
    We will mostly use crossed modules of Lie groups in order to describe Lie 2-groups. Some background material and further pointers to the literature on higher groups and bundles are found in \cref{app:higher_groups} and \cref{app:higher_principal_bundles}.
    
    \paragraph{The 2-group $\underline{\sfTD}_n$.} We regard the abelian group $\sfU(1)^{2n}$ as the quotient $\IR^{2n}/\IZ^{2n}$ and extend the corresponding action groupoid by a factor of $\sfU(1)$, leading to the Lie groupoid
    \begin{subequations}\label{eq:TD_n_2-group}
        \begin{equation}
            \begin{gathered}
                \begin{tikzcd}
                    \IR^{2n}\times \IZ^{2n}\times \sfU(1)\arrow[r,shift left] 
                    \arrow[r,shift right] & \IR^{2n}
                \end{tikzcd}~,~
                \begin{tikzcd}[column sep=1.5cm,row sep=large]
                    \phantom{\sft(h)} \xi & \xi-m_1\arrow[l,bend left,swap,out=-20,in=200]{}{(\xi,m_1,\phi_1)}& \xi-m_1-m_2\arrow[l,bend left,swap,out=-20,in=200]{}{(\xi-m_1,m_2,\phi_2)}\arrow[ll,bend right,out=20,in=-200]{}{(\xi,m_1+m_2,\phi_1+\phi_2)}
                \end{tikzcd},
                \\
                \sfid_\xi\coloneqq (\xi,0,0)~,~~~(\xi,m,\phi)^{-1}\coloneqq (\xi-m,-m,-\phi)~,
            \end{gathered}
        \end{equation}    
        which becomes the (strict) Lie 2-group\footnote{See \cref{app:higher_groups} for a review of higher groups and a summary of our conventions.}  $\underline{\sfTD}_n$ together with the monoidal structure and inverse functor defined by
        \begin{equation}
            \begin{aligned}
                (\xi_1,m_1,\phi_1)\otimes (\xi_2,m_2,\phi_2)&\coloneqq(\xi_1+\xi_2,m_1+m_2,\phi_1+\phi_2-\langle \xi_1,m_2\rangle)~,
                \\
                \sfinv(\xi,m,\phi)&\coloneqq (-\xi,-m,-\phi-\langle \xi,m\rangle)
            \end{aligned}
        \end{equation}    
    \end{subequations}    
    for $\xi,\xi_{1,2}\in \IR^{2n}$, $m,m_{1,2}\in \IZ^{2n}$ and $\phi,\phi_{1,2}\in \sfU(1)$. We will always use additive notation for elements in $\IR/\IZ\cong \sfU(1)$. The binary bracket $\langle-,-\rangle$ is defined as 
    \begin{equation}\label{eq:FOOBARBAZ}
        \langle \xi_1,\xi_2\rangle=\xi^\rmT _1
        \begin{pmatrix} 
            0 & 0 \\ 
            \unit_n &0 
        \end{pmatrix}
        \xi_2
        ~~~\mbox{or}~~~
        \left\langle
        \begin{pmatrix}
            \hat \xi_1
            \\
            \check \xi_1
        \end{pmatrix},
        \begin{pmatrix}
            \hat \xi_2
            \\
            \check \xi_2
        \end{pmatrix}\right\rangle=\check \xi_1 \hat \xi_2
    \end{equation}
    for all $\xi_{1,2}\in \IR^{2n}$. This Lie 2-group corresponds to a crossed module of Lie groups,
    \begin{equation}\label{eq:def_TD_n}
        \begin{gathered}
            \sfTD_n~\coloneqq~\big(\IZ^{2n}\times\sfU(1) \xrightarrow{~\sft~}\IR^{2n}\big)~,
            \\
            \sft(m,\phi)\coloneqq m~,
            \\
            \xi\acton(m,\phi)\coloneqq (m,\phi-\langle \xi,m\rangle)
        \end{gathered}
    \end{equation}
    with the group products in $\IZ^{2n}\times\sfU(1)$ and $\IR^{2n}$ abelian and evident.
    
    \paragraph{Lie 2-algebra of $\underline{\sfTD}_n$.} We will also need the infinitesimal description of $\underline{\sfTD}_n$ in terms of the Lie 2-algebra $\frtd_n$. The latter is given as the crossed module of Lie algebras
    \begin{equation}\label{eq:def_TD_n_alg}
        \begin{gathered}
            \frtd_n~=~\big(\IR\xrightarrow{~\sft~}\IR^{2n}\big)~,
            \\
            \sft(y)=0~,~~~
            \xi\acton y=0
        \end{gathered}
    \end{equation}
    for $y\in \IR$ and $\xi\in \IR^{2n}$. Weak morphisms of Lie 2-groups describing automorphisms of $\underline{\sfTD}_n$ then translate to invertible Lie 2-algebra morphisms $\phi\colon\frtd_n\rightarrow \frtd_n$ as defined in \cref{app:higher_Lie_algebras}. 
    
    \subsection{Automorphisms of \texorpdfstring{$\underline{\sfTD}_n$}{TDn}}
    
    In~\cite{Nikolaus:2018qop}, the authors announced the result that the group of isomorphism classes of objects $\underline{\pi_0}(\sfAut(\underline{\sfTD}_n))$ in the 2-group of automorphisms given by crossed intertwiners $\sfAut(\underline{\sfTD}_n)$ is isomorphic to the group $\sfG\sfO(n,n;\IZ)$ defined in~\eqref{eq:def_GO}; the proof was given recently in~\cite{Waldorf:2022lzs}. In the following, we identify a subset of these automorphisms that will be suitable for our purposes.
    
    \paragraph{Automorphism 2-functors.} Weak morphisms $\Phi\colon\underline{\sfTD}_n\rightarrow \underline{\sfTD}_n$ can be equivalently regarded as weak 2-functors between the corresponding one-object 2-groupoids $\Phi\colon\sfB\underline{\sfTD}_n\rightarrow \sfB\underline{\sfTD}_n$ as defined in~\cref{app:2-groupoid_basics}. Such a 2-functor consists of a functor $\Phi_1\colon\underline{\sfTD}_n\rightarrow \underline{\sfTD}_n$ and a natural transformation given by a map $\Phi_2\colon\IR^{2n}\times \IR^{2n}\rightarrow \IR^{2n}\times\IZ^{2n}\times \sfU(1)$ that satisfies the naturality and coherence conditions listed in~\eqref{eq:weak_morphism}. We start from the most general ansatz 
    \begin{equation}
        \begin{aligned}
            \Phi_1(\xi,m,\phi)&=(\Phi_1^0(\xi,m,\phi),\Phi_1^1(\xi,m,\phi),\Phi_1^2(\xi,m,\phi))~,
            \\
            \Phi_2(\xi_1,\xi_2)&=(\Phi_2^0(\xi_1,\xi_2),\Phi_2^1(\xi_1,\xi_2),\Phi_2^2(\xi_1,\xi_2))~,
        \end{aligned}
    \end{equation}
    where all components are evident smooth maps. The naturality condition reads as
    \begin{equation}\label{eq:TD_naturality}
        \begin{aligned}
            \Phi_2(\xi_1,\xi_2)\circ &(\Phi_1(\xi_1,m_1,\phi_1)\tildeotimes \Phi_1(\xi_2,m_2,\phi_2))
            \\
            &=\Phi_1(\xi_1+\xi_2,m_1+m_2,\phi_1+\phi_2-\langle \xi_1,m_2\rangle)\circ \Phi_2(\xi_1-m_1,\xi_2-m_2)~,
        \end{aligned}
    \end{equation}
    and the coherence condition is
    \begin{multline}\label{eq:TD_coherence}
            \Phi_2(\xi_1+\xi_2,\xi_3)\tilde \circ(\Phi^0_2(\xi_1,\xi_2)+\Phi_1^0(\xi_3),\Phi^1_2(\xi_1,\xi_2),\Phi^2_2(\xi_1,\xi_2))
            \\
            =
            \Phi_2(\xi_1,\xi_2+\xi_3)\tilde \circ (\Phi_1^0(\xi_1)+\Phi_2^0(\xi_2,\xi_3),\Phi_2^1(\xi_2,\xi_3),\Phi_2^2(\xi_2,\xi_3)\\-\langle \Phi_1^0(\xi_1),\Phi_2^1(\xi_2,\xi_3)\rangle).
    \end{multline}
    Because $\Phi_1$ is a functor, we have 
    \begin{equation}
        \begin{aligned}
            \Phi_1^0(\xi,m,\phi)&=\Phi_1^0(\xi)~,
            \\
            \Phi_1^1(\xi,m,\phi)&=\Phi_1^0(\xi)-\Phi_1^0(\xi-m)~,
            \\
            \Phi_1^2(\xi,0,0)&=0~,
            \\
            \Phi_1^2(\xi,m_1+m_2,\phi_1+\phi_2)&=\Phi^2_1(\xi,m_1,\phi_1)+\Phi^2_1(\xi-m_1,m_2,\phi_2)~.
        \end{aligned}
    \end{equation}    
    Applying the target map to both sides of~\eqref{eq:TD_naturality} implies that $\Phi_2^0(\xi_1,\xi_2)=\Phi_1^0(\xi_1+\xi_2)$. The composition on the left-hand side of~\eqref{eq:TD_naturality} implies that $\Phi^1_2$ measures the failure of $\Phi_1^0$ to be additive:
    \begin{equation}
        \Phi^1_2(\xi_1,\xi_2)=\Phi_1^0(\xi_1+\xi_2)-\Phi_1^0(\xi_1)-\Phi_1^0(\xi_2)~.
    \end{equation}
    The other composition is automatically satisfied, and the sources of both sides of~\eqref{eq:TD_naturality} match. 
    
    We now restrict ourselves to a particular class of morphisms in which $\Phi_1^0$ is a group isomorphism on objects. This amounts to
    \begin{equation}
        \Phi_1^0(\xi)=g\xi~,~~~\Phi_1^1(\xi,m,\phi)=gm~,~~~\Phi_2^1=0
    \end{equation}
    for $g\in \sfGL(2n;\IZ)$, and we directly restrict $g$ further to be an element in $\sfGO(n,n;\IZ)$. We note that the restriction to crossed intertwiners in~\cite{Nikolaus:2018qop} certainly implies this restriction. The coherence condition then reduces to
    \begin{equation}
        \Phi_2^2(\xi_1,\xi_2)+\Phi_2^2(\xi_1+\xi_2,\xi_3)=\Phi_2^2(\xi_1,\xi_2+\xi_3)+\Phi_2^2(\xi_2,\xi_3)~,
    \end{equation}
    which implies 
    \begin{equation}
        \Phi_2^2(\xi_1,0)=\Phi_2^2(0,\xi_2)=\Phi_2^2(0,0)~.
    \end{equation}
    Also, the naturality condition~\eqref{eq:TD_naturality} for $\xi_1=m_1=\phi_2=0$ reduces to
    \begin{equation}
        \Phi_1^2(\xi_2,m_2,\phi_1)=\Phi_1^2(\xi_2,m_2,0)+\Phi_1^2(0,0,\phi_1)~,
    \end{equation}
    allowing us to split $\Phi_1^2$ into two components,
    \begin{equation}
        \Phi_1^2(\xi_2,m_2,\phi_1)=\Phi_1^{21}(\xi_2,m_2)+\Phi_1^{22}(\phi_1)
    \end{equation}
    with $\Phi_1^{21}(0,0)=0$. Naturality for $\xi_1=m_1=0$ then implies linearity of $\Phi_1^{22}$.
    
    Further following~\cite{Nikolaus:2018qop}, we restrict to morphisms with $\Phi_1^{21}(\xi,m)=\Phi_2^2(m,\xi)$ and assume $\Phi_2^2$ to be bilinear. This now completely solves the coherence relation~\eqref{eq:TD_coherence}. We further set $\Phi_1^2(\phi)=(-1)^{|g|}\phi$, reducing the naturality condition~\eqref{eq:TD_naturality} to
    \begin{equation}
        \Phi_2^2(\xi_1,m_2)-\Phi_2^2(m_2,\xi_1)
        =\langle g \xi_1,g m_2\rangle-(-1)^{|g|}\langle \xi_1,m_2\rangle~.
    \end{equation}
    For an element $g$ parameterised as in~\eqref{eq:g-parameterisation}, we have
    \begin{equation}
        \begin{aligned}
            \Phi_2^2(\xi,m)-\Phi^2_2(m,\xi)&=\xi^\rmT 
            \begin{pmatrix}
                C^\rmT A & C^\rmT  B 
                \\
                D^\rmT A -(-1)^{|g|}\unit_n & D^\rmT  B
            \end{pmatrix}m
            \\
            &=\xi^\rmT 
            \begin{pmatrix}
                C^\rmT A & C^\rmT  B 
                \\
                (-1)^{|g|}\unit_n-(-1)^{|g|}\unit_n-B^\rmT C & D^\rmT  B
            \end{pmatrix}m~.
        \end{aligned}
    \end{equation}
    For convenience, we define the antisymmetric matrix
    \begin{equation}
        \begin{aligned}
            \rho(g)&\coloneqq 
            g^\rmT \begin{pmatrix} 0_n & 0_n \\ \unit_n & 0 \end{pmatrix}g
            -(-1)^{|g|}\begin{pmatrix} 0_n & 0_n \\ \unit_n & 0 \end{pmatrix}
            \\
            &=\begin{pmatrix}
                C^\rmT A & C^\rmT  B 
                \\
                D^\rmT A -(-1)^{|g|}\unit_n & D^\rmT  B
            \end{pmatrix}=
            \begin{pmatrix}
                C^\rmT A & C^\rmT  B 
                \\
                -B^\rmT C & D^\rmT  B
            \end{pmatrix}
            \\
            &=-\rho^\rmT (g)~.
        \end{aligned}
    \end{equation}
    We further decompose this matrix into its lower triangular part $\rho_L(g)$ and its transpose,
    \begin{equation}\label{eq:def_rho_L}
        \rho(g)=\rho_L(g)-\rho_L(g)^\rmT ~.
    \end{equation}    
    We then have 
    \begin{equation}
        \Phi_2^2(\xi_1,\xi_2)\coloneqq \xi_1^\rmT \big(\rho_L(g)+\zeta\big) \xi_2
    \end{equation}
    for $\zeta$ an element of $\sfSym(2n;\IZ)$, the additive group of $2n\times 2n$-dimensional symmetric matrices. Altogether, we have identified a subset of automorphisms of $\underline{\sfTD}_n$ which is parameterised by $\sfGO(n,n;\IZ)\times \sfSym(2n;\IZ)$ according to
    \begin{equation}\label{eq:automorphisms_TD_n}
        \begin{aligned}
            \Phi^{g,\zeta}_1(\xi,m,\phi)&=\big(g\xi,gm,(-1)^{|g|}\phi+m^\rmT (\rho_L(g)+\zeta)\xi\big)~,
            \\
            \Phi^{g,\zeta}_2(\xi_1,\xi_2)&=(g(\xi_1+\xi_2),0,\xi_1^\rmT (\rho_L(g)+\zeta)\xi_2)~.
        \end{aligned}
    \end{equation}
    
    \paragraph{The group $\sfGO(n,n;\IZ)\times \sfSym(2n;\IZ)$.} We note that
    \begin{equation}
        g^\rmT _2\rho(g_1)g_2+(-1)^{|g_1|}\rho(g_2)=\rho(g_1g_2)~,
    \end{equation}
    and we measure the failure of the same relation to hold for the lower triangular part by the symmetric integer-valued matrix
    \begin{equation}\label{eq:def_sigma_L}
        \sigma_L(g_1,g_2)\coloneqq g^\rmT _2\rho_L(g_1)g_2+(-1)^{|g_1|}\rho_L(g_2)-\rho_L(g_1g_2)\in \sfSym(2n;\IZ)~.
    \end{equation}
    This matrix-valued function satisfies 
    \begin{equation}\label{eq:sigma_relation}
        -g_3^\rmT \sigma_L(g_1,g_2)g_3+\sigma_L(g_1,g_2g_3)+(-1)^{|g_1|}\sigma_L(g_2,g_3)-\sigma_L(g_1g_2,g_3)=0~.
    \end{equation}
    
    Composition of the automorphisms~\eqref{eq:automorphisms_TD_n} now induces a group structure on the space $\sfGO(n,n;\IZ)\times \sfSym(2n;\IZ)$ with the (associative) product given by 
    \begin{equation}\label{eq:auto_horiz_comp}
        \begin{aligned}
            (g_1,\zeta_1)\times(g_2,\zeta_2)=\big(g_1g_2,g_2^\rmT \zeta_1g_2+(-1)^{|g_1|}\zeta_2+\sigma_L(g_1,g_2)\big)~.
        \end{aligned}
    \end{equation}    
    One can lift this group to a 2-group by adding natural 2-transformations to these weak 2-endofunctors. For the purposes of this thesis, however, we can work with a smaller 2-group acting on $\underline{\sfTD}_n$.
    
    \subsection{2-group action on \texorpdfstring{$\underline{\sfTD}_n$}{TDn}}\label{ssec:2-group_action}
    
    \paragraph{The 2-group $\scGO(n,n;\IZ)$.} The action of a 2-group on a 2-group is readily defined, see \cref{app:higher_groups}. An action of $\sfGO(n,n;\IZ)$, trivially regarded as a (strict) 2-group, on $\underline{\sfTD}_n$ that is an extension of the weak 2-functor \eqref{eq:automorphisms_TD_n} with $\zeta=0$ does not exist directly, but the calculations involved in showing this make it evident that the slightly enlarged 2-group $\scGO(n,n;\IZ)$ does allow for a 2-group action. This 2-group has underlying Lie groupoid
    \begin{subequations}\label{eq:def_scGOnnZ}
        \begin{equation}
            \begin{gathered}
                \begin{tikzcd}
                    \sfGO(n,n;\IZ)\times \IZ^{2n}\arrow[r,shift left] 
                    \arrow[r,shift right] & \sfGO(n,n;\IZ)
                \end{tikzcd}~,~~
                \begin{tikzcd}[column sep=2.0cm,row sep=large]
                    g & g \arrow[l,bend left,swap,out=-20,in=200]{}{(g,z_1)}& g\arrow[l,bend left,swap,out=-20,in=200]{}{(g,z_2)}\arrow[ll,bend right,out=20,in=-200]{}{(g,z_1+z_2)}
                \end{tikzcd}~,
                \\
                \sfid_g=(g,0)~,~~~(g,z)^{-1}=(g,-z)~,
            \end{gathered}
        \end{equation}   
        and the monoidal product and corresponding inverse are given by 
        \begin{equation}
            \begin{aligned}
                (g_1,z_1)\otimes (g_2,z_2)&=(g_1g_2,z_1+g_1z_2)
                \eand
                \sfinv(g,z)&=(g^{-1},-g^{-1}z)
            \end{aligned}
        \end{equation}
        for all $g,g_{1,2}\in \sfGO(n,n;\IZ)$ and $z,z_{1,2}\in \IZ^{2n}$. The associator allowing for the existence of the action reads as
        \begin{equation}\label{eq:assoc_GOnnZ}
            \begin{aligned}
                \MoveEqLeft
                \sfa(g_1,g_2,g_3)\\
                &=\Big(g_1g_2g_3,\frac{(-1)^{|g_1g_2g_3|}}{2}g_1g_2g_3\eta\times\big(g_3^\rmT \rmdiag(\sigma_L(g_1,g_2))
                \\
                &\hspace{3cm}+\rmdiag\big(\sigma_L(g_1g_2,g_3)-\sigma_L(g_1,g_2g_3)-(-1)^{|g_1|}\sigma_L(g_2,g_3)\big)\Big)
                \\
                &=\Big(g_1g_2g_3,\frac{(-1)^{|g_1g_2g_3|}}{2}g_1g_2g_3\eta~\big(g_3^\rmT \rmdiag(\sigma_L(g_1,g_2))-\rmdiag(g_3^\rmT \sigma_L(g_1,g_2)g_3)\big)\Big)~,
            \end{aligned}
        \end{equation}
        where we have used~\eqref{eq:sigma_relation} in the computations.
    \end{subequations}    
    Here, $\rmdiag\colon\sfSym(2n;\IZ)\rightarrow \IZ^{2n}$ extracts the diagonal vector of a square matrix, and $\eta$ is the usual $\sfO(n,n;\IZ)$-invariant metric defined in~\eqref{eq:Onn-metric}. We note that indeed\footnote{This follows from all terms proportional to off-diagonal elements of $\zeta$ appearing twice, and all terms proportional to diagonal elements appearing with the even factor of the form $(g_3)_{ii}((g_3)_{ii}-1)$.}
    \begin{equation}
        \tfrac12(g_3^\rmT \rmdiag(\zeta)-\rmdiag(g_3^\rmT \zeta g_3))\in \IZ^{2n}
    \end{equation}
    for any symmetric matrix $\zeta\in\sfSym(2n;\IZ)$, and the associator is well-defined. The associator is fully encoded in a normalised cocycle $\sfGO(n,n;\IZ)^3\rightarrow \IZ^{2n}$: in particular, $\sfa(1,g_2,g_3)$, $\sfa(g_1,1,g_3)$, and $\sfa(g_1,g_2,1)$ are trivial, and 
    \begin{equation}
        \begin{aligned}
            \sfid_{g_1}\otimes \sfa(g_2,g_3,g_4)+\sfa(g_1,g_2g_3,g_4)+\sfa(g_1,g_2,g_3)=\sfa(g_1,g_2,g_3g_4)+\sfa(g_1g_2,g_3,g_4)~.
        \end{aligned}
    \end{equation}   
    
    The 2-group $\scGO(n,n;\IZ)$ is thus a special Lie 2-group in the sense of~\cite{Baez:0307200}. In particular, it is skeletal, i.e.~isomorphic objects are equal. Moreover, it was shown in~\cite{Waldorf:2022lzs} that the automorphism 2-group of $\underline{\sfTD}_n$ based on crossed intertwiners is equivalent to a 2-group with the same underlying groupoid as $\scGO(n,n;\IZ)$; this 2-group is strictly equivalent to $\scGO(n,n;\IZ)$.
    
    \paragraph{Action $\scGO(n,n;\IZ)\curvearrowright\underline{\sfTD}_n$.} The action $\scGO(n,n;\IZ)\curvearrowright\underline{\sfTD}_n$ is now given by the following data: the unital bifunctor
    \begin{subequations}\label{eq:GO-action}
        \begin{equation}
            \begin{aligned}
                \acton\colon\scGO(n,n;\IZ)\times \underline{\sfTD}_n&\rightarrow \underline{\sfTD}_n~,
                \\
                (g,z)\times(\xi,m,\phi)&\mapsto (g\xi,gm,(-1)^{|g|}\phi+m^\rmT \rho_L(g)\xi+z^\rmT \eta g \xi)~,
            \end{aligned}
        \end{equation}
        the natural transformation
        \begin{equation}
            \begin{aligned}
                \Upsilon_{\scGO(n,n;\IZ)}\colon (g_1g_2)\acton \xi &\xrightarrow{~\cong~} g_1\acton(g_2\acton \xi)~,
                \\
                \Upsilon_{\scGO(n,n;\IZ)}(g_1,g_2,\xi)&\coloneqq (g_1g_2\xi,0,\tfrac12 \xi^\rmT \sigma_L(g_1,g_2)\xi+\tfrac12\rmdiag(\sigma_L(g_1,g_2))^\rmT \xi)~,
            \end{aligned}
        \end{equation}
        and the natural transformation
        \begin{equation}
            \begin{aligned}
                \Upsilon_{\sfTD_n}\colon g\acton(\xi_1+\xi_2)&\xrightarrow{~\cong~} (g\acton \xi_1)+(g\acton \xi_2)~,
                \\
                \Upsilon_{\sfTD_n}(g,\xi_1,\xi_2)&\coloneqq (g(\xi_1+\xi_2),0,-\xi_1^\rmT \rho_L(g)\xi_2)~.
            \end{aligned}
        \end{equation}
    \end{subequations}    
    One can check that these data satisfy all the relations required for a 2-group action. In particular, the functors $\Upsilon_{\scGO(n,n;\IZ)}$ and $\Upsilon_{\sfTD_n}$ satisfy indeed the required coherence relations found in~\cite[Prop.~3.2]{Garzn:2001aa}. In the underlying computations, we have to use the fact that
    \begin{equation}
        \tfrac12 m^\rmT \zeta m+\tfrac12\rmdiag(\zeta)^\rmT m
    \end{equation}
    is an integer\footnote{This is clear from the fact that in the first term each summand involving off-diagonal components of $\zeta$ appears with a factor of $2$ due to the symmetry of $\zeta$, and the second term then corrects the diagonal sum to an integer.} for all $\zeta\in \sfSym(2n;\IZ)$ and $m\in \IZ^{2n}$.
    
    \paragraph{Action of subgroups.} Let us briefly go through the various subgroups of $\sfGO(n,n;\IZ)$ we introduced in~\cref{ssec:review_diff_refine} and list their action on $\underline{\sfTD}_n$.
    \begin{subequations}
        \begin{itemize}
            \item[$A)$]  $A$-transformations are parameterised by  elements $A\in \sfGL(n;\IZ)\subset \sfO(n,n;\IZ)$, and the corresponding automorphisms are strict:
            \begin{equation}
                \begin{gathered}
                    \Phi^A_1(\xi,m,\phi)\coloneqq \left(\left(\begin{smallmatrix} A & 0 \\ 0 & (A^\rmT )^{-1}  \end{smallmatrix}\right) \xi ~,~ \left(\begin{smallmatrix} A & 0 \\ 0 & (A^\rmT )^{-1}  \end{smallmatrix}\right) m~,~\phi\right)~,
                    \\
                    \Phi^A_2(\xi_1,\xi_2)\coloneqq \left(\left(\begin{smallmatrix} A & 0 \\ 0 & (A^\rmT )^{-1}  \end{smallmatrix}\right) (\xi_1+\xi_2)~,~0~,~0\right)~.
                \end{gathered}
            \end{equation}
            \item[$B)$] $B$-transformations are parameterised by an antisymmetric, integer-valued matrix $B$. The corresponding automorphisms read as
            \begin{equation}
                \begin{gathered}
                    \Phi^B_1(\xi,m,\phi)\coloneqq \left(\left(\begin{smallmatrix} \unit_n & -B  \\ 0 & \unit_n \end{smallmatrix}\right) \xi ,\left(\begin{smallmatrix} \unit_n & -B  \\ 0 & \unit_n \end{smallmatrix} \right) m,\phi-\check m^\rmT  B_L \check \xi\right),
                    \\
                    \Phi^B_2(\xi_1,\xi_2)\coloneqq \left(\left(\begin{smallmatrix} \unit_n & -B  \\ 0 & \unit_n \end{smallmatrix}\right) (\xi_1+\xi_2)~,~0~,~-\check \xi^\rmT _1 B_L \check \xi_2\right)~,
                \end{gathered}
            \end{equation}
            where $B=B_L-B_L^\rmT $ and where \(\check\xi\) was defined in \eqref{eq:FOOBARBAZ}. This action was also defined, with minor differences in $\Phi_2$, in~\cite[Sect.~4.1]{Nikolaus:2018qop}.
            \item[$\beta)$] $\beta$-transformations are parameterised by an antisymmetric, integer-valued matrix $\beta$. The corresponding automorphisms read as
            \begin{equation}
                \begin{gathered}
                    \Phi^\beta_1(\xi,m,\phi)\coloneqq \left(\begin{psmallmatrix} \unit_n & 0  \\ -\beta & \unit_n \end{psmallmatrix} \xi ~,~ \begin{psmallmatrix} \unit_n & 0  \\ -\beta & \unit_n \end{psmallmatrix} m~,~\phi+\hat m^\rmT  \beta_L \hat \xi\right)~,
                    \\
                    \Phi^\beta_2(\xi_1,\xi_2)\coloneqq \left(\begin{psmallmatrix} \unit_n & 0  \\ -\beta & \unit_n \end{psmallmatrix} (\xi_1+\xi_2)~,~0~,~\hat \xi^\rmT _1 \beta_L \hat \xi_2\right)~,
                \end{gathered}
            \end{equation}
            where $\beta=\beta_L-\beta_L^\rmT $ and where \(\hat\xi\) was defined in \eqref{eq:FOOBARBAZ}.
            \item[$T_k$)] The generators of factorised dualities are parameterised by $k\in\{1,\ldots,n\}$ and a sign. The corresponding automorphisms read as
            \begin{equation}\label{eq:Tk-morphisms}
                \begin{gathered}
                    \Phi^{T_k}_1(\xi,m,\phi)\coloneqq \left( \begin{psmallmatrix} \unit_n-1_k & \pm 1_k \\ \pm 1_k & \unit_n-1_k \end{psmallmatrix} \xi ~,~ \begin{psmallmatrix} \unit_n-1_k & \pm 1_k \\ \pm 1_k & \unit_n-1_k \end{psmallmatrix} m,\phi-\langle m ,1_k \xi\rangle\right)~,
                    \\
                    \Phi^{T_k}_2(\xi_1,\xi_2)\coloneqq \left(\begin{psmallmatrix} \unit_n-1_k & \pm 1_k \\ \pm 1_k & \unit_n-1_k \end{psmallmatrix} (\xi_1+\xi_2)~,~0~,~-\langle \xi_1,1_k\xi_2\rangle\right)~.
                \end{gathered}
            \end{equation}
            \item[$G$)] The $G$-transformations are parameterised by $s_{1,2}\in\{\pm1\}$, and the corresponding automorphisms are strict:
            \begin{equation}
                \begin{gathered}
                    \Phi^G_1(\xi,m,\phi)\coloneqq \left( \begin{psmallmatrix} s_1\unit_n & 0 \\ 0 & s_2\unit_n \end{psmallmatrix} \xi ~,~ \begin{psmallmatrix} s_1\unit_n & 0 \\ 0 & s_2\unit_n \end{psmallmatrix} m,s_1s_2\phi\right)~,
                    \\
                    \Phi^G_2(\xi_1,\xi_2)\coloneqq \left(\begin{psmallmatrix} s_1\unit_n & 0 \\ 0 & s_2\unit_n \end{psmallmatrix} (\xi_1+\xi_2)~,~0~,~0\right)~.
                \end{gathered}
            \end{equation}
        \end{itemize}
    \end{subequations}    
    
    \paragraph{The Lie 2-group $\underline{\sfTD}^\ltimes_n$.} In order to capture T-dualities involving general affine torus bundles, we have to extend the Lie 2-group $\underline{\sfTD}_n$ by the action of $\sfGL(n;\IZ)\subset \sfGO(n,n;\IZ)$ to the semidirect product
    \begin{equation}
        \underline{\sfTD}^\ltimes_n\coloneqq \sfGL(n;\IZ)\ltimes \underline{\sfTD}_n~.
    \end{equation}
    Here, the action of $\sfGL(n;\IZ)$ is that of the geometric subgroup on $\sfGO(n,n;\IZ)$. We note that T-duality indeed just allows for an extension by $\sfGL(n;\IZ)\subset \sfGO(n,n;\IZ)$ and not the perhaps expected group $\sfGL(2n;\IZ)$. This amounts to a link between the orientations of the torus bundles $\check P$ and $\hat P$ in~\eqref{eq:T-duality_correspondence}.
    
    Note that the group $\sfGL(n;\IZ)$ indeed acts on $\underline{\sfTD}_n$ without any need for a further extension, which is due to $\rho_L(g)=0$ for $g\in \sfGL(n;\IZ)\subset \sfGO(n,n;\IZ)$. The 2-group $\underline{\sfTD}^\ltimes_n$ thus has underlying Lie groupoid
    \begin{subequations}\label{eq:TD_n_hat}
        \begin{equation}
            \begin{gathered}
                \begin{tikzcd}
                    \sfGL(n;\IZ)\times \IR^{2n}\times \IZ^{2n}\times \sfU(1)\arrow[r,shift left] 
                    \arrow[r,shift right] & \sfGL(n;\IZ)\times \IR^{2n}
                \end{tikzcd}~,
                \\
                \begin{tikzcd}[column sep=2.0cm,row sep=large]
                    (g,\xi) & (g,\xi-m_1)\arrow[l,bend left,swap,out=-20,in=200]{}{(g,\xi,m_1,\phi_1)}& (g,\xi-m_1-m_2)\arrow[l,bend left,swap,out=-20,in=200]{}{(g,\xi-m_1,m_2,\phi_2)}\arrow[ll,bend right,out=20,in=-200]{}{(g,\xi,m_1+m_2,\phi_1+\phi_2)}
                \end{tikzcd}~,
                \\
                \sfid_{(g,\xi)}=(g,\xi,0,0)~,~~~(g,\xi,m,\phi)^{-1}=(g,\xi-m,-m,-\phi)
            \end{gathered}
        \end{equation}    
        and monoidal structure and inverse functor
        \begin{equation}
            \begin{aligned}
                (g_1,\xi_1,m_1,\phi_1)\otimes (g_2,\xi_2,m_2,\phi_2)&\coloneqq(g_1g_2,\xi_1+g_1\xi_2,m_1+g_1m_2,\phi_1+\phi_2-\langle \xi_1,g_1m_2\rangle)~,
                \\
                \sfinv(\xi,m,\phi)&\coloneqq (g^{-1},-g^{-1}\xi,-g^{-1}m,-\phi-\langle \xi,g^{-1}m\rangle)
            \end{aligned}
        \end{equation}    
    \end{subequations}    
    for $g\in \sfGL(n;\IZ)\subset \sfO(n,n;\IZ)$, $\xi,\xi_{1,2}\in \IR^{2n}$, $m,m_{1,2}\in \IZ^{2n}$, and $\phi,\phi_{1,2}\in \sfU(1)$. Like $\underline{\sfTD}_n$, $\underline{\sfTD}^\ltimes_n$ is a strict Lie 2-group, and the corresponding crossed module of Lie groups is 
    \begin{equation}\label{eq:def_TD_n_hat}
        \begin{gathered}
            \sfTD^\ltimes_n~\coloneqq~\big(\IZ^{2n}\times\sfU(1) \xrightarrow{~\sft~}\sfGL(n;\IZ)\ltimes\IR^{2n}\big)~,
            \\
            \sft(m,\phi)=(1,m)~,
            \\
            (g,\xi)\acton(m,\phi)=(gm,\phi-\langle \xi,gm\rangle)~.
        \end{gathered}
    \end{equation}
    The associated crossed module of Lie algebras is the same as $\frtd_n$.
    
    \section{Geometric T-duality with principal 2-bundles}\label{sec:geometric_t-duality}
    
    \subsection{Topological T-duality correspondences as principal 2-bundles}
    
    In the following, we give the description of geometric T-duality correspondences in terms of principal 2-bundles, extending the description found in~\cite{Nikolaus:2018qop} to T-dualities involving affine torus bundles.
    
    \paragraph{Span of principal 2-bundles.} As mentioned above, it has been shown in~\cite{Nikolaus:2018qop} that T-duality correspondences can be formulated as spans of principal 2-bundles $\check \scP$, $\hat \scP$, and $\scP_\rmC$ over $X$,
    \begin{equation}\label{eq:geometric_T_duality_span}
        \begin{tikzcd}[column sep=1cm, row sep=0.8cm]
            & \arrow[ld,"\check \sfp",swap] \scP_\rmC \arrow[rd,"\hat \sfp"]& & \\
            \check \scP & & \hat \scP
        \end{tikzcd}
    \end{equation}
    which are induced by correspondences of Lie 2-groups. In the following, we review the cocycle description of the above principal 2-bundles as well as the projections $\check \sfp$ and $\hat \sfp$ between them. We will always consider principal 2-bundles subordinate to a surjective submersion $Y\rightarrow X$.
    
    \paragraph{The principal 2-bundle $\scP_\rmC$.}
    The structure Lie 2-group of $\scP_\rmC$ is the crossed module of Lie groups $\sfTD^\ltimes_n$ defined in~\eqref{eq:def_TD_n_hat}. Correspondingly, the general cocycle relations for principal 2-bundle~\eqref{eq:adjusted_cocycles} specialise as follows:
    \begin{subequations}\label{eq:TD-top-cocycles}
        \begin{equation}
            \begin{gathered}
                h=(m_{ijk},\phi_{ijk})\in C^\infty(Y^{[3]},\IZ^{2n}\times\IR/\IZ)~,\\
                g=(g_{ij},\xi_{ij})\in C^\infty(Y^{[2]},\sfGL(n;\IZ)\times\IR^{2n})
            \end{gathered}
        \end{equation}
        with $(ij)\in Y^{[2]}$ and $(ijk)\in Y^{[3]}$, which satisfy
        \begin{equation}
            \begin{aligned}
                \phi_{ikl}+\phi_{ijk}&=\phi_{ijl}+\phi_{jkl}-\langle \xi_{ij},g_{ij}m_{jkl}\rangle~,
                \\
                m_{ikl}+m_{ijk}&=m_{ijl}+g_{ij}m_{jkl}~,
                \\
                g_{ik}&=g_{ij}g_{jk}~,
                \\
                \xi_{ik} &= m_{ijk}+\xi_{ij}+g_{ij}\xi_{jk}
            \end{aligned}
        \end{equation}
        on $Y^{[4]}$ and $Y^{[3]}$, respectively.
    \end{subequations}    
    
    Two such cocycles $(g,h)$ and $(\tilde g,\tilde h)$ are considered equivalent if they are related by a coboundary consisting of maps 
    \begin{subequations}\label{eq:TD-top-coboundaries}
        \begin{equation}
            \begin{gathered}
                b=(m_{ij},\phi_{ij})\in C^\infty(Y^{[2]},\IR/\IZ\times \IZ^{2n})~,\\
                a=(g_i,\xi_i)\in C^\infty(Y,\sfGL(n;\IZ)\times\IR^{2n})~,
            \end{gathered}
        \end{equation}
        that link the cocycles by the relations
        \begin{equation}
            \begin{aligned}
                \phi_{ik}+\phi_{ijk}&=\tilde \phi_{ijk}-\langle \xi_i,g_i\tilde m_{ijk}\rangle+\phi_{ij}+\phi_{jk}-\langle \xi_{ij},g_{ij}m_{jk}\rangle~,
                \\
                m_{ik}+m_{ijk}&=g_i\tilde m_{ijk}+m_{ij}+g_{ij}m_{jk}~,
                \\
                \tilde g_{ij}&=g_i^{-1}g_{ij}g_j~,
                \\
                \xi_i +g_i\tilde \xi_{ij}&=m_{ij}+\xi_{ij}+g_{ij}\xi_j~,
            \end{aligned}
        \end{equation}
        over $Y^{[3]}$ and $Y^{[2]}$, cf.~the general formulas~\eqref{eq:adjusted_coboundaries}.
    \end{subequations}    
    
    \paragraph{The flip morphism.} The \emph{flip morphism}~\cite{Nikolaus:2018qop} is simply the action of the concatenation of all factorised dualities $T^+_1\circ \cdots \circ T^+_n$, i.e.~the $\sfGO(n,n;\IZ)$-transformation $g=\eta$. The corresponding automorphism reads as
    \begin{equation}
        \begin{gathered}
            \Phi_1(g,\xi,m,\phi)\coloneqq \left( \eta g\eta~,~\eta \xi ~,~ \eta m~,~\phi-\langle \xi ,m\rangle \right)~,
            \\
            \Phi_2(g_1,\xi_1;g_2,\xi_2)\coloneqq \left(\eta (\xi_1+\xi_2)~,~0~,~\langle \xi_2,\xi_1\rangle\right)~,
        \end{gathered}
    \end{equation}
    cf.~\eqref{eq:Tk-morphisms}. At the level of the crossed module of Lie algebras, $\Phi$ induces the following endomorphism $\phi$ on $\frtd_n$:
    \begin{equation}\label{eq:infinitesimal_flip_morphism}
        \begin{gathered}
            \phi^\text{flip}_0(\xi)=
            \begin{pmatrix} 
                0 & \unit_n
                \\
                \unit_n & 0         
            \end{pmatrix}
            \xi
            ~,~~~
            \phi^\text{flip}_1(y)= y~,
            \\
            \phi^\text{flip}_2(\xi_1,\xi_2)=\langle \xi_2,\xi_1\rangle-\langle \xi_1,\xi_2\rangle~.
        \end{gathered}
    \end{equation}
    
    As explained in \cref{app:higher_principal_bundles}, this automorphism  gives rise to a morphism of principal $\sfTD_n$-bundles, which we also call the \emph{flip morphism}. Moreover, it is clear that this automorphism is not an inner automorphism; therefore, the morphism of principal $\sfTD_n$-bundles is not an isomorphism of principal 2-bundles.
    
    \paragraph{The principal 2-bundles $\check \scP$ and $\hat \scP$.} The structure 2-group of $\check\scP$ and $\hat \scP$ is the crossed module of Lie groups
    \begin{equation}\label{eq:def_TB_F2}
        \begin{gathered}
            \sfTB^\ltimes_n~=~\big(\IZ^n\times C^\infty(\IT^n,S^1)\xrightarrow{~\sft~}\sfGL(n;\IZ)\ltimes\IR^n\big)~,
            \\
            \sft(m,f)=m~,
            \\
            (g,\xi)\acton(m,f)=(gm,c\mapsto f(c-g\xi))=(m,f\circ \sfs_{g\xi})
        \end{gathered}
    \end{equation}
    for all $g\in \sfGL(n;\IZ)$, $\xi\in \IR^n$, $m\in \IZ^n$, and $f\in C^\infty(\IT^n,S^1)$, where $\sfs_\xi$ denotes the function $\sfs_\xi\colon t\mapsto t-\xi$. The crossed module $\sfTB^{\mathrm{F2}}_n$ defined in~\cite{Nikolaus:2018qop} is obtained by restricting to $g=\unit$. 
    
    The higher bundles $\check\scP$ and $\hat \scP$ are correspondingly described by cocycles consisting of maps 
    \begin{subequations}\label{eq:TBF2-cocycles}
        \begin{equation}
            \begin{gathered}
                h=(m_{ijk},f_{ijk})\in C^\infty\big(Y^{[3]},\IZ^{n}\times C^\infty(\IT^n,S^1)\big)~,\\
                g=(g_{ij},\xi_{ij})\in C^\infty(Y^{[2]},\sfGL(n;\IZ)\times\IR^{n})~,
            \end{gathered}
        \end{equation}
        which satisfy
        \begin{equation}
            \begin{aligned}
                f_{ikl}+f_{ijk}&=f_{ijl}+f_{jkl}\circ \sfs_{g_{ij}\xi_{ij}}~,
                \\
                m_{ikl}+m_{ijk}&=m_{ijl}+g_{ij}m_{jkl}~,
                \\
                g_{ij} &= g_{ij}g_{jk}~,
                \\
                \xi_{ik} &= m_{ijk}+\xi_{ij}g_{ij}+\xi_{jk}~.
            \end{aligned}
        \end{equation}
    \end{subequations}    
    Two such cocycles $(g,h)$ and $(\tilde g,\tilde h)$ are considered equivalent if they are related by a coboundary consisting of maps 
    \begin{subequations}\label{eq:TBF2-coboundaries}
        \begin{equation}
            \begin{gathered}
                b=(m_{ij},f_{ij})\in C^\infty(Y^{[2]},\IZ^n\times C^\infty(\IT^n,S^1))~,
                \\
                a=(g_i,\xi_i)\in C^\infty(Y,\sfGL(n;\IZ)\times\IR^{n})~,
            \end{gathered}
        \end{equation}
        that link the cocycles by the relations
        \begin{equation}
            \begin{aligned}
                f_{ik}+f_{ijk}&=\tilde f_{ijk}\circ \sfs_{g_i\xi_i}+f_{ij}+f_{jk}\circ\sfs_{g_{ij}\xi_{ij}}~,
                \\
                m_{ik}+m_{ijk}&=g_i\tilde m_{ijk}+m_{ij}+g_{ij}m_{jk}~,
                \\
                \tilde g_{ij}&=g_i^{-1}g_{ij}g_j~,
                \\
                \xi_i +g_i\tilde \xi_{ij}&=m_{ij}+\xi_{ij}+g_{ij}\xi_j~.
            \end{aligned}
        \end{equation}
    \end{subequations}    
    
    \paragraph{The two projections.} It remains to specify the projections $\check \sfp\colon\scP_\rmC\rightarrow \check \scP$ and $\hat \sfp\colon\scP_\rmC\rightarrow \hat \scP$ in~\eqref{eq:geometric_T_duality_span} which establish the span relating geometrically T-dual T-backgrounds to each other. The projection $\check \sfp$ is called the \emph{right-leg projection} in~\cite{Nikolaus:2018qop}\footnote{Note that we interchanged right and left as compared to~\cite{Nikolaus:2018qop}, so that the right-leg projection is the left projection in diagram~\eqref{eq:geometric_T_duality_span}.}, and it is given by a 
    morphism of Lie 2-groups that induces the bundle map
    \begin{equation}\label{eq:formula_geometric_pi-check}
        \begin{aligned}
            \check\sfp\colon(g_{ij},\xi_{ij},m_{ijk},\phi_{ijk})&=\left(
            \begin{pmatrix} 
                \hat g_{ij}\\ \check g_{ij}\end{pmatrix},
            \begin{pmatrix} 
                \hat \xi_{ij}\\ \check \xi_{ij}\end{pmatrix},
            \begin{pmatrix} 
                \hat m_{ijk}\\ \check m_{ijk}
            \end{pmatrix},\phi_{ijk}
            \right)
            \\
            &\mapsto (\check g_{ij},\check \xi_{ij},\check m_{ijk},c\mapsto \phi_{ijk}+\hat m_{ijk}^\rmT  \check g_{ij}c)~.
        \end{aligned}
    \end{equation}
    The projection $\hat \sfp$ is obtained by concatenating the flip morphism from above with the right-leg projection. Explicitly, this amounts to the map 
    \begin{equation}\label{eq:formula_geometric_pi-hat}
        \begin{aligned}
            \hat\sfp\colon(g_{ij},\xi_{ij},m_{ijk},\phi_{ijk})&=\left(
            \begin{pmatrix} 
                \hat g_{ij}\\ \check g_{ij}\end{pmatrix},
            \begin{pmatrix} 
                \hat \xi_{ij}\\ \check \xi_{ij}\end{pmatrix},
            \begin{pmatrix} 
                \hat m_{ijk}\\ \check m_{ijk}
            \end{pmatrix},
            \phi_{ijk}\right)
            \\
            &\mapsto \left(
            \hat g_{ij},
            \hat \xi_{ij}
            ,
            \hat m_{ijk}
            ,c\mapsto \phi_{ijk}+\check m_{ijk}^\rmT \hat g_{ij}c+\check \xi_{ij}^\rmT\hat \xi_{jk}   
            \right)~.
        \end{aligned}
    \end{equation}
    This completes the formulation of topological geometric T-correspondences in terms of spans of higher principal bundles: two principal $\sfTB^\ltimes_n$-bundles $\check \scP$ and $\hat \scP$ form T-dual pairs if there is a double fibration~\eqref{eq:geometric_T_duality_span} with projections~\eqref{eq:formula_geometric_pi-check} and~\eqref{eq:formula_geometric_pi-hat}.
    
    \paragraph{Recovering T-backgrounds.} Consider the image of the right leg projection $\check \sfp$ as given in~\eqref{eq:formula_geometric_pi-check}, defining the principal $\sfTB^\ltimes_n$-bundle $\check \scP$ subordinate to a surjective submersion $\sigma\colon Y\rightarrow X$. The triple $(\check g_{ij},\check \xi_{ij},\check m_{ijk})$ clearly defines an affine torus bundle, which we regard as a principal fiber bundle with fibers $\sfGL(n;\IZ)\times (\IR/\IZ)^n$ over $X$. This principal bundle is given by the quotient 
    \begin{equation}
        \check P \coloneqq (V/\IZ^n)/\sim
        ~~~\mbox{with}~~~
        V\coloneqq Y\times \sfGL(n;\IZ)\times \IR^n~,
    \end{equation}
    where two points $(y,s,t)$, $(y',s',t')\in V$ are equivalent if and only if 
    \begin{equation}
        \sigma(y)=\sigma(y')~,~~~s=\check g(y,y')s'~,\eand t-t'=\check g(y,y')\check \xi(y,y')~.
    \end{equation}
    Correspondingly, we may cover $\check P$ by the implied surjective submersion $V\rightarrow \check P$. Consider now cocycles describing the pullback 2-bundle $\check \pi^* \check\scP$ subordinate to $V\rightarrow \check P$, where $\check \pi$ is the projection defined in~\eqref{eq:T-duality_correspondence}, and recall that any bundle naturally trivialises when pulled back over itself. We note that 
    \begin{equation}
        \begin{aligned}
            V^{[2]}&\coloneqq V\times_{\check P} V
            \\
            &=\{(y_i,y_j,s_i,s_j,t_i,t_j)\in Y^2\times\IR^{2n}\mid
            \\
            &\hspace{3cm}
            \sigma(y_i)=\sigma(y_j),\;
            s_i=\check g_{ij}s_j,\;
            t_i-t_j-\check g_{ij}\check \xi_{ij}\in \IZ^n
            \}~.
        \end{aligned}
    \end{equation}
    Correspondingly, we can define a coboundary $(g_i,\xi_i,m_{ij},f_{ij})$ as in~\eqref{eq:TBF2-coboundaries} by
    \begin{equation}
        \begin{aligned}
            g_i\coloneqq s_i~,~~~\xi_i\coloneqq t_i~,~~~m_{ij}\coloneqq t_i-t_j-\check g_{ij}\check \xi_{ij}~,~~~f_{ij}\coloneqq 0~.
        \end{aligned}
    \end{equation}
    This coboundary induces a 2-bundle isomorphism which trivialises the $\check P$ part in the cocycle:
    \begin{equation}\label{eq:recovering_gerbe_data_top}
        \begin{aligned}
            (\check g_{ij},\check \xi_{ij},\check m_{ijk}, f_{ijk})~~\xrightarrow{~\cong~}~~\big(\unit,0,0,c \mapsto f_{ijk}+\check m_{ijk}^\rmT g_i(c-t_i)\big)~,
        \end{aligned}
    \end{equation}
    and the $\sfU(1)$-cocycle given by the part $(f_{ijk}-\check m_{ijk}^\rmT g_i t_i)$ constant in $c$ defines an abelian gerbe subordinate to the cover $V\rightarrow \check P$. This is the abelian gerbe $\check \scG$ that, together with $\check P$, forms the toric T-background captured by $\check \scP$.
    
    \paragraph{Equivalence of principal 2-bundles.} We note that~\cite[Thm.~3.4.5]{Nikolaus:2018qop} shows that the left leg projection yields a bijection between isomorphism classes of principal $\sfTD_n$-bundles and principal $\sfTB^\text{F2}_n$-bundles. This bijection evidently extends to a bijection between isomorphism classes of principal $\sfTD_n^\ltimes$-bundles and $\sfTB_n^\ltimes$-bundles. In this sense, it is clear that no information is gained or lost by choosing to work with either $\check \scP$ or $\scP_\rmC$, at least for principal torus bundles. This point will be important below as a differential refinement only exists on $\scP_\rmC$.
    
    \subsection{Differential refinement of \texorpdfstring{$\scP_\rmC$}{PC}}
    
    \paragraph{Adjustment data for $\sfTD^\ltimes_n$.} In order to define a non-flat connection on a principal 2-bundle, one generically needs to lift the conventional definition in the literature to that of an adjusted one, see \cref{app:higher_principal_bundles}. As noted there, an adjustment for a crossed module of Lie groups $\caG=(\sfH\xrightarrow{~\sft~}\sfG,\acton)$ is really an algebraic datum, given by a map $\kappa\colon\sfG\times \frg\rightarrow \frh$ linear in $\frg$, where $\frg$ and $\frh$ are the Lie algebras of $\sfG$ and $\sfH$, respectively. To qualify as an adjustment, the map $\kappa$ has to satisfy the condition
    \begin{equation}\label{eq:adjustment_condition_2}
        \begin{aligned}
            (g_2^{-1}g_1^{-1})\acton (h^{-1}(X\acton h))
            &+g_2^{-1}\acton\kappa(g_1,X)
            \\
            &+\kappa(g_2,g_1^{-1}X g_1-\sft(\kappa(g_1,X)))-\kappa(\sft(h)g_1g_2,X)=0
        \end{aligned}
    \end{equation}
    for all $g_{1,2}\in \sfG$, $h\in \sfH$, and $X\in \frg$. 
    
    For the crossed module of Lie groups $\sfTD^\ltimes_n$, a valid choice is 
    \begin{equation}
        \kappa\colon (g,\xi;X)\mapsto (0,-\langle X,\xi\rangle)
    \end{equation}
    for $g\in \sfGL(n;\IZ)$ and $\xi,X\in \IR^{2n}$, as one can verify by straightforward computation.
    
    \paragraph{No adjustment for $\sfTB^\text{F2}_n$ or $\sfTB^\ltimes_n$.} For the general theory of principal 2-bundles with connection, it is interesting to note that there is, in fact, no adjustment for the crossed module of Lie groups $\sfTB^\text{F2}_n$. The $C^\infty(\IT^n,\sfU(1))$ part of condition~\eqref{eq:adjustment_condition_2} simplifies to
    \begin{equation}\label{eq:aux_adjustment_condition}
        \begin{aligned}
            f(c-X+\xi_1+\xi_2)&-f(c+\xi_1+\xi_2)
            +\kappa_1(\xi_1,X)(c+\xi_2)
            \\
            &+\kappa_1(\xi_2,X-\kappa_0(\xi_1,X))(c)-\kappa_1(m+\xi_1+\xi_2,X)(c)=0~,
        \end{aligned}
    \end{equation}
    where we have changed notation to $g=\xi$ and split $h=(m,f)$ and $\kappa=(\kappa_0,\kappa_1)$. For simplicity, consider the special case where $f$ is an affine function,
    \begin{equation}
    	f(x) \coloneqq \phi + m'\cdot x
    \end{equation}
    with $m'\in \IZ$. For generic $X$, the first two terms in~\eqref{eq:aux_adjustment_condition} lead to a non-vanishing, $m'$-dependent term, while the remaining terms are independent of $m'$. Thus, there is clearly no fixed $\kappa$ that can satisfy~\eqref{eq:aux_adjustment_condition} for arbitrary $f$. This problem persists for the Lie 2-group $\sfTB^\ltimes_n$.
    
    One can now speculate that the absence of an adjustment is due to the disconnected nature of the components of the 2-group. Regarding the situation at the level of Lie 2-algebras as done in~\cite{Saemann:2019dsl} and in particular in~\cite{Borsten:2021ljb}, we would still conjecture that any crossed module of connected Lie groups admits an adjustment.

    \paragraph{Differential refinement of $\scP_\rmC$.} We begin with the differential refinement of the principal $\sfTD^\ltimes_n$-bundle $\scP_\rmC$ in the geometric T-duality span~\eqref{eq:geometric_T_duality_span}, making the adjusted cocycle formulas~\eqref{eq:adjusted_cocycles} explicit. Beyond the topological cocycle data $(g_{ij},\xi_{ij},m_{ijk},\phi_{ijk})$, cf.~\eqref{eq:TD-top-cocycles}, we have the 1- and 2-forms
    \begin{subequations}\label{eq:diff_refined_cocycles}
        \begin{equation}
            \Lambda\in \Omega^1(Y^{[2]})~,~~~
            A\in \Omega^1(Y,\IR^{2n})~,~~~
            B\in \Omega^2(Y)
        \end{equation}
        satisfying the gluing relations
        \begin{equation}
            \begin{aligned}
                \Lambda_{ik} &= \Lambda_{jk} + \Lambda_{ij}+\rmd \phi_{ijk}-\langle A_i,m_{ijk}\rangle~,
                \\
                A_j &= g_{ij}^{-1}A_i + g_{ij}^{-1}\rmd \xi_{ij}~,
                \\
                B_j &= B_i + \rmd \Lambda_{ij} + \langle\rmd A_i,\xi_{ij}\rangle~.
            \end{aligned}
        \end{equation}
    \end{subequations}        
    The adjusted curvature of this connection on $\scP_\rmC$ is given by locally defined 2- and 3-forms
    \begin{equation}
        F=\rmd A\in \Omega^2(Y,\IR^{2n})
        \eand
        H=\rmd B+\langle\rmd A,A\rangle\in \Omega^3(Y)~.
    \end{equation}
    
    Two differentially refined cocycles $(g,h,A,\Lambda,B)$ and $(\tilde g,\tilde h,\tilde A,\tilde \Lambda,B)$ are equivalent if there is a differentially refined coboundary between them. Such a coboundary is given by the data $(\xi,m,\phi)$ of a topological coboundary, cf.~\eqref{eq:TD-top-coboundaries}, together with a 1-form 
    \begin{equation}
        \lambda\in \Omega^1(Y)
    \end{equation}
    such that
    \begin{equation}
        \begin{aligned}
            \tilde \Lambda_{ij}&= \Lambda_{ij}+\lambda_j-\lambda_i-\rmd \phi_{ij}-\langle A_i,m_{ij}\rangle~,
            \\
            \tilde A_i&=g_i^{-1}A_i+g_i^{-1}\rmd \xi_i~,
            \\
            \tilde B_i&=B_i +\mathrm d\lambda_i
            +\langle \rmd A_i,\xi_i\rangle~.
        \end{aligned}
    \end{equation}
    
    \paragraph{T-duality.} Note that, when applying the flip morphism to a differentially refined cocycle describing a principal $\sfTD^\ltimes_n$-bundle with connection, we can simply use~\eqref{eq:cocycles_under_morphisms} to identify the images of the cocycles describing the principal 2-bundle $\scP_\rmC$ under flips. Explicitly, we have the following maps: 
    \begin{equation}\label{eq:images_of_flip}
        \begin{aligned}
            \xi_{ij}&\mapsto \tilde \xi_{ij}=
            \begin{pmatrix}
                \hat{\tilde{\xi}}_{ij}
                \\
                \check{\tilde{\xi}}_{ij}
            \end{pmatrix}=
            \begin{pmatrix}
                \check \xi_{ij}
                \\
                \hat \xi_{ij}
            \end{pmatrix}~,
            ~~~&
            h_{ijk}&\mapsto \tilde h_{ijk}=h_{ijk}-\check \xi_{ij}\cdot \hat \xi_{ij}~,
            \\
            g_{ij}&\mapsto \tilde g_{ij}=
            \begin{pmatrix}
                \hat{\tilde{g}}_{ij}
                \\
                \check{\tilde{g}}_{ij}
            \end{pmatrix}=
            \begin{pmatrix}
                \check g_{ij}
                \\
                \hat g_{ij}
            \end{pmatrix}~,
            \\
            A_i&\mapsto \tilde A_i=
            \begin{pmatrix}
                \hat{\tilde{A}}_i
                \\
                \check{\tilde{A}}_i
            \end{pmatrix}=
            \begin{pmatrix}
                \check A_i
                \\
                \hat A_i
            \end{pmatrix}~,
            ~~~&
            \Lambda_{ij}&\mapsto \tilde\Lambda_{ij}=\Lambda_{ij}~,
            \\
            B_i&\mapsto \tilde B_i=B_i+\check A_i\wedge\hat A_i~.
        \end{aligned}
    \end{equation}
    As a consequence, the curvatures are mapped to
    \begin{equation}
        F_i\mapsto \tilde F_i=\begin{pmatrix}
            \hat{\tilde{F}}_i
            \\
            \check{\tilde{F}}_i
        \end{pmatrix}
        =\begin{pmatrix}
            \rmd \check A_i
            \\
            \rmd \hat A_i
        \end{pmatrix}
        \eand 
        H_i\mapsto \tilde H_i=\rmd \tilde B_i+(\rmd \check A_i)\wedge \hat A_i~,
    \end{equation}
    and we note that the 3-form part of the curvature remains invariant: $H_i=\tilde H_i$.
    
    Due to the absence of an adjustment for $\sfTB^\ltimes_n$, however, we do not have analogues of the left- or right-leg projections that map to the principal 2-bundles $\check \scP$ and $\hat\scP$. Instead, we have to work over the correspondence space, but these cocycles contain all the required information, as we will show next.
    
    \paragraph{Recovering the differentially refined T-backgrounds.} Because we do not have the bundles $\check \scP$ or $\hat \scP$ at our disposal, we work with the bundle $\scP_\rmC$. The cocycle data $(\check g_{ij},\check \xi_{ij},\check m_{ijk}^\IZ,\check A_i)$ and $(\hat g_{ij},\hat \xi_{ij},\hat m_{ijk}^\IZ,\hat A_i)$ contained in a cocycle describing $\scP_\rmC$ describe two affine torus bundles $\check P$ and $\hat P$ over $X$ equipped with connections. It remains to recover the two gerbes $\hat \scG$ and $\check \scG$.
    
    To this end, we pull back $\scP_\rmC$ along $\Pi=\check \pi\circ \check \sfp=\hat \pi\circ \hat \sfp$, cf.~\eqref{eq:T-duality_correspondence}, so that the part corresponding to the affine torus bundles trivialises. Let $\sigma\colon Y\rightarrow X$ be a surjective submersion. Then the correspondence space forms an affine torus bundle, which we regard as a fiber bundle with fibers $\sfGL(n;\IZ)\times \IR^{2n}$. This bundle can be identified with
    \begin{equation}
        \check P\times_X\hat P=(V/\IZ^{2n})/\sim
        ~~~\mbox{with}~~~
        V\coloneqq Y\times \sfGL(n;\IZ)\times \IR^{2n}~,
    \end{equation}
    where two points $(y,s,t)$, $(y',s',t')\in V$ are equivalent if and only if 
    \begin{equation}
        \sigma(y)=\sigma(y')~,~~~~s=g(y,y')s'~,\eand t-t'=g(y,y')\xi(y,y')~.
    \end{equation}
    The fibered product over the correspondence space is then given by 
    \begin{equation}
        \begin{aligned}
            V^{[2]}=\{(y_i,y_j,s_i,s_j,t_i,t_j)&\in Y^2\times\sfGL(n;\IZ)^2\times \IR^{4n}\mid
            \\
            &\sigma(y_i)=\sigma(y_j),\;
            s_i=g_{ij}s_j,\;
            t_i-t_j-g_{ij}\xi_{ij}\in \IZ^n
            \}~,
        \end{aligned}
    \end{equation}
    and we introduce the coboundary $(g_i,\xi_i,m_{ij},\phi_{ij})$ with
    \begin{equation}
        \begin{aligned}
            g_i\coloneqq s_i~,~~~
            \xi_i\coloneqq t_i~,~~~m_{ij}\coloneqq t_i-t_j-g_{ij}\xi_{ij}~,~~~\phi_{ij}\coloneqq 0~.
        \end{aligned}
    \end{equation}
    This coboundary mostly trivialises the cocycle describing $\scP_\rmC$:
    \begin{equation}\label{eq:recovering_gerbe_data}
        \begin{aligned}
            (\xi_{ij},m_{ijk},\phi_{ijk})~~\xrightarrow{~\cong~}~~\big(\unit,0,\phi_{ijk}+\langle \xi_{ij},t_j-t_k-g_{ij}\xi_{jk}\rangle\big)~.
        \end{aligned}
    \end{equation}
    The latter expression contains the cocycle $(\phi_{ijk}+\langle \xi_{ij},t_j-t_k-g_{ij}\xi_{jk}\rangle)$ that defines an abelian gerbe subordinate to the cover $\check P\times_X\hat P$. We note that this expression does not depend on $\check t$; therefore, it is the pullback of a gerbe $\hat\scG$ on $\hat \scP$ along the map $\hat \pi$ in~\eqref{eq:T-duality_correspondence}.
    
    Let us now consider the differential refinement of the \v Cech cocycle $(\unit,0,0,\phi_{ijk}+\langle \xi_{ij},t_j-t_k-\xi_{jk}\rangle)$. The data $(\Lambda,A,B)\in \Omega^1(V^{[2]})\oplus\Omega^1(V,\IR^{2n})\oplus\Omega^2(V)$ satisfy
    the gluing relations
    \begin{equation}\label{eq:recovery_differential_refinement}
        \Lambda_{ik} = \Lambda_{jk} + \Lambda_{ij}+\rmd \phi_{ijk}~,~~~A_j = A_i~,~~~B_j = B_i + \rmd \Lambda_{ij} ~,
    \end{equation}
    and we note that $(\hat \Lambda,\hat B)$ form the differential refinement of the gerbe $\hat \pi^*\hat \scG$. In order to recover the gerbe $\check \scG$, we apply first the flip morphism to the cocycle data and then go through the same procedure.    
    
    \subsection{Example: Geometric T-duality with nilmanifolds}\label{ssec:ex_top_T_nilmanifolds}
    
    \paragraph{Topological T-duality with nilmanifolds.}
    An instructive example of geometric T-duality is that of geometric T-duality between three-dimensional nilmanifolds with $H$-fluxes, i.e.~abelian gerbes with 3-form curvature $H$. Recall that a three-dimensional nilmanifold $N_k$ is a principal circle bundle over $\IT^2$ characterised by its first Chern number $k\in \rmH^2(\IT^2,\IZ)\cong\IZ$. We can describe it as a quotient of $\IR^3$ with coordinates $(x,y,z)$ by the relations 
    \begin{equation}\label{eq:def_nil_1}
        (x,y,z)\sim(x,y+1,z)\sim(x,y,z+1)\sim(x+1,y,z-ky)~,
    \end{equation}
    where $x$ and $y$ are coordinates on the base and $z$ is the fiber coordinate. Subordinate to the surjective submersion $\IR^2\rightarrow \IT^2$, we can describe the principal circle bundle $N\rightarrow \IT^2$ by a Čech cocycle given by the map $g\colon(\IR^2)^{[2]}\rightarrow \IR$ with\footnote{We note that $(\IR^2)^{[2]}=\IR^2\times_{\IT^2}\IR^2\cong \IR^2\times \IZ^2$ and $(\IR^2)^{[3]}=\IR^2\times_{\IT^2}\IR^2\times_{\IT^2}\IR^2\cong \IR^2\times \IZ^2\times \IZ^2$.}    
    \begin{equation}\label{eq:nilmanifold_bundle_cocycle}
        g(x,y;x',y')\coloneqq k(x'-x)y~.
    \end{equation}

    A gerbe $\scG_\ell$ on the nilmanifold $N$ is characterised by its Dixmier--Douady class $\ell\in \rmH^3(N,\IZ)\cong\IZ$; subordinate to the surjective submersion $\IR^3\rightarrow N$, we can describe it by a Čech cocycle $h\colon(\IR^2)^{[3]}\rightarrow \IR$ with
    \begin{equation}\label{eq:nilmanifold_gerbe_cocycle}
        h(x,y,z;x',y',z';x'',y'',z'')\coloneqq\ell x(y-y')(z'-z'')~.
    \end{equation}
    
    It is well-known that the T-background $(\IT^2,N_k,\scG_\ell)$ is an $F^2$-background, and (topological) geometric T-duality corresponds to the duality
    \begin{equation}\label{eq:top_T-duality}
        (\IT^2,N_k,\scG_\ell)~~\xleftrightarrow{~~~}~~(\IT^2,N_\ell,\scG_k)~.
    \end{equation}
    
    \paragraph{Differential refinement.} The connection on the principal circle bundle $N_k\rightarrow \IT^2$ is given by $1$-forms
    \begin{subequations}\label{eq:def_nil_2}
        \begin{equation}
            A(x,y)\coloneqq k x\,\rmd y \in \Omega^1(\IR^2)~,
        \end{equation}
        which are local pullbacks of the global 1-form $\rmd z+kx\,\rmd y$ on $N_k$. This leads to the Kaluza--Klein metric
        \begin{equation}
            g(x,y,z) = \rmd x^2+\rmd y^2+(\rmd z+kx\,\rmd y)^2
        \end{equation}
        on the total space of $N_k$. Moreover, the gerbe $\scG_\ell$ over $N_k$ is endowed with a connective structure given by the 2-form connection and the 1-form 
        \begin{equation}
            \begin{aligned}
                B(x,y,z) &= \ell x\,\rmd y\wedge\rmd z\in \Omega^2(\IR^3)~,
                \\
                \Lambda(x,y,z;x',y',z') &=\ell(x-x')y \rmd z\in \Omega^2(\IR^3)~,
            \end{aligned}
        \end{equation}
        and these data satisfy the cocycle conditions~\eqref{eq:adjusted_cocycles} for the 2-group $\sfB\sfU(1)=(\sfU(1)\rightarrow *)$ in additive notation. The curvature of the gerbe $\scG_\ell$ is the image of its Dixmier--Douady class in de~Rham cohomology,
        \begin{equation}
            H=\ell \rmd x\wedge \rmd y\wedge \rmd z~.
        \end{equation}
    \end{subequations}

    \paragraph{Higher bundle description.} The topological T-duality~\eqref{eq:top_T-duality} can be described by a principal $\sfTD^\ltimes_1$-bundle\footnote{We note that the structure group can be reduced to $\sfTD_1$.} $\scP_\rmC$ over $\IT^2$ subordinate to the submersion $\IR^2\rightarrow\IT^2$, and the cocycles~\eqref{eq:TD-top-cocycles} specialise as follows:
    \begin{subequations}\label{eq:nilmanifold_cocycle}
        \begin{equation}
            \begin{aligned}
                g&=\begin{pmatrix}
                    \hat g,~\hat \xi
                    \\
                    \check g,~\check \xi
                \end{pmatrix}~,~~~\begin{aligned}
                    \hat g(x,y;x',y') &= \unit~, ~~~&\hat \xi(x,y;x',y') &= \ell(x'-x)y~, \\
                    \check g(x,y;x',y') &= \unit~,~~~&\check \xi(x,y;x',y') &= k(x'-x)y~,
                \end{aligned}
                \\
                m&= \begin{pmatrix}
                    \hat m
                    \\
                    \check m
                \end{pmatrix}
                ~,~~~
                \begin{aligned}
                    \hat m(x,y;x',y';x'',y'') &= -\ell(x''-x')(y'-y)~,\\
                    \check m(x,y;x',y';x'',y'') &= -k(x''-x')(y'-y)~,
                \end{aligned}
                \\
                \phi(x,y;x',y';x'',y'') &= \tfrac12k\ell\left(y'(xx''-xx'-x'x'')-(x''-x')(y'^2-y^2)x\right)~.
            \end{aligned}
        \end{equation}
        The differential refinement~\eqref{eq:diff_refined_cocycles} is given by 
        \begin{equation}
            \begin{aligned}
                A&=\begin{pmatrix}
                    \check A \\
                    \hat A
                \end{pmatrix}=
                \begin{pmatrix}
                    k x\,\mathrm dy
                    \\
                    \ell x\,\mathrm dy
                \end{pmatrix}~,
                \\
                B(x,y) &= 0~,
                \\
                \Lambda(x,y;x',y') &= \tfrac12k\ell(xx'\,\rmd y+(xy+x'y'+y^2(x'-x))\,\rmd x)~.
            \end{aligned}
        \end{equation}
    \end{subequations}    
    We note that under the flip homomorphism, the roles of $k$ and $\ell$ are interchanged, as expected.
    
    \paragraph{Span of principal 2-bundles.} Focusing on the topological part, the image of the projection $\check \sfp\colon\scP_\rmC\rightarrow \check \scP$ in~\eqref{eq:geometric_T_duality_span} is obtained from formula~\eqref{eq:formula_geometric_pi-check}. It is given by the $\sfTB^{\rm F2}_n$-bundle $\check P$ over $\IT^2$ which, subordinate again to the surjective submersion $\IR^2\rightarrow \IT^2$, is described by the cocycle
    \begin{equation}\label{eq:top_part_nilmanifold_higher_cocycle}
        \begin{aligned}
            f(x,y;x',y';x'',y'') &=\Big(c\mapsto\tfrac12k\ell\left(y'(xx''-xx'-x'x'')-(x''-x')(y'^2-y^2)x\right)\\
            &\hspace{3cm}-\ell(x''-x')(y'-y)c\Big)~,
            \\
            m(x,y;x',y';x'',y'') &=-k(x''-x')(y'-y)~,
            \\
            \xi(x,y;x',y') &=k(x'-x)y~.
        \end{aligned}
    \end{equation}
    On the other hand, we have the projection $\hat \sfp\colon\scP_\rmC\rightarrow \hat \scP$ in~\eqref{eq:geometric_T_duality_span}, whose image is 
    \begin{equation}
        \begin{aligned}
            f(x,y;x',y';x'',y'') &=\Big(c\mapsto\tfrac12k\ell\left(y'(xx''-xx'-x'x'')-(x''-x')(y'^2-y^2)x\right)\\
            &\hspace{3cm}-k(x''-x')(y'-y)c+k\ell(x'-x)y(x''-x')y'\Big)~,
            \\
            m(x,y;x',y';x'',y'') &=-\ell(x''-x')(y'-y)~,
            \\
            \xi(x,y;x',y') &=\ell(x'-x)y~.
        \end{aligned}
    \end{equation}
    
    \paragraph{Recovering the full T-backgrounds.} From the cocycle~\eqref{eq:nilmanifold_cocycle}, we readily extract the cocycle data for the two circle bundles $N_k$ and $N_\ell$ described by $(\check \xi, \check m, \check A)$ and $(\hat \xi, \hat m, \hat A)$, respectively. 
    
    The gerbe cocycles are also extracted from~\eqref{eq:nilmanifold_cocycle} using the formulas~\eqref{eq:recovering_gerbe_data} and~\eqref{eq:recovery_differential_refinement}. In the case of the gerbe $\scG_\ell$ over $N_k$, we obtain the 2-form
    \begin{equation}
        B=\langle \rmd A,a\rangle=k\rmd x\wedge \rmd y~\hat t~.
    \end{equation}
    Identifying $\hat t$ with $z$, this potential 2-form has curvature
    \begin{equation}
        H=k\rmd x\wedge \rmd y\wedge \rmd z~,
    \end{equation}
    which is indeed the curvature of the gerbe $\hat \scG_k$ on $\hat P=N_\ell$.
    
    \section{T-duality with T-folds and principal 2-groupoid bundles}\label{sec:T-folds}
    
    The first step in generalising geometric T-backgrounds is to consider T-folds, i.e.~T-backgrounds which are locally geometric but globally glued together by general elements of the T-duality group. As before, we will restrict ourselves to affine torus bundles so that the T-duality group is $\sfGO(n,n;\IZ)$. We start by considering the mathematical description of Kaluza--Klein reduction that corresponds to double dimensional reduction.
    
    \subsection{Kaluza--Klein reductions yield principal 2-groupoid bundles}
    
    From our above discussion, it is clear that T-duality is intimately related to Kaluza--Klein reduction: the definition of topological T-duality via the Gysin sequence relies on fiber integration, and the metric on the total space of the principal torus bundle is defined as the Kaluza--Klein metric. For a more detailed discussion of this point with regards to double field theory, see also~\cite{Berman:2019biz}. In order to push our analysis to the non-geometric situation, let us first motivate in a non-rigorous fashion  the origins of the 2-group $\sfTD_n$ based on Kaluza--Klein reductions. For a related but slightly different perspective that interprets double field theory as a Kaluza--Klein theory, see also~\cite{Alfonsi:2019ggg,Alfonsi:2020nxu,Alfonsi:2021ymc}.
    
    \paragraph{Kaluza--Klein reduction.} By a Kaluza--Klein reduction we mean the reduction of geometric structures on a principal torus bundle $P$ over a manifold $X$ to geometric structures on the manifold $X$. Mathematically, most geometric structures we want to reduce (as e.g.~Riemannian metrics, gerbes, and principal bundles with connections) are given by functors that are represented by a classifying space $\caC$. An example would be principal $\sfG$-bundles for $\sfG$ some topological group, which are maps from $P$ to $\caC=\sfB\sfG$. In the following, we will focus on topological aspects, which suffices for the present section.
    
    If $P=X\times \IT^n$ is topologically trivial, then we have the usual currying relation
    \begin{equation}
        C^0(X\times \IT^n,\caC)=C^0(X,C^0(\IT^n,\caC))~,
    \end{equation}
    where $C^0(A,B)$ denotes the space of continuous maps from $A$ to $B$.\footnote{For this to hold, one must work in a Cartesian-closed category of topological spaces. The category of all topological spaces and continuous maps fails to be Cartesian-closed, but there are well-known fixes for this, e.g.\ working with compactly generated weakly Hausdorff spaces. For physical purposes, one might want to work with smooth maps rather than continuous ones; in that case one can use e.g.\ diffeological spaces, cf.\ \cite{Baez:0807.1704} for a detailed discussion. Here, we neglect such technical details.
    } This is due to the functors $\IT^n\times-$ and $C^0(\IT^n,-)$ forming an adjunction in a Cartesian-closed category. Taking homotopy classes,
    \begin{equation}
        [X\times \IT^n,\caC]=[X,C^0(\IT^n,\caC)]~,
    \end{equation}
    we see that $C^0(\IT^n,\caC)$ classifies $\caC$-objects on a trivial $n$-torus bundle.
    Note that, for $n=1$, we obtain the maps from $X$ into the free loop space $L\caC\coloneqq C^0(S^1,\caC)$ of $\caC$.
    
    If $P$ is non-trivially fibered over $X$, then the above discussion holds only locally. In particular, the fibers can only be identified with $\IT^n$ up to an action of $\sfU(1)^n$, and we replace the mapping space $[\IT^n,\caC]$ with the homotopy quotient space\footnote{This is not an ordinary quotient since the action of $\sfU(1)^n$ has fixed points (on constant maps). Technically, such homotopy quotients can be realised in the category of topological spaces by using (topologically realised) classifying spaces to remove such fixed points. These details do not concern us, however: for our purposes, it is much more natural to model them in terms of higher group(oid)s as we explain.}
    \begin{equation}\label{eq:gen_cyclic_loop_space}
        C^0(\IT^n,\caC)\sslash\sfU(1)^n~.
    \end{equation}
    In the case $n=1$, this quotient is also called the \emph{cyclic loop space}, and the mapping
    \begin{equation}\label{eq:reduction_functor}
        [P,\caC]\rightarrow [X,C^0(\IT^n,\caC)\sslash\sfU(1)^n]
    \end{equation}
    is also called \emph{double dimensional reduction}; see~\cite{Fiorenza:2016ypo,Fiorenza:2016oki} and also the corresponding $n$Lab page\footnote{\url{https://ncatlab.org/nlab/show/geometry+of+physics+--+fundamental+super+p-branes}} for further details. We note that there is again an adjunction between the reduction functor~\eqref{eq:reduction_functor} and the corresponding oxidation functor. 
    
    \paragraph{The group $\sfTD_1$.} In our double dimensional reduction, we will have to restrict ourselves to the zero modes along the fibers. This restriction, however, allows for much computational simplification, as we will see in the following.
    
    For one-dimensional T-duality, we are interested in the case $n=1$ and $\caC=\sfB\sfB\sfU(1)$, the classifying space for abelian bundle gerbes\footnote{This is simply the strict 2-category with a single object, a single 1-cell and $\sfU(1)$ as its 2-cells.}. Recall that for any higher group $\sfG$, there is a homotopy equivalence between $L\sfB\sfG$ and the homotopy quotient $\sfB\sfG\sslash\sfG$, which can be modeled by the corresponding action groupoid. In the case of $\sfB\sfB\sfU(1)$, we thus identify
    \begin{equation}
        \begin{aligned}
            L\sfB\sfB\sfU(1)&\cong \sfB\sfB\sfU(1)\times \sfB\sfU(1)
            \\
            &\cong \big(\sfU(1)\times \sfU(1)\rightrightarrows \sfU(1)\rightrightarrows *\big)~.
        \end{aligned}
    \end{equation}
    The cyclic loop space $L\sfB\sfB\sfU(1)\sslash\sfU(1)$ is again a homotopy quotient, and we arrive at
    \begin{equation}
        \begin{aligned}
            \big(L\sfB\sfB\sfU(1)\sslash\sfU(1)\big)&\cong \sfB\sfU(1)\times L\sfB\sfB\sfU(1)
            \\
            &\cong \big(\sfU(1)\times\sfU(1)\times \sfU(1)\rightrightarrows \sfU(1)\times \sfU(1)\rightrightarrows *\big)~.
        \end{aligned}
    \end{equation}
    We note that the latter space is the classifying space of a smooth Lie 2-group $\scG$,
    \begin{equation}
        \big(L\sfB\sfB\sfU(1)\sslash\sfU(1)\big)\cong \sfB\scG~,
    \end{equation}
    where 
    \begin{equation}
        \scG=\sfB\sfU(1)\times \sfU(1)\times \sfU(1)~.
    \end{equation}
    Replacing the groups $\sfU(1)$ with 2-groups $\IR\times \IZ\rightrightarrows \IR$ and taking the resulting crossed module of Lie groups, we arrive at the complex
    \begin{equation}
        \sfTD_1=\big(\sfU(1)\times \IZ^2\xrightarrow{~\sft~} \IR^2,\acton \big)~,
    \end{equation}
    which underlies the 2-group $\sfTD_1$.
    We note that the action $\acton$ has to be inferred to be that for $\sfTD_1$.
    
    \paragraph{Recovering the group $\sfTD_n$.} We now readily iterate the above procedure. Because $\scG$ is a Cartesian product of groups, we can do this for each factor separately. Above we saw that, in each step,
    \begin{equation}
        \sfB\sfB\sfU(1)\rightarrow \sfB(\sfB\sfU(1)\times \sfU(1)\times \sfU(1))~.
    \end{equation}
    Similarly, it is easy to see that
    \begin{equation}
        \sfB\sfU(1)\rightarrow \sfB\sfU(1)\times \sfU(1)\times\sfB\sfU(1)~,
    \end{equation}
    where the last factor, coming from the $\sfU(1)$-action of the cyclification, acts on the second, producing an image in the first. This is no longer the classifying space of a Lie group but a Lie groupoid whose set of objects is $\sfU(1)$. We can consistently truncate to the first factor $\sfB\sfU(1)$ in order to retain a group. Iterating this procedure $n$ times and replacing $\sfU(1)$-factors with $\IZ\rightarrow \IR$, we arrive at 
    \begin{equation}
        \sfTD_n=\big(\sfU(1)\times \IZ^{2n}\xrightarrow{~\sft~} \IR^{2n},\acton \big)~.
    \end{equation}
    Again, deriving the correct action $\acton$ requires substantially more work.
    
    \paragraph{Lie Groupoids.} In the iteration procedure above, we truncated the part obtained from $\sfB\sfU(1)$-factors to preserve the 2-group structure. After two dimensional reductions, however, we ought to keep these groupoid parts. This is intuitively clear as a 2-form $B$-field, dimensionally reduced twice, will give rise to scalar fields, which should take values in the space of objects of this groupoid. We will develop this point in the following section. A further dimensional reduction can then be captured by an augmented groupoid, and we will discuss this later in \cref{sec:non_geometric}.
    
    \subsection{Lie 2-groupoid for T-duality with T-folds}\label{ssec:def_TDn_groupoid}
    
    \paragraph{Narain moduli space.} The Lie 2-groupoids arising in the dimensional reduction come with a manifold of objects, which will be the target space of additional scalar fields produced by the dimensional reduction. These have a 1-form field strength and correspond to $0$-branes from a string theory perspective.
    
    As is well-known, the moduli of the Riemannian metric and the Kalb--Ramond $B$-field on $\IT^n$ are given by the \emph{Narain moduli space}~\cite{Narain:1985jj}
    \begin{equation}
        M_n=\sfO(n,n;\IZ)~\backslash~\sfO(n,n;\IR)~/~\big(\sfO(n;\IR)\times \sfO(n;\IR)\big)\eqqcolon \sfO(n,n;\IZ)~\backslash~Q_n~,
    \end{equation}
    which has dimension $n^2$. As argued above, we have to replace the T-duality group $\sfO(n,n;\IZ)$ by $\sfGO(n,n;\IZ)$ to allow for general torus bundles. Correspondingly, we will work with the scalar manifold
    \begin{equation}
        GM_n=\sfGO(n,n;\IZ)~\backslash~\sfO(n,n;\IR)~/~\big(\sfO(n;\IR)\times \sfO(n;\IR)\big)\eqqcolon \sfGO(n,n;\IZ)~\backslash~Q_n~.
    \end{equation}

    \paragraph{Lie 2-groupoid $\scTD_n$.} We can replace the quotient $GM_n$ again by its action groupoid,
    \begin{equation}
        \sfGO(n,n;\IZ)\ltimes Q_n~\rightrightarrows~Q_n~.
    \end{equation}
    The advantage of this replacement is that $GM_n$ generically has non-trivial 1-cycles, while $Q_n$ is contractible\footnote{This is due to $\sfO(n;\IR)\times \sfO(n;\IR)$ being a maximal compact subgroup of $\sfO(n,n;\IR)$; the topology of a Lie group is essentially that of its maximal compact subgroup.}. We can then combine this Lie groupoid with the Lie 2-group $\sfTD_n$, and it is evident how to do this: we need to extend $\sfGO(n,n;\IZ)$ in the action groupoid by $\scGO(n,n;\IZ)$ and have it act diagonally on both $Q_n$ and $\sfTD_n$.
    
    This results in the Lie 2-groupoid $\scTD_n$ with the following 2-, 1-, and 0-cells:\footnote{Note that $\scTD_n$ can be thought of as (i.e.~it is equivalent to) a bundle of 2-groups with fiber $\sfTD_n$ on the orbifold $Q_n/\sfGO(n,n;\IZ)$, up to the additional copy of $\IZ^{2n}$ in $(\scTD_n)_2$. Around a non-contractible cycle labeled by $g\in\sfGO(n,n;\IZ)$, the fiber $\sfTD_n$ undergoes a monodromy given by $g$. This construction is the direct categorified analogue of a bundle of groups on an orbifold $\Sigma/\Gamma$, which arises e.g.~in Yang--Mills-matter theories with gauge group $\sfG$ with a scalar field taking values in a manifold $\Sigma$, for which a discrete subgroup $\Gamma\subset\sfAut(\sfG)$ that acts on $\Sigma$ has been gauged.}
    \begin{equation}
        \begin{aligned}
            (\scTD_n)_2&=\sfGO(n,n;\IZ)\times \IZ^{2n}\times\IR^{2n}\times \IZ^{2n}\times \sfU(1)\times Q_n~,
            \\
            (\scTD_n)_1&=\sfGO(n,n;\IZ)\times \IR^{2n}\times Q_n~,
            \\
            (\scTD_n)_0&=Q_n~.
        \end{aligned}
    \end{equation}
    The 2- and 1-morphisms read as
    \begin{equation}
        \begin{gathered}
            (g,\xi,q)\xLeftarrow{~(g,\xi,z,m,\phi,q)~}(g,\xi-m,q)~,
            \\
            (q)\xleftarrow{~(g,\xi,q)~}(g^{-1}q)~,
        \end{gathered}
    \end{equation}
    and they compose vertically and horizontally according to
    \begin{equation}
        (g,z_1,\xi,m_1,\phi_1,q)\circ(g,z_2,\xi-m_1,m_2,\phi_2,q)\coloneqq(g,z_1+z_2,\xi,m_1+m_2,\phi_1+\phi_2,q)
    \end{equation}
    and
    \begin{equation}
        \begin{aligned}
            &(g_1,z_1,\xi_1,m_1,\phi_1,q)\otimes(g_2,z_2,\xi_2,m_2,\phi_2,g_1^{-1}q)
            \\
            &\hspace{2cm}\coloneqq\Big(g_1g_2,z_1+g_1z_2,\xi_1+g_1\xi_2,m_1+g_1m_2,
            \\
            &\hspace{3cm}\phi_1+(-1)^{|g_1|}\phi_2-\langle \xi_1,g_1m_2\rangle+m_2^\rmT \rho_L(g_1)\xi_2+z_1^\rmT \eta g_1\xi_2,q\Big)~.
        \end{aligned}
    \end{equation}
    Due to the expression in the last component, horizontal composition is no longer associative, and we have the associator
    \begin{equation}
        \begin{aligned}
            &\sfa(g_1,\xi_1,q_1;g_2,\xi_2,q_2;g_3,\xi_3,q_3)
            \\
            &\hspace{0.5cm}=\Big(\sfa(g_1,g_2,g_3),(\sfid_{\xi_1}\otimes\Upsilon_{\sfTD_n}^{-1}(g_1,\xi_2,g_2\xi_3))\circ(\sfid_{\xi_1}\otimes\sfid_{g_1\xi_2}\otimes \Upsilon_{\scGO(n,n;\IZ)}(g_1,g_2,\xi_3))\Big)
            \\
            &\hspace{0.5cm}=\Big(\sfa(g_1,g_2,g_3),
            (\xi_1+g_1(\xi_2+g_2\xi_3),0,\xi_2^\rmT \rho_L(g_1)g_2\xi_3)
            \\
            &\hspace{2.5cm}\circ(\xi_1+g_1\xi_2+g_1g_2\xi_3,0,\tfrac12\xi_3^\rmT \sigma_L(g_1,g_2)\xi_3+\tfrac12\rmdiag(\sigma_L(g_1,g_2))^\rmT \xi_3),q_1\Big)
            \\
            &\hspace{0.5cm}=\Big(\sfa(g_1,g_2,g_3),\xi_1+g_1(\xi_2+g_2\xi_3),0,
            \\
            &\hspace{3.5cm}\xi_2^\rmT \rho_L(g_1)g_2\xi_3+\tfrac12\xi_3^\rmT \sigma_L(g_1,g_2)\xi_3+\tfrac12\rmdiag(\sigma_L(g_1,g_2))^\rmT \xi_3),q_1\Big)~,            
        \end{aligned}
    \end{equation}
    where $\sfa(g_1,g_2,g_3)$ is the associator in $\scGO(n,n;\IZ)$ defined in~\eqref{eq:assoc_GOnnZ}, and $q_{i+1}=g_i^{-1} q_i$, cf.~\eqref{eq:associator_semidirect_product}. Note that horizontal composition is still unital.
    
    \paragraph{T-duality as a gauge symmetry.} In our Lie 2-groupoid $\scTD_n$, the T-duality group $\sfGO(n,n;\IZ)$ appears explicitly on par with the gauge group $\IR^{2n}/\IZ^{2n}$. The T-duality group therefore is to be regarded as a gauge group. Because the group is discrete, there are no associated gauge potentials, but there are associated 2-groupoid bundle isomorphisms, effectively quotienting the space of inequivalent principal 2-groupoid bundles, while at the same time giving rise to new, topologically non-trivial bundles.        
    
    \subsection{T-duality correspondences involving T-folds}
    
    A T-duality correspondence between T-folds is now a principal $\scTD_n$-bundle, and we develop the cocycle description of such a bundle in the following.
    
    \paragraph{Generalities on higher groupoid bundles.} Just as in the case of principal 2-bundles, higher groupoid bundles are conveniently described in terms of cocycles, i.e.\ functors from the Čech groupoid of a surjective submersion to the higher groupoid itself.
    
    Let $M$ be a manifold and $\sigma\colon Y\rightarrow M$ a surjective submersion, and let $\check\scC(Y\rightarrow M)$ be the corresponding Čech groupoid, cf.~\eqref{eq:def_Cech_groupoid}, trivially regarded as a (strict) higher Lie groupoid. Let $\scG$ be a (higher) Lie groupoid, which we call the \emph{structure groupoid}. A \emph{(higher) groupoid bundle} over $M$ subordinate to the surjective submersion $\sigma$ with structure groupoid $\scG$ is then an (appropriately defined) higher functor 
    \begin{equation}
        \Phi\colon\check\scC(Y\rightarrow M)\rightarrow \scG~.
    \end{equation}
    Groupoid bundle isomorphisms are given by (higher) natural transformations between two such functors, and the higher isomorphisms are then identified with modifications and higher transfors. Note that a groupoid bundle whose structure groupoid is the delooping $\sfB\sfG$ of a Lie group $\sfG$ is just a principal $\sfG$-bundle.
    
    The definition of higher functors for Lie $n$-groupoids for $n>2$ is technically very involved, and it is a good idea to switch to the perspective of quasi-groupoids defined in terms of Kan simplicial manifolds, cf.~\cref{app:quasi-groupoids} and e.g.~\cite{Jurco:2016qwv}. The same holds for the definition of a differential refinement.
    
    We remark that from a physical perspective, (1-)groupoid bundles are the geometric structures underlying gauged sigma models.
    
    \paragraph{Cocycle description of $\scTD_n$-bundles.} We now specialise the above abstract discussion to the case of the structure groupoid $\scTD_n$. This leads to the groupoid extension of the discussion in~\cite{Jurco:2014mva}. A $\scTD_n$-bundle over a manifold $M$ subordinate to the surjective submersion $\sigma\colon Y\rightarrow M$ is then a weak 2-functor\footnote{see~\cref{app:2-groupoid_basics} for definitions} $\Phi\colon\check\scC(Y\rightarrow M)\rightarrow \scTD_n$. Such a functor is encoded in the data
    \begin{equation}
        \begin{aligned}
            (g,z,\xi,m,\phi,q)
            &\in C^\infty(Y^{[3]},\sfGO(n,n;\IZ)\times \IZ^{2n}\times\IR^{2n}\times \IZ^{2n}\times \sfU(1)\times Q_n)~,
            \\
            (g,\xi,q)&\in C^\infty(Y^{[2]},\sfGO(n,n;\IZ)\times \IR^{2n}\times Q_n)~,
            \\
            q&\in C^\infty(Y,Q_n)~,
        \end{aligned}
    \end{equation}
    which define the natural isomorphism $\Phi_2$, the functor $\Phi_1$, and the function $\Phi_0$, respectively. On $Y^{[2]}$, we then have 
    \begin{equation}
        q_j=g_{ij}^{-1}q_i
    \end{equation}
    with $(ij)\in Y^{[2]}$, while on $Y^{[3]}$, we deduce that 
    \begin{equation}
        e_{ijk}\circ (\sfid_{d_{ij}}\otimes \sfid_{d_{jk}})=\sfid_{d_{ik}}\circ e_{ijk}
    \end{equation}
    for $d_{ij}=(g_{ij},\xi_{ij},q_{i})$, $e_{ijk}=(g_{ik},z_{ijk},\xi_{ik},m_{ijk},\phi_{ijk},q_{i})$, and $(ijk)\in Y^{[3]}$ or, equivalently,
    \begin{equation}\label{eq:groupoid_cocycle_relations_Y3}
        \begin{aligned}
            g_{ik}&=g_{ij}g_{jk}~,
            \\
            \xi_{ik}&=m_{ijk}+\xi_{ij}+g_{ij}\xi_{jk}~.
        \end{aligned}
    \end{equation}    
    Finally, on $Y^{[4]}$, we have
    \begin{equation}
        e_{ikl}\circ (e_{ijk}\otimes \sfid_{d_{kl}})=e_{ijl}\circ (\sfid_{d_{ij}}\otimes e_{jkl})\circ \sfa(d_{ij},d_{jk},d_{kl})
    \end{equation}
    or, equivalently,
    \begin{equation}
        \begin{aligned}
            z_{ijk}+z_{ikl} &= z_{ijl}+g_{ij}z_{jkl}
            \\
            &\hspace{0.5cm}+\frac{(-1)^{|g_{ik}|}}{2}g_{ik}\eta\,\rmdiag(\sigma_L(g_{ij},g_{jk}))
            +\frac{(-1)^{|g_{il}|}}{2}g_{il}\eta\,\rmdiag(\sigma_L(g_{ik},g_{kl}))
            \\
            &\hspace{0.5cm}-\frac{(-1)^{|g_{il}|}}{2}g_{il}\eta\,\rmdiag(\sigma_L(g_{ij},g_{jl}))
            -\frac{(-1)^{|g_{jl}|}}{2}g_{il}\eta\,\rmdiag(\sigma_L(g_{jk},g_{kl}))~,
            \\
            m_{ijk}+m_{ikl} &= m_{ijl}+g_{ij}m_{jkl}~,
            \\
            \phi_{ijk} + \phi_{ikl} &=\phi_{ijl} + (-1)^{|g_{ij}|}\phi_{jkl}-\langle \xi_{ij}, g_{ij}m_{jkl}\rangle+
            m_{jkl}^\rmT \rho_L(g_{ij})\xi_{jl} +
            \xi_{jk}^\rmT \rho_L(g_{ij})g_{jk}\xi_{kl}
            \\
            &\hspace{0.5cm}+\tfrac12 \xi_{kl}^\rmT \sigma_L(g_{ij},g_{jk})\xi_{kl}+\tfrac12\rmdiag(\sigma_{L}(g_{ij},g_{jk}))^\rmT \xi_{kl}-z_{ijk}^\rmT \eta g_{ik}\xi_{kl}~,
        \end{aligned}
    \end{equation}
    where the functions $\rho_L$ and $\sigma_{L}$ were defined in~\eqref{eq:def_rho_L} and~\eqref{eq:def_sigma_L}, respectively. Note that the second equation is automatically satisfied due to~\eqref{eq:groupoid_cocycle_relations_Y3}.
    
    Let us now differentially refine this topological\footnote{We note that the ``scalar part'' of a topological groupoid bundle can already be considered as a part of the differential refinement. This is certainly more sensible from a physical perspective, where scalar fields arise from dimensionally reducing gauge potentials.} cocycle data. The adjusted cocycle data only seem to exist if 
    \begin{subequations}\label{eq:diff_refined_cocycles2}
        \begin{equation}\label{eq:z-rel}
            z_{ijk}=\frac{(-1)^{|g_{ik}|}}{2}g_{ik}\eta\,\rmdiag(\sigma_L(g_{ij},g_{jk}))~.
        \end{equation}
        In this case, we have 1-~and 2-forms
        \begin{equation}
            \Lambda\in \Omega^1(Y^{[2]})~,~~~
            A\in \Omega^1(Y,\IR^{2n})~,~~~
            B\in \Omega^2(Y)~,
        \end{equation}
        which satisfy the gluing relations
        \begin{equation}
            \begin{aligned}
                \Lambda_{ik} &= (-1)^{|g_{ij}|}\Lambda_{jk} + \Lambda_{ij}+\rmd \phi_{ijk}-\langle A_i,m_{ijk}\rangle
                \\
                &\hspace{1cm}+\tfrac12 \rmd \xi_{jk}\rho(g_{ij})\xi_{jk}-A_k^\rmT \eta g_{kj}\eta \rho(g_{ij})\xi_{jk}+\tfrac12\rmd\left(\xi_{jk}^\rmT \rho_L(g_{ij})\xi_{jk}\right)
                \\
                &= (-1)^{|g_{ij}|}\Lambda_{jk} + \Lambda_{ij}+\rmd \phi_{ijk}-\langle A_i,m_{ijk}\rangle
                \\
                &\hspace{1cm}+\rmd \xi_{jk}\rho_L(g_{ij})\xi_{jk}-A_k^\rmT \eta g_{kj}\eta \rho(g_{ij})\xi_{jk}~,
                \\
                A_j &= g_{ij}^{-1}A_i + g_{ij}^{-1}\rmd \xi_{ij}~,
                \\
                (-1)^{|g_{ij}|}B_j &= B_i + \rmd \Lambda_{ij} + \langle\rmd A_i,\xi_{ij}\rangle-\tfrac12A_j^\rmT \rho(g_{ij})A_j~.
            \end{aligned}
        \end{equation}
    \end{subequations}        
    We note that, for $g_{ij}\in \sfGL(n;\IZ)\subset \sfO(n,n;\IZ)$, the relation~\eqref{eq:z-rel} is automatically satisfied for $z_{ijk}=0$, and we recover the cocycle relations for differentially refined principal $\sfTD^\ltimes_n$-bundles.
    
    \subsection{Example of a T-fold}
    
    \paragraph{3-dimensional example.}
    Let us consider again the example of the three-dimensional nilmanifold defined in~\eqref{eq:def_nil_1} and~\eqref{eq:def_nil_2} with $k=0$ and T-dualise along the $y$- and $z$-directions. In the T-correspondence, our base manifold $X$ is then simply the circle parameterised by $x$, and we consider a principal $\scTD_n$-bundle over $S^1$ subordinate to the cover $\IR\rightarrow S^1$. 
    
    Due to dimensionality, there can be no non-trivial triple intersections, and we can thus set $z_{ijk}=m_{ijk}=\phi_{ijk}=0$. Since principal torus bundles over a circle are topologically trivial, we can trivialise $\check g$ and $\hat g$. This leaves us with the scalars $(q_i)\colon Y\to Q_2$ and the transition functions $(g_{ij})\colon Y^{[2]}\rightarrow \sfGO(2,2;\IZ)$ with the only non-trivial cocycle conditions being
    \begin{equation}
        g_{ij}\acton q_j=q_i\eand g_{ij}g_{jk}=g_{ik}
    \end{equation}
    for all $(ij)\in Y^{[2]}$ and $(ijk)\in Y^{[3]}$, together with the evenness condition
    \begin{equation}\label{eq:g-even}
        g_{ik}\eta\,\rmdiag(\sigma_L(g_{ij},g_{jk}))\in2\IZ^4
    \end{equation}
    required for adjustment.
    Modulo this evenness condition, these data are the same as those defining a groupoid bundle over $S^1$ with structure groupoid the action groupoid corresponding to the action $\sfGO(2,2;\IZ)\curvearrowright Q_2$. This, in turn, is the same as a map of orbifolds $q\colon S^1\rightarrow GM_2$.
    
    Suppose now that the map $q$ factors through a map 
    \begin{equation}
        q'\colon S^1\rightarrow \fro(2;\IZ) \setminus Q_2~,
    \end{equation}
    where $\fro(2;\IZ)$ denotes the (abelian) subgroup of $\beta$-transformations in $\sfGO(2,2;\IZ)$. In terms of cocycle data, the maps $q_i$ are then glued together by transformations $g_{ij}\in \fro(2,\IZ)$. In this case, \eqref{eq:g-even} holds automatically.
    
    We recall that $Q_2\coloneqq\sfO(2,2;\IR)/\sfO(2;\IR)^2$ is contractible and, furthermore, diffeomorphic to $\IR^4$. This manifold is identified with the four scalars arising from the dimensional reduction of the metric and the Kalb--Ramond field:
    \begin{equation}
        \phi_{g_{yy}}=0~,~~~
        \phi_{g_{xy}}=0~,~~~
        \phi_{g_{zz}}=1~,
        \eand
        \phi_B=\ell x~.
    \end{equation}
    The $\beta$-transformation then corresponds to the matrix
    \begin{equation}
        g_{x+1,x}=\begin{pmatrix}
            1 & 0 & 0 & 0 
            \\
            0 & 1 & 0 & 0 
            \\
            0 & \ell & 1 & 0
            \\
            -\ell & 0 & 0 & 1
        \end{pmatrix}~,
    \end{equation}
    and it acts on $\phi_B$ according to $\phi_B\rightarrow \phi_B+\ell$, leaving the other scalars unmodified.
    
    After a T-duality transformation along both directions, the $\beta$-transformation turns into a $B$-transformation, and the structure of the bundle can be encoded in an ordinary abelian gerbe over the nilmanifold with the transition $1$-forms $\Lambda$ encoding the gluing.
    
    \paragraph{Special $\scTD_n$-bundles.} Consider a $\scTD_n$-bundle that is isomorphic to a $\scTD_n$-bundle described by a cocycle whose underlying cocycle is such that the $(g_{ij})$ are purely $\beta$-transformations. We will call such a bundle a \emph{special $\scTD_n$-bundle}. In this case, the cocycle relations simplify in that $\sigma_L(g_1,g_2)$ vanishes so that $z_{ijk}$ can be put to zero.\footnote{The same holds when the $(g_{ij})$ are purely $B$-transformations, purely $A$-transformations, purely factorised dualities, or purely $G$-transformations.} Moreover, the cocycle relations for the components $\hat \xi_{ij}$ and $\hat A_i$ in the cocycle data are precisely the same as those in the case of trivial $g_{ij}$. We thus recover an ordinary principal $\sfU(1)^n$-bundle with connection over the base manifold.
    
    \paragraph{General picture.} In general, a $\scTD_n$-groupoid bundle $\scP$ describes a T-duality correspondence between two T-folds. If the $\scTD_n$-groupoid bundle is special, then one of the T-backgrounds in the correspondence will descend from a $\scTD_n$-groupoid bundle in which all $g_{ij}$ are merely $B$-transformations and thus fully geometrical. Generically, however, the T-dual of a T-fold does not have to be geometric.
    
    
    \subsection{Half-geometric T-correspondences as principal \texorpdfstring{$\sfTD^{\text{$\tfrac12$geo}}_n$}{TDnhalfgeometric}-bundles}
    
    Let us briefly compare our construction with that of~\cite{Nikolaus:2018qop}, where the topological part of T-correspondences involving geometric T-backgrounds and T-folds was described in terms of higher geometry.
    
    \paragraph{Half-geometric T-correspondences.} We follow the nomenclature of~\cite{Nikolaus:2018qop} and call a T-correspondence involving a geometric background of type $F^1$ and a T-fold background \emph{half-geometric}. Topological T-backgrounds of type $F^1$ have been shown to correspond to functors represented by the 2-group $\sfTB^\text{F1}_n$~\cite{Nikolaus:2018qop}. This 2-group is given by the semidirect product of the strict 2-group $\sfTB^\text{F2}_n$ with the abelian subgroup of $\beta$-transformations $\fro(n;\IZ)\subset \sfGO(n,n;\IZ)$,
    \begin{equation}
        \sfTB^\text{F1}_n\coloneqq \fro(n;\IZ)\ltimes \sfTB^\text{F2}_n~.
    \end{equation}
    For the definition of the action and the resulting semidirect product, see~\cite[App.~A.4]{Nikolaus:2018qop}; the corresponding cocycles of (topological) principal $\sfTB^\text{F1}_n$-bundles were also given in~\cite{Nikolaus:2018qop}.
    
    A half-geometric T-correspondence can then be described by a principal $\sfTD^{\text{$\tfrac12$geo}}_n$-bundle, where the structure 2-group is defined as 
    \begin{equation}\label{eq:TD_half_geo}
        \sfTD^{\text{$\tfrac12$geo}}_n\coloneqq \frso(n;\IZ)\ltimes \sfTD_n ~.
    \end{equation}
    The left-leg projection is equivariant in a particular sense and induces a map $\hat p\colon\sfTD^{\text{$\tfrac12$geo}}_n\rightarrow \sfTB^\text{F1}_n$. The main result of~\cite{Nikolaus:2018qop} is that this map is a bijection and that every $F^1$-background is the image of the left-leg projection of a principal $\sfTD^{\text{$\tfrac12$geo}}_n$-bundle.
    
    \paragraph{Comparison to $\scGO(n,n;\IZ)\ltimes \sfTD_n$.} In our approach, the group $\sfTD^{\text{$\tfrac12$geo}}_n$ should be compared to the semidirect product 2-group
    \begin{equation}
        \scGO(n,n\IZ)\ltimes \sfTD_n~,
    \end{equation}
    where the relevant group action was defined in~\eqref{eq:GO-action}. This 2-group is the group of morphisms with source a specific object $q\in Q_n$ in the Lie 2-groupoid $\scT\scD_n$. There is now an evident inclusion
    \begin{equation}
        \sfTD^{\text{$\tfrac12$geo}}_n\coloneqq \frso(n;\IZ)\ltimes \sfTD_n\hookrightarrow\scGO(n,n;\IZ)\ltimes \sfTD_n~.
    \end{equation}
    
    We note that a differential refinement of an $F^1$-background automatically induces scalar fields; in the presence of these, the 2-group $\sfTD^{\text{$\tfrac12$geo}}_n$ is not large enough to accommodate all required transformations. Conversely, however, it is clear that any topological $\sfTD^{\text{$\tfrac12$geo}}_n$-bundle embeds into a topological $\scTD_n$-bundle. By~\cite[Thm.~4.2.2.]{Nikolaus:2018qop}, our construction certainly provides freedom to capture the topological part of any $F^1$-T-background, and there is a left-leg projection from principal $\sfTD^{\text{$\tfrac12$geo}}_n$-bundles to principal $\sfTB^\text{F1}_n$-bundles, cf.~again~\cite{Nikolaus:2018qop}.

    \section{T-duality with non-geometric backgrounds}\label{sec:non_geometric}
    
    We can now complete the final step, which consists in generalising our description of T-duality to non-geometric T-backgrounds, i.e.~T-backgrounds that are not even locally geometric. It is well known that Kaluza--Klein reductions with duality twists or on doubled twisted tori gives rise to $R$-flux~\cite{Dabholkar:2002sy,Hull:2009sg}. In order to describe general non-geometric backgrounds, we will thus have to incorporate non-trivial $R$-fluxes into our picture.
    
    \subsection{From Kaluza--Klein reduction to the tensor hierarchy}
    
    \paragraph{Adjustments and tensor hierarchies.} As mentioned in~\cref{app:higher_principal_bundles}, connections on higher principal bundles require an adjustment, which consists of an additional datum on the gauge group. This additional datum is readily obtained in the case in which the higher gauge algebra is derived from the differential graded vector space underlying a \emph{tensor hierarchy}, cf.~\cite{Borsten:2021ljb}. Of interest to us is the fact that Kaluza--Klein reductions lead to gauged supergravities in which non-geometric fluxes essentially define (at least parts of) the embedding tensor~\cite{Aldazabal:2011nj,Geissbuhler:2011mx,Grana:2012rr}. A general tensor hierarchy for double field theory was then defined in~\cite{Hohm:2013nja}. We note that also~\cite{Alfonsi:2020nxu} argues that the gauge potential forms arising in a Kaluza--Klein interpretation of double field theory should be arranged into a tensor hierarchy.
    
    \paragraph{Tensor hierarchy for $\sfGO(n,n;\IZ)$.} Recall from Chapter~5 that  the \emph{embedding tensor} is a linear map
    \begin{equation}
        \Theta\colon V_{-1}\rightarrow V_0
    \end{equation}
    into the Lie algebra $V_0=\frg$ of the global symmetry group $\sfG$ from a representation $V_{-1}$
    satisfying the quadratic closure constraint
    \begin{equation}\label{eq:closure_constraint}
        [\Theta(v_1),\Theta(v_2)]=\Theta(\Theta(v_1)v_2)~,
    \end{equation}
    as well as the representation constraint.
    
    In our case, we set $\sfG=\sfGO(n,n;\IZ)$ and $V_{-1}=\IR^{2n}$, the space in which the gauge potential 1-forms take values. Correspondingly, the embedding tensor is a map $\Theta\colon\IR^{2n}\rightarrow \fro(n,n;\IR)$, which exponentiates to a map
    \begin{equation}
        \bar\Theta\colon\IR^{2n}\rightarrow \sfGO(n,n;\IR)~.
    \end{equation}
    The map $\Theta$ needs to satisfy the closure constraint~\eqref{eq:closure_constraint} as well as the representation constraint. We note that a generic map $\Theta$ is an element in the tensor product of the fundamental and the adjoint representation of $\fro(n,n;\IZ)$, which decomposes as follows:
    \begin{equation}
        \tyng(1,1) \otimes \tyng(1) = \tyng(1)\oplus \tyng(1,1,1) \oplus \tyng(2,1)~.
    \end{equation}
    Usually, the representation constraint is selected by requiring supersymmetry. Here, we have good heuristic reasons to impose the condition
    \begin{equation}\label{eq:representation_constraint}
        \Theta\in \tyng(1,1,1)
    \end{equation}
    as we will argue now. Recall that $R$-flux corresponds to the 3-form flux $H$ wrapped around three of the $2n$ directions in the $2n$ torus fiber directions, subject to additional constraints reflecting the fact that the $2n$ coordinates cannot be regarded as geometric simultaneously. Under dimensional reduction, we expect the $R$-form fluxes for $(n+1)$-dimensional torus fibers to originate from the $R$- and $Q$-fluxes of $n$-dimensional torus fibers, and this branching rule essentially fixes~\eqref{eq:representation_constraint}. We then have the following association of fluxes to $\sfGO(n,n;\IZ)$-representations:
    \begin{equation}
        \mbox{$f$-flux}~\leftrightarrow~\tyng(1)~,~~~\mbox{$Q$-flux}~\leftrightarrow~\tyng(1,1)~,~~~\mbox{$R$-flux}~\leftrightarrow~\tyng(1,1,1)~.
    \end{equation}
    Here, ``$f$-flux'' refers to the curvatures of $\hat A$ and $\check A$, taking values in $\IZ^{2n}$. It corresponds to $H$ wrapped around one of the $2n$ directions in the $2n$-torus fibers and forms the fundamental representation, while the $Q$-flux corresponds to $H$ wrapped around two of the $2n$ directions in the $2n$-torus fibers and, hence, to the adjoint representation, namely the linearisation of the non-linear adjoint representation of $\sfGO(n,n;\mathbb Z)$ on itself.
    
    A further check of our choice~\eqref{eq:representation_constraint} comes from considering the embedding of T-duality into U-duality, for which supersymmetric arguments fully determine the representations. For the embedding $\sfE_{7(7)}\supset\sfSO(6,6)\times\sfSL(2)$, the representation $\mathbf{912}$ of the embedding tensor for $\sfE_{7(7)}$ decomposes as
    \begin{equation}
        \textbf{912} \to (\textbf{12},\textbf2)\oplus (\textbf{220},\textbf2)\oplus \dotsb~,
    \end{equation}
    and we only find the representations
    \begin{equation}
        \tyng(1,1,1)=\textbf{220}\eand \tyng(1)=\textbf{12}
    \end{equation}
    of $\sfSO(6,6)$, but not
    \begin{equation}
        \tyng(2,1)=\textbf{560}~.
    \end{equation}
    
    Note that we also have a second representation space $V_{-2}=\sfU(1)\times \IZ^{2n}$, and compatibility with this representation requires that the images of integer vectors should be a symmetry of the gauge 2-groupoid $\sfTD_{2n}$. In particular, we need to impose that
    \begin{equation}
        \rmim(\bar\Theta)\subset \sfGO(n,n;\IZ)~.
    \end{equation}
    We will denote the set of embedding tensors satisfying the quadratic and linear representation constraints by $\bar R_n$ and the subset that further satisfies this integrality constraint by $R_n\subset\bar R_n$.
    
    Exponentiation of the map $\Theta$ implies that the image of $\bar\Theta$ lies in $\sfSO^+(n,n;\IZ)$, the connected component of $\sfGO(n,n;\IZ)$ containing the identity.
    
    Finally, we will restrict ourselves in this thesis to the case of ungauged (super)gravity theories, i.e.~T-background configurations in which the 1-form potentials are purely abelian. This implies that the image of the embedding tensor $\Theta$ is an abelian Lie algebra. Hence, in the following, $\Theta$ and $\bar\Theta$ will always be Lie algebra and Lie group homomorphisms, respectively, with domains the abelian Lie algebra $\IR^{2n}$ and Lie group $\IR^{2n}$, respectively.
    
    \paragraph{Relation to $R$-fluxes.} 
    
    We can now identify the discrete moduli in $\bar\Theta$ and relate them to $R$-fluxes. For $n=0$ and $n=1$, there are unique group homomorphisms \(\bar\Theta\colon\IZ^0\to\sfSO^+(0,0;\IZ)\cong1\) and $\bar\Theta\colon\IZ^2\to\sfSO^+(1,1;\IZ)\cong1$. Hence, there are no $R$-fluxes in either case.
    
    For $n=2$, there exist group homomorphisms $\bar\Theta\colon\IZ^4\to\sfSO^+(2,2;\IZ)$, for which we must check the representation constraint. Infinitesimally, $\Theta\colon\IR^4\to\fro(2,2;\IR)$ forms a Lie algebra homomorphism. The image of this linearisation is an abelian Lie subalgebra of \(\fro(2,2;\IR)\), whose dimension can be at most $2$. If the dimension of the image is \(0\), this corresponds to trivial $R$-charge. If the dimension is \(1\), it is straightforward to check that the representation constraint fails. If the dimension is \(2\), the image corresponds to a pair of mutually commuting translations in \(\IR^{2,2}\); the requirement that the exponentiated rotations be integral implies that one of them can be taken to be along a space-like 2-plane and the other along a time-like 2-plane. Thus, the putative $R$-flux can be put in a standard form, and the representation constraint can then be easily checked to fail. Hence, for \(n=2\) there are no non-trivial $R$-fluxes either.
    
    In dimensions $n\ge3$, however, non-trivial $R$-fluxes exist. One sufficient ansatz is to consider group homomorphisms
    \begin{equation}
        \IZ^n\to\fro(n;\IZ)~,
    \end{equation}
    where \(\fro(n;\IZ)\) is the abelian group of \(n\times n\) antisymmetric integer matrices, that are given by pairing with an \(n\)-dimensional totally antisymmetric integer 3-tensor, i.e.\ an element of
    \begin{equation}
        \{f\colon\{1,\dotsc,n\}^3\to\mathbb Z\mid f_{ijk}=-f_{jik}=-f_{ikj}\}\cong\IZ^{\binom n3}~,	
    \end{equation}
    in which each of the \(n\) generators map to linearly independent elements of \(\fro(n;\IZ)\); this then defines a group homomorphism
    \begin{equation}
        \IZ^{2n}\to\fro(n;\IZ)\subset\sfGO(n,n;\IZ)
    \end{equation}
    in which elements of the other \(\IZ^n\) simply map to zero, and which manifestly satisfies the representation constraint.
    Thus, in $n$ dimensions the set of  $R$-fluxes contains at least \(\IZ^{\binom n3}\), which corresponds to the geometric cohomology 3-classes. Of course, there are many possible choices of the embedding $\fro(n;\IZ)\hookrightarrow\sfGO(n,n;\IZ)$ by conjugations; these correspond to the T-duality orbits of the geometric cohomology 3-classes.
    
    \subsection{The complete groupoid of T-duality}
    
    \paragraph{Motivation: augmented higher groupoids.} For $n\le2$, the dimensional reduction of the Kalb--Ramond $B$-field and its field strength $H$ creates 1-forms and scalars with corresponding curvature 2- and 1-forms. These are accounted for in the gauge Lie 2-groupoid $\scTD_n$. As evident from~\eqref{eq:direct_product_cohomology_decomposition}, dimensional reduction of $H$ with $n\ge3$ will produce $0$-forms, which, however, clearly cannot be seen as curvatures of non-existing $(-1)$-forms.
    
    The groupoid picture, however, suggests a resolution. We note that 2-forms and their 3-form curvatures are essentially encoded by the 2-cells of $\scTD_n$, while the 1-forms and their 2-form curvatures correspond to the 1-cells. The scalars are then encoded by the $0$-cells. We can regard the Lie 2-groupoid $\scTD_n$ as a Lie 2-quasi-groupoid, i.e.~a simplicial manifold satisfying the relevant Kan condition, cf.~\cref{app:quasi-groupoids}. In this context, there is the notion of augmented groupoid, which allows us to include $(-1)$-cells as the image of a single face map. Such an augmentation is quite natural: consider for example the Čech groupoid of a surjective submersion $\sigma\colon Y\rightarrow M$, cf.~\cref{app:higher_principal_bundles}. We can augment the nerve of the Čech groupoid by $M$ and obtain the augmented quasi-groupoid
    \begin{equation}\label{eq:augmented_Cech_nerve}
        \check\scC_\text{aug}(Y\rightarrow M)\ \coloneqq\ \left(\begin{tikzcd}
            \ldots \arrow[r,shift left,shift left,shift left] 
            \arrow[r,shift right,shift right,shift right]
            \arrow[r,shift left] 
            \arrow[r,shift right] & Y^{[3]} \arrow[r,shift left,shift left] 
            \arrow[r,shift right,shift right]
            \arrow[r] &
            Y^{[2]}\arrow[r,shift left] 
            \arrow[r,shift right] & Y \arrow[r] & M
        \end{tikzcd}\right)~.
    \end{equation}
    
    In order to capture all aspects of non-geometric T-duality, we evidently have to augment the Lie 2-groupoid $\scTD_n$ by $\bar R_n$, resulting in the augmented Lie 2-quasi-groupoid $\scTD^\text{aug}_n$. 
    
    We note that the sequence of reductions indeed terminates here: $0$-form curvatures do not reduce any further. Thus, the picture of augmented Lie groupoids is indeed sufficient for arbitrary $n$.
    
    \paragraph{Augmented Lie 2-quasi-groupoid $\scTD^\text{aug}_n$.} In a first step, we enlarge the space of 0-cells in the Lie 2-groupoid $\scTD_n$ from $Q_n$ to $Q_n\times R_n$ to incorporate the $0$-form field strengths.\footnote{Similar to $\scTD_n$, this 2-groupoid can be thought of as a bundle of 2-groups, with fiber $\sfTD_n$, on the disconnected orbifold $(Q_n\times R_n)/\sfGO(n,n;\IZ)$.} The only additional datum we need is an action of the semidirect product group $\sfGO(n,n;\IZ)\ltimes \IR^{2n}$ on the now enlarged space of scalars $Q_n\times R_n$, and we define\footnote{The semidirect product $\sfGO(n,n;\IZ)\ltimes \IR^{2n}$ is the usual one, i.e.~$(g_1,\xi_1)(g_2,\xi_2)=(g_1g_2,\xi_1+g_1\xi_2)$ for $g_{1,2}\in\sfGO(n,n;\IZ)$ and $\xi_{1,2}\in\IR^{2n}$.}
    \begin{equation}
        \begin{aligned}
            \acton\colon(\sfGO(n,n;\IZ)\ltimes \IR^{2n})\times (Q_n\times R_n)&\rightarrow (Q_n\times R_n)~,
            \\
            (g,\xi;q,r)&\mapsto\big((g\acton r)(\xi)gq,g\acton r\big)=\big(gr(g^{-1}\xi)q,g\acton r\big)~,
        \end{aligned}
    \end{equation}
    where the action \(g\acton r\) is defined as
    \begin{equation}\label{eq:action_on_r}
        (g\acton r)(\zeta)\coloneqq gr(g^{-1}\zeta)g^{-1}
    \end{equation}
    for \(\zeta\in\IZ^{2n}\). This is indeed a group action as one readily verifies by direct computation:
    \begin{equation}
        \begin{aligned}
            (g_1,\xi_1)\acton ((g_2,\xi_2)\acton (q,r))&=(g_1,\xi_1)\acton(g_2r(g_2^{-1}\xi_2)q,g_2\acton r)
            \\
            &=(g_1(g_2\acton r)(g_1^{-1}\xi_1)g_2r(g_2^{-1}\xi_2)q,g_1g_2\acton r)
            \\
            &=(g_1g_2r(g_2^{-1}g_1^{-1}\xi_1)r(g_2^{-1}\xi_2)q,g_1g_2\acton r)
            \\
            &=\Big(g_1g_2r\big((g_1g_2)^{-1}(\xi_1+g_1\xi_2)\big)q,g_1g_2\acton r\Big)
            \\
            &=((g_1,\xi_1)(g_2,\xi_2))\acton(q,r)~,
        \end{aligned}
    \end{equation}
    where we used that $r$ is a group homomorphism.
    
    To incorporate ``$(-1)$-form potentials,'' we then turn it into a Lie 2-quasi-groupoid by constructing its Duskin nerve~\cite{Duskin02simplicialmatrices}. The full augmented Lie 2-quasi-groupoid $\scTD^\text{aug}_n$ is then obtained by augmenting it with the space $\bar R_n$ representing the ``$(-1)$-form potentials.'' The underlying simplicial manifold is given by 
    \begin{equation}
        \scTD^\text{aug}_n\coloneqq\ \left(\begin{tikzcd}[cramped,column sep=small]
            \cdots \arrow[r,shift left,shift left,shift left] 
            \arrow[r,shift right,shift right,shift right]
            \arrow[r,shift left] 
            \arrow[r,shift right] & (\scTD^\text{aug}_n)_2\arrow[r,shift left,shift left] 
            \arrow[r,shift right,shift right]
            \arrow[r] &
            (\scTD^\text{aug}_n)_1\arrow[r,shift left] 
            \arrow[r,shift right] & (\scTD^\text{aug}_n)_0 \arrow[r] & (\scTD^\text{aug}_n)_{-1}
        \end{tikzcd}\right)
    \end{equation}
    with
    \begin{equation}
        (\scTD^\text{aug}_n)_0\coloneqq Q_n\times R_n
        \eand
        (\scTD^\text{aug}_n)_{-1}\coloneqq \bar R_n~.
    \end{equation}
    The remaining sets $(\scTD^\text{aug}_n)_i$ with $i\geq 1$ are those given by the Duskin nerve construction. This construction becomes rather technical; fortunately this intuitive picture of $\scTD^\text{aug}_n$ is sufficient for all our purposes.
    
    \subsection{\texorpdfstring{$\scTD^\text{aug}_n$}{TDnhat}-bundles}\label{ssec:TDnhat-bundles}
    
    \paragraph{Topological cocycles.} Because of the simplicity of the augmentation, we can directly extend the topological cocycles of $\scTD_n$-bundles to cocycles of $\scTD^\text{aug}_n$-bundles. These are encoded in augmented simplicial maps from the augmented Čech 2-quasi-groupoid to the augmented Lie 2-quasi-groupoid $\scTD^\text{aug}_n$. Explicitly, we have the data
    \begin{subequations}
        \begin{equation}
            \begin{aligned}
                (g,z,\xi,m,\phi,q,r)
                &\in C^\infty(Y^{[3]},\sfGO(n,n;\IZ)\times \IZ^{2n}\times\IR^{2n}\times \IZ^{2n}\times \sfU(1)\times Q_n\times R_n)~,
                \\
                (g,\xi,q,r)&\in C^\infty(Y^{[2]},\sfGO(n,n;\IZ)\times \IR^{2n}\times Q_n\times R_n)~,
                \\
                (q,r)&\in C^\infty(Y,Q_n\times R_n)~,
                \\
                r&\in C^\infty(M,\bar R_n)~,
            \end{aligned}
        \end{equation}
        which satisfy the relations
        \begin{equation}\label{eq:augmented_cocycle}
            \begin{aligned}
                r_i&=r~,
                \\
                (q_j,r_j)&=(g_{ij}^{-1},-g_{ij}^{-1}\xi_{ij})\acton(q_i,r_i)~,~~~(q_{ijk},r_{ijk})=(q_{ij},r_{ij})=(r_i,q_i)~,
                \\
                g_{ik}&=g_{ij}g_{jk}~,~~~g_{ijk}=g_{ik}~,
                \\
                \xi_{ik}&=\xi_{ij}+g_{ij}\xi_{jk}+m_{ijk}~,~~~\xi_{ijk}=\xi_{ik}~,
                \\
                m_{ijk}+m_{ikl} &= m_{ijl}+g_{ij}m_{jkl}~,
                \\
                \phi_{ijk} + \phi_{ikl} &=\phi_{ijl} + (-1)^{|g_{ij}|}\phi_{jkl}-\langle \xi_{ij}, g_{ij}m_{jkl}\rangle+
                m_{jkl}^\rmT \rho_L(g_{ij})\xi_{jl} +
                \xi_{jk}^\rmT \rho_L(g_{ij})g_{jk}\xi_{kl}
                \\
                &\hspace{0.5cm}+\tfrac12 \xi_{kl}^\rmT \sigma_L(g_{ij},g_{jk})\xi_{kl}~,
            \end{aligned}
        \end{equation}
        as well as the following condition required for adjustment:
        \begin{equation}\label{eq:z-rel-2}
            z_{ijk}=\frac{(-1)^{|g_{ik}|}}{2}g_{ik}\eta\,\rmdiag(\sigma_L(g_{ij},g_{jk}))~.
        \end{equation}
        
        \paragraph{Differential refinement.} As we have only added discrete structures to our gauge Lie 2-groupoid that do not affect the continuous cocycles, the differential refinement (excluding the scalar fields) is the same as that of $\scTD_n$. That is, we have 1-~and 2-forms
        \begin{equation}
            \Lambda\in \Omega^1(Y^{[2]})~,~~~
            A\in \Omega^1(Y,\IR^{2n})~,~~~
            B\in \Omega^2(Y)~,
        \end{equation}
    \end{subequations}    
    satisfying the relations~\eqref{eq:diff_refined_cocycles2}.
    
    \paragraph{Compatibility of $Q$- and $R$-fluxes.} Notice that, even though the $(-1)$-form potentials $r$ are a~priori valued in the smooth space $\bar R_n$ (similar to all other potentials), they are constrained to be quantised as elements of $R_n$ by the cocycle condition $r=r_i\in R_n$. This accords with the fact that $(-1)$-form potentials do not encode independent local degrees of freedom, unlike potentials of higher form degrees.
    
    The condition~\eqref{eq:augmented_cocycle} implies a compatibility condition between the $Q$-flux and the $R$-flux: a generic $Q$-flux cannot coexist with a generic $R$-flux. If the $Q$-flux is given by $g_{ij}$ taking values in some subgroup $\Gamma\subset\sfGO(n,n;\IZ)$, then the $R$-fluxes $r$ are constrained to take values in the stabiliser subgroup of $\Gamma$ under the group action~\eqref{eq:action_on_r}. In particular, a vanishing $Q$-flux is compatible with arbitrary $R$-flux in $R_n$, whereas a generic $Q$-flux is compatible only with the trivial $R$-flux that is the constant map in $R_n$.
    
    This has the following physical interpretation. The $R$-flux $r$, regarded as the embedding tensor $r\colon\IR^{2n}\to\sfGO(n,n;\IR)$, identifies the abelian gauge group \(\IR^{2n}\) of the 1-forms with a subgroup of \(\sfGO(n,n;\IR)\) given by the image of $r$. In the presence of non-trivial Q-flux, as specified by $g_{ij}\in\sfGO(n,n;\IZ)$, however, this identification holds only locally, since \(g_{ij}\) acts non-trivially on the $r_i$. In order for this identification to be globally well-defined, one requires that $r$ be equivariant under the action of $g_{ij}$, which is implied by~\eqref{eq:augmented_cocycle}. This renders the $R$-flux globally well-defined for each connected component of space-time (in the complement of domain walls).
    
    It is illustrative to examine the compatibility between $Q$-flux and $R$-flux in the purely geometric case, i.e.\ when all fluxes correspond to the Kalb--Ramond 3-form flux $H$ wrapped around the $n$ geometric directions in an $n$-torus. The $Q$-flux being geometric in this sense corresponds to the ansatz $g_{ij}\in\fro(n;\IZ)\subset\sfGO(n,n;\IZ)$, and similarly $r_i$ must be a map $\IZ^{2n}\to\sfGO(n,n;\IZ)$ whose image lies in $\fro(n;\IZ)$. In such a case, since the adjoint action of $\fro(n;\IZ)$ on itself is trivial, the condition that $g_{ij}^{-1}\acton r = r$ corresponds to the condition that $r$ be constant on the orbits of the $\fro(n;\IZ)$-action on the domain $\IZ^{2n}$, i.e.
    \begin{equation}
        r(\hat a,\check a)=r(\hat a,\check a+g\hat a)
    \end{equation}
    for $(\hat a,\check a)\in\IZ^{2n}$ and arbitrary $g\in\fro(n;\IZ)$. The classification of such orbits is non-trivial. However, it is sufficient that $r$ satisfy the stronger condition
    \begin{equation}
        r(\hat a,\check a)=\hat r(\hat a)
    \end{equation}
    for some function $\hat r\colon\IZ^n\to\fro(n;\IZ)$. After additionally imposing the representation constraint, such $R$-fluxes correspond to elements of $\IZ^{\binom n3}$. Hence, we see that the purely geometric case is consistent with the constraints imposed by~\eqref{eq:augmented_cocycle}.

    \paragraph{Reductions of $\scTD^\text{aug}_n$.} Let us stress the obvious point that the augmented Lie quasi-groupoid $\scTD^\text{aug}_n$ naturally restricts to the higher Lie groupoids and higher Lie groups relevant to T-duality correspondences without $R$-fluxes or without $Q$- and $R$-fluxes, as expected. In particular, if the scalar fields are fixed to be constant, then $\scTD^\text{aug}_n$ effectively reduces to the 2-group $\sfTD^\ltimes_n$. At the other extreme, one may set the 1-form and 2-form fields to be trivial. Then one obtains a sigma model on the Narain moduli space $\sfGO(n,n;\IZ)\setminus(Q_n\times R_n)$ with constraints on the superselection sectors coming from the augmentation as discussed above; if we further turn off $R$-fluxes completely, this reduces to a sigma model on $GM_n=\sfGO(n,n;\IZ)\setminus Q_n$ subject to the minor constraint \eqref{eq:g-even} on discrete moduli.
    
    \paragraph{Classification of branes.} The augmented Lie 2-quasi-groupoid $\scTD^\text{aug}_n$ that we obtained also leads to a natural classification of branes that appear in toroidal compactifications of string theory.
    
    In general, a codimension~\(k\) brane can be stable if it couples to a \((k-2)\)-form potential magnetically, so that branes can be classified by classifying the corresponding \((k-2)\)-form potentials; a codimension~\(k\) brane can also be stable if it carries a non-trivial topological charge, i.e.\ one in which the scalar field exhibits a non-trivial monodromy around the \((k-1)\)-sphere around the brane; such codimension~\(k\) branes are classified by the homotopy group \(\pi_{k-1}(\Sigma)\) of the manifold (or orbifold) that the scalars take values in. A higher gauge groupoid \(\scG\), describing both \(p\)-forms for \(p>0\) as well as scalars, unifies both of these conditions, such that codimension~\(k\) branes are uniformly classified by \(\pi_{k-1}(\scG)\); the fact that \(\pi_{k-1}(-)\) is abelian for \(k\ge3\) corresponds to the fact that three codimensions suffice to exclude anyonic statistics and, hence, non-abelian charges.
    
    In the case of $\scTD^\text{aug}_n$, we have the following result:
    \begin{equation}
        \begin{aligned}
            \pi_0(\scTD^\text{aug}_n) &= R_n~,
            \\
            \pi_1(\scTD^\text{aug}_n) &= \sfGO(n,n;\IZ)~,
            \\
            \pi_2(\scTD^\text{aug}_n) &= \IZ^{2n} \times \IZ^{2n}~,
            \\
            \pi_3(\scTD^\text{aug}_n) &= \IZ ~.
        \end{aligned}
    \end{equation}
    These deviate from the expectations somewhat due to the complications of adjustment, and we comment on each case in the following.
    
    Codimension~$1$ branes, or domain walls, are labeled by the change in $R$-flux across them, which is thus labeled by an element of $R_n$. When the $R$-flux belongs to $\fro(n;\IZ)\subset\sfGO(n,n;\IZ)$, the domain wall corresponds to an NS-brane wrapped around $n-3$ directions of the $n$-torus. Note that the presence of generic domain walls may be incompatible with the presence of generic defect branes as explained above.
    
    Codimension~$2$ branes, or defect branes~\cite{Bergshoeff:2011se}, are labeled by a non-trivial monodromy of the $Q_n/\sfGO(n,n;\IZ)$-valued scalar field around it, hence by $\sfGO(n,n;\IZ)$. Upon dimensional uplift, these correspond to NS-branes wrapped around $n-2$ directions of the $n$-torus, or to KK-branes, or to bound states of both, depending on the element of $\sfGO(n,n;\IZ)$. Somewhat unexpectedly, the condition for the existence of an adjustment implies that the monodromy $g$ must satisfy, for every pair of integers $m_1,m_2$, the condition that
    \begin{equation}
        g^{m_1+m_2}\eta\,\rmdiag(\sigma_L(g^{m_1},g^{m_2}))\in2\IZ^{2n}~,
    \end{equation}
    which is the special case of~\eqref{eq:g-even} for a codimension~2 brane. This condition always holds for $n=1$, but it fails for generic $g\in\sfGO(n,n;\IZ)$ for $n\ge2$. However, in case $g$ belongs to one of the special subgroups discussed in \cref{ssec:2-group_action} --- namely, the $\sfGL(n;\IZ)$ subgroup of $A$-transformations, the $\fro(n;\IZ)$ subgroup of $B$-transformations, the $\fro(n;\IZ)$ subgroup of $\beta$-transformations, the subgroup of factorised dualities, or the $\IZ_2\times\IZ_2$ subgroup generated by $\rmdiag(s_1,\dotsc,s_1,-s_2,\dotsc,-s_2)$ for $s_1,s_2=\pm1$ --- the condition always holds. Thus, such ``ordinary'' codimension~2 branes do exist. The fact that the permitted $Q$-fluxes do not form a subgroup of $\sfGO(n,n;\IZ)$ means that such defect branes are mutually non-local: the presence of one defect brane may forbid the presence of another defect brane somewhere else.
    
    One expects a $2n$-plet of codimension~3 branes, corresponding to a single $\sfGO(n,n;\IZ)$ orbit consisting of NS-branes wrapped around $n-1$ directions and KK-branes. Hence, the presence of the additional copy of $\IZ^{2n}$, which ultimately comes from the non-trivial 2-group structure of $\scGO(n,n;\IZ)$ in~\eqref{eq:def_scGOnnZ}, comes as a surprise. However, the adjustment condition~\eqref{eq:z-rel} requires that this spurious charge be fixed by the $Q$-fluxes, such that the actual possible set of codimension~2 brane charges is simply labeled by $\IZ^{2n}$ as expected.
    
    The unique codimension~4 brane corresponds to an NS-brane fully wrapped around the $n$-torus, coupling magnetically to the Kalb--Ramond field.
    
    \subsection{Example: \texorpdfstring{$R$}{R}-space}
    
    \paragraph{Generic $R$-space.} A generic $R$-space is described as (part of) a full $\scTD^\text{aug}_n$-bundle over $X$. To render this data manageable, we can restrict ourselves to an $R$-space in which all fields are set to zero except for the scalar fields $(q,r)\in C^\infty(Y,Q_n\times R_n)$, the monodromies $(g,\xi)\in C^\infty(Y^{[2]},\sfGO(n,n;\IZ)\times\IR^{2n})$, and $z\in C^\infty(Y^{[3]},\IZ^{2n})$. By~\eqref{eq:z-rel}, the \(g_{ij}\) define a \(\sfGO(n,n;\IZ)\)-principal bundle that is even in the sense of \eqref{eq:g-even}.
    By the same equation, the $g_{ij}$ also fix the higher monodromies $z_{ijk}$. Note that we can set the connection data \((\Lambda,A,B)\) consistently to zero. In this case, the data \((g_{ij},q_i,r_i)\) are equivalent (modulo the condition \eqref{eq:g-even}) to that defining a cocycle of the action groupoid of $\sfGO(n,n;\IZ)$ on $Q_n\times R_n$.
    
    Thus, each connected component of the base manifold $X$ is associated to an element of $R_n$ --- the $R$-flux. Within each connected component, then, one has additional $Q$-flux around non-contractible cycles; if the $R$-flux vanishes for a connected component $X_i$, then $Q$-flux is simply valued in terms of a group homomorphism
    \(\pi_1(X_i)\to\sfGO(n,n;\IZ)\) satisfying \eqref{eq:g-even}.
    
    \paragraph{Nilmanifold example.} Let us again consider the example of the nilmanifold, this time T-dualised completely, so that the base space is merely a point. Here, we can set all higher fields to zero except for $r$ and $q$, and these specify elements $r\in R_n$ and $q\in Q_n$.


\chapter{Conclusion and outlook}

Throughout this thesis, we have seen that adjusted nonabelian higher gauge theories offer a viable nonperturbative extension to the usual world of gauge theories and that they appear naturally in the description of many physical phenomena from various corners of high-energy physics. In a sense this is not surprising: since Yang and Mills we have known that nonabelian gauge symmetries pervade nature, and since Kaluza and Klein we have known that higher dimensions, in which higher-degree form fields naturally arise, offer a natural geometrisation of physics; nonabelian higher gauge theories are simply one point in the confluence of these two currents.

Furthermore, we should mention that higher gauge theories naturally fit into a homotopy-algebraic framework for perturbative physics, including recursion rules \cite{Macrelli:2019afx}, colour--kinematics duality \cite{Borsten:2021hua,Borsten:2021rmh,Borsten:2022aa}, and string theory \cite{Kajiura:0410291,Doubek:2020rbg}.

Nevertheless, a number of challenges and opportunities for future work remain.
\begin{description}
\item[Parallel transport] Our discussion extends in principle straightforwardly to higher dimensions, except that one should use simplicial models of the required higher path groupoids and higher Lie groups, as~in~e.g.~\cite{Jurco:2016qwv}; the technicalities of higher coherence laws will otherwise overwhelm.
\item [\(E_2L_\infty\)-algebras] 
There are three main questions that remain or arise from our work. First, it would be certainly very interesting to explore further the relationship of our constructions to ones existing in the literature. We feel that e.g.~$\opEilh$-algebras should have appeared in other algebraic contexts; for example, the deformed Leibniz rule arising in $\ophLie$- and $\opEilh$-algebras is very similar to the formula in~\cite[Theorem 5.1]{Steenrod:1947aa} for Steenrod's cup products.\footnote{We thank Jim Stasheff for pointing out this potential link.}
    Second, the formulation of $\ophLie$ and, consequently $\opEilh$, only features two levels of binary brackets: \(\epsilon^0_2\) and \(\epsilon^1_2\). Thus, while the symmetry property of the bracket \(\epsilon^0_2\) is relaxed up to homotopy by the alternator \(\epsilon^1_2\), the symmetry property of \(\epsilon^1_2\) is not relaxed up to homotopy by a putative second-order alternator \(\epsilon^2_2\). In technical terms, the resulting notion of \(E_2L_\infty\)-algebras fails to be a cofibrant resolution of the Lie operad over general commutative rings (rather than over fields of characteristic~\(0\)). The notion of \(E_2L_\infty\)-algebras suffices for applications to tensor hierarchies, but a more mathematically natural formulation should extend it to a notion of \(EL_\infty\)-algebras in which all symmetry properties are relaxed up to homotopy.
    Third, most of our applications of $E_2L_\infty$-algebras involved them only in their hemistrict form, namely as $\ophLie$-algebras. This is due to the fact that we were only able to refine the derived bracket construction to a construction of an $\ophLie$-algebra from a differential graded Lie algebra. As we explain in \ref{ssec:gen_tensor_hierarchies}, there is a clear indication that some tensor hierarchies originate from $E_2L_\infty$-algebras that are not $\ophLie$-algebras but that can be obtained from $L_\infty$-algebras. This suggests a much wider generalization of the derived bracket construction, which would be certainly very useful to have. In particular, it would allow us to characterize a very large class of $E_2L_\infty$-algebras for which the problem of defining the kinematical data of higher adjusted gauge theories, such as the data arising in the tensor hierarchies, is fully under control. Altogether, we believe that the general picture, both in generalized geometry and in higher gauge theory, will ultimately require using fully fledged $E_2L_\infty$-algebras.
\item[T-duality]
    There are a few open questions arising from our constructions. First of all, we observe that the T-duality group $\sfGO(n,n;\IZ)$ does not act on the 2-group $\sfTD_n$, while the extension $\scGO(n,n;\IZ)$ does. This leads to additional moduli in our description, which are then canceled by condition~\eqref{eq:z-rel} arising from demanding the existence of adjusted curvatures. Similarly, our cocycles for principal $\scTD^\text{aug}_n$-bundles impose topological restrictions on the set of $Q$- and $R$-fluxes. It would be useful to understand both from a physical perspective. Second, it would be important to link our description of T-duality for non-geometric spaces to the descriptions available in the literature, in particular to~\cite{Mathai:2004qq,Bouwknegt:2004ap}. Third, it would be very interesting to relate our constructions much more closely to double field theory, in particular to the global constructions of~\cite{Deser:2018flj} based on the formalism of~\cite{Deser:2016qkw}. Fourth, it may be possible to use our framework to make progress with the definition and the understanding of non-abelian T-duality as well as Poisson--Lie T-duality; an interesting perspective on the latter has recently been given in~\cite{Arvanitakis:2021lwo}. The issue here is that with the inclusion of non-abelian gauge groups, the relevant 2-groups including the gauge potential become more and more complicated, cf.~\cite{Rist:2022hci}. In a related vein, while in this thesis we restrict to the case of ungauged (super)gravities, it may be feasible to generalize our results to the gauged case with more non-trivial tensor hierarchies. Finally, all our constructions lift, in principle, readily to U-duality, and this is currently the focus of our attention~\cite{Borsten:2022ab}; see also~\cite{Alfonsi:2021ymc,Sati:2021rsd} for related work.
\end{description}


\bibliographystyle{alphaurl}
\bibliography{bigone}


\end{document}